\documentclass[]{spie}  

\usepackage{amsmath,amsfonts,amssymb}
\usepackage{graphicx}
\usepackage[colorlinks=true, allcolors=blue]{hyperref}

\usepackage{wrapfig}
\usepackage{sidecap}

\title{Polarization Modeling and Predictions for DKIST Part 5: Impacts of enhanced mirror and dichroic coatings on system polarization calibration.} 

\author[a]{David M. Harrington}
\author[a]{Stacey R. Sueoka}
\author[a]{Amanda J. White}
\affil[a]{National Solar Observatory, 8 Kiopa'a Street, Ste 201 Pukalani, HI 96768, USA}

\authorinfo{Further author information: (Send correspondence to D.M.H) \\
D.M.H.: E-mail: dharrington@nso.edu, Telephone: 1 808 572 6888}

\begin{document} 
\maketitle

\begin{abstract}
The Daniel K. Inouye Solar Telescope (DKIST) is designed to deliver accurate spectro-polarimetric calibrations across a wide wavelength range and large field of view for solar disk, limb and coronal observations. DKIST instruments deliver spectral resolving powers up to 300,000 in multiple cameras of multiple instruments sampling nanometer scale bandpasses. We require detailed knowledge of optical coatings on all optics to ensure we can predict and calibrate the polarization behavior of the system. Optical coatings can be metals protected by many dielectric layers or several-micron thick dichroics. Strong spectral gradients up to 60$^\circ$ retardance per nanometer wavelength and several percent diattenuation per nanometer wavelength are observed in such coatings. Often, optical coatings are not specified with spectral gradient targets for polarimetry in combination to both average and spectral threshold type specifications. DKIST has a suite of interchangeable dichroic beam splitters using up to 96 layers. We apply the Berreman formalism in open-source Python scripts to derive coating polarization behavior. We present high spectral resolution examples on dichroics where transmission can drop 10\% with associated polarization changes over a 1 nm spectral bandpass in both mirrors and dichroics. We worked with a vendor to design dichroic coatings with relatively benign polarization properties that pass spectral gradient requirements and polarization requirements in addition to reflectivity. We now have the ability to fit multi-layer coating designs which allow us to predict system level polarization properties of mirrors, anti-reflection coatings and dichroics at arbitrary incidence angles, high spectral resolving power and on curved surfaces through optical modeling software packages. Performance predictions for polarization at large astronomical telescopes requires significant metrology efforts on individual optical components combined with systems-level modeling efforts. We show our custom-built laboratory spectropolarimeter and metrology efforts on protected metal mirrors, anti-reflection coatings and dichroic mirror samples. 
\end{abstract}


\clearpage

\tableofcontents

\clearpage

\section{DKIST Optics \& Polarization Models for Calibration}
\label{sec:intro}  

The Daniel K. Inouye Solar Telescope (DKIST) on Haleakal\={a}, Maui, Hawai'i is presently under construction with operations beginning around 2020. The telescope has a 4.2 m off-axis F/ 2 primary mirror (4.0 m illuminated) and a suite of polarimetric instrumentation in a coud\'{e} laboratory \cite{2014SPIE.9145E..25M, Keil:2011wj, Rimmele:2004ew}. Many of the proposed science cases rely on high spectral resolution polarimetry with imaging capabilities from scanning or tilting the instrument. Optics allow for stepping of spectrograph slits, scanning through wavelengths with Fabry-Perot interferometers and using imaging fiber bundles to create imaging spectropolarimetric capability over visible and near infrared wavelengths covering wide fields of view. Many science cases require strictly simultaneous observation of several spectral lines with multiple instruments. DKIST can operate up to 8 polarimetric cameras simultaneously to achieve these goals.

DKIST uses seven mirrors to collect and relay light to a rotating coud\'{e} lab to provide flexible capabilities \cite{Marino:2016ks, McMullin:2016hm,Johnson:2016he,2014SPIE.9147E..0FE, 2014SPIE.9147E..07E, 2014SPIE.9145E..25M}. Operations involve four polarimetric instruments presently spanning the 380 nm to 5000 nm wavelength range. We also have two high speed imagers covering visible and near infrared wavelengths. A sequence of dichroic beam splitters (and optionally windows or mirrors) called the Facility Instrument Distribution Optics (FIDO) allows for changing of instrument configurations on a timescale of less than half an hour. The FIDO optics allow simultaneous operation of three polarimetric instruments optimized for 380nm to 1800nm while all using the adaptive optics system for correction \cite{2014SPIE.9147E..0FE, 2014SPIE.9147E..07E, 2014SPIE.9147E..0ES, SocasNavarro:2005bq}. Another instrument (CryoNIRSP) can receive all wavelengths to 5000 nm but without use of the adaptive optics system. We refer the reader to recent papers outlining the various capabilities of the first-light instruments \cite{McMullin:2016hm, 2014SPIE.9147E..07E, 2014SPIE.9145E..25M, 2014SPIE.9147E..0FE, Rimmele:2004ew}. 

This paper is part of a series investigating polarization performance expectations for the DKIST instrument suite. In HS17\cite{2017JATIS...3a8002H} we outlined the DKIST optical layout and properties of a very simple enhanced silver mirror coating model. This coating recipe was used in Zemax to estimate the field of view and beam footprint variation of the combined system optics to ViSP and Cryo-NIRSP. We also showed the predicted Mueller matrix for the DKIST primary and secondary mirrors, mounted ahead of our calibration optics. In H17a\cite{Harrington:2017eja} we showed polarization calibrations of a night time telescope and system calibrations with a visible spectropolarimeter using the daytime sky. In H18\cite{2017JATIS...3d8001H}, we applied the Berreman calculus\cite{1972JOSA...62..502B, 2014btfp.book.....M} to polarization fringes formed in multi-layer crystals with predictions and data collected in the lab and at a solar telescope.  We then extended this calculus in HS18 \cite{Harrington:2018bt} to converging and diverging beams. Fringes were measured at various focal ratios and compared to simulations in converging beams at solar and night time telescopes as well as in the laboratory.  We also showed thermal models for the DKIST retarders along with thermal perturbation models for the polarization fringes \cite{Harrington:2018bt}. We recently have investigated spatial variation of retardance across multi-layer retarders made of polished crystals, stretched polycarbonate and ferro-electric liquid crystals in HS18b\cite{Harrington:2018jb}.  This variation was then included in the DKIST optical model to show polarization calibration errors as functions of field angle and wavelength. We used a definition of calibration efficiency to show how we can use a single calibration retarder to simultaneously and efficiently calibrate all DKIST instruments from 380 nm to 1650 nm, representing the entire first-light AO corrected suite. In this paper, we extend the coating model efforts of HS17\cite{2017JATIS...3a8002H} to many vendors, highly enhanced metals, hundred-layer dichroics and our system of beam splitters. We present measurements and coating models for all optics presently coated in the DKIST telescope and most of the first light instrument suite along with system-level predictions for polarization performance. We show some issues with complex coating formulas particularly for our high spectral resolving power instrument suite. We introduce tolerance analysis at the system level given new measurements of spatial and shot-to-shot coating variability.

\begin{figure}[htbp]
\begin{center}
\vspace{-1mm}
\hbox{
\hspace{-0.0em}
\includegraphics[height=7.9cm, angle=0]{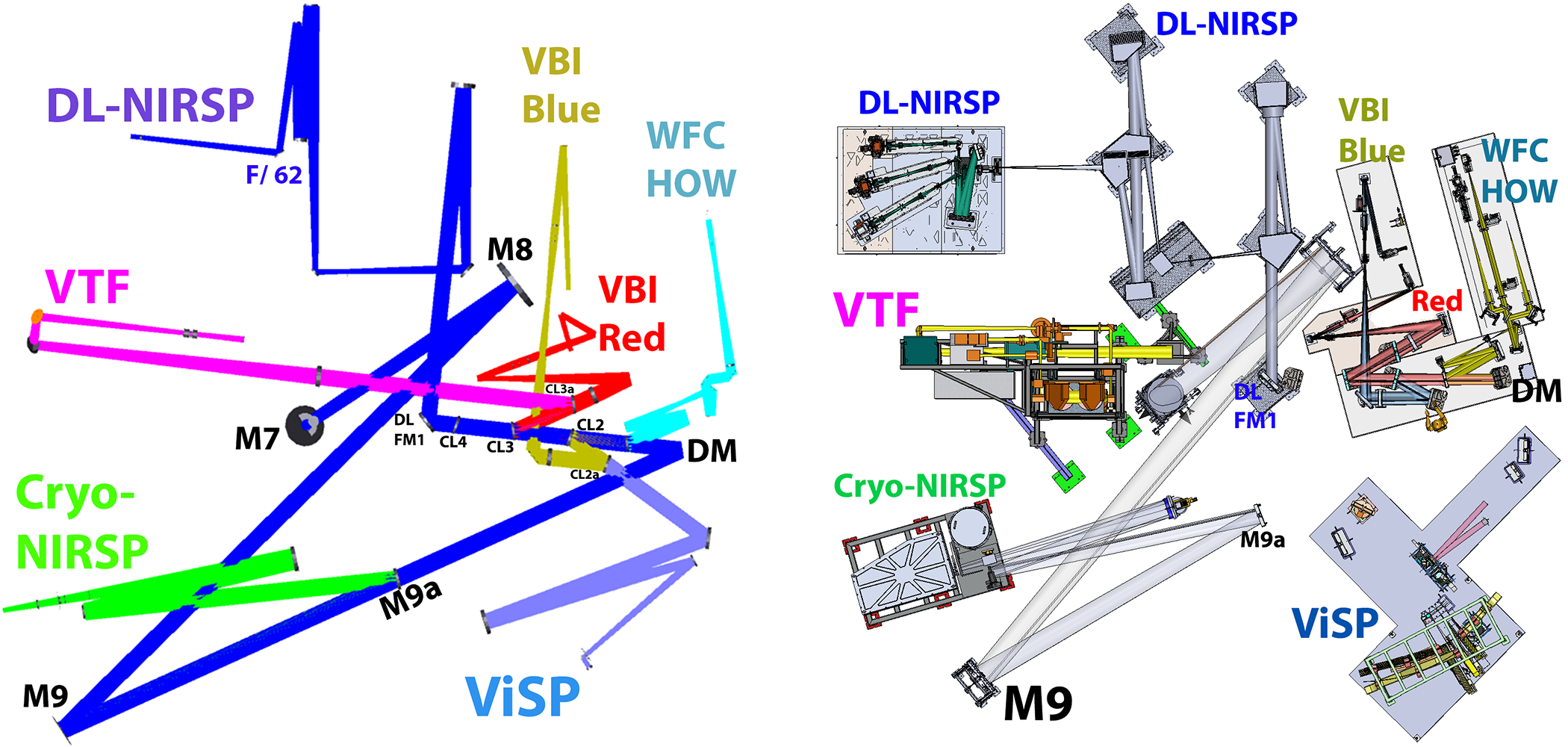}
}
\caption[] 
{ \label{fig:optical_telescope_model} DKIST coud\'{e} lab optical Zemax model at left and mechanical model at right.  The telescope feeds the beam at F/ 53 to the coud\'{e} lab.  M7 folds the beam level with the lab floor.  M8 collimates the beam.  M9 directs the beam towards various instruments. The FIDO system of dichroics sends selectable wavelength ranges simultaneously towards various combinations of instruments  (ViSP, VTF, DL-NIRSP, VBI).  Each instrument has multiple cameras used simultaneously.  The optional M9a directs the entire beam towards Cryo-NIRSP.  }
\vspace{-5mm}
\end{center}
\end{figure}

We show in Figure \ref{fig:optical_telescope_model} the optical and mechanical layout of the instruments in the coud\'{e} laboratory. The left hand graphic shows the optical Zemax model of the instruments with the beam propagating to the various focal planes close to where polarization modulation occurs.  The dark blue beam shows the Diffraction Limited Near Infrared Spectropolarimeter (DL-NIRSP) which uses an imaging fiber bundle to create spectropolarimetric images on three separate cameras. The light blue beam shows the Visible Spectropolarimeter (ViSP) which uses a slit to scan the field of view while imaging simultaneously with three separate cameras. The magenta beam shows the Visible Tunable Filter (VTF) imaging through a Fabry-Perot type spectropolarimeter.  These three instruments represent the available post-adaptive optics (AO) polarimetric instrumentation.  There are also two high speed imaging systems collectively called the Visible Broadband Imager (VBI) with Red and Blue channels. Additional instrumentation is associated with the adaptive optics system low order wavefront sensor and high order wavefront sensor.  All AO instruments see the first beam splitter associated with the wavefront sensor (WFS-BS1) in transmission. The high order wavefront sensor is fed by the Fresnel reflection off the uncoated surface of WFS-BS1 as seen by the cyan colored rays in Figure \ref{fig:optical_telescope_model}.

\begin{wrapfigure}{r}{0.55\textwidth}
\centering
\vspace{-3mm}
\begin{tabular}{c} 
\hbox{
\hspace{-1.3em}
\includegraphics[height=6.1cm, angle=0]{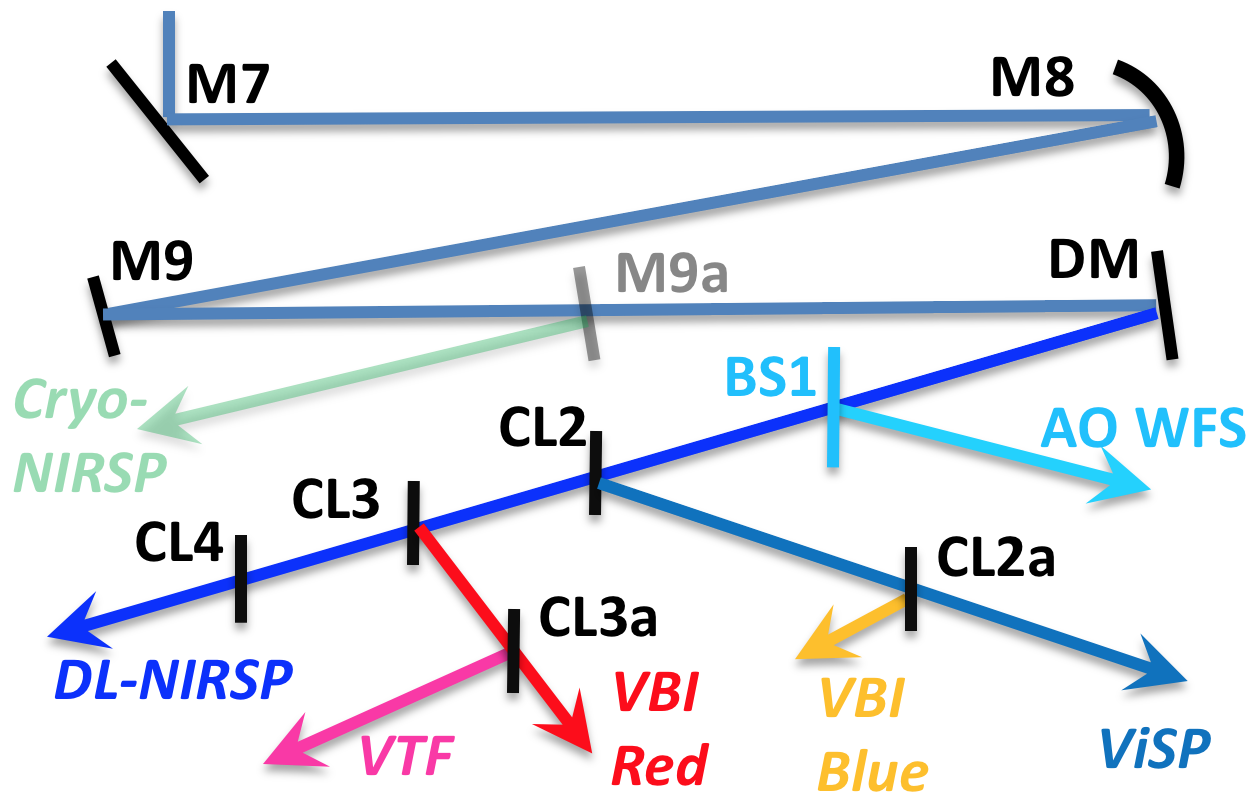}
}
\end{tabular}
\caption[] 
{\label{fig:optics_schematic}  The coud\'{e} lab cartoon layout beginning with DKIST mirrors M7 thorough the deformable mirror M10 (DM). FIDO optic stations CL2, CL2a, CL3, CL3a and CL4 can be configured using a suite of mirrors, windows and dichroics to optimize the configuration for particular use cases. The first beam splitter is permanently mounted and feeds the wavefront sensor (WFS). }
\vspace{-3mm}
\end{wrapfigure}

Complex polarization modulation and calibration strategies are required for such a mulit-instrument system \cite{2014SPIE.9147E..0FE,2014SPIE.9147E..07E, Sueoka:2014cm, 2015SPIE.9369E..0NS, deWijn:2012dd}. At present design, three different retarders are in fabrication for use in calibration near the Gregorian focus \cite{2014SPIE.9147E..0FE,Sueoka:2014cm,Sueoka:2016vo}. The planned 4 m European Solar Telescope (EST), though on-axis, will also require similar calibration considerations \cite{SanchezCapuchino:2010gy, Bettonvil:2010cj,Collados:2010bh}. The upcoming Chinese Giant Solar Telescope is exploring segmented designs and modeling coating variability between segments \cite{Yao:2018ej,Yuan:2016en}. Many solar and night-time telescopes have performed polarization calibration of complex optical pathways \cite{DeJuanOvelar:2014bq, Joos:2008dg, Keller:2009vj,Keller:2010ig, Keller:2003bo, Rodenhuis:2012du, Roelfsema:2010ca, 1994A&A...292..713S, 1992A&A...260..543S,  1991SoPh..134....1A, Schmidt:2003tz, Snik:2012jw,  Snik:2008fh, Snik:2006iw, SocasNavarro:2011gn, SocasNavarro:2005jl, SocasNavarro:2005gv, Spano:2004ge, Strassmeier:2008ho, Strassmeier:2003gt, Tinbergen:2007fd, 2005A&A...443.1047B, 2005A&A...437.1159B}.

As part of the coud\'{e} laboratory, there are multiple interchangeable dichroic beam splitters, windows and mirrors collectively called the Facility Instrument Distribution Optics (FIDO). The FIDO optics are configurable to distribute various wavelengths to different instruments as observers require. The optics are mounted in stations labeled following a Coud\'{e} Lab (CL) station numbering system CL2, CL3, etc.  As all AO-assisted instruments see the first beamsplitter WFS-BS1 in transmission, we start the numbering system at 2.  The optics are designed such that the wedge angles are matched in each optic, and every instrument sees either 2 or 4 beamsplitters in transmission to compensate for the wedge and associated wavelength variation in beam deflection. The two stations CL2 and CL3, use a reflection off the optic.  There are two optical stations after these reflections named CL2a and CL3a respectively. We show a cartoon layout of the coud\'{e} laboratory optics in Figure \ref{fig:optics_schematic}.  We follow the same color convention as in Figure \ref{fig:optical_telescope_model}.  The FIDO mirrors, dichroics and windows interchangeably used in the CL stations will be discussed in later sections. There is a separate instrument that does not use the adaptive optics but instead uses a seeing limited, all-reflective beam path.  The Cryogenic Near Infrared Spectropolarimeter (Cryo-NIRSP) covers wavelengths out to 5000 nm by inserting a pickoff mirror before the adaptive optics system. 

Each of the instruments are built by different teams with a variety of mirror coatings often using multiple formulas from multiple vendors in several independent coating chamber shots.  The DKIST instrument mirrors can be just one or two protective layers up to very complex enhanced protected type coatings with 29 dielectric layers over the metal and a coating thickness over of over 3 $\mu$m. The interchangeable FIDO mirrors, windows and dichroics are provided by DKIST and also have a diversity of coatings. The dichroic coating designs include up to nearly one hundred dielectric layers and thickness of 9 $\mu$m. These coatings can produce spectrally diverse and complex behavior requiring a detailed treatment as outlined in this paper.

DKIST is designed as a multi-decade lifespan facility supporting a diverse array of use cases with a suite of polarimeters. The slit spectropolarimeter, integral field fiber-fed imaging spectropolarimeter and Fabry-Perot imaging spectropolarimeter described above step or scan across solar features on disk, limb and the corona. Often multiple cameras within multiple instruments work in concert with both active and adaptive optics locking onto nearby solar features for wavefront correction and pointing stabilization. The suite of instruments is designed to be flexible in configuration, upgradable, stable in calibration and support an incredibly diverse range of science objectives.

Within the current DKIST science planning process, there are already hundreds of proposed observing cases spanning near ultraviolet wavelengths (0.393 $\mu$m) to thermal infrared (4.6$\mu$m) often with several cameras on several instruments operating simultaneously. The expected flux levels from on-disk observations at visible wavelengths to coronal observations in the thermal infrared range in amplitude by at least factors of millions. Very large changes are also anticipated in spatial and spectral sampling, camera frame rates, modulation strategies and time to noise limits. Some use cases are seeing limited without adaptive optics and are sampled coarsely to achieve very high sensitivity. Other cases use the adaptive optics system to achieve diffraction limited performance with sampling at the highest spatial and spectral powers delivered. The DKIST AO system has 1600 actuators and is anticipated to deliver Strehl ratios of 0.3 at 500 nm wavelength in median seeing conditions \cite{Marino:2016ks, Marino:2012jq, Johnson:2016he}. The DKIST upgrade to multi-conjugate AO already in progress should push delivered high-Strehl performance to wider fields \cite{Schmidt:2017jh, Schmidt:2016bj, Schmidt:2016jz}. Polarization modulation speeds can span multiple orders of magnitude (e.g. the ferro electric liquid crystal modulator in VTF is capable of kHz rates versus discrete modulation on timescales $<$0.01 Hz for coronal observations with DL-NIRSP or Cryo-NIRSP) \cite{2014SPIE.9147E..0ES, 2016arXiv160706767S}. 

Expected solar magnetic field strengths can range from a fraction of a Gauss spatially unresolved below the DKIST diffraction limit to several kiloGauss covering the entire instantaneous field of view of a DKIST instrument. Translation of an error in magnetic field to an error in measured Stokes vectors implies some numerical techniques to relate changes in modeled field properties to errors in a measurement through some kind of atmospheric model and inversion process. A general, instrument-unspecific framework to systematically relate the error bars in a Stokes measurement to the error bars in a magnetic field through an inversion code does not exist. Often, specific measured Stokes vectors can be perturbed by changing assumed instrument calibrations to assess field errors such as in Appendix E of Jaeggli 2011\cite{Jaeggli:2011vg}. Signal to noise estimates and simple models for instabilities of individual components can also be related to field uncertainties  in specific cases \cite{delToroIniesta:2012hq}. 

Detailed performance predictions are useful to assess the inaccuracies that can arise when observing and calibrating across such a wide range of instrument configurations. We also find the system level performance predictions useful to estimate the magnitude of variations, and assess the calibration techniques required. We require knowledge of the optical surfaces, the coating behavior across field angle and pupil position and the modes of intended use (slit scanning, pupil steering, temporal sampling, etc). For calibrations to be accurate, we must assess the magnitude of expected errors and the stability of all components in the optical system. 

Often, most science cases call for continuum polarization stability so that the zero point and length of the recovered Stokes vector is comparable between different instrument pointings (sometimes called the image mosaic, field mosaic or other terms of varying applicability). The requirements on the orientation of the vector (which manifest from retardance spatial variation and other coordinate geometry issues) are often less stringent as a few degrees of orientation variation in a recovered Stokes vector does not directly impact comparison of differing solar atmosphere models. Depolarization is often ignored as it is usually a small fraction of a percent change in the magnitude of a recovered Stokes vector, and it can be modeled with proper system level tools (e.g. for DKIST\cite{2017JATIS...3a8002H}). 

With this paper, we extend our prior modeling efforts to include several parameters of optical coatings required for a realistic system level modeling. Any altitude-azimuth telescope with a non-zero field of view has temporal dependence across the field as mirror groups change their relative geometry. Realistic coatings are not identical between coating runs and are only within manufacturing tolerances of nominal designs. Polarization properties of real coated mirrors never cancel perfectly. The transmission, diattenuation, polarizance and retardance parameters can be individually perturbed by field rotation, spatially non-uniform coatings and optical instabilities. By knowing the wavelength dependence of the system performance at high spectral resolving power, we can effectively plan for calibrations at appropriate configurations (e.g. spectral resolution, field scanning, spatial sampling, temporal averaging). We also design coatings that do not cause undue calibration challenges, such as coatings changing retardance by more than 1 wave in a few nm of wavelength. This work updates performance models that will inform limits to the accuracy of calibration techniques when we decide how wide of a field we can step, modes for field scanning, available spectral binning, wavelength interpolation, etc. Some of these updates will be included in the instrument performance calculators used to plan observations (published online at https://dkist.nso.edu/CSP/instruments).

\subsection{Mirror Grouping Models, Polarization \& System Mueller Matrices}
\label{sec:sub_Group_Model}

We create a polarization model for the telescope and the suite of instruments using our knowledge of the optical coatings and substrates. We showed in HS17\cite{Harrington:2017dj} some predictions for the polarization behavior of the telescope and instrument feed optics using nominal coating formulas derived with the Zemax-provided refractive indices. We showed the mirrors introduce some slight field of view dependence for the polarization calibration as well as a very mild depolarization from our off-axis primary and secondary mirrors. The magnitude of field-dependent variation shown in our prior work \cite{Harrington:2017dj} is not changed by our work presented here.  In this paper, we consider only the on-axis beam at the nominal instrument bore-sight. Field of view considerations represent a significant complication.  A common technique for simplifying the system polarization models is to group mirrors together that maintain a fixed orientation with respect to each other. We call this the {\it group model}. For DKIST systems engineering, we also need to predict the Mueller matrix of the system while accounting for polarization properties of the many mirrors mounted in front of the polarization modulators in each instrument. The basic calibration plan is to use the DKIST calibration optics to simultaneously fit for the telescope group model, the modulation matrix of the instruments and certain properties of the calibration optics. We show here the mathematics behind some of the simplifications assumed in the group model.  We then asses how to predict these terms in later sections of the paper.

In this work, we denote the Stokes vector as {\bf S} = $[I,Q,U,V]^T$. In this formalism, $I$ represents the total intensity, $Q$ and $U$ the linearly polarized intensity along polarization position angles $0^\circ$ and $45^\circ$ in the plane perpendicular to the light beam, and $V$ is the right-handed circularly polarized intensity. The typical convention for astronomical polarimetry by the International Astronomical Union is for the +$Q$ electric field vibration direction to be aligned to celestial North-South, while +$U$ has the electric field vibration direction aligned to North-East and South-West. The propagation axis points towards the observer.  In laboratory settings frequently $+Q$ is defined as horizontal or vertical. For solar studies, a common definition is to have +Q parallel to the solar equator or in the positive right ascension direction.\cite{2008SoPh..249..233I, Thompson:2006gl, Skumanich:1997eh}

\begin{wrapfigure}{l}{0.36\textwidth}
\vspace{-8mm}
\begin{equation}
{\bf M}_{ij} =
 \left ( \begin{array}{rrrr}
 II   	& QI		& UI		& VI		\\
 IQ 	& QQ	& UQ	& VQ	\\
 IU 	& QU	& UU	& VU		\\
 IV 	& QV	& UV		& VV		\\ 
 \end{array} \right ) 
\label{eqn:MM}
\end{equation}
\vspace{-8mm}
\end{wrapfigure}

The Mueller matrix is the 4x4 matrix that transfers Stokes vectors \cite{1992plfa.book.....C, Chipman:2014ta, Chipman:2010tn}. Each element of the Mueller matrix is denoted as the transfer coefficient \cite{Chipman:2010tn, 2013pss2.book..175S}. For instance the coefficient [0,1] in the first row transfers $Q$ to $I$ and is denoted $QI$. The first row terms are denoted $II$, $QI$, $UI$, $VI$. The first column of the Mueller matrix elements are $II$, $IQ$, $IU$, $IV$. In this paper we will use the notation in Equation \ref{eqn:MM}.  The output Stokes vector is related to the input vector via a simple transfer equation ${\bf S}_{i_{output}} = {\bf M}_{ij} {\bf S}_{i_{input}}$.  With this formalism, the Stokes vector from some patch of solar atmosphere would be transferred by the Mueller matrix of each optic between the sun and the sensor.

We adopt a notation where a rotation is denoted as ${\bf \mathbb{R}}$. We note that a rotation of a Mueller matrix must include rotations on both sides of the matrix to preserve input coordinate systems: ${\bf \mathbb{R}}(-\theta)$ ${\bf M}$ ${\bf \mathbb{R}}(-\theta)$. There are three main coordinate rotations in DKIST.  The elevation axis is between M4 and M5  (${\bf \mathbb{R}}_{El}$).  The azimuth axis is between M6 and M7 (${\bf \mathbb{R}}_{Az}$).  The DKIST coud\'{e} laboratory is on a rotating platform so there is a separate rotational degree of freedom with the coud\'{e} angle in addition to the azimuth of the target (${\bf \mathbb{R}}_{TA}$).

\begin{wrapfigure}{r}{0.50\textwidth}
\vspace{-6mm}
\begin{equation}
{\bf S}_{coude} = {\bf \mathbb{R}}_{TA}   {\bf \mathbb{R}}_{Az}  {\bf M}_6 {\bf M}_5  {\bf \mathbb{R}}_{El} {\bf M}_4 {\bf M}_3   {\bf M}_2  {\bf M}_1 {\bf S}_{input} 
\label{eqn:MM_group_model_raw}
\end{equation}
\begin{equation}
{\bf S}_{Cryo} = {\bf M}_{CFM}{\bf M}_{CSM}{\bf M}_{9a} {\bf M}_9 {\bf M}_8  {\bf M}_7 {\bf S}_{coude} 
\label{eqn:MM_group_model_cryo}
\end{equation}
\begin{equation}
{\bf S}_{thruAO} = {\bf M}_{BS1b} {\bf M}_{BS1f} {\bf M}_{DM} {\bf M}_9 {\bf M}_8  {\bf M}_7 {\bf S}_{coude} 
\label{eqn:MM_group_model_toAO}
\end{equation}
\vspace{-6mm}
\end{wrapfigure}

There are six mirrors that collect the solar flux and relay the beam to the coud\'{e} laboratory.  The seventh mirror (M7) folds the beam onto the coud\'{e} laboratory floor.  The eighth mirror (M8) is an off axis collimating mirror and the ninth mirror (M9) is a coma correcting fold mirror with a specific figure.  With this notation, we can explicitly compute the transfer equations to see how Stokes vectors will behave at various locations along the optical path.  In Equation \ref{eqn:MM_group_model_raw} we show the input Stokes vector ${\bf S}_{input}$ being transferred from the telescope primary mirror to the coud\'{e} laboratory ${\bf S}_{coude}$ just before reflection off M7.  

\begin{wrapfigure}{r}{0.59\textwidth}
\vspace{-7mm}
\begin{equation}
{\bf S}_{toDL} = {\bf M}_{CL4b} {\bf M}_{CL4f} {\bf M}_{CL3b} {\bf M}_{CL3f} {\bf M}_{CL2b}{\bf M}_{CL2f}  {\bf S}_{thruAO} 
\label{eqn:MM_group_model_DLfido}
\end{equation}
\begin{equation}
{\bf S}_{DL} = {\bf M}_{FM4} {\bf M}_{FM24} {\bf M}_{FSM} {\bf M}_{FM2} {\bf M}_{FM2} {\bf M}_{OAM1} {\bf M}_{FM1} {\bf S}_{toDL} 
\label{eqn:MM_group_model_DLinstrument}
\end{equation}
\vspace{-7mm}
\end{wrapfigure}

The instrument Cryo-NIRSP does not use the AO system. The system uses a pickoff flat mirror called M9a at 9$^\circ$ incidence angle that directs light to this instrument. The next mirror in the system is a flat pupil steering mirror we denote CSM working at 4$^\circ$ incidence angle. This is followed by the off axis mirror focusing the beam at F/ 18 using a 1.1$^\circ$ fold angle denoted CFM. The beam passes through the polarization modulator to the spectrograph entrance slit. Equation \ref{eqn:MM_group_model_cryo} shows the Mueller matrices transferring the coude lab Stokes vector to the Cryo-NIRSP modulator. The Cryo-NIRSP feed optics and modulator properties are described in our prior references\cite{Harrington:2018jb,Harrington:2018bt,Harrington:2017jh,Harrington:2017dj}.

\begin{wrapfigure}{l}{0.46\textwidth}
\centering
\vspace{-3mm}
\begin{tabular}{c} 
\hbox{
\hspace{-1.2em}
\includegraphics[height=7.8cm, angle=0]{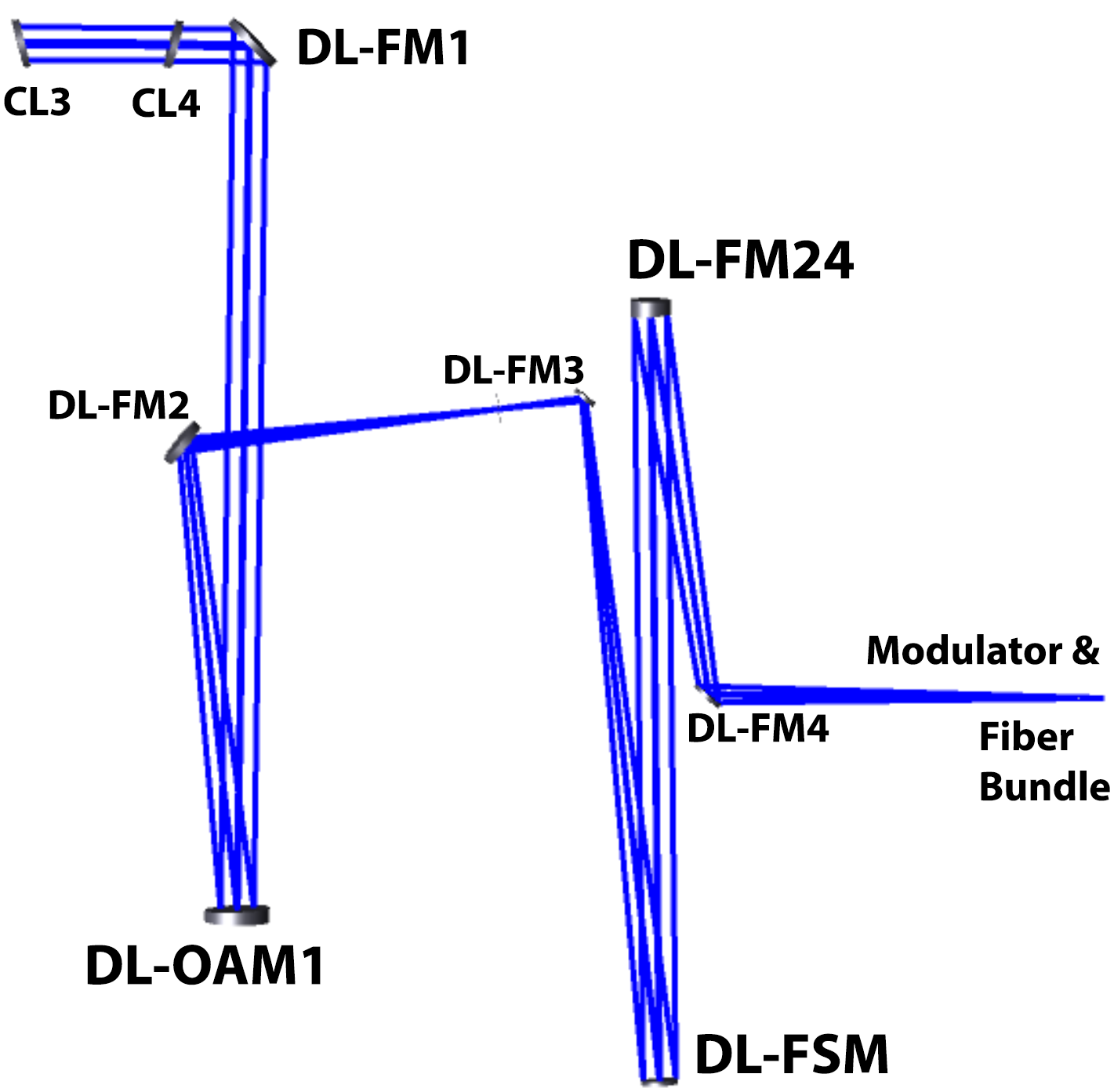}
}
\end{tabular}
\caption[]  {\label{fig:dl_optics}  The DL-NIRSP F/ 24 optical Zemax model from the CL3 FIDO optic station through the 7 instrument reflections to the modulator and fiber bundle.   }
\vspace{-7mm}
 \end{wrapfigure}

For the rest of the polarimetric instruments, the beam must propagate through the AO system as shown in Equation \ref{eqn:MM_group_model_toAO}. The DL-NIRSP, ViSP and VTF all see M7, M8 and M9 as well as the tenth mirror as the adaptive optics system deformable mirror (DM = M10). All instruments using the AO system must also account for the WFS-BS1 transmission and diattenuation from both the uncoated front surface and broad-band anti-reflection coated wedged back surface we denote in Equation \ref{eqn:MM_group_model_toAO} as BS1f and BS1b respectively.

As an example of the Mueller matrix calculation, we show in Equation \ref{eqn:MM_group_model_DLfido} the dichroic coating front surface reflection and broad-band anti-reflection coated back surface reflections off the interchangeable FIDO optics feeding the DL-NIRSP instrument. We explicitly call out the optical stations CL2, CL3 and CL4 along with the separate Mueller matrices for the front and back surface reflections. We show later both theoretical models and polarimetric measurements for several of the FIDO dichroic coatings used to compute the system Mueller matrices. 

The mirrors included in the DL-NIRSP relay optics also need to be included to compute the expected Mueller matrix of the system to the modulator.  In Equation \ref{eqn:MM_group_model_DLinstrument} we show the optics transferring the Stokes vector exiting the last FIDO optic through the DL-NIRSP mirrors in the F/ 24 configuration to the modulator mounted in front of the imaging fiber bundle. The Zemax optical model for the on-axis field angle is seen in Figure \ref{fig:dl_optics}.

\begin{wrapfigure}{r}{0.53\textwidth}
\vspace{-6mm}
\begin{equation}
{\bf S}_{group} =  {\bf M}_{Mod} {\bf \mathbb{R}}_{TA} {\bf \mathbb{R}}_{Az} {\bf M}_{5,6}  {\bf \mathbb{R}}_{El} {\bf M}_{3,4} {\bf M}_{1,2} {\bf S}_{input} 
\label{eqn:MM_group_model}
\end{equation}
\vspace{-8mm}
\end{wrapfigure}

The fundamental assumption of the {\it group model} is that the static optics can all be multiplied together and fit with a greatly reduced number of variables. For the present modeling efforts, we also ignore all field of view dependence though we can calculate magnitudes and make the models more complex as needed. We show the {\it group model} in Equation \ref{eqn:MM_group_model}. 

The primary and secondary mirrors are ahead of the calibration optics so they are fit separately using a variety of techniques. The third mirror is near a focal plane and is actively pointed to maintain optical alignment by small amounts. The third and fourth mirrors are modeled together as a group, ignoring the small angular offsets. The first four mirrors are upstream of the elevation axis. Similarly, M6 is near a pupil plane and is also tilted by small amounts to maintain optical alignment.  The fifth and sixth mirrors are modeled together as a group.  After the rotations about the azimuth and coud\'{e} table axes, all optics are fixed and become part of the Mueller matrix for all optics ahead of the modulator, denoted as ${\bf M}_{Mod}$.  This Mueller matrix is expected to be computed for all field angles and wavelengths on every sensor and it includes all optics in the relay optics, AO system, FIDO and within the instruments.

\begin{wrapfigure}{l}{0.41\textwidth}
\vspace{-5mm}
\begin{equation}
 \left ( \begin{array}{rrrr}
 II   		& QI/II	& UI/II	& VI/II	\\
 IQ/II 	& QQ/II	& UQ/II	& VQ/II	\\
 IU/II 	& QU/II	& UU/II	& VU	/II	\\
 IV/II	 	& QV/II	& UV	/II	& VV	/II	\\ 
\end{array} \right ) 
\label{eqn:MM_IntensNorm}
\end{equation}
\vspace{-5mm}
\end{wrapfigure}

We showed an example for DL-NIRSP with the F/ 24 configuration in Equations \ref{eqn:MM_group_model_toAO}, \ref{eqn:MM_group_model_DLfido} and \ref{eqn:MM_group_model_DLinstrument}. The Mueller matrix combines polarization behavior of 6 surfaces through the feed optics and AO system, 6 surfaces with complex dichroic coatings in FIDO and another 7 mirror surfaces inside DL-NIRSP ahead of the modulator.  For any configuration change in FIDO or an instrument, the Mueller matrix must be calculated. This Mueller matrix will be computed for several example FIDO configurations later in this paper.  Later in this paper, we provide an assessment of the sensitivity to mirror coating properties on each mirror. The assumptions underlying the simplifications in the {\it group model} will also be assessed.

We also will adopt an astronomical convention for displaying Mueller matrices where we normalize every element by the $II$ element to remove the influence of transmission on the other matrix elements as seen in Equation \ref{eqn:MM_IntensNorm}. Thus subsequent Figures will display a matrix that is not formally a Mueller matrix but is convenient for displaying the separate effects of transmission, retardance and diattenuation in simple forms.  The transmission is shown in the [0,0] element while all other elements are normalized by transmission for convenient interpretation.

\subsection{Summary: Optical Path \& Predicting System Polarization}

We outlined the DKIST optical path to the coud\'{e} laboratory through the instruments to the modulating retarders. We provide a model for polarization of the various optics in the system by functional groups called the {\it group model}. The 6 mirrors between the sky and the coud\'{e} lab are combined into 3 groups that rotate with respect to one another. We then showed how the Mueller matrices of the coud\'{e} relay optics, the adaptive optics system, the Facility Instrument Distribution Optics (FIDO) dichroics and internal instrument optics were combined into a single Mueller matrix representing all optics ahead of the polarization modulator. We showed an example of the DL-NIRSP instrument in the F/ 24 configuration and described the 19 optical surfaces that will impact this system Mueller matrix. The ViSP and Cryo-NIRSP instruments are described in a prior reference \cite{2017JATIS...3a8002H}. Next, we show how we can measure polarization properties of coated optics in reflection and transmission. We examine mirror data sets and repeatability in Section \ref{sec:nlsp_data}. Section \ref{sec:mirror_coating_fits} shows examples of fitting mirror coating models to data sets along with a comparison of metrology for retardance and diattenuation from multiple instruments. We show broad band anti-reflection coatings in Section \ref{sec:wbbar1} used in DKIST beam splitters and calibration optics. Models for FIDO Dichroic coatings and examples from our vendor are shown in Section \ref{sec:dichroics}. We discuss transmission and polarization artifacts only seen at high spectral resolving power. We then combine this coating information to show predictions of the {\it group model} parameters and the Mueller matrix from the derived optical coating properties of all optics between the sky and the polarimeter. The combined spectral behavior of the mirrors, beam splitters and windows is assessed in a few anticipated DKIST observing configurations in Section \ref{sec:system_group_model}.

\clearpage

\section{NSO Laboratory Spectro-Polarimeter \& Performance}
\label{sec:nlsp_data}

\begin{wrapfigure}{l}{0.60\textwidth}
\centering
\vspace{-3mm}
\begin{tabular}{c} 
\hbox{
\hspace{-1.0em}
\includegraphics[height=7.0cm, angle=0]{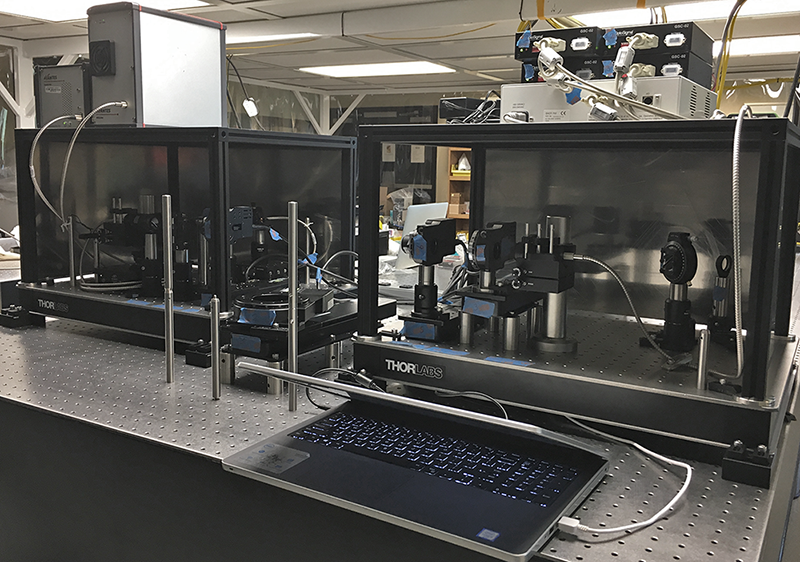}
}
\end{tabular}
\caption[]  {\label{fig:nlsp_lab_setup}  The NLSP installed in the Boulder laboratory.   }
\vspace{-2mm}
 \end{wrapfigure}

The National Solar Observatory Laboratory Spectro-Polarimeter (NLSP) uses two spectrographs to simultaneously measure polarized spectra with a wire grid polarizing beam splitter. We use a collimated optical setup for polarization measurement. A fiber coupled light source is collimated by an achromatic doublet lens and then stopped to a circular beam of 4 mm diameter using laser cut masks. This provides a narrow-field, uniform, collimated light source. A polarization state generator consists of a rotating wire grid polarizer and rotating third-wave achromatic linear retarder mounted upstream of the sample location. After the sample, a rotating third-wave linear retarder is mounted as a modulator. The final optic is a fixed orientation analyzing wire grid polarizer also used as a polarizing beam splitter.  As detectors, we use visible and near-infrared spectrographs from Avantes. The visible spectrograph covers 380 nm to 1200 nm while the NIR spectrograph covers 900 nm to 1650 nm wavelength.

The beam transmitted through the wire grid polarizer feeds the visible spectrograph via filters, aperture stops and a lens. At the lens focus, a fiber couples light to the spectrograph. The beam reflected off the wire grid polarizer is passed through a separate set of filter, aperture stop and lens optics into the near infrared (NIR) spectrograph. This NIR arm has an additional polarizer with wires parallel to the analyzer to remove the fresnel reflection component off the glass and maintain high contrast.  Figure \ref{fig:nlsp_lab_setup} shows a picture of the system with the polarization state generator half on the right and the spectropolarimeter half on the left.  The sample under test goes on the rotation and translation stage in between the two sides.

\begin{wraptable}{l}{0.64\textwidth}
\vspace{-4mm}
\caption{DKIST Enhanced Ag Mirror Samples}
\label{table:mirror_sample_list}
\centering
\begin{tabular}{l l l  |  l l l}
\hline
\hline
Name				& Run			& N			& Name				& Run		& N		\\
\hline
\hline
DKIST M2				& 13BE18 		&  2			& DKIST Eval 1			& Unknwn		& 1			\\
DKIST M3				& 14BE04			& 1			& DKIST Eval 2			& Unknwn		& 1			\\
DKIST M4$_{pre}$		& 15BA34			& 1			& DKIST Eval 3			& Unknwn		& 1			\\
DKIST M4				& 15BA35			& 2			& Cryo-NIRSP 1 		& 16BB07		& 1			\\
DKIST M5				& 12BD18			& 1			& Cryo-NIRSP 2 		& 16BB21		& 1			\\
DKIST M5$_s$			& 12BD19			& 1			& Cryo-NIRSP P 		& 16BD15		& 2			\\
DKIST M6				& 14BE05			& 1			& DL-NIRSP 1			& 16BE16		& 1			\\			
DKIST M6$_s$			& 14BE04			& 1			& DL-NIRSP 1a			& 16BE17		& 1			\\
DKIST M7				& 16BD16			& 1			& DL-NIRSP 1b			& {\it 16BE17}	& 1		\\
DKIST M10			& 15BA23			& 4			& DL F00-207			& 16BB22		& 1			\\
\hline
\hline
\end{tabular}
\vspace{-4mm}
\end{wraptable}

We have a reflective configuration for NLSP where a sample can reflect the beam to an additional set of optics. As this reflective setup must always have an optic in the sample location, we cannot calibrate absolute reflectivity. But we can use our calibration of the system optics to measure the retardance and diattenuation of the sample. Presently, we have only calibrated this channel at a fixed incidence angle (AOI) of 45$^\circ$, but the system is capable of a much wider range of angles. The sample is mounted on a rotation stage controlling AOI which itself is mounted on a translation stage. 

The mirrors for the DKIST optics as well as the coud\'{e} instruments are provided by various instrument partners who use a range of commercial vendors. There is a great diversity of enhanced protected, protected and bare metal mirrors in the path between the sun and any DKIST camera. Additionally, many-layer dielectric dichroic coatings are part of the Facility Instrument Distribution Optics (FIDO) system. The ViSP will see one dichroic coating in reflection and another in transmission with the appropriate anti-reflection coating at 15$^\circ$ incidence. The VTF will see one dichroic in transmission and two in reflection. The DL-NIRSP will see three dichroics in transmission with the three anti-reflection coatings. All these instruments will see the wavefront sensor beam splitter in transmission with a broad-band anti-reflection coating on the back side. 

\begin{wraptable}{l}{0.40\textwidth}
\vspace{-3mm}
\caption{Other Ag \& Al Mirror Samples}
\label{table:mirror_sample_list_alt}
\centering
\begin{tabular}{l l l}
\hline
\hline
Name				& Run			& N		\\
\hline
\hline
DKIST M1				& AFRL Bare Al	 	& 2		\\
DKIST M1 spare		& AFRL Bare Al	 	& 2		\\
\hline
FIDO  Samples			&  		& 		\\
\hline
IOI Enh. ProtAg 		& EAg-300 5-5033	& 1		\\
IOI Enh. ProtAg 		& EAg-700 8-6282	& 1		\\
IOI Enh. ProtAg 		& EAg1-450 8-6898	& 1		\\
\hline
FIDO C-M1$_{pre}$		& EAg1-420 6-7759	& 2		\\	
FIDO C-M1			& EAg1-420 6-7766	& 3		\\
ViSP Re-coat			& EAg1-420 6-7767	& 3		\\
DKIST M9				& EAg1-450 9-3095	& 3		\\
\hline
Other Samples			& 				& 		\\
\hline
DL-NIRSP EMF		& Protected Ag		& 1		\\
DKIST M8 EMF			& Protected Ag99b	& 3		\\
ViSP Very-EAg	 		& 29 Layer + Ag	& 2		\\
ViSP RMI EAg			& Prot. Enh. Ag		& 2		\\	
Zygo	 M9a Samp.		& Prot. Enh. Ag		& 4		\\
Zygo DL-FM1			& Prot. Enh. Ag		& 1		\\ 
\hline
BBSO Newport 		& Prot. Enh. Ag		& 2		\\
GREGOR 			& Protected Ag		& 2		\\
\hline
Thor Labs				& Protected Ag 		& 3		\\
Edmund Optics$^{vbi}$	& Protected Ag		& 1		\\
Edmund Optics			& Protected Ag		& 4		\\
\hline
\hline
\end{tabular}
\vspace{-3mm}
\end{wraptable}

In Table \ref{table:mirror_sample_list} we list many samples of the DKIST enhanced protected silver coating we tested. We show the many witness samples collected from coating shots of the {\it identical formula} procured for the DKIST telescope optics, as well as two of the coud\'{e} instrument:  DL-NIRSP and Cryo-NIRSP and preliminary evaluation samples from an unknown run. These witness samples represent what should be identical coatings spatially across the chamber over many repeated coating shots. The N column shows how many witness samples we have from each run representing different spatial positions within the coating chamber.  The run 13BE18 is noted as both the mirror and witness sample substrates are SiC.  We also note that run 14BE04 contained both DKIST mirrors M3 and the spare M6.  We received a sample from the coating shot just before M4 which we denote M4$_{pre}$.  We also have a spare 5$^{th}$ mirror denoted M5$_s$.  For the specific case of the DL-NIRSP mirror DL F00-207 we did not receive a sample, but this is a small flat optic so we tested the optic itself directly. One of the samples from the DL-NIRSP team was labeled 16BE15/17 but the sample itself had 16BE17 marked on the side so we make the assumption this is a separate spatial position for 16BE17 that we denote 1b shown as italics in Table \ref{table:mirror_sample_list}. 

We show several alternate coating samples in Table \ref{table:mirror_sample_list_alt}. We have the bare aluminum coatings from the 4m primary mirror as well as the spare {\it commissioning mirror} both done by the Air Force Research Labs (AFRL) on Haleakala, adjacent to DKIST. We tested enhanced protected silver mirror samples from Infinite Optics, Inc. (IOI) during FIDO mirror and dichroic coating assessments. We were given coating designs, reflectivity and diattenuation along with samples for Enhanced Silver (EAg) formula names 300, 700, 1-420 and 1-450. We are partly through coating several other DKIST mirros  (M9 and the FIDO mirrors) with an IOI coating.  We also note that three of the ViSP mirrors were originally a very enhanced 29-layer silver coating that was subsequently stripped and re-coated with the IOI coating EAg1-420.

The bottom section of Table \ref{table:mirror_sample_list_alt} shows witness samples we've tested from commercial sources. Some will be used in DKIST while others are used at other solar telescopes including the Big Bear Solar Observatory (BBSO) Goode Solar Telescope (GST, formerly the New Solar Telescope, NST) and the GREGOR solar telescope. The two samples from the very-enhanced ViSP mirrors were tested. There is an additional large flat from Edmund Optics procured for the DKIST VBI-blue instrument we tested and denote in Table \ref{table:mirror_sample_list_alt}. This mirror is likely similar to some of the BBSO mirrors from Edmund Optics with a catalog protected silver coating. We later bought 4 small samples from Edmunds nominally with the  same protected silver coating to assess variability. We also include the three Thor Labs protected silver mirrors that we use at NSO for laboratory testing. These samples were all procured at one time but without any guarantee of being from the same coating shot. On some DL-NIRSP mirrors, the nominal DKIST-specified silver formula was not procured.  We also show measurements of this alternate protected silver coating by Dynasil's Evaporated Metal Films (EMF). The team did not receive witness samples for this alternate coating so we tested the DL-NIRSP spectrograph flat optic F00-201 directly in NLSP. We also coated DKIST M8 at EMF with a blue-enhanced version of the Ag99 coating we label Ag99b. Zygo Corporation provided four samples of their nominal enhanced protected silver coating as part of the FIDO project coating the removable mirror feeding Cryo-NIRSP (M9a). Zygo also coated the fold mirror to DL-NIRSP after the sequence of beamsplitters (DL-FM1). We also obtained a sample from the two coating runs of the enhanced protected silver used for the ViSP internal fold mirrors. One of these folds is after the coated slit mask and impacts the system polarization modulation matrix.

We demonstrate here the use of the Berreman calculus to fit coating data from NLSP as well as to compare polarization performance of various coated mirrors and dielectrics. We use the common equation for a Mueller matrix derived from a single flat fold. Details of the Mueller matrix equations used are in Appendix \ref{sec:app_Mueller}. The reflectivities parallel and perpendicular to the plane of incidence ($R_p$ and $R_s$) can be used to derive the transmission and diattenuation terms. In the normalized Mueller matrix the $\frac{IQ}{II}$ and $\frac{QI}{II}$ terms are a normalized reflectivity difference ratio (R$_s$-R$_p$)/(R$_s$+R$_p$). The retardance ($\delta$) is a term in the UV rotation matrix in the lower right quadrant.

\begin{wrapfigure}{r}{0.57\textwidth}
\centering
\vspace{-3mm}
\begin{tabular}{c} 
\hbox{
\hspace{-1.0em}
\includegraphics[height=7.2cm, angle=0]{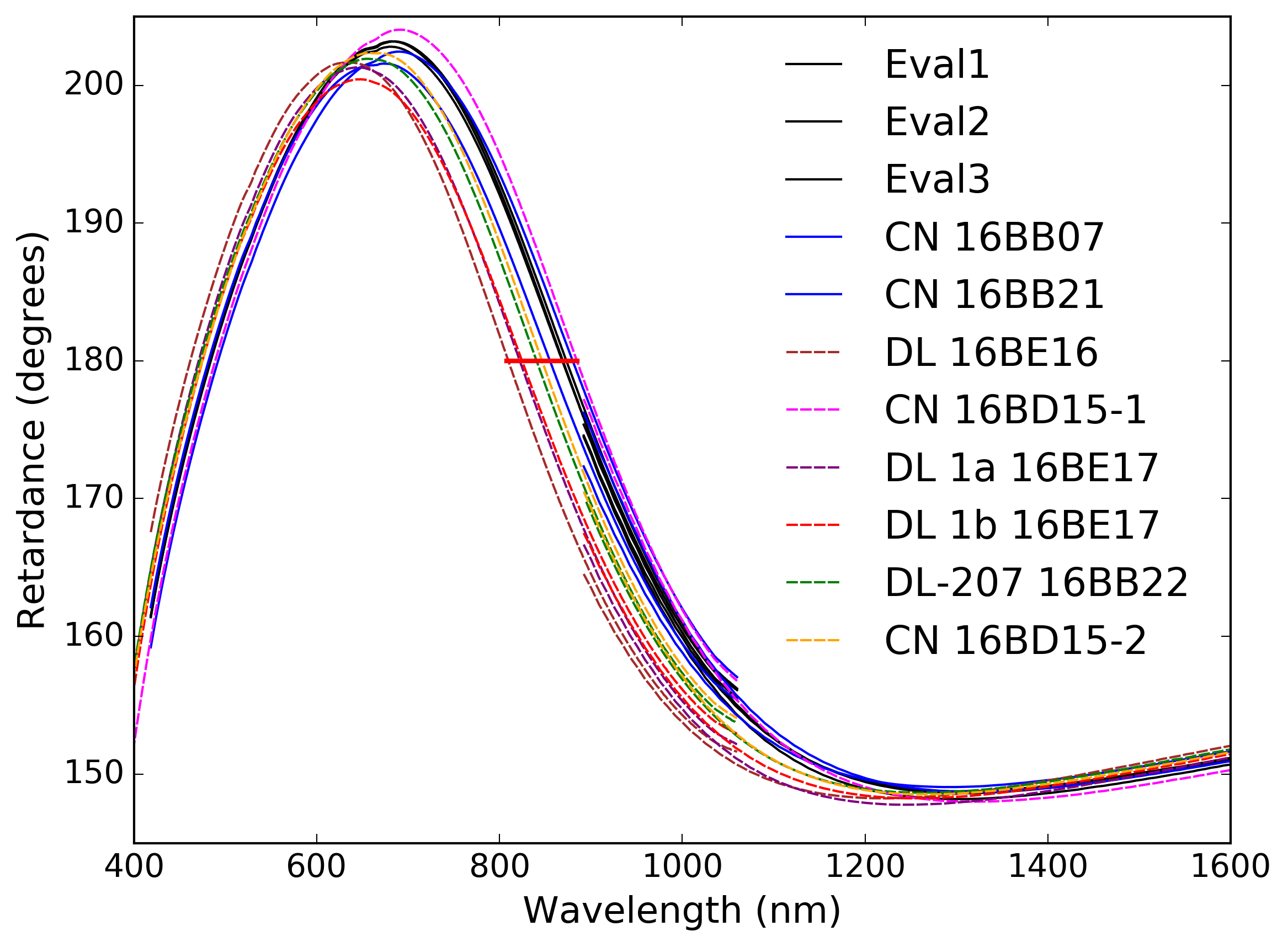}
}
\end{tabular}
\caption[] 
{\label{fig:quantum_compiled_nlsp_data}  The retardance computed from the Mueller matrix elements measured with the reflective arm of NLSP at 45$^\circ$ incidence angle. The measured retardance is near the theoretical unperturbed 180$^\circ$ at two wavelengths, around 460 nm and 810 nm. The short horizontal bar shows a 60 nm spread in wavelengths for 180$^\circ$ retardance in the NIR. We separately show the visible and near infrared spectrograph data sets with good agreement in the 900 nm to 1100 nm wavelength overlap range. }
\vspace{-3mm}
\end{wrapfigure}

With NLSP, we are able to characterize the coatings as applied to DKIST telescope and instrument optics. In any manufacturing process, there will be variation. We show in Figure \ref{fig:quantum_compiled_nlsp_data} witness samples from a nominal coating formula applied by a vendor well within manufacturing tolerances to several DKIST optics and to the DL-NIRSP and Cryo-NIRSP instrument feed optics. The curves show retardance varying from under 150$^\circ$ to just over 200$^\circ$. The different color curves highlight how the coating applied to the DL-NIRSP instrument gives significantly different results in retardance from the DKIST telescope optics and the Cryo-NIRSP feed optics. The solid red horizontal line in Figure \ref{fig:quantum_compiled_nlsp_data} shows how the wavelength of zero net mirror retardance varies from 810 nm to 885 nm depending on the coating. 

These retardance changes can be caused by only a few nanometers variation in thickness of a dielectric layer in a multi-layer coating. These variations are well within typical manufacturing tolerances and are entirely expected, particularly when retardance or diattenuation targets are not included in the coating specifications. The reflectivity of all our mirrors easily pass the DKIST reflectance criteria. The NLSP measured retardance differences are orders of magnitude larger than the NLSP sensitivity and we show later in this paper how these changes impact the system polarization performance predictions.


\section{Mirror Coating Models \& Fits to Polarization}
\label{sec:mirror_coating_fits}

We begin assessing the DKIST coatings by following our previously published technique \cite{Harrington:2017ejb, 2017JATIS...3a8002H} to select a simple one or two layer coating model. We identify the best fit of the model retardance to the NLSP measurements using only the thickness of the dielectric coating layers as variables. The refractive indices of the layer materials are interpolated from lookup tables derived from public references such as refractiveindex.info. This simple method is quite limited in that only retardance is fit, not diattenuation or reflectivity. An additional limitation is that only a single incidence angle is used in the fit. However, having a model coating that reasonably reproduces the spectral behavior and magnitude of retardance, diattenuation and reflectivity allows us to estimate the magnitude of several types of polarization artifacts possible in DKIST. We provide more details below and in Appendix \ref{sec:refractive_index_and_fitting}.

\begin{figure}[htbp]
\begin{center}
\vspace{-1mm}
\hbox{
\hspace{-1.1em}
\includegraphics[height=6.49cm, angle=0]{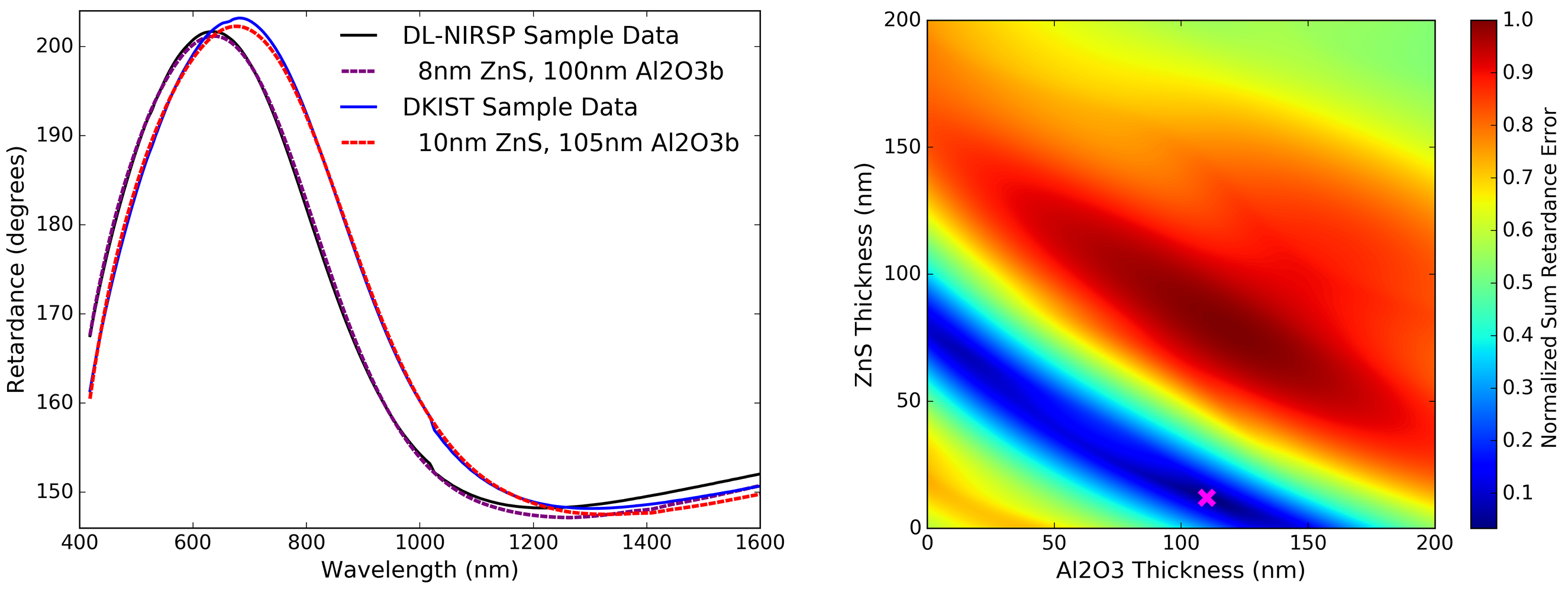}
}
\caption[] 
{ \label{fig:quantum_retardance_change_may31_to_july14} The left plot shows measured and modeled retardance curves for two enhanced protected silver samples. The solid curves show NLSP measurements and dashed curves show best fit two layer coating models. The right hand plot shows the wavelength averaged retardance error normalized from 0 to 1 for a two-layer protective coating of ZnS over Al$_2$O$_3$ on top of the silver base layer. }
\vspace{-5mm}
\end{center}
\end{figure}

Any imperfections in knowledge of the materials, number of layers or material refractive index will degrade the model fit.  A grid of coating retardance models are computed for each thickness of the dielectric material. The predicted retardance is then subtracted from the NLSP data to create an error curve. These retardance error curves are squared and summed to create a single wavelength averaged error metric which is minimized to determine the fit. In the left hand graphic of Figure \ref{fig:quantum_retardance_change_may31_to_july14} we show two example retardance curves for the DKIST enhanced protected silver samples. The solid black line shows measurements of a witness sample for the DKIST instrument DL-NIRSP.  The dashed dark blue line shows an example fit model coating of ZnS at 8 nm thickness coated over 100 nm of Al$_2$O$_3$ using the Boidin et al \cite{Boidin:2016kx} refractive index formula on top of silver. This coating has the theoretical 180$^\circ$ retardance upon reflection at wavelengths of 462 nm and 810 nm.

\begin{wrapfigure}{r}{0.57\textwidth}
\centering
\vspace{-3mm}
\begin{tabular}{c} 
\hbox{
\hspace{-1.0em}
\includegraphics[height=7.2cm, angle=0]{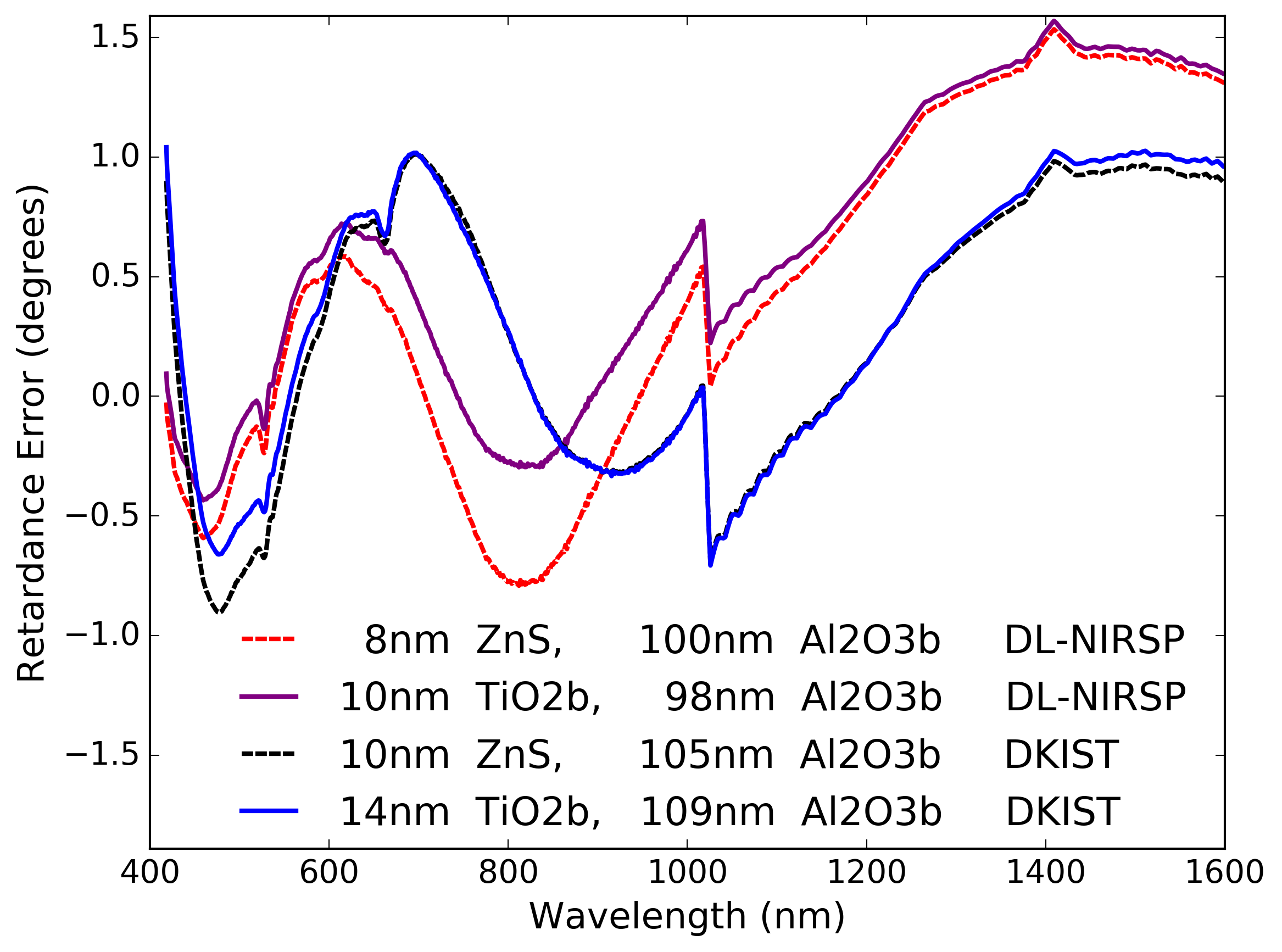}
}
\end{tabular}
\caption[] 
{\label{fig:metal_index_change_quantum_pars}  The difference between retardance data and models for a reflection off a DKIST enhanced protected silver sample following the left graphic of Figure \ref{fig:quantum_retardance_change_may31_to_july14}. Residuals are roughly 1$^\circ$ retardance but are orders of magnitude larger than our statistical limits.  The step in error at 1020 nm wavelength is caused by the change between visible and near infrared spectrographs.  }
\vspace{-3mm}
\end{wrapfigure}

The solid blue line of Figure \ref{fig:quantum_retardance_change_may31_to_july14} shows another DKIST witness sample of the same coating formula but from a different shot than the DL-NIRSP instrument shown in black. There is some normal and expected shot to shot variation in the manufacturing process. The dashed red line in the left hand graphic shows an example fit model coating of ZnS at 10 nm thickness coated over 105 nm of Al$_2$O$_3$ using the Boidin et al \cite{Boidin:2016kx} refractive index formula on top of silver. This coating has the theoretical 180$^\circ$ retardance upon reflection at wavelengths of 483 nm and 870 nm, which corresponds to a change of 21 nm and 60 nm wavelength respectively from the prior coating shot. This coating model fit is different from the DL-NIRSP sample by about 2 nm in the top ZnS layer and 5 nm in the bottom Al$_2$O$_3$ layer. The retardance fits match the appropriate data set well and are very significantly different from each other.

This technique uses a simple brute force search of the possible coating formulas and is limited to a very small number of variables. Essentially two thicknesses are fit but the refractive index wavelength dependence for each dielectric must be specified as well as the wavelength range over which to compute errors.  The right hand plot of Figure \ref{fig:quantum_retardance_change_may31_to_july14} shows the wavelength averaged error value on a color scale for the DKIST sample against thickness for the ZnS and Al$_2$O$_3$ layers. Red shows high error values while dark blue into black shows the error minimum values.  The magenta cross marks the coating model solution identified with this simple brute-force method.

Figure \ref{fig:metal_index_change_quantum_pars} shows the difference between a few best-fit models and the NLSP data sets. We follow same color scheme as Figure \ref{fig:quantum_retardance_change_may31_to_july14} for coating models made with the various ZnS and Al$_2$O$_3$ refractive index models. We also now include a new model with the TiO$_2$ refractive index equation from Boidin as the top coating layer.  The best fit coating layer thickness only changes by a few nanometers when using these different refractive index equations. The retardance error changes magnitude slightly, with larger errors at the extreme wavelength ends of the data set. We can clearly see spectral oscillations in retardance error of 1.5$^\circ$ peak to peak.  In essence, both two-layer models are incomplete at reproducing the measured spectral complexity at levels below $\pm$1$^\circ$ retardance. This $\pm$1$^\circ$ retardance error is orders of magnitude above our NLSP metrology statistical noise limits and indicates the limitations of the simple two-layer thickness fitting technique. With NLSP, our data is now of sufficient quality that we must add more variables to include variable refractive index dependence on wavelength, the metal refractive index values and to also include reflectivity and diattenuation in the fitting metric. However, this model is sufficient for estimating DKIST system calibration behavior and showing how real coatings may impact DKIST polarization models.

\begin{figure}[htbp]
\begin{center}
\vspace{-3mm}
\hbox{
\hspace{-0.0em}
\includegraphics[height=17.99cm, angle=0]{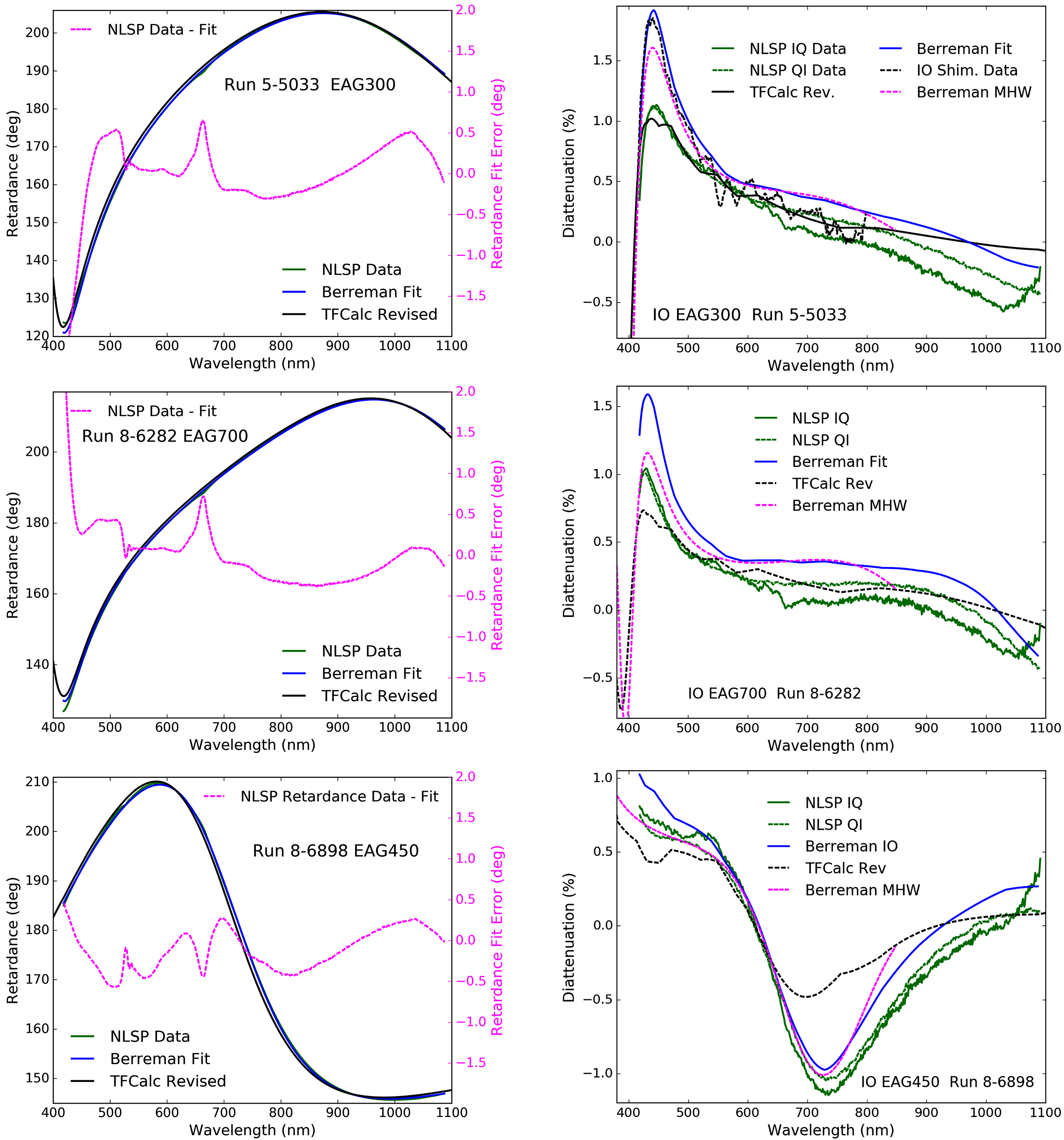}
}
\caption[Retardance] 
{ \label{fig:retardance_fits_to_mirror_samples}  A comparison of retardance (left) and diattenuation (right) data and models.  Each panel has the NLSP data recorded with the IOI mirror sample in reflection at 45$^\circ$ incidence shown in green.  For the diattenuation data, we have two independent estimates of diattenuation from both the QI and IQ elements of the NLSP measured Mueller matrix. The best fit Berreman model was found via a simple search for dielectric thicknesses in a grid with 1 nm step size. The best-fit Berreman model (to retardance only) is plotted in each panel in blue. The difference in retardance between the NLSP data and the best-fit Berreman model is shown in magenta using the right hand y-axis of the left hand retardance plots. The Berreman model reproduces the data to roughly $\pm$2$^\circ$ in retardance with a very repeatable mis-match function with wavelength. The left hand graphics show agreement between data and models to roughly fraction of a percent magnitudes, though diattenuation is not included in the fit. The dashed magenta line in the right hand diattenuation plots shows the model using an alternate refractive index for the silver from MHW\cite{2014btfp.book.....M}. An IOI diattenuation spectra from their Shimadzu system is shown by the dashed black line in the upper right graph.}
\vspace{-8mm}
\end{center}
\end{figure}

The materials deposited and their refractive index upon deposition is one of the major limitations to this modeling. When we have vendor-provided refractive index data or can adjust the refractive index wavelength interpolation, we achieve significantly better fits. Figure \ref{fig:retardance_fits_to_mirror_samples} shows measurements and associated retardance fits to three enhanced protected silver mirror samples from Infinite Optics, Inc. The mirrors represent different coating formulas, materials and design choices.

The default materials and refractive we used in fitting the DKIST mirrors above are provided in standard software packages such as the Thin Film Calculator (TFCalc), Zemax optical design studio or in text books such as McCall, Hodgkinson and Wu (MHW) \cite{2014btfp.book.....M}. With vendor-provided refractive index data and a coating designs (in TFCalc files or as Zemax coating recipes), we follow the same coating layer thickness fit to match the retardance values. We then revised the model to use the as-built thicknesses to show the diattenuation prediction. We did not revise the refractive index of the dielectric materials or the silver or allow any variation of material refractive index. As shown above, the diattenuation is very sensitive to the complex refractive index of the silver coating so we do not expect good fits for these parameters.

The green lines in Figure \ref{fig:retardance_fits_to_mirror_samples} show the NLSP data. Left shows retardance and right shows diattenuation. Most retardance curves in green are nearly invisible as the model fits nearly perfectly overlay the retardance data set. For diattenuation, NLSP measures the IQ and QI elements of the Mueller matrix independently, so both are shown to demonstrate systematic and statistical error limits. The blue curves show our Python-script calculations using the Berreman calculus when using the vendor provided refractive indices where given, and nominal TFCalc values otherwise. Typical fits are now within 0.5$^\circ$ retardance error peak to peak and a much smaller RMS for the IOI samples.  The fit was significantly worse for the DKIST EAg sample shown above in Figures \ref{fig:quantum_retardance_change_may31_to_july14} and \ref{fig:metal_index_change_quantum_pars}. This mismatch is likely caused by both varying refractive index of coating materials and additional coating layers not included in our fit as we have no vendor information on this coating. We can clearly see an artifact in the NLSP data set around 680 nm wavelength corresponding to a 0.5$^\circ$ narrow spike. Our fits also fail significantly at shorter wavelengths (where the refractive index interpolations are worse).

With the TFCalc model revisions to the dielectric thicknesses, we can also compare NLSP diattenuation measurements to various theoretical calculations. In the right hand graphics of Figure \ref{fig:retardance_fits_to_mirror_samples} we show the diattenuation and a measurement provided by IOI in the upper right.  The IOI diattenuation measurements shown in dashed black are recorded with orthogonal polarizers in their Shimadzu spectrophotometer to measure reflectivity of S and P polarization states. 

The TFCalc silver default refractive index values are used in this dashed black line. The fits are generally within 0.5\% diattenaution but with high spectral variation in the quality of the fit. We ran another Berreman model using the MHW\cite{2014btfp.book.....M} refractive index for silver to demonstrate the impact of a change in assumed metal optical properties.This alternate model matches the data better at different wavelengths. We note the MHW indices are only fit across visible wavelengths and are only plotted where the materials properties are valid. As metallic coatings deposited in coating chambers can have variation in properties with deposition, a mis-match between literature values and actual as-coated data is expected.

\clearpage
\subsection{Refractive Indices, Materials Parameters \& Coating Model Interpretation}
\label{sec:mirror_fit_limitations}

\begin{wraptable}{l}{0.18\textwidth}
\vspace{-3mm}
\caption{Indices}
\label{table:effective_refractive_indices}
\centering
\begin{tabular}{c c}
\hline\hline
Material		& Index 			\\
\hline
\hline
ZnS			& 2.31	\\
TiO$^t_2$		& 2.31 	\\
TiO$^b_2$	& 2.18	\\
TiO$_2$		& 2.04	\\
HfO$_2$		& 1.98	\\
SiO			& 1.92	\\
\hline
Al$_2$O$^b_3$& 1.67	\\
Al$_2$O$_3$	& 1.55	\\
SiO$_2$		& 1.45	\\
MgF$_2$		& 1.37	\\
\hline
\hline
\end{tabular}
\vspace{-3mm}
\end{wraptable}

The refractive index values used in various software modeling packages varies. In common software packages such as TFCalc or Zemax, the basic software package provides some default materials and refractive index values as lookup tables or simple equations. This nominal is useful from a modeling perspective as it provides rough guides to actual coating materials. However, there are several factors that impact the actual refractive index of deposited materials in the coating process. These include the coating deposition process, temperature, final material density, varying growth styles, impurity concentrations and others. Thus, detailed information is required from a vendor about their specific materials and known process before having confidence in modeling coatings using refractive index data. In most cases, we do not need to know and likely will not be told the actual materials and all the proprietary details of a coating. However, a useful performance model for mirror properties can be created using simple parametric curves from the public materials data as representative of common coating types. We do not need to know specific coating formulation details to estimate the polarization performance of the system. As a modeling approach for this paper, we simply assume that the refractive index formulas may change substantially, in some cases over 15\%. We list examples in Table \ref{table:effective_refractive_indices} for a wavelength of 850 nm drawn from TFCalc, Zemax, MHW \cite{2014btfp.book.....M} and other references. The b superscript denotes the Boidin reference and the t denotes the TFCalc default.  For instance, crystal TiO$_2$ would have an index of 2.3 while in some coating literature the index is in the range 2.0 to 2.2. Thus a refractive index curve fit in our study is not indicative of an actual material, but of an effective refractive index that may be similar to materials commonly used in the coating process.  

\begin{figure}[htbp]
\begin{center}
\vspace{-2mm}
\hbox{
\hspace{-0.4em}
\includegraphics[height=5.5cm, angle=0]{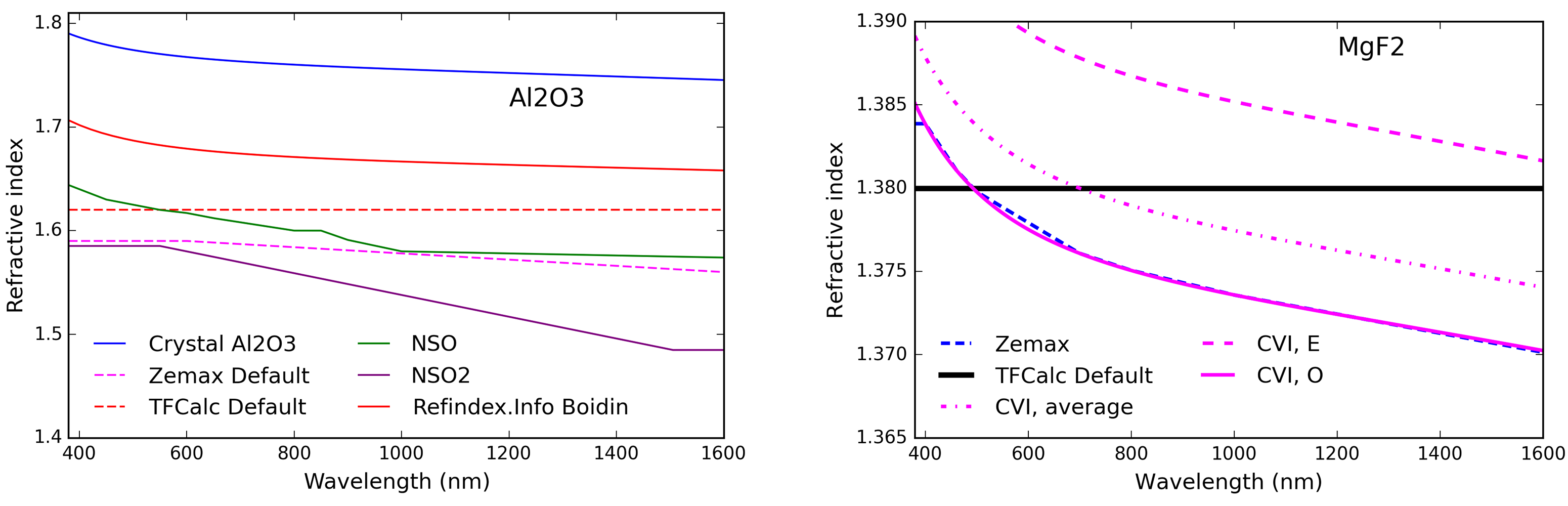}
} 
\caption[Refractive Index] 
{ \label{fig:refractive_index}  The refractive indices published for common coating materials. The RefractiveIndex.info website data is shown along with data from the Zemax coating file provided with version 16.5, 2016 and the TFCalc default material files where applicable. Left shows Al$_2$O$3$ and right shows MgF$_2$ crystal and coating values. The CVI Melles Griot catalog equation is used for crystal MgF$_2$. The difference in refractive index between amorphous coated materials and crystalline materials can be several percent, giving strong changes in predicted coating performance.}
\vspace{-8mm}
\end{center}
\end{figure}

Figure \ref{fig:refractive_index} shows example curves taken from a common website (RefractiveIndex.info), the default coating file in the software packages TFCalc and Zemax, along with vendor catalog values. On the left side we can see crystal sapphire (Al$_2$O$_3$) having an index around 1.75. When used in a coating, TFCalc gives a value slightly above 1.6 constant for all wavelengths as the dashed red curve while Zemax uses a value slightly below 1.6 that falls with wavelength as the dashed magenta curve. The website RefractiveIndex.info cites Boidin et al. \cite{Boidin:2016kx} in the red curve from just above 1.7 falling to 1.65 at long wavelengths. We have internal DKIST engineering reports that also use slightly higher and lower values as seen by the green and purple curves.

Slightly better agreement is seen between values for a common coating material, MgF$_2$.  The right hand graphic of Figure \ref{fig:refractive_index} shows the CVI Laser Optics \& Melles Griot catalog equation for crystalline MgF$_2$ as the solid and dashed magenta lines.  The magenta dot-dashed line represents the weighted average of an amorphous version simply computed as twice the ordinary and once the extraordinary equations combined. Zemax provides a point-wise linearly interpolated version seen as the dashed blue line in Figure \ref{fig:refractive_index} that essentially tracks the CVI catalog equation for the ordinary index of crystal MgF$_2$.  The TFCalc software package uses a wavelength independent value of 1.38.  

\begin{wrapfigure}{r}{0.61\textwidth}
\centering
\vspace{-5mm}
\begin{tabular}{c} 
\hbox{
\hspace{-1.0em}
\includegraphics[height=7.4cm, angle=0]{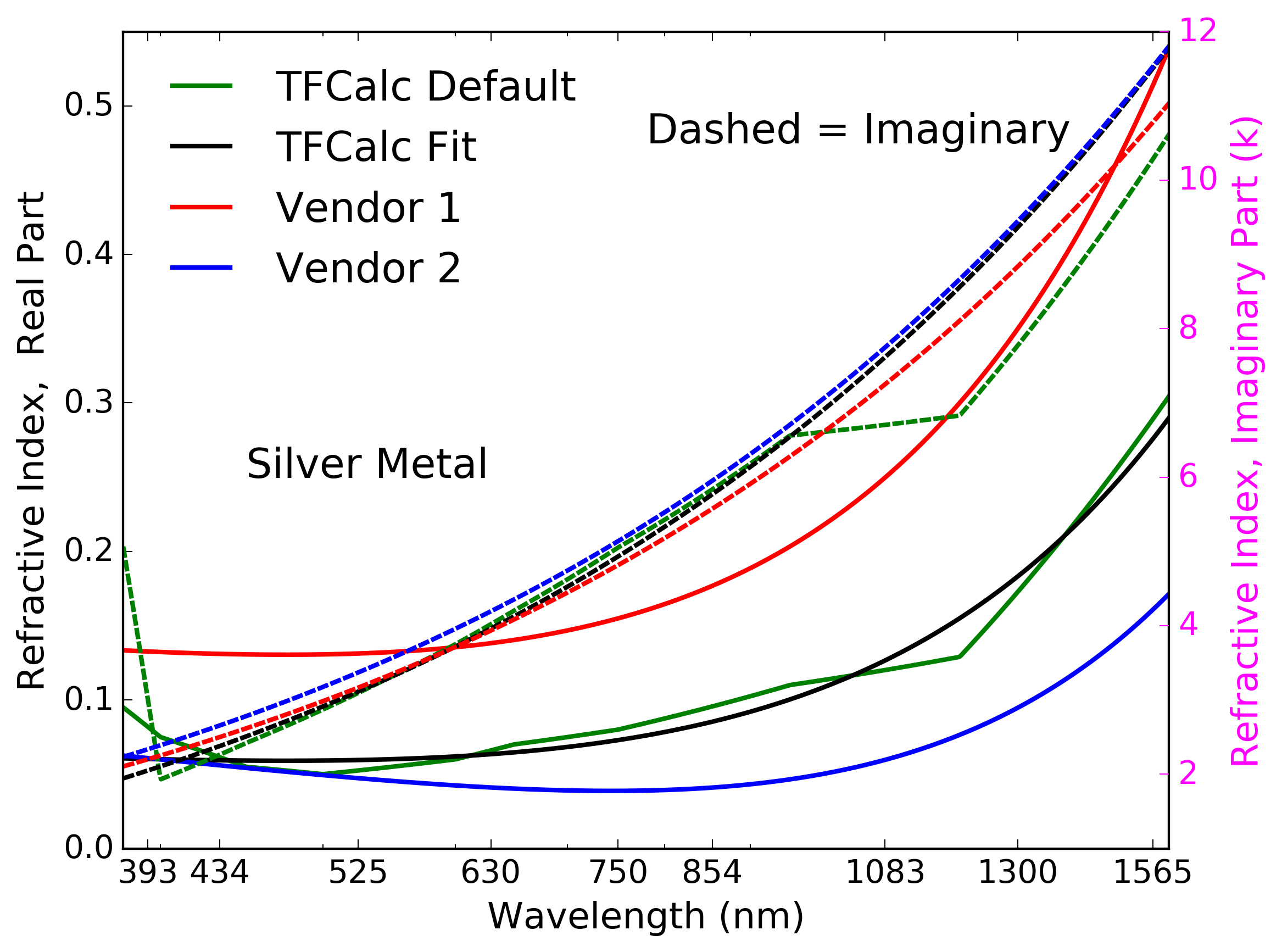}
}
\end{tabular}
\caption[] 
{\label{fig:silver_refractive_index}  The real and imaginary refractive index components of silver. The real component (n) is shown as the solid lines on the left hand y axis with values ranging from 0.05 to 0.5. The green lines show the default TFCalc refractive indices.  Note some discontinuities at 400 nm and 1000 nm wavelengths. Different colors show simple polynomial model fits to TFCalc default values and models provided by vendors in various example coatings. The dashed lines show the imaginary component (k) in exponential form (n-ki) on the right hand y axis. The values of k reflect orders of magnitude change in the wave penetration depth into the metal. }
\vspace{-3mm}
\end{wrapfigure}

The refractive index of the silver metal has a strong influence on polarization properties of the coatings. Literature and vendor values vary widely. Figure \ref{fig:silver_refractive_index} shows an example of linearly interpolated TFCalc default values in green with limited data points and some strong changes in spectral behavior around 400 nm and 1000 nm wavelength.  We also show simple functional fits to the real and imaginary components of silver metal used in various coating models. The solid lines show simple polynomial models for the real part of the refractive index (n) with values ranging from 0.05 up to 0.5. 

The imaginary part (k) is shown as dashed lines in Figure \ref{fig:silver_refractive_index} using a Zemax-style convention where the index is modeled as (n-ki). This imaginary index is exponential in behavior and the electromagnetic wave does not penetrate more than a few nanometers into the metal layer when the complex index is greater than 2 or 3. For the silver metal coating model in Figure \ref{fig:silver_refractive_index}, we see values of roughly 3 at blue wavelengths rising linearly to 11 or 12 at wavelengths of 1600 nm. Often the vendor curves are discontinuous as the green line of Figure \ref{fig:silver_refractive_index} shows for the default TFCalc values. 

\begin{wrapfigure}{l}{0.53\textwidth}
\centering
\vspace{-3mm}
\begin{tabular}{c} 
\hbox{
\hspace{-1.4em}
\includegraphics[height=5.85cm, angle=0]{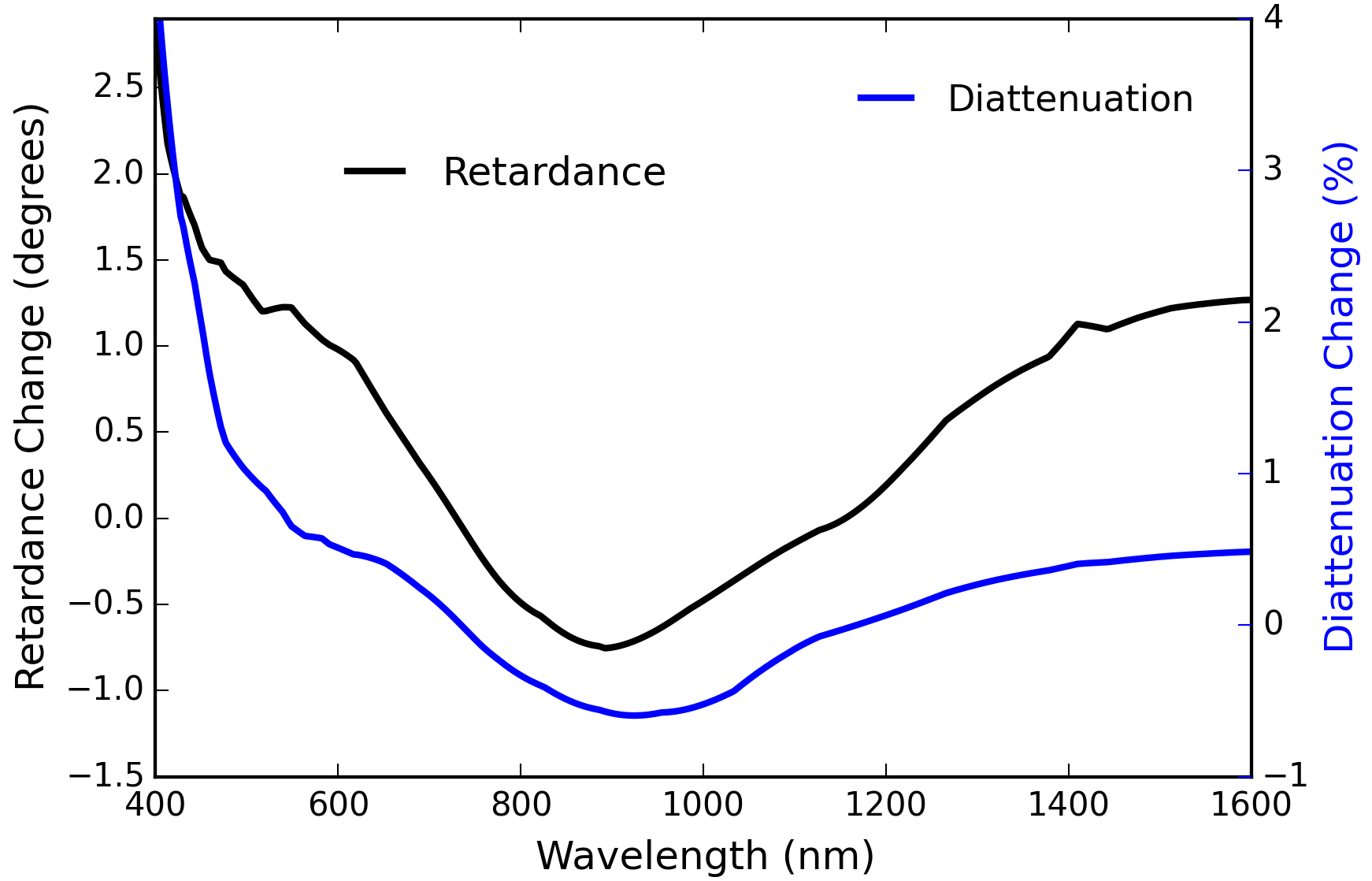}
}
\end{tabular}
\caption[] 
{\label{fig:metal_index_perturb}  The difference between retardance and diattenuation parameters computed using two different refractive index values for silver metal in the coating for a reflection off a DKIST enhanced protected silver mirror formula at 45$^\circ$ incidence. }
\vspace{-2mm}
 \end{wrapfigure}

In prior works\cite{Harrington:2017ejb,2017JATIS...3a8002H}, we only allowed the thickness of one or two dielectric coating layers to vary as a simple two variable optimization problem. However, the metal layer thickness and refractive index both have large polarimetric impact. We show retardance and diattenuation changes in Figure \ref{fig:metal_index_perturb} when using two different formulations for the silver metal indices. In both models we use 10 nm of ZnS with indices from RefractiveIndex.info coated over 105 nm of Al$_2$O$_3$ with refractive indices using the Boidin et al \cite{Boidin:2016kx} values on RefractiveIndex.info. We change only the refractive index of the silver to show the impact. 

The black curve with the left hand Y axis shows that retardance can change by over 2$^\circ$ peak-to-peak for a coating with 20$^\circ$ to 40$^\circ$, a 10\% effect. Much larger changes are seen in diattenuation. The diattenuation change shown in blue using the right hand Y axis is up to 4\% at short wavelengths and roughly half a percent in the visible to near infrared wavelengths. The diattenuation for this coating is only 1.5\% magnitude peak to peak. This metal index difference changes the diattenuation over 300\%.  The silver metal properties can dominate the fit of diattenuation.  Vendors sometimes provide coating performance predictions. Infrequently, this may be be accompanied by names of materials used, such as: {\it coating X is SiO$_2$ protecting the Ag}.  Usually, the layer thickness is not disclosed and the refractive index values for the as-coated material differ significantly from literature values. The retardance is much more sensitive to the dielectric material thicknesses as well as refractive indices. The diattenuation is very sensitive to the real and imaginary refractive index components of the metal as well as the dielectrics.

\begin{wrapfigure}{r}{0.60\textwidth}
\centering
\vspace{-3mm}
\begin{tabular}{c} 
\hbox{
\hspace{-1.0em}
\includegraphics[height=7.6cm, angle=0]{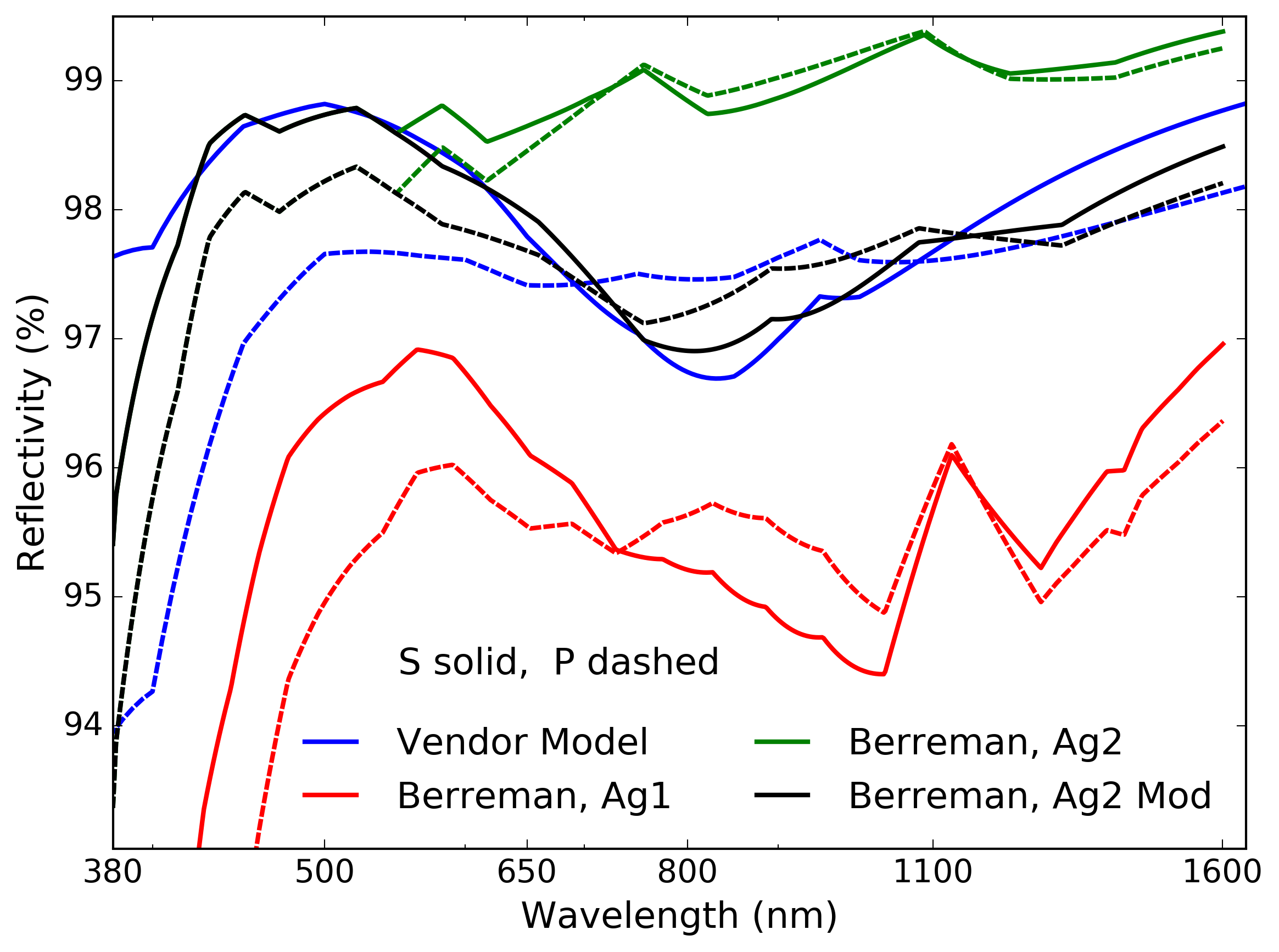}
}
\end{tabular}
\caption[] 
{\label{fig:Compare_Quantum_Reflectivity}  The reflectivity variation when linearly interpolating tabulated data for refractive index of the silver metal layer. The dashed lines show parallel polarization state (P) while the solid lines show the perpendicular polarization state (S) at a 45$^\circ$ incidence angle. Blue shows a vendor prediction while black, red and green show Berreman models using various refractive index curves. Discontinuities come from interpolation between refractive index table values.}
\vspace{-1mm}
 \end{wrapfigure}

In Figure \ref{fig:Compare_Quantum_Reflectivity} we show the reflectance predicted for this same coating model but we use a few different variations of the silver metal complex refractive index. The reflectivity data shows clear discontinuities where TFCalc as well as our Berreman code perform a linear interpolation between points in a table of refractive index values. Solid lines shows the S- polarization state while dashed lines show the P- polarization state. This particular coating has diattenuation change sign in the 700 nm to 1050nm wavelength range. The blue curve shows a vendor-provided TFCalc model. Each vendor may have their own internal materials databases, sometimes where refractive indices are adjusted to their results or other times modified versions of literature values. We expect significant variation between any historical literature values, the vendor models and actual coatings. This is especially true when our simple models likely do not correspond to actual materials and may not include all the layers in the actual coating. The green curve shows one of our Berreman models of silver refractive indices that over-estimates reflectivity at near infrared wavelengths by 1.5\%.  The red curve shows a separate Berreman prediction using lower real refractive index components for the silver that under-estimates the reflectivity by over 2\%.  The black curve is a by-hand modification of the green curve at infrared wavelengths to show that reflectivity can be met, but the diattenuation prediction is still significantly in error. The various look-up tables of silver refractive indices presented can change the reflectivity by over 3\%. As we currently do not perform a simultaneous fit to reflectivity, diattenuation and retardance, we expect our models to contain errors as presented in this section.  This becomes a major limitation of polarization performance modeling, requiring future development for simultaneous fitting of many variables.

\begin{figure}[htbp]
\begin{center}
\vspace{-1mm}
\hbox{
\hspace{-0.4em}
\includegraphics[height=6.3cm, angle=0]{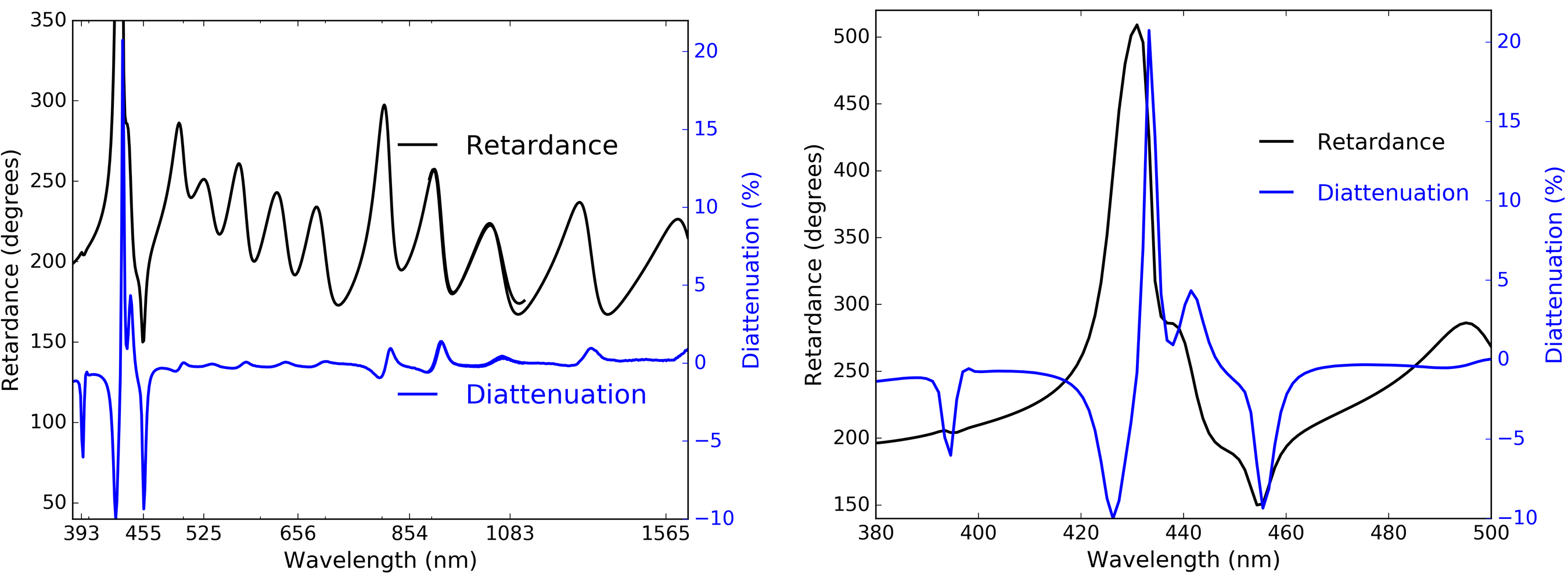}
}
\caption[] 
{ \label{fig:nutek_mirrors_nlsp_dia_ret} Retardance in black and diattenuation in blue measured with NLSP for the ViSP feed mirrors coating witness sample at 45$^\circ$ incidence. The left hand graphic shows all NLSP wavelengths from 380 nm to 1650 nm.  The right hand graphic shows the 380 nm to 520 nm wavelength range where the mirrors show a very strong change in retardance of almost a complete wave over 20 nm bandpass. }
\vspace{-7mm}
\end{center}
\end{figure}

\subsection{DKIST ViSP: Enhanced Protected Silver Mirrors With 29 Dielectric Layers}
\label{sec:sub_visp_ioi_mirrors}

We show in this section a significantly more complex enhanced silver mirror coating that was nominally going to be used in DKIST by one of our partner instruments. The Visible Spectropolarimeter (ViSP) team chose to use a many-layer enhancement on the feed mirrors between the FIDO dichroics and their modulating retarder after the spectrograph entrance slit. This coating is nominally 29 layers of dielectric with an oscillating high-low refractive index design. This represents roughly 3 $\mu$m of dielectric coated on top of the silver metal layer. Given the many layers, significant spectral variation is expected along with the presence of narrow spectral features. Though the team has stripped and re-coated their mirrors with an alternate coating, we include this coating metrology here to show the impact of many-layer enhanced designs and consideration of manufacturing issues. 

Figure \ref{fig:nutek_mirrors_nlsp_dia_ret} shows the ViSP mirror witness sample polarization properties measured in NLSP at 45$^\circ$ incidence. The left hand graphic shows retardance in black and diattenuation in blue for the full NLSP measurement range. The right hand graphic shows the shorter visible wavelengths where very rapid but well measured spectral changes are seen. The retardance changes by over 300$^\circ$ in 20 nm wavelength range, giving a spectral gradients up to 60$^\circ$ per 1 nm wavelength. This bandpass is comparable to the full spectral range measured across the ViSP sensors. Diattenuation similarly changes from -10\% to +20\% in a very narrow band pass. This kind of mirror coating has impact for DKIST as the modulation matrix must be wavelength dependent with the assumption of variation at these magnitudes. We note that requirements against strong spectral gradient in retardance and diattenuation were not included in any specifications. Given the metrology results, the ViSP team has already stripped and re-coated their mirrors. Behavior for the ViSP feed optics had they kept these coatings will be explored in later sections of this paper.  This case is a good example of what happens when many-layer coatings are specified giving rise to complex, large and spectrally narrow polarization properties.

\subsection{Summary of Coating Model Fits to Reflectivity, Diattenuation \& Retardance}
\label{sec:sub_mirror_coating_summary}

We presented examples of two-layer coating models fit to NLSP retardance measurements in this Section. We showed how the dielectric layer thickness and material refractive index impacts fitting measured retardance curves. In Section \ref{sec:mirror_fit_limitations}, we showed how the complex refractive index of the metal layer strongly influences the reflectivity and diattenuation. When fitting a coating formula to measured data, the refractive indices of all components need to be assessed. For DKIST, the retardance values are critical as they determine the field dependence of the cross-talk and ultimately drive requirements on how DKIST calibrates instruments and with what model for the mirror retardance. When attempting to fit retardance, diattenuation and reflectivity simultaneously, all refractive index values become critical. In Section \ref{sec:mirror_fit_limitations} we showed how public literature values for refractive indices may roughly approximate coating behavior, but a detailed fit to high accuracy in all performance parameters requires substantially more detailed knowledge of the coating materials properties. Examples of coating model fits and witness sample measurements were provided for three Infinite Optics, Inc. samples where we have much better refractive index information. Measurements of polarization for a more complex mirror coating with 29 dielectric layers over silver initially planned for use in the DKIST Visible Spectropolarimeter were shown in Section \ref{sec:sub_visp_ioi_mirrors}. In Appendix \ref{sec:refractive_index_and_fitting} we show more examples of mirror polarization properties from commercial sources used in DKIST and the Goode Solar Telescope (formerly the New Solar Telescope) at the Big Bear Solar Observatory, along with examples of coating models and refractive index variation impacts on predicted behavior. We have samples from the GREGOR solar telescope and DKIST Visible Tunable Filter (VTF) instrument shown in Appendix Section \ref{sec:sub_VTF_GREGOR_mirrors}. We move on from many-layer dielectric protected mirrors to many-layer dielectric coatings used as anti-reflection coatings in Sections \ref{sec:wbbar1} and dichroic beam splitters in \ref{sec:dichroics}. Techniques for fitting properties of such coatings also become more complex.

\clearpage

\section{WBBAR1 for DKIST WFS-BS1, FIDO \& Calibration Optics}
\label{sec:wbbar1}

\begin{wrapfigure}{r}{0.58\textwidth}
\centering
\vspace{-3mm}
\begin{tabular}{c} 
\hbox{
\hspace{-1.0em}
\includegraphics[height=7.3cm, angle=0]{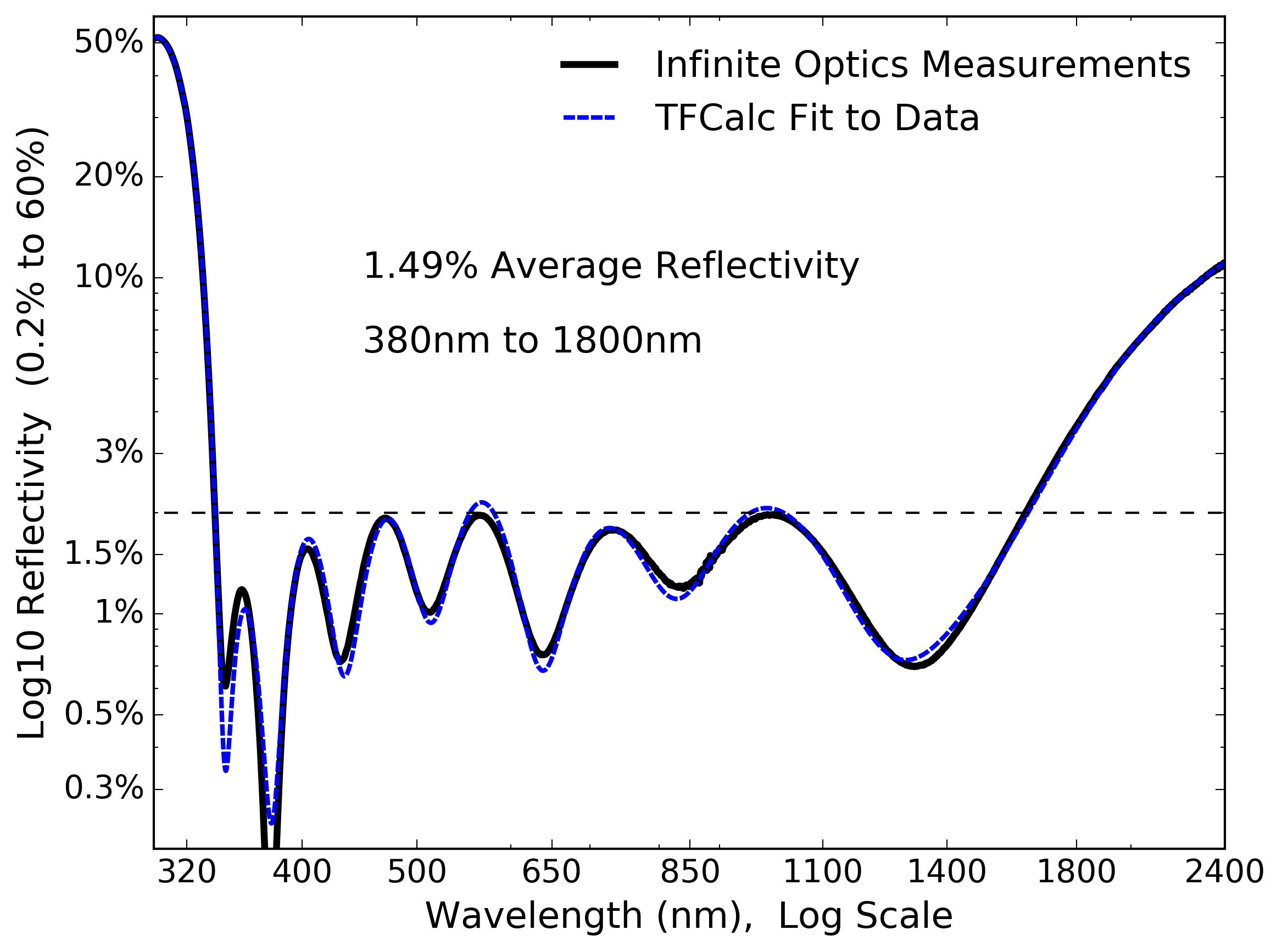}
}
\end{tabular}
\caption[] 
{\label{fig:IO_BBAR_Wide_BS1}  The Infinite Optics broad-band anti-reflection coating on a Heraeus Infrasil 301 fused silica substrate. Black shows spectrophotometric measurements from Infinite Optics in unpolarized light. The dashed blue curve shows a TFCalc model where all layers were adjusted in thickness to fit the transmission model to the measurements. An incidence angle of 8$^\circ$ was used in the model and for the data collection. }
\vspace{-4mm}
 \end{wrapfigure}

Anti-reflection coatings applied to various windows and beam splitters have potential for polarization impact on the DKIST optical train. The nominal suite of visible-light instruments fed by the AO system and dichroics includes paths with between one and four beam splitter transmissions. The behavior of any anti-reflection coating will thus be multiplied by one to more than five surfaces. As these optics must cover our entire instrument suite, the wavelength range and performance requirements are stringent. 

Infinite Optics, Inc designed a wide wavelength range anti-reflection coating to cover the 380 nm to 1800 nm wavelength region.  The nominal design includes a thin strippable layer, then fourteen layers oscillating between SiO$_2$ and HfO$_2$ of roughly 350 nm total physical thickness for each material at very roughly 50 nm per layer.  The final outer layer is roughly 130 nm of MgF$_2$. The coating is 0.86 $\mu$m total physical thickness and 16 layers.  We call this coating {\bf WBBAR1}. Several tests to date are listed in Table \ref{table:wbbar1_samples} showing chamber and run number along with the optic coated. This WBBAR1 formula is essentially a dichroic coating with high UV reflectance and good visible to near infrared transmission.  We also have a slightly modified design optimized for the 620 nm to 1800 nm wavelength range called {\bf WBBAR2} that will be used on some of the dichroics described later that work in transmission only at longer wavelength ranges. All instruments post-AO are fed in transmission through the wavefront sensor beam splitter WFS-BS1 at 15$^\circ$ incidence angle. The wedged FIDO dichroic beam splitters also have an anti-reflection coating on the back surfaces.

\begin{wraptable}{l}{0.28\textwidth}
\vspace{-4mm}
\caption{WBBAR1 Samples}
\label{table:wbbar1_samples}
\centering
\begin{tabular}{l l}
\hline\hline
Run		& Description 	\\
\hline
\hline
7-4246	& Prelim Test 	\\
10-0095	& Final Test 	\\
12-6267   	& PA\&C Win S1 \\
12-6268   	& PA\&C Win S2 \\
67 \& 67  	& Infrasil S1 \& S2 \\
10-0231	& WFC-BS1 Test	\\
10-0233	& WFC-BS1	\\
\hline
\hline
\end{tabular}
\vspace{-5mm}
\end{wraptable}

Using TFCalc, we can adjust the individual layer thicknesses to fit spectrophotometric data allowing us to create an as-built coating design. Figure \ref{fig:IO_BBAR_Wide_BS1} shows an example.  The black curve shows spectrophotometry from Infinite Optics on a coating test run. The baseline TFCalc design model includes four thin layers that are important to achieve performance but increase the design sensitivity to manufacturing tolerances. The dashed blue curve shows a best-fit TFCalc model to the as-built spectrophotometric measurements. Note that in this model we used refractive index data provided by Infinite Optics. Only the material thickness was allowed to vary in the fit, not the refractive index data for each material.   

\begin{figure}[htbp]
\begin{center}
\vspace{-1mm}
\hbox{
\hspace{-0.8em}
\includegraphics[height=6.4cm, angle=0]{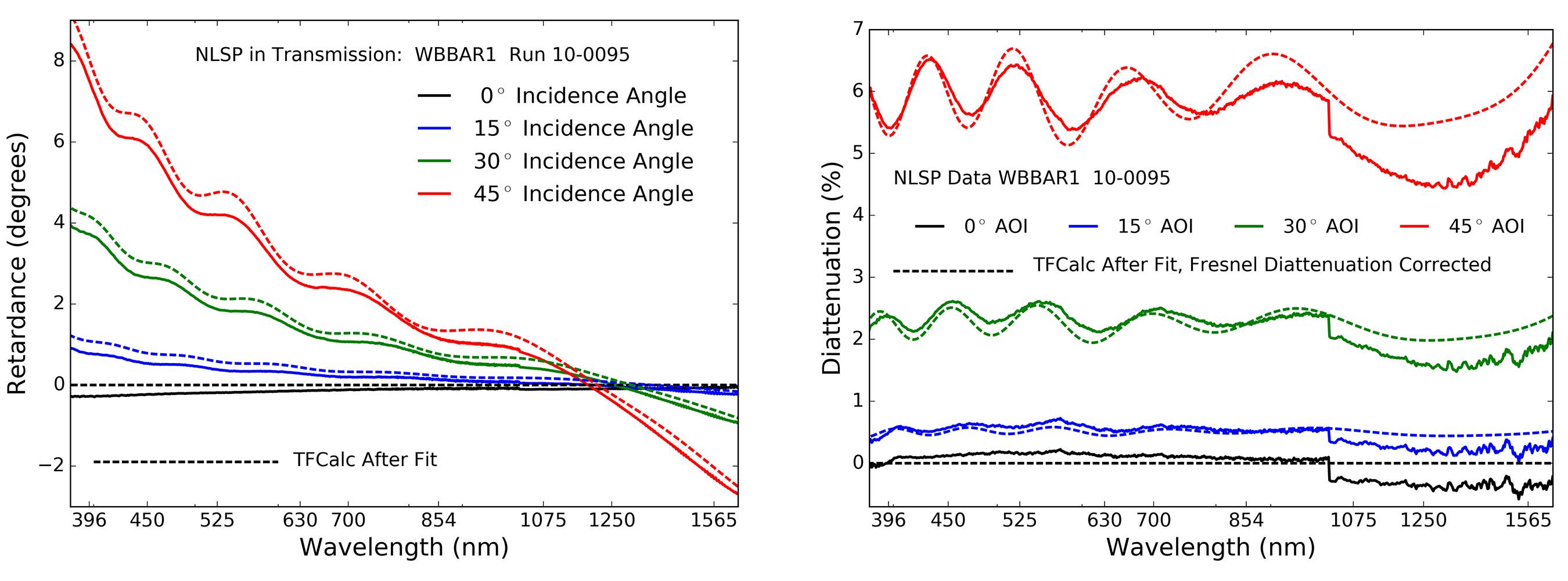}
}
\caption[] 
{ \label{fig:IO_BBAR_Transmission_Retardance_Diattenuation} Transmission retardance (left) and diattenuation (right) of Infinite Optics WBBAR1 sample 10-0095. Measurements are solid lines and are compared to TFCalc model predictions as dashed lines. TFCalc models were created from the nominal design file modified by fitting the IOI unpolarized transmission spectrum at AOI=8$^\circ$.   }
\vspace{-6mm}
\end{center}
\end{figure}

The TFCalc polarization predictions are compared to NLSP measurements in Figure \ref{fig:IO_BBAR_Transmission_Retardance_Diattenuation}. The sample from chamber 10, run 0095 was measured with NLSP in transmission through a fused silica witness sample at 0$^\circ$, 15$^\circ$, 30$^\circ$ and 45$^\circ$ incidence angle.  The retardance data is shown on the left while diattenuation data is seen on the right.  The retardance data is smooth with magnitudes of less than 1$^\circ$ at 15$^\circ$ incidence angles, as will be used in the DKIST optics WFS-BS1 and the FIDO beam splitters.  This retardance is nearly negligible. Similarly, the diattenuation is less than 0.6\% at 15$^\circ$ incidence angles. In the diattenuation data, the step in the measurements occurs when data between the visible (VIS) and near infrared (NIR) spectrographs is spliced together at 1020 nm wavelength.  For the diattenuation data set, we had to account for the additional diattenuation of the Fresnel reflection off the uncoated back surface of the substrate.  This adds roughly 4.5\% diattenuation at 45$^\circ$ incidence angle.

\subsection{Anti-reflection \& dichroic summary: Fitting Many-layer TFCalc Models}
\label{sec:wbbar1_dichroic_summary}

In this section we showed how simple anti-reflection coating designs can be adjusted to fit the as-built layer thicknesses by using unpolarized transmission spectra. With these adjusted models, we are able to then predict transmission, diattenuation and retardance that closely matches our NLSP measurements and vendor metrology.  A wide-wavelength range anti-reflection coating design with 14 layers of SiO$_2$ and HfO$_2$ and two additional layers (top and bottom) was fit to unpolarized transmission data. Subsequent predictions were made for NLSP retardance and diattenuation with agreement better than 1$^\circ$ retardance and a fraction of a percent diattenuation. Errors increased with higher incidence angles due to the exacerbated optical misalignments in NLSP caused by the tilted glass substrate. In Appendix \ref{sec:BBAR_Coatings}, we show coating repeatability and measurements of 1-side and 2-side coated samples. These samples are identical in coating design to three DKIST windows within FIDO called Coud\'{e} Windows 1 2 and 3 (e.g. denoted C-W1).  The AO wave front beam splitter (WFS-BS1) is only back-side coated with WBBAR1 with FIDO C-W2 being delivered similarly. The FIDO C-W1 and C-W3 will be both-side coated similar to the two-side coated windows in the DKIST calibration optic (CalPol2 window). Now that we have successfully shown this example of repeatably manufacturing and fitting relatively simple, thin coatings in transmission, we move on to the much thicker FIDO dichroic coatings working both in transmission and in reflection.  \\

\clearpage

\section{Dichroic Coatings: Polarization Performance \& FIDO Designs}
\label{sec:dichroics}

\begin{wraptable}{r}{0.34\textwidth}
\vspace{-4mm}
\caption{Dichroic Specs}
\label{table:dichroic_spec_list}
\centering
\begin{tabular}{l l l l}
\hline \hline
Clear aperture 290 mm		\\
Angle of Incidence 15$^\circ$	\\
Coating thk. $\pm$1\% spatial var. \\
Diattenuation $<$ 2\% \\
Wedge angle 0.5$^\circ$ $\pm$ 1 arcsec \\
10.5 nm RMS refl WFE, pwr rem \\
48 nm RMS refl WFE power \\
7.5nm RMS trans WFE \\
\hline
\hline
\end{tabular}
\vspace{-3mm}
\end{wraptable}

The Facility Instrument Distribution Optics (FIDO) contain a set of interchangeable mirrors, dichroic beam splitters and windows used to send various wavelengths to the suite of DKIST post-AO coud\'{e} instruments. The optics are mounted in the collimated beam after the adaptive optics system and require a 290 mm clear aperture to accommodate the diverging 2.83 arc minute field of view at station CL4 with a tolerance. 

The optics can cause non-common path wavefront errors as they are mounted in a collimated beam after the wavefront sensor and near a pupil. The wavefront qualities are critical to delivering diffraction limited performance for each instrument and stress from coatings is a major design consideration. The substrates are Heraeus Infrasil 302 at 43 mm thickness. 

\begin{wraptable}{l}{0.34\textwidth}
\vspace{-3mm}
\caption{Dichroic Samples}
\label{table:dichroic_sample_list}
\centering
\begin{tabular}{l l l l}
\hline
\hline
Name		& Lyr		& Thk	& Run		\\
			& 		& $\mu$m	& Nmbr		\\
\hline \hline
C-BS-465		& 25		& 1.5		& TBD		\\
WBBAR1		& 16		& 0.9		& TBD		\\
\hline
C-BS-555		& 48		& 3.1		& TBD		\\
WBBAR1		& 16		& 0.9		& TBD		\\
\hline
C-BS-643		& 52		& 3.5		& TBD		\\
WBBAR2		& 10		& 0.8		& TBD		\\
\hline
C-BS-680		& 52		& 3.8		& TBD		\\
WBBAR2		& 10		& 0.8		& TBD		\\
\hline
C-BS-950		& 96		& 8.8		& TBD		\\
WBBAR2		& 10		& 0.8		& TBD		\\
\hline
\hline
Dich. A		& 61		& 4.1		& 8-5652		\\
Dich. B		& 84		& 6.9		& 7-2802		\\
Dich. C		& 21		& 1.0		& lpw1-400	\\
\hline
\hline
\end{tabular}
\vspace{-3mm}
\end{wraptable}

The optical specifications are quite demanding with significant impact on the allowable coatings. Table \ref{table:dichroic_spec_list} shows some of the highlights. These are large parts that must be interchangeable without disturbing other optics.  As such, the wavefront error in both transmission and reflection must be incredibly flat after coating, including power in the transmitted wavefront. Stress in the coating must be compensated by pre-dishing the Infrasil substrates so coatings must be both low stress and repeatable in stress and WFE to ensure each dichroic beam splitter is interchangeable. The wedge angle of each substrate must also be identical to better than one arcsecond for a wedge of half degree.  As most coatings are highly reflective at 633 nm, testing must be done for transmission at long wavelengths. The guaranteed intrinsic stress birefringence of $<$5 nm per cm of optical path at 633 nm wavelength ensures that there is minimal variable wavefront error and retardance.  As the optics are near a pupil plane, birefringence will spatially average significantly and the resulting mild depolarization is not a concern.  

The optics are at 15$^\circ$ incidence angle which does create polarization through the complex dichroic coatings on the front surfaces and also the broad-band anti-reflection coatings on the back surfaces. We modeled polarization fringes in two recent papers \cite{Harrington:2018cx,2017JATIS...3d8001H} and do not expect to observe significant fringing given the 0.5$^\circ$ wedge angle in each of the FIDO beam splitters.

\begin{wrapfigure}{r}{0.50\textwidth}
\centering
\vspace{-4mm}
\begin{tabular}{c} 
\hbox{
\hspace{-1.0em}
\includegraphics[height=6.45cm, angle=0]{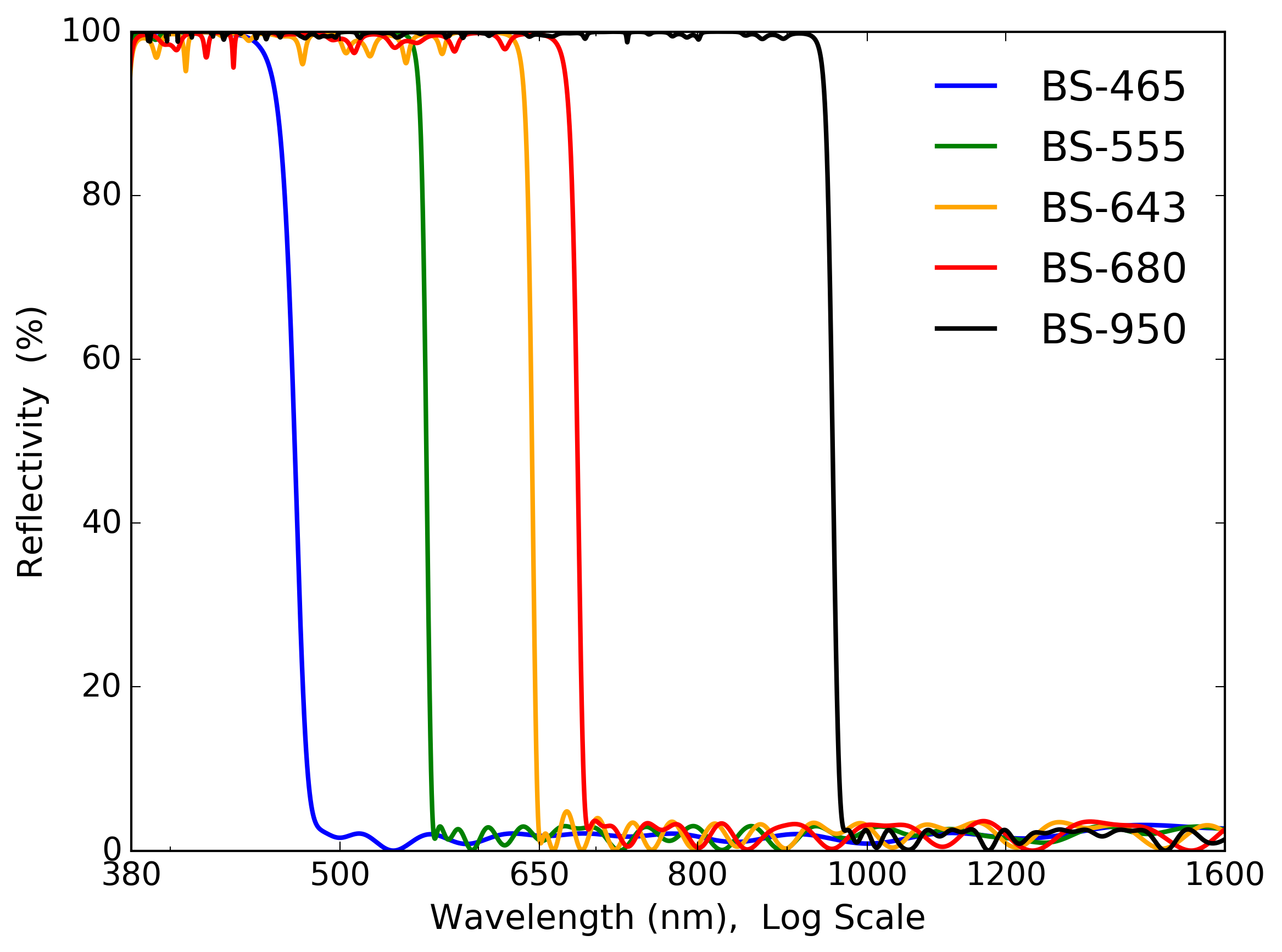}
}
\end{tabular}
\caption[] 
{\label{fig:FIDO_AOI15_reflectivity} Design reflectivity for the FIDO optics.  }
\vspace{-4mm}
 \end{wrapfigure}

There are presently five dichroics planned for fabrication as shown in Table \ref{table:dichroic_sample_list}. The naming convention denotes the wavelength where the beam splitter switches from reflective to transmissive with a 50\% value. As an example, the Coud\'{e} Beam Splitter reflecting wavelengths short of 465 nm while transmitting wavelengths longer than 465 nm is denoted C-BS-465. The dichroic suite is thus 465, 555, 643, 680 and 950.  We also show in Table \ref{table:dichroic_sample_list}, the appropriate anti-reflection coating intended for the back surface. This will either be WBBAR1 described above or the related coating WBBAR2.  

We are in the process of verifying the coating performance, repeatability, spatial uniformity, stress, wavefront error, etc so we list TBD in Table \ref{table:dichroic_sample_list}.  We also list three dichroic samples used in early evaluation and design from Infinite Optics that we label Dich. A, B and C.

We show example reflectivity curves of the planned dichroic coatings at 15$^\circ$ incidence angle in Figure \ref{fig:FIDO_AOI15_reflectivity}.  The designs have fairly sharp transitions with most coatings switching from 80\% reflective to 80\% transmissive in less than 10 nm wavelength. We chose this style of coating as they are low stress evaporative coatings with reasonable repeatability and achievable 1\% physical thickness uniformity across the entire aperture.

\begin{figure}[htbp]
\begin{center}
\vspace{-1mm}
\hbox{
\hspace{-0.6em}
\includegraphics[height=6.3cm, angle=0]{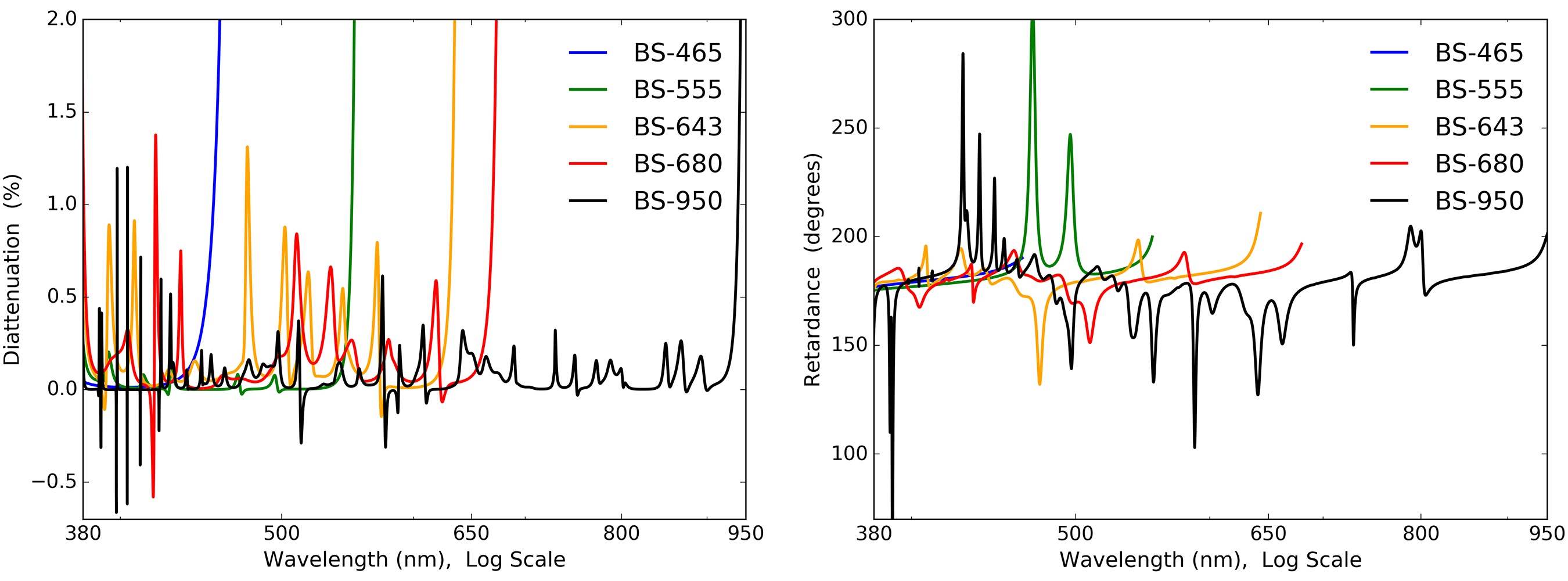}
}
\caption[] 
{ \label{fig:FIDO_AOI15_reflectivity_retardance} The diattenuation (left) and retardance (right) for the FIDO optics from the TFCalc design for the beam reflected at 15$^\circ$ incidence angle. Designs were constrainted to have diattenuation below 2\%.  }
\vspace{-5mm}
\end{center}
\end{figure}

The retardance of the reflected beam at 15$^\circ$ incidence angle is shown in the right hand graphic of Figure \ref{fig:FIDO_AOI15_reflectivity_retardance}. The thickest coating is C-BS-950 with 96 layers and an 8.8 $\mu$m physical thickness. There are strong spectral variations expected in retardance curves for dichroics with tens of layers. The theoretical retardance derivative has several bandpasses with gradients of 10$^\circ$ per nanometer wavelength with a few specific narrow features up to magnitudes of 100$^\circ$ retardance per nm wavelength. We show in this section that these rapid swings in retardance are observable and can also be mitigated by design. Similarly, these narrow spectral features in dichroic coatings are also coincident with significant diattenuation as well as strong changes in transmission.  All of these narrow spectral features of dichroics have impact for DKIST calibration as we can possibly anticipate strong spectral changes across the $\sim$nm instrument bandpasses in the modulation matrix through retardance and diattenuation, in addition to substantial throughput changes.  Through designing coatings with minimal diattenuation, we were able to achieve retardance values as in Figure \ref{fig:FIDO_AOI15_reflectivity_retardance} without many waves wrapping and excessively sharp spectral features.

Though the retardance curves appear to have many {\it spikes}, there are no multiple-wave wrapping wavelengths, reflectivity is above 96\% for all wavelengths in the reflection band, and the reflected diattenuation values are all below 2\% for any wavelength in reflection.  We show diattenuation for the reflected beam in each design in the left hand graphic of Figure \ref{fig:FIDO_AOI15_reflectivity_retardance}.  Values are always well below 2\% magnitude though there are some narrow spectral features that are sensitive to coating manufacturing tolerances.

The FIDO beamsplitter coating designs are currently undergoing a significant uniformity, stress and polarization testing process. As part of DKIST systems engineering, we needed to verify the spectral predictions of TFCalc against polarization measurements for various coating samples simultaneous with repeatability of coating stress, wavefront error and other relevant performance parameters. Here we present testing of Infinite Optics dichroic samples as well as some preliminary design predictions for DKIST.

\clearpage

\subsection{FIDO Dichroic C-BS-465: Coating Spatial Uniformity \& Model Fitting}
\label{sec:FIDO_spatial_variation}
In this section, we make a detailed analysis of the thinnest FIDO dichroic coating.  We assess two separate coating shots for spatial uniformity, spectral performance and variability of the individual coating layers. This BS-465 design uses 24 layers, has a 1502 nm physical thickness with the thinnest layer at 17.0 nm in the design. The coating follows a common design with a strippable layer as the base then alternating SiO$_2$ and TiO$_2$ layers.  A thicker SiO$_2$ outer layer is the air interface.  

\begin{wraptable}{l}{0.54\textwidth}
\vspace{-3mm}
\caption{Dichroic Coating 465 Testing}
\label{table:dichroic_465_test_list}
\centering
\begin{tabular}{l l l }
\hline
\hline
Name		& Run		& Meas		\\
\hline
\hline
Test1		& 10-0150		& S\&P 15$^\circ$, Unif 0$^\circ$ 9 samp.		\\
Test2		& 10-0153		& S\&P 15$^\circ$, Unif 0$^\circ$ 9 samp.		\\
Filter1		& 10-0154		& \%T 0$^\circ$ Design: (HL)$^3$ 2H (LH)$^3$   \\
Filter2		& 10-0156		& \%T 0$^\circ$ Design: (HL)$^3$ 2H (LH)$^3$   \\
\hline
\hline
\end{tabular}
\vspace{-3mm}
\end{wraptable}

\begin{figure}[htbp]
\begin{center}
\vspace{-0mm}
\hbox{
\hspace{-0.6em}
\includegraphics[height=6.3cm, angle=0]{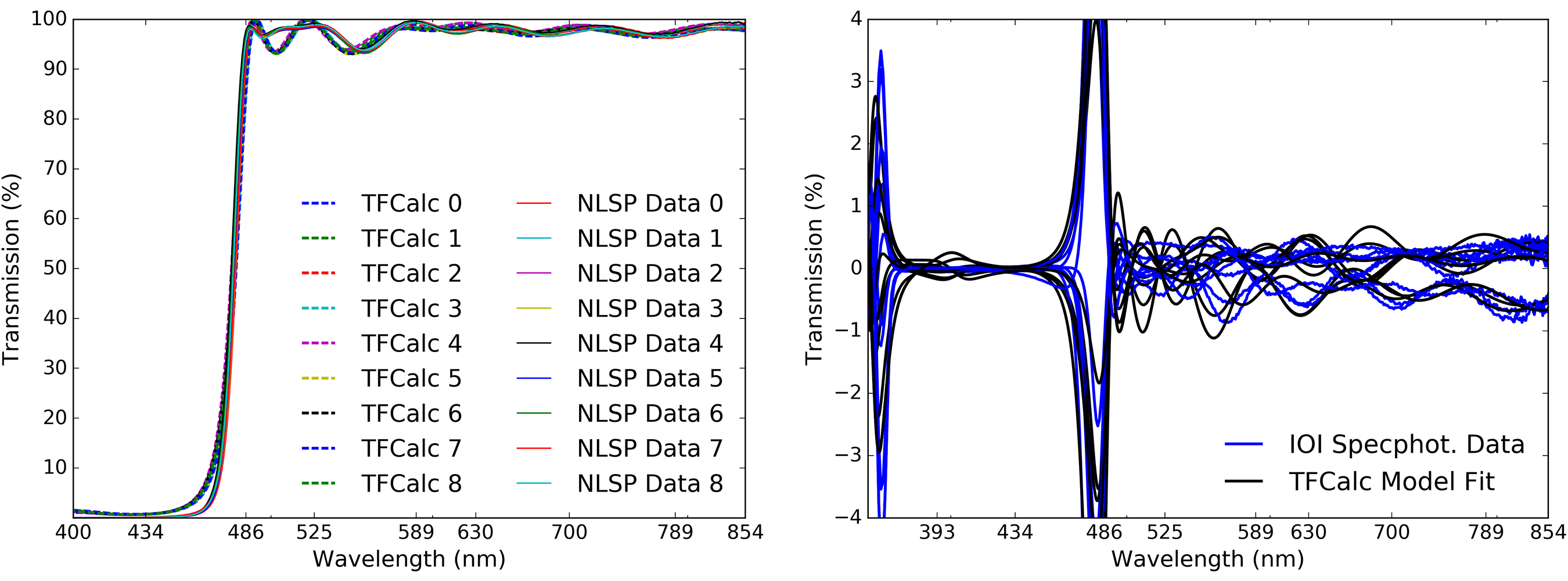}
}
\caption[] 
{\label{fig:dichroic_465_thickness_variation_errors} The left graphic show TFCalc models after fitting the IOI measurements, along with the uniformity sample measurements. The right graphic shows the differences between the average and each individual curve to highlight spatial variation. Blue shows spatial variation in measurements while black shows spatial variation derived from the TFCalc models. Dichroic 465 test run 10-1053 had transmission measured on 9 witness samples at AOI=0$^\circ$. Transmission curves are adjusted for a theoretical $\sim$3.8\% reflection loss from the uncoated sample back surface. }
\vspace{-6mm}
\end{center}
\end{figure}

Table \ref{table:dichroic_465_test_list} shows the test runs for this dichroic coating.  We received uniformity measurements at 15$^\circ$ incidence angle on 9 samples coated in run 10-0153 around April 21, 2018.  The transmission and reflection measured by IOI with their spectrograph is shown in the left hand graphic Figure \ref{fig:dichroic_465_thickness_variation_errors} as the solid lines. We then used this spectral data and selected wavelengths from 350 nm to 1200 nm to perform coating model fitting in TFCalc. 

After this uniformity testing, additional tests of the chamber for spatial uniformity and refractive index of the deposited materials were done. The tests used a standard 13-layer narrow band filter design (HL)$^3$ 2H (LH)$^3$. Table \ref{table:dichroic_465_test_list} shows these two tests in the same chamber (10) but with coating shot 0154 immediately after the dichroic and another test six days later in shot 0156.  With the nominal wavelength set around 480 nm for this test, the quarter-wave optical thickness is 82 nm for the Low index material SiO$_2$ and 55 nm for the High index material TiO$_2$. The variance of the filter central wavelength was used as a statistical measure of the layer by layer variation, showing that we did indeed pass a 1\% variation of the physical thickness across the aperture.  With these updated refractive indices in hand, we could fit the TFCalc models to the C-BS-465 dichroic sample data. The left hand graphic of Figure \ref{fig:dichroic_465_thickness_variation_errors} shows the best fit TFCalc models as dashed lines. Some of the spectral oscillations are not well fit, but the transition wavelengths and general behavior are reproduced.  We can expect sptaial variation in both reflection and transmission across the beam following the magnitudes shown in Figure \ref{fig:dichroic_465_thickness_variation_errors}.

The right hand graphic of Figure \ref{fig:dichroic_465_thickness_variation_errors} shows the variation between the mean and the nine individual spatial samples.  The TFCalc model variation is shown in black and measurement residual errors are shown in blue.  As expected, there are larger errors near wavelengths where spectral gradients are strongest. We also do not reproduce a somewhat larger spectral oscillation around 475 nm wavelength with a depth of a few percent. Note that we have adjusted the transmission to account for the $\sim$3.8\% Fresnel reflection loss from the uncoated sample back surface using the Fresnel equations and the refractive index data for Heraeus Infrasil provided by the manufacturer.

\begin{figure}[htbp]
\begin{center}
\vspace{-2mm}
\hbox{
\hspace{-0.8em}
\includegraphics[height=6.3cm, angle=0]{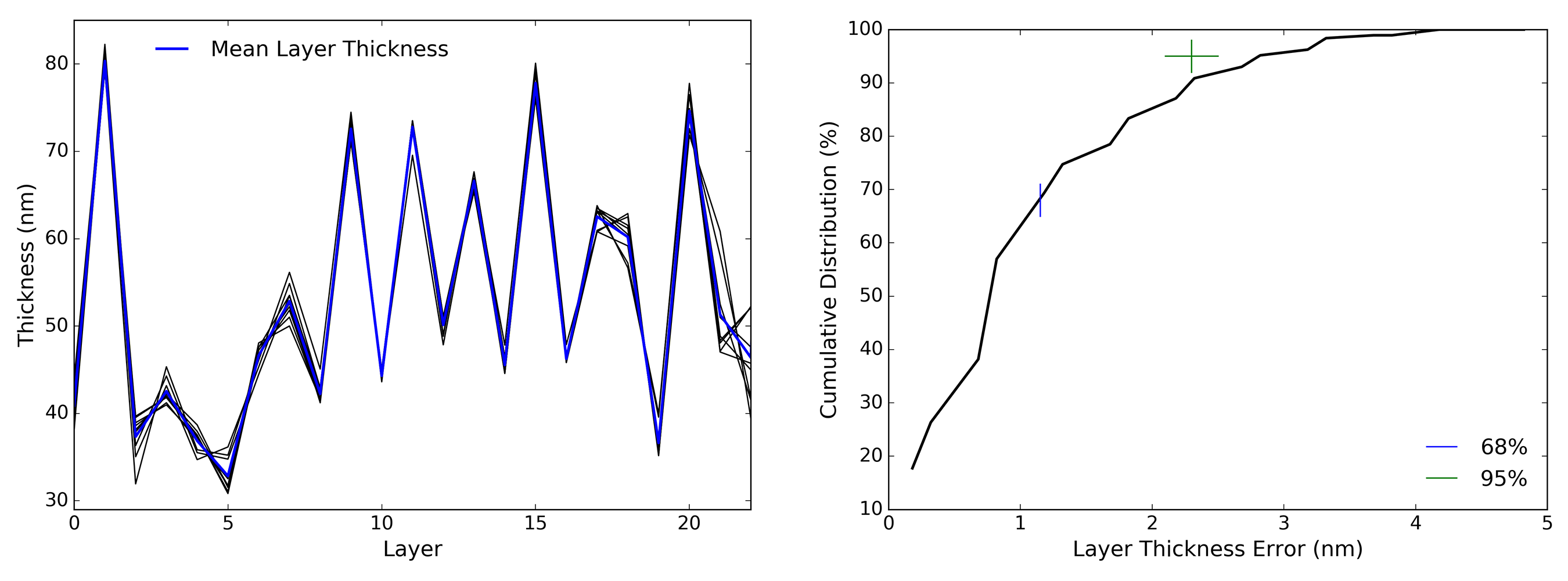}
}
\caption[] 
{\label{fig:dichroic_465_thickness_variation_TFCalc_Fitting} TFCalc fits to the nine uniformity sample measurements for Dichroic 465 test run 10-1053.  Left shows all nine of the TFCalc models.  Some layers are computed to vary more than others, especially the outer layer.  The right graphic shows the cumulative distribution of errors for each layer in each model with respect to the average. }
\vspace{-7mm}
\end{center}
\end{figure}

The statistics of layer variation in these coatings show that we pass a 1\% physical coating thickness variation across the clear aperture, and that the variations essentially follow Gaussian statistics across the aperture. In the left hand graphic of Figure \ref{fig:dichroic_465_thickness_variation_TFCalc_Fitting} we show the variation between layers in the TFCalc fitting process. The baseline design thickness is shown as the blue curve with each layer of alternating SiO$_2$ and TiO$_2$ having thicknesses between roughly 30 nm and 80 nm. Each of the 9 colored curves shows a fit to the individual spectra.

\begin{figure}[htbp]
\begin{center}
\vspace{-0mm}
\hbox{
\hspace{-0.8em}
\includegraphics[height=6.3cm, angle=0]{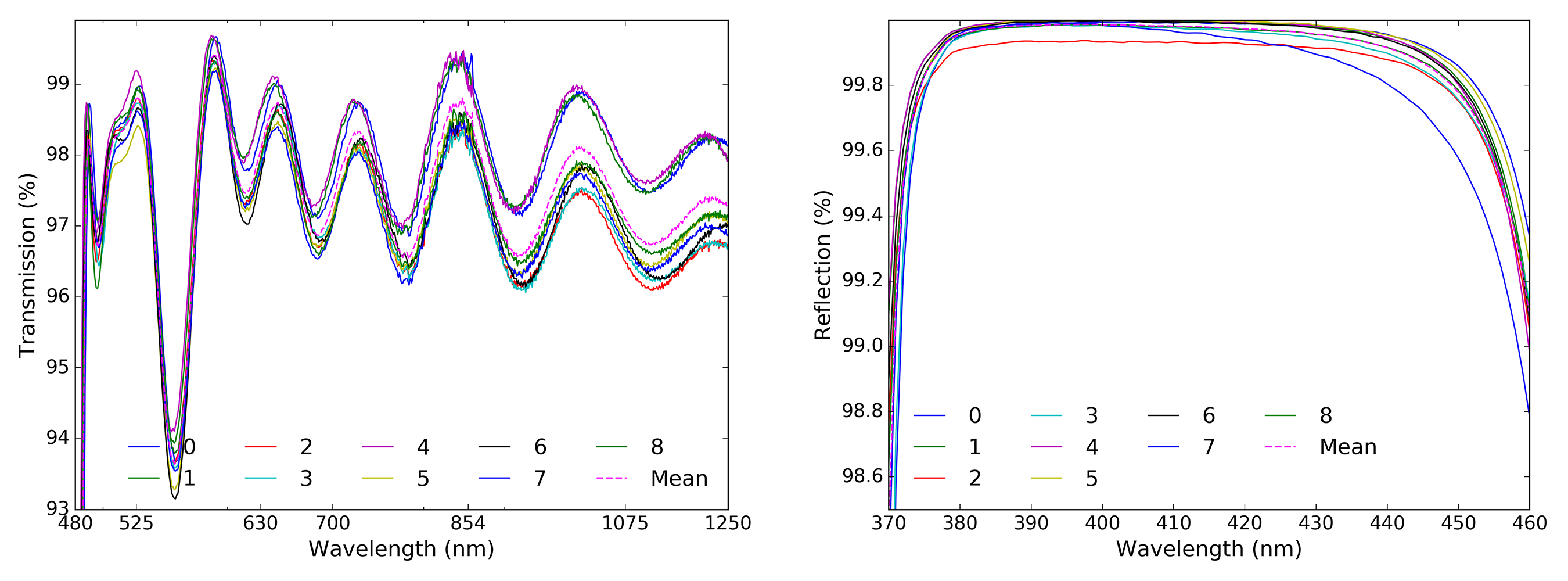}
}
\caption[] 
{\label{fig:dichroic_465_refl_trans_specific_regions}  The transmission bandpass (left) and reflection bandpasses (right) for IOI measurements of the 9 samples throughout the chamber for the C-BS-465 dichroic coating uniformity test run 10-1053 at AOI=0$^\circ$.}
\vspace{-6mm}
\end{center}
\end{figure}

In the right hand graphic of Figure \ref{fig:dichroic_465_thickness_variation_TFCalc_Fitting} we show the statistics of the variation between layers. The cumulative distribution of errors in each individual layer against the nominal design is shown as the solid black line. We did not see any evidence of any layer being particularly thicker or thinner as a function of aperture radius, or layer depth. A few layers show more variation than others, but the statistics are essentially the same as other layers with radius and depth. The crosses show 1-sigma 68\% and 2-sigma 95\% errors for a Gaussian distribution fit to the histogram.  More than 68\% of the layers are within 1.15 nm layer thickness of the average.  For 95\% of the points, the layer variation is less than 2.9 nm.

Figure \ref{fig:dichroic_465_refl_trans_specific_regions} shows the transmission and reflection data in the appropriate bandpass intended for feeding the DKIST instruments. Reflectivity is over 99\% for the range 380 nm to 450 nm.  Similarly, after compensating for the Fresnel reflection, the transmission is over 96\% for all wavelengths between 480 nm and the long wavelength cutoff of the metrology, except for a small bandpass around 550 nm. The mean transmission is over 97\% for this coating.  The actual FIDO dichroic will have the WBBAR1 coating on the back side, minimizing the back surface reflection losses.

\begin{wrapfigure}{r}{0.53\textwidth}
\centering
\vspace{-4mm}
\begin{tabular}{c} 
\hbox{
\hspace{-1.1em}
\includegraphics[height=6.6cm, angle=0]{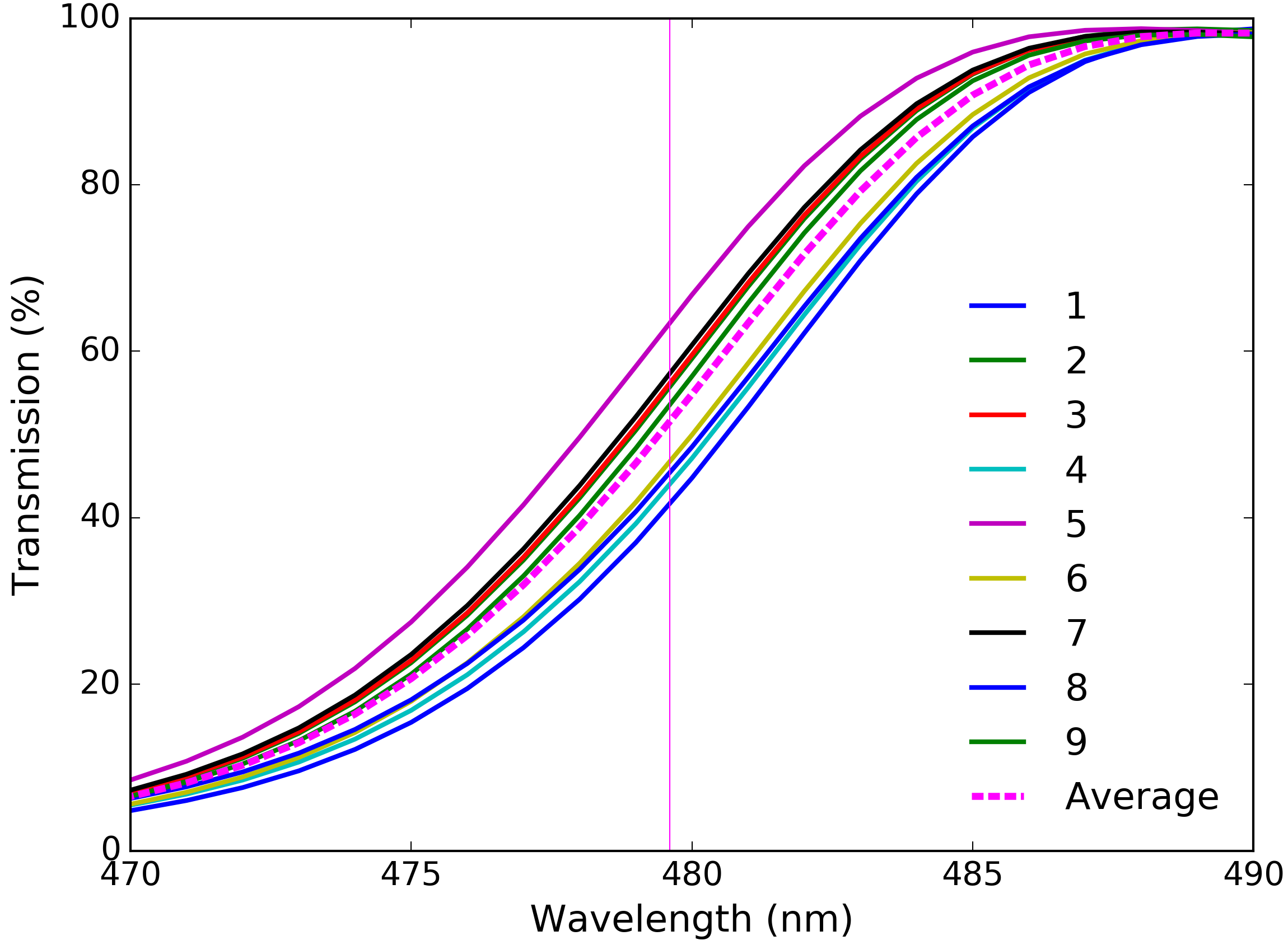}
}
\end{tabular}
\caption[] 
{\label{fig:dichroic_465_transmission_test1_aoi0}  Transmission measurements at AOI=0$^\circ$ of the 9 uniformity samples for C-BS-465 dichroic run 10-1053.  }
\vspace{-2mm}
 \end{wrapfigure}

Figure \ref{fig:dichroic_465_transmission_test1_aoi0} shows the IOI transmission data around the transition wavelength at normal incidence (0$^\circ$) corrected for the back surface Fresnel losses. The transition wavelength between reflection and transmission is slightly longer than the nominal 50\% transmission at 465 nm wavelength.  For this design there is also roughly 5 nm wavelength shift to the blue when used at the nominal 15$^\circ$ FIDO orientation. The transition wavelength is noted by the straight black line 479.3 nm along with a spread of $\pm$1.3 nm wavelength at 50\% transmission across the aperture. This 20\% transmission spatial variation across the aperture would raise calibration concerns if using this optic at the transition wavelength.  However, the nominal dichroic coating specification of reflectivity $>$90\% for wavelengths short of 440 nm and $>$90\% transmissive for wavelengths longer than 490 nm is easily met with this test. The design is also easily adjusted to slightly shorter transition wavelength following these test results.

\begin{figure}[htbp]
\begin{center}
\vspace{-1mm}
\hbox{
\hspace{-0.8em}
\includegraphics[height=6.3cm, angle=0]{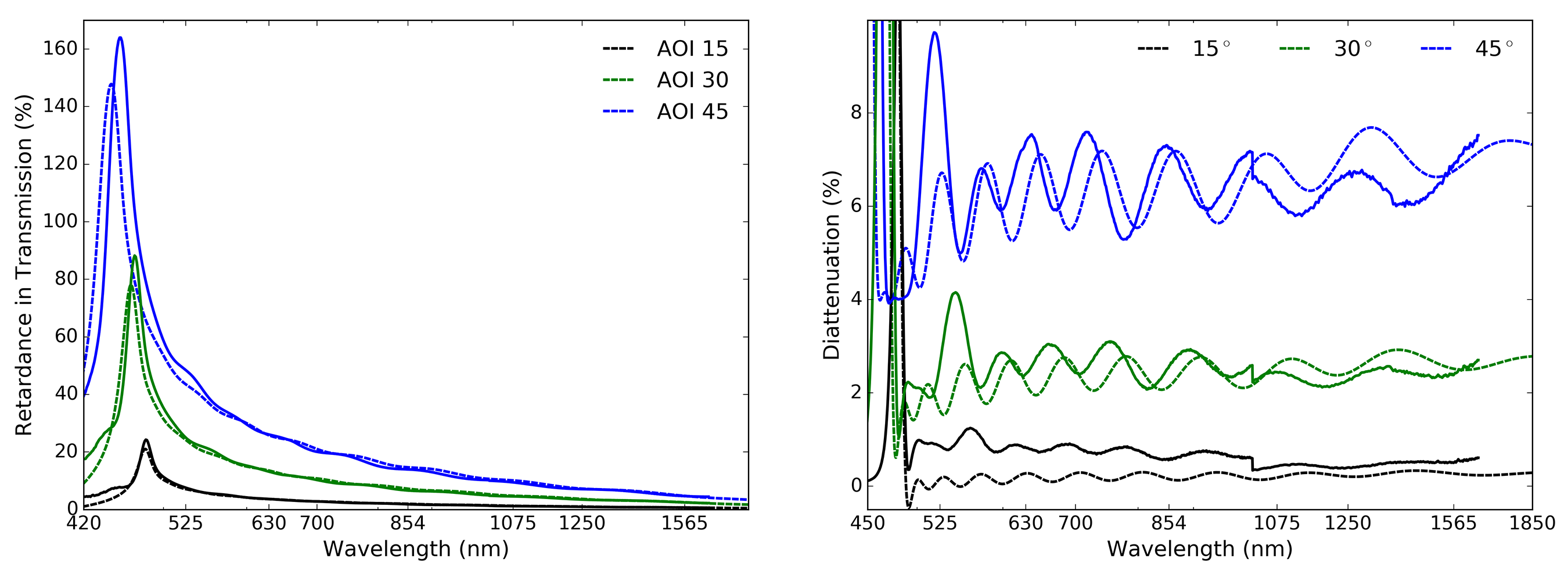}
}
\caption[] 
{ \label{fig:IO_465_Transmission_Retardance_Diattenuation_NLSP} Transmission retardance (left) and diattenuation (right) measured with NLSP at various incidence angles for the C-BS-465 dichroic samples. The solid lines show the data and dashed lines allow comparison with the best-fit TFCalc model predictions. }
\vspace{-5mm}
\end{center}
\end{figure}

We received all 9 samples and performed testing in NLSP for polarization. Figure \ref{fig:IO_465_Transmission_Retardance_Diattenuation_NLSP} shows the NLSP-measured retardance and diattenuation as a function of incidence angle. We tilted the sample between 0$^\circ$ and 45$^\circ$. The retardance has strong spectral gradients near the transition wavelengths, as expected.  Diattenuation is spectrally stable in transmission with mild oscillation in the transition band.  For the beam at 15$^\circ$ incidence, the diattenuation is less than 1\%.

We note that the NLSP reflective arm currently only measures samples at 45$^\circ$ incidence so we do not have a direct measurement of polarization performance at the FIDO incidence angle of 15$^\circ$.  The TFCalc models closely match the predictions, and this dichroic has essentially negligible polarization impact in the reflection band for the reflected beam.  Retardance is less than 5$^\circ$ across the entire 380 nm to 460 nm wavelength range.  Diattenuation is less than 0.2\% across that same range.  The IOI measurements of reflected diattenuation in this region are also consistent with zero given their measurement uncertainty.  This will of course not be true for the more complex dichroics discussed later, but we conclude here that we have a valid model supported by several data sets for this simple FIDO dichroic.

\subsection{Infinite Optics Dichroic A:  Reflection 380 nm to 580 nm}
\label{sec:sub_IO_dichroicA}

When fitting significantly more complex coating designs, there is significant degeneracy. A completely free fit could adjust the thickness of every layer and the refractive index of each material to match the data. Fitting every individual spectral oscillation is usually quite difficult, resulting in unrealistic layer thickness and fits that are not close to the actual deposited coating. Often, there are direct measurements and estimates of layer thicknesses recorded during the coating process providing the ability to constrain the fit.  We demonstrate some constrained fitting here.   

\begin{wrapfigure}{l}{0.55\textwidth}
\centering
\vspace{-3mm}
\begin{tabular}{c} 
\hbox{
\hspace{-1.0em}
\includegraphics[height=6.2cm, angle=0]{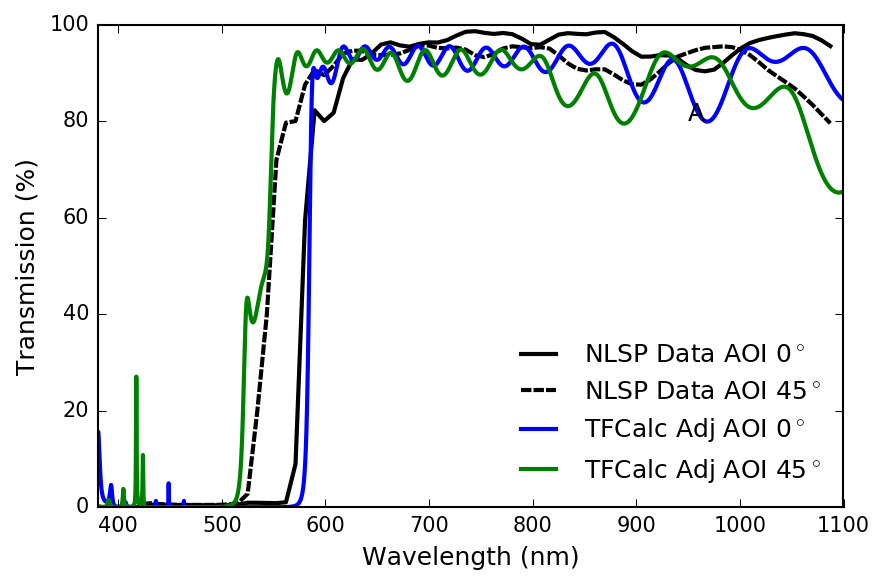}
}
\end{tabular}
\caption[] 
{\label{fig:IO_dichroicA_NLSP_measured_transmission_vs_predicted_AOI}   NLSP transmission spectra are shown in black along with TFCalc design predictions as blue and green for Infinite Optics, Inc. Dichroic A run number 8-5652. }
\vspace{-4mm}
\end{wrapfigure}

We received an Infinite Optics sample from run number 8-5652 we label Dichroic A. This dichroic reflects wavelengths shorter than roughly 580 nm wavelength. This design was {\it not optimized for polarization} or for use at high incidence angles and is expected to have strong polarization artifacts. The design uses 61 layers for a total 4.00$\mu$m thickness. Most of the layers are alternating SiO$_2$ and TiO$_2$ at roughly quarter wave thickness for a 510nm reference wavelength (87nm / 59nm per layer).  We can compare the theoretical TFCalc design files with our NLSP measurements in transmission.  Figure \ref{fig:IO_dichroicA_NLSP_measured_transmission_vs_predicted_AOI} shows the measured $II$ Mueller matrix element from NLSP along with the TFCalc predictions at the appropriate AOI. The two NLSP measurements are shown in black with normal incidence as the solid line and 45$^\circ$ incidence as the dashed line. The blue and green curves show the TFCalc design files for incidence angles 0$^\circ$ and 45$^\circ$ respectively.

We took the TFCalc design and adjusted the transmission to account for the Fresnel reflection off the uncoated back surface. Note that we had lower resolving power of the NLSP setup for these measurements. A 9 nm FWHM instrument profile reduced the magnitude of some of the narrow, low magnitude spectral transmission features seen in the lower left reflection band. It also mildly reduces some of the spectral ripple magnitudes at the shortest wavelengths.  But there are significant amplitude differences and the instrument profile does not reduce the spectral ripple significantly. As seen in Figure \ref{fig:IO_dichroicA_NLSP_measured_transmission_vs_predicted_AOI}, the rough shape of the measured transmission curve is matched by the nominal design as is the 50/50 transition wavelength, but the stack up of thickness variation in 61 coating layers combined with variations in refractive index can cause the detailed spectral dependence to change significantly in spectral ripple as well as transition wavelength.  Fortunately, we can use TFCalc to constrain a fit of the design to the NLSP measured transmission, diattenuation and retardance.  Depending on the limits set for the allowable variation of each layer and on the refractive index values chosen, matching all spectral oscillations in a single transmission spectrum can drive the coating model far away from the actual layer thicknesses estimated by other means.

\begin{figure}[htbp]
\begin{center}
\vspace{-3mm}
\hbox{
\hspace{-0.7em}
\includegraphics[height=6.3cm, angle=0]{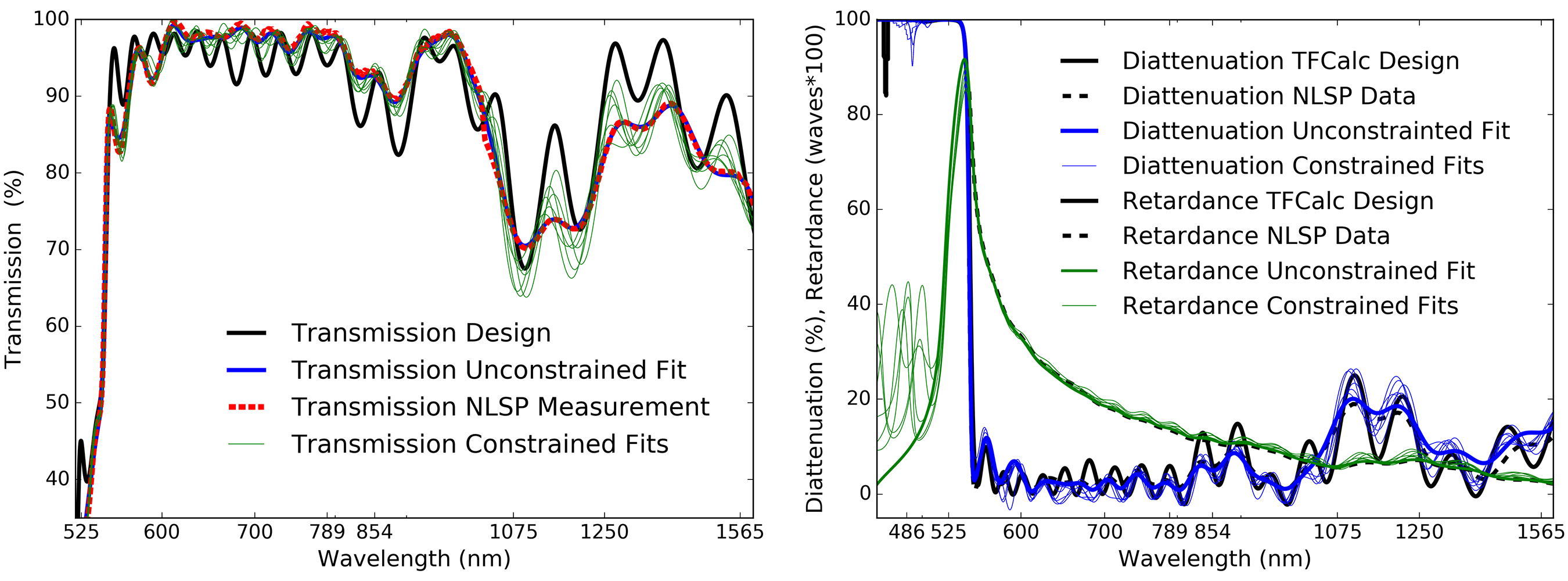}
}
\caption[] 
{ \label{fig:IO_dichroicA_derived_transmission_retardance_diattenuation} Left shows the NLSP measured transmission for the Dichroic A, IO run number 8-5652 as dashed red.  The TFCalc design is solid back, the TFCalc unconstrained fit is solid blue (matching the NLSP data well) and the constrained TFCalc fits are thin green lines.  Right shows the NLSP measured retardance and diattenuation in transmission as black dashed lines with the TFCalc designs as solid lines.  The Y axis is both \% diattenuation and retardance in 100ths of waves.  The various blue lines show TFCalc diattenuation models and green lines show retardance models.    }
\vspace{-8mm}
\end{center}
\end{figure}

An example of the fitting process is shown in Figure \ref{fig:IO_dichroicA_derived_transmission_retardance_diattenuation} for the transmitted beam at an incidence angle of 45$^\circ$. The left graphic shows the measured transmission function along with various fits. The right hand graphic shows retardance in green and measured diattenuation in blue as well as the same series of constrained fits. Each of the individual fitted models is computed with ever wider range allowed for the variation of each layer thickness against the design.  The fits improve to this single spectrum, but some layer thickness become quite different from the estimates derived during the coating process.  The retardance is nearly a full wave in the transition band from short wavelength reflection to long wavelength transmission. The diattenuation is similarly close to 100\%. In the transmission band, the retardance smoothly decays from a full wave around 600 nm wavelength to under 0.1 waves (36$^\circ$) for long wavelengths.  The diattenuation however has a more complex spectral pattern. This filter was nominally designed to have high transmission only around 700 nm wavelength. Diattenuation contains many spectral oscillations of a few percent in the high transmission bandpass but up to 20\% at near infrared wavelengths.  




\subsection{Infinite Optics Dichroic B: Reflection $<$680nm on High Refractive Index Glass}
\label{sec:sub_IO_dichroicB}

\begin{wrapfigure}{l}{0.57\textwidth}
\centering
\vspace{-3mm}
\begin{tabular}{c} 
\hbox{
\hspace{-1.0em}
\includegraphics[height=6.2cm, angle=0]{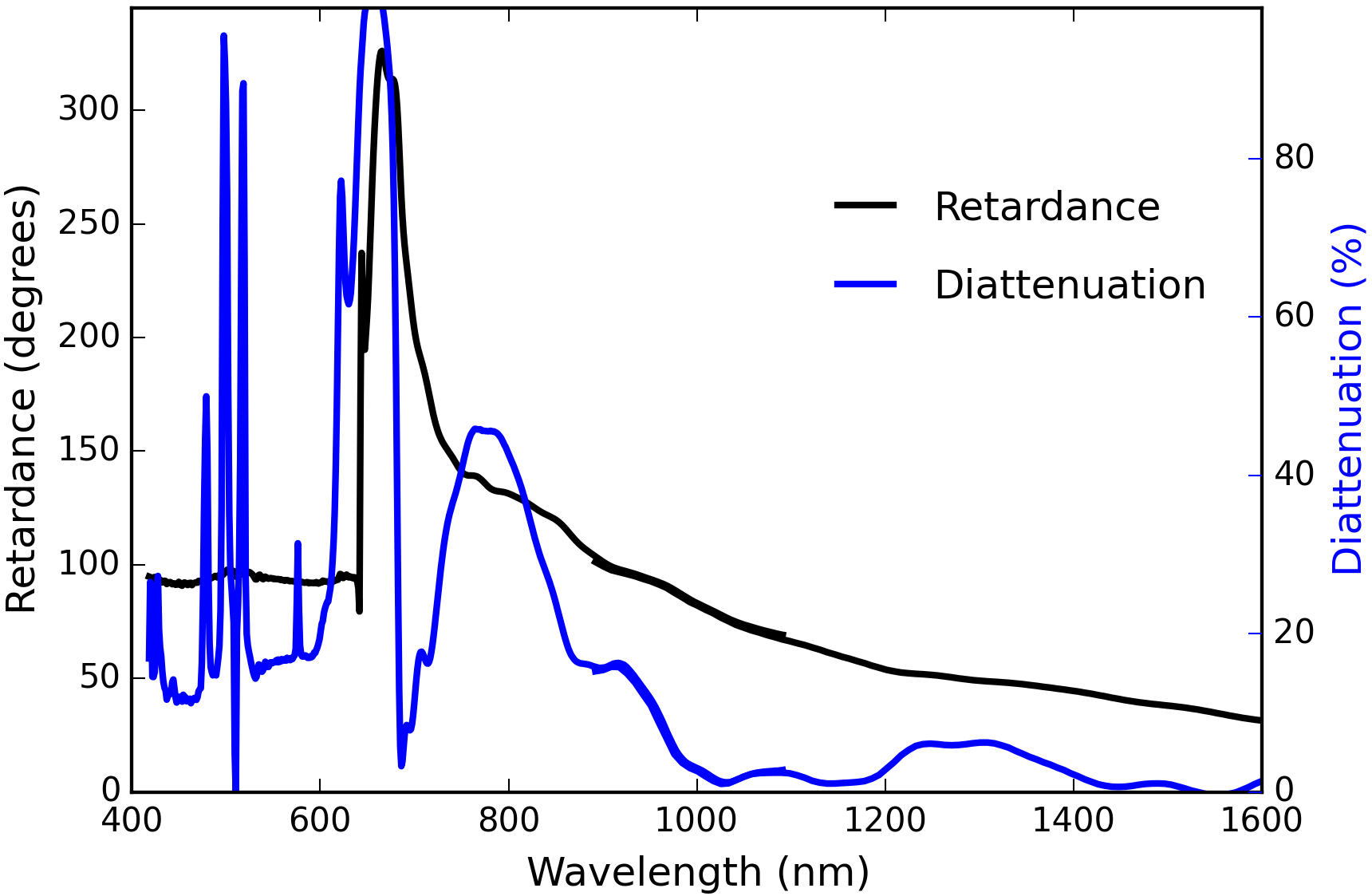}
}
\end{tabular}
\caption[] 
{\label{fig:IO_dichroicB_transmitted_pol}  The NLSP measured polarization properties in transmission for the Infinite Optics Dichroic B sample at 45$^\circ$ incidence with run number 7-2802. Black shows retardance on the left hand y axis.  Blue shows diattenuation on the right hand y axis. }
\vspace{-4mm}
\end{wrapfigure}

We received another Infinite Optics sample from run number 7-2802 we label Dichroic B. This dichroic reflects wavelengths shorter than roughly 680 nm wavelength and is coated on a high refractive index SF11 glass substrate. This design was {\it not optimized for polarization} or for use at high incidence angles and is expected to have strong polarization artifacts and spectrally narrow features. The design uses 84 layers for a total 6.59$\mu$m thickness. Most of the layers are alternating SiO$_2$ and TiO$_2$, the same as the DKIST FIDO dichroic designs. In Figure \ref{fig:IO_dichroicB_transmitted_pol}, we show the polarization properties of dichroic B in transmission at 45$^\circ$ incidence derived from the NLSP measured Mueller matrix. The retardance is shown in black using the left hand y axis running from 0$^\circ$ to just over 300$^\circ$. The blue curve shows diattenuation using the right hand y axis running from 0\% to 100\%. As this dichroic is highly reflective for wavelengths shorter than 680 nm, the transmitted flux is very low. However, we do reproduce stable polarization measurements with NLSP even with transmission less than 1\%. We have typical SNRs over 10,000 for a transmissive optic so we still can achieve SNRs in the range of 1000 with 1\% transmission. There are narrow spectral features in diattenuation of the transmitted beam that are entirely real as we detail in the next section.  As this coating was not optimized for transmission, strong diattenuation is seen for the transmitted beam even in the transmission bandpass.

\begin{wrapfigure}{r}{0.60\textwidth}
\centering
\vspace{-3mm}
\begin{tabular}{c} 
\hbox{
\hspace{-1.0em}
\includegraphics[height=7.7cm, angle=0]{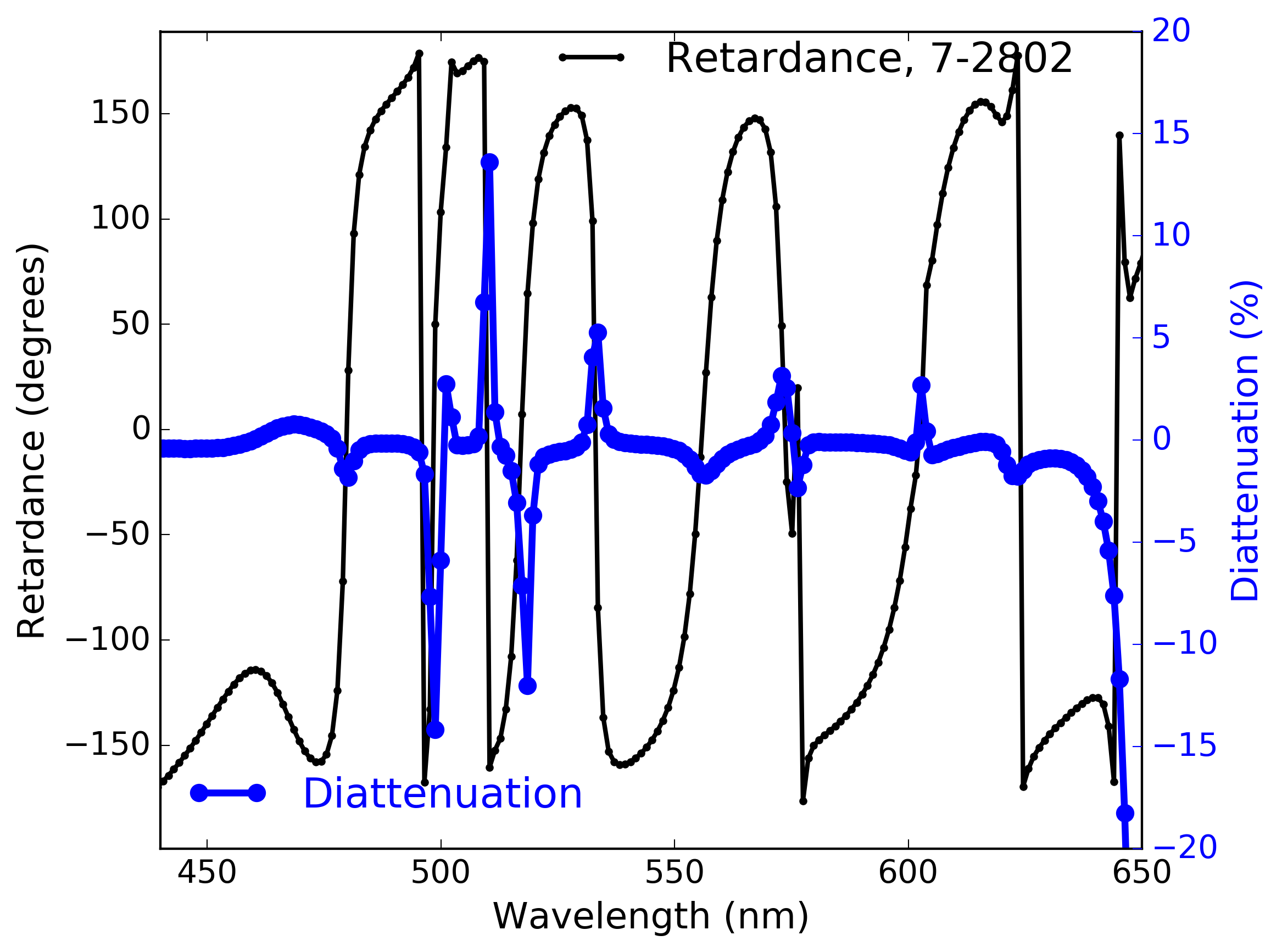}
}
\end{tabular}
\caption[] 
{\label{fig:IO_dichroicB_reflected_pol}  The NLSP measured retardance in black and diattenuation in blue for the Infinite Optics Dichroic B sample run 7-2802 at 45$^\circ$ incidence in reflection. The 440 nm to 655 nm wavelength region was selected as the efficient reflection band pass. }
\vspace{-4mm}
\end{wrapfigure}

This coating was designed only for high reflectivity at low incidence angles. For coatings designed without specification of polarization, the coating can have very strong rates of spectral change in addition to large magnitudes of both retardance and diattenuation.  In Figure \ref{fig:IO_dichroicB_reflected_pol}, we show the NLSP measured retardance and diattenuation of Dichroic B in reflection at an incidence angle of 45$^\circ$. 

The retardance is shown in black with multiple wavelength regions showing strong spectral changes wrapping through several waves of retardance. The diattenuation is shown in blue with large antisymmetric swings in the same wavelength regions where retardance changes quickly. Spectral changes are over $\pm$10\% in just a few nanometers wavelength, comparable to the spectral resolving power of NLSP.  We show this dichroic as an example of how optimizing a design for only reflectivity can create narrow spectral regions where the transmission and polarization performance pose calibration challenges.

\clearpage
\subsection{Narrow Spectral Features: Spectral Calibration with Many-Layer Coatings}
\label{sec:sub_dichroic_spikes}

Coating models predict narrower spectral features and stronger wavelength gradients as the number of layers and coating thickness increase. Features in common astronomical dichroics can exist where the reflectivity can drop over 10\% from the $>$99.9\% nominal performance along with strong polarization dependent response. 

\begin{wrapfigure}{r}{0.60\textwidth}
\centering
\vspace{-3mm}
\begin{tabular}{c} 
\hbox{
\hspace{-1.0em}
\includegraphics[height=7.4cm, angle=0]{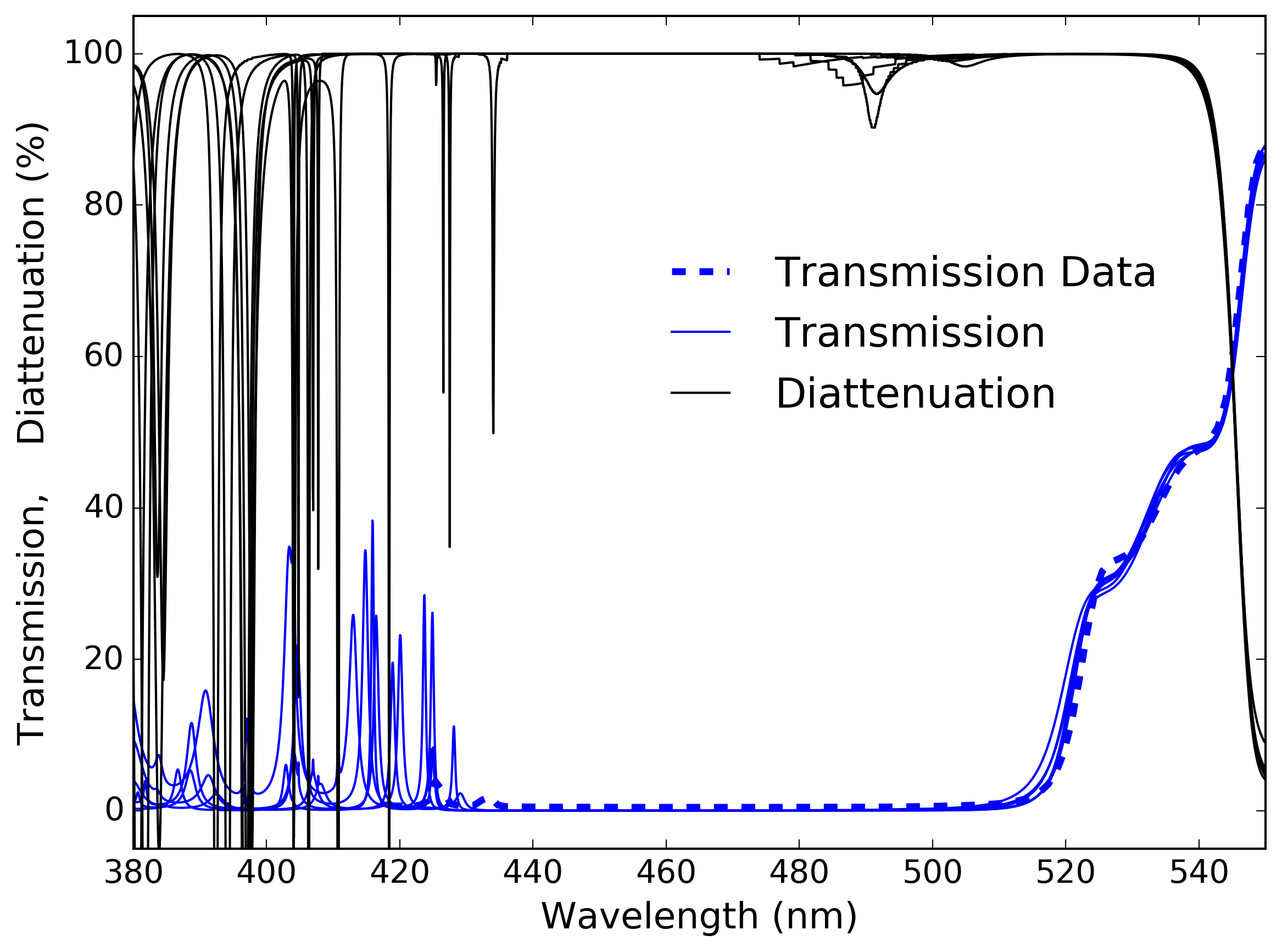}
}
\end{tabular}
\caption[] 
{\label{fig:IO_dichroicA_TFCalc_Spikes}  The TFCalc models of Dichroic A run 8-5652 from a constrained fit to the NLSP data.  Narrow, strong features are predicted at short wavelengths with high sensitivity to the specific constraints applied. Blue shows transmission, black shows diattenuation.  The thick dashed blue line is the NLSP measured transmission spectrum. }
\vspace{-3mm}
\end{wrapfigure}

As an example, Figure \ref{fig:IO_dichroicA_TFCalc_Spikes} shows the constrained fit TFCalc models to the NLSP data sets from above. Blue shows the transmission spectrum and black shows the diattenuation for a 45$^\circ$ incidence angle. We note that in the reflection band around 450 nm to 500 nm, this design shows almost no narrow spectral features and minimal sensitivity to the varying fit constraints.  However, at shorter wavelengths of 380 nm to 440 nm, there are transmission features predicted at amplitudes of 5\% to almost 40\% depending on the model.  As NLSP does not have significant signal for this dichroic at short wavelengths, we can only plot the measured transmission spectrum as the thick blue line of Figure \ref{fig:IO_dichroicA_TFCalc_Spikes}.

We measured transmission spectra for Infinite Optics Dichroics A, run 8-5652 and B, run 7-2802 over the 370 nm to 600 nm wavelength range at 4 pm per step with the Meadowlark Spex instrument we previously described. \cite{Harrington:2018cx} In the Meadowlark Spex system, their double-grating CT spectrograph used a 30 $\mu$m wide slit setting and a photo-multiplier as the sensor.  Spex had an instrument profile of roughly 25 pm optical FWHM over visible wavelengths. Sampling was 4 picometers wavelength per step or about 6 samples per optical FWHM. Spectral resolving power is thus roughly 20,000 sampled at one part in 120,000 \cite{Harrington:2018cx}.

\begin{figure}[htbp]
\begin{center}
\vspace{-1mm}
\hbox{
\hspace{-1.0em}
\includegraphics[height=5.5cm, angle=0]{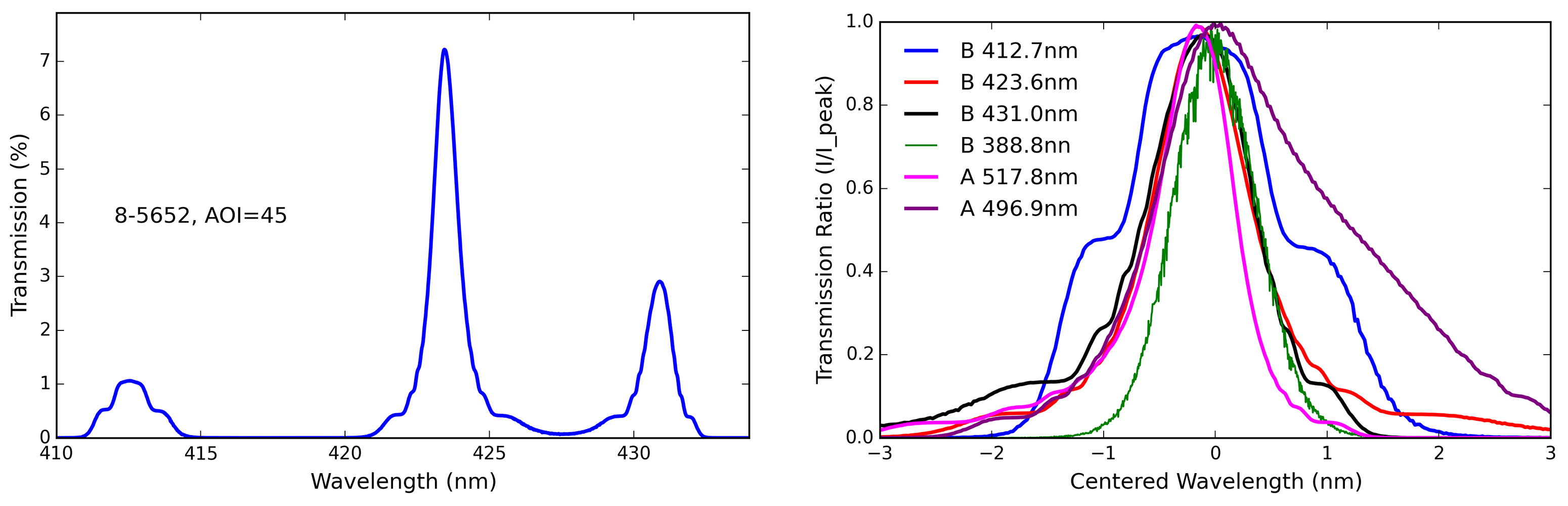}
}
\caption[] 
{ \label{fig:IO_dichroic_spikes} Transmission measurements of Infinite Optics dichroics in the reflection bandpass.  The left panel shows transmission for the run 8-5652 Dichroic A.  Three narrow features at amplitudes of 1\% to 7\% transmission are seen in this bandpass and are predicted in the TFCalc model.  The right panel shows a compilation of several narrow features in both run 7-2802 Dichroic B and run 8-5652 Dichroic A.  Some features are Gaussian in spectral profile while others are complex.  Amplitudes range from a fraction of a percent to 7\%.  }
\vspace{-5mm}
\end{center}
\end{figure}

Figure \ref{fig:IO_dichroic_spikes} shows examples of the narrow spectral features measured in the Infinite Optics Dichroics A and B. The left hand graphic shows a 25 nm bandpass measuring transmission for Dichroic A run 8-5652 in the wavelength region where this optic is highly reflective. The nominal reflectivity is over 99.9\% as designed but there are narrow spectral features that reach up to 7\% transmission. In the right hand graphic of Figure \ref{fig:IO_dichroic_spikes} we collected several spectral features from both Dichroics A and B. We centered them on a relative scale by the central wavelength seen in the legend. We then normalized them by the peak transmission. The wavelength dependent profile of each feature is complex, but the widths are roughly 1 nm to 2 nm FWHM.

\subsection{Summary of Dichroic Coatings: Spectral Features and Optimization}
\label{sec:sub_summarize_dichroics}

In this section we showed the nominal DKIST designs for the Facility Instrument Distribution Optics (FIDO).  The FIDO dichroic coating designs were shown using 25 to 96 layers and thicknesses from 1.5 $\mu$m to 8.8 $\mu$m. These designs were compared with as-built samples of many-layer coatings. We showed NLSP spectropolarimetric data, the nominal thin film models and fits of those models to our as-built data. The first example dichroic (A, run 8-5652) shown in Section \ref{sec:sub_IO_dichroicA} had 61 layers and showed spectrally smooth behavior in both reflection and transmission with high reflectivity and a sharp transition. An alternate dichroic (B, run 7-2802) example was shown in Section \ref{sec:sub_IO_dichroicB} with very high reflectivity but spectrally narrow bands of high diattenuation along with retardance oscillations of over 5 waves in the reflection bandpass of 420 nm to 680 nm. We then showed these very narrow features are quite real, measurable and expected in various designs in Section \ref{sec:sub_dichroic_spikes}. We presented spectrophotometric measurements with a resolving power of 20,000 on the Meadowlark SPEX system and example TFCalc design files showing how different dichroic designs can be quite sensitive to manufacturing errors and create these narrow but quite strong spectral features.

Given these measurements of several dichroic styles, DKIST optimized our designs not only for high reflectivity but low diattenuation and spectrally smooth retardance while also being constrained by the minimum stress and wavefront error requirements. The C-BS-950 design presents challenges given the nominal 96 layers, but the sensitivity was minimized and the retardance should maintain spectral gradients of less than 10$^\circ$ per nm of bandpass for almost all of the required bandpass with particular emphasis on key DKIST spectral wavelengths.  We are presently performing a thorough coating design verification and repeatability study to ensure DKIST achieves the required optical performance on the FIDO optics.

We showed how the thin film design tools such as TFCalc can be used to verify that dichroic coatings fall within tolerances and also to predict the presence and magnitude of narrow spectral features.  Many astronomical instruments are now observing simultaneous with other instruments, often covering narrow bandpass at high spectral resolving power. To ensure success in designing and calibrating these instruments, the properties of all the many-layer coatings in the system must be known. By controlling the design, benign performance can be assured.  For DKIST, we showed mirrors, broad-band anti-reflection coatings and dichroics in the above sections.  We now combine the optics into their appropriate groups and create a system-level model for DKIST with all coatings on all optics.

\clearpage
\section{A DKIST System Model for Polarization and Throughput}
\label{sec:system_group_model}

We now use the coating analysis and metrology to predict the mirror Mueller matrix and {\bf group model} parameters for every optic ahead of each instrument modulator. With this group model, we can predict the polarization response of the system at any azimuth, elevation or coud\'{e} table angle\cite{2017JATIS...3a8002H}. We can also make estimates of expected uncertainties in coating performance and show how a tolerance analysis could be done to inform mirror coating choices. In the system calibration process we are fitting the group model parameters of diattenuation and retardance for every mirror group. These terms also combine with the derivation of a modulation matrix across the field of view of each instrument. Given our spectral smoothness requirements on the coatings, we can ensure that sub nanometer wavelength variation between instrument bandpasses does not degrade the calibration of the telescope between instruments. Depending on the calibration fitting recipes used, a wide value search space can lead to degeneracies and unphysical results in the fit to mirror properties. By showing this tolerance analysis and our fitting errors, we show that we can significantly restrict the range allowed for fitting the telescope mirror group parameters. We also gain speed in computation when we can use simple gradient-minimization techniques as opposed to wide area search techniques (such as differential evoloution). For DKIST with many simultaneous wavelengths and a 24 hour calibration time requirement, the computational gains are expected to be significant. With our Berreman code we can make assumptions about the range of coating properties and directly compute the reflectivity, diattenuation and retardance of the mirrors under several kinds of model perturbations.  Knowing the range of expected polarization properties also helps validate any DKIST-derived calibrations when using our custom retarders we have mapped spatially and assessed for thermal instabilities\cite{Harrington:2018jb,Harrington:2018bt,2017JATIS...3d8001H}.

\begin{figure}[htbp]
\begin{center}
\vspace{-1mm}
\hbox{
\hspace{-0.3em}
\includegraphics[height=8.2cm, angle=0]{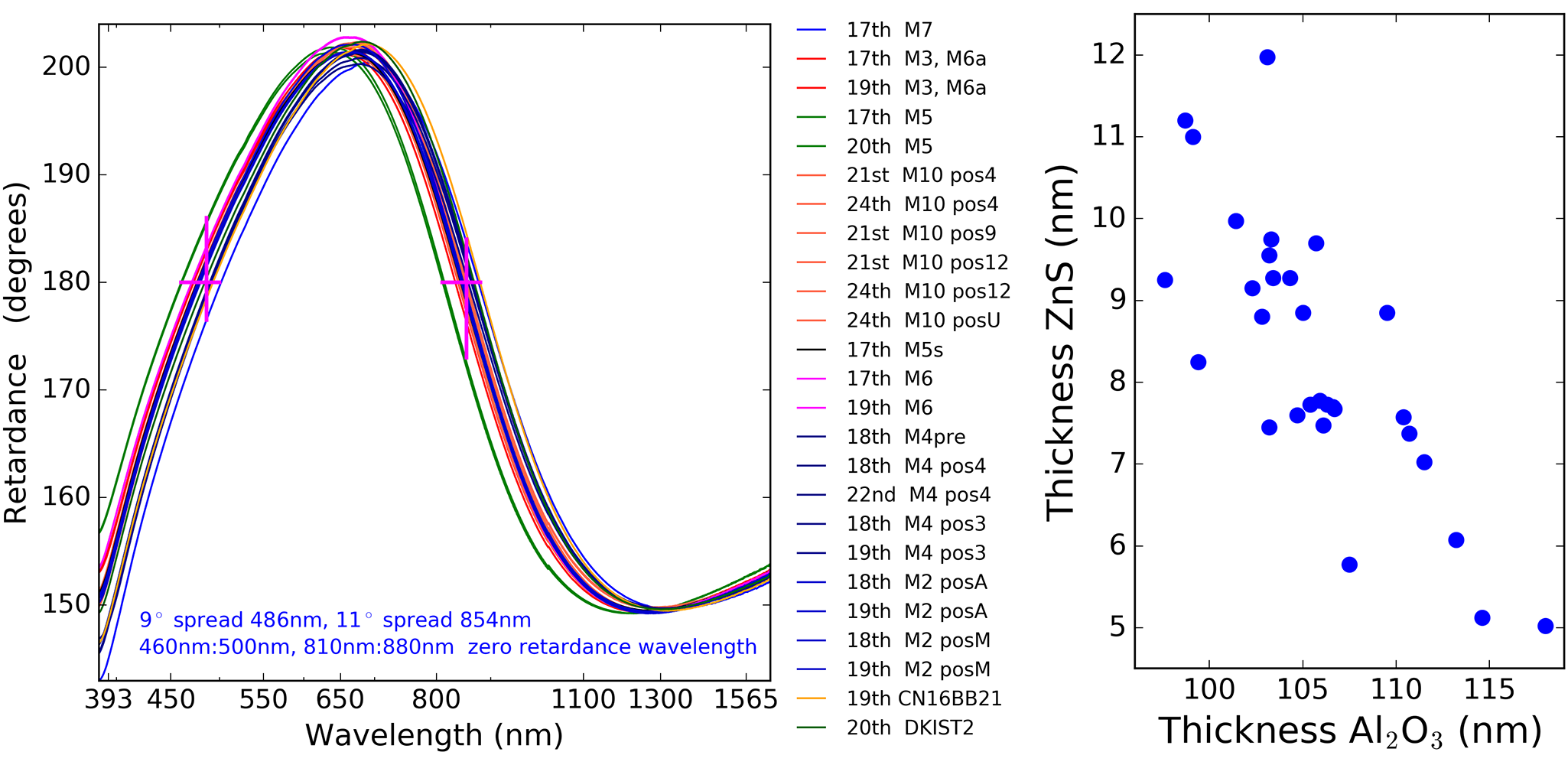}
}
\caption[] 
{\label{fig:group_model_nlsp_retardance} Left shows the retardance derived from NLSP Mueller matrix measurements in September 2018 for the DKIST telescope mirrors along with comparison to some instrument partner witness samples. Right shows the best-fit thickness of the simple two-layer ZnS over Al$_2$O$_3$ model using the direct search brute force method. }
\vspace{-5mm}
\end{center}
\end{figure}

In September of 2018, we had acquired witness samples for every telescope mirror in the DKIST path that has been coated to date. This includes the optics M2 through M10. We received samples from M8 and M9 in late 2018 (as detailed in Appendix \ref{sec:sub_DKIST_M9_ViSP_FiDO}). In several cases we received multiple witness samples from different spatial positions in the coating chamber. Figure \ref{fig:group_model_nlsp_retardance} shows the retardance measurements from this September 2018 campaign in the left hand graphic. In several cases, we re-mounted and re-measured samples to assess our procedures over timescales $>$1 year. The repeatability and an assessment of the errors is included in Appendix \ref{appendix:nlsp_cal}.  As noted in the Figure, there is a spread of up to $\sim$10$^\circ$ retardance at individual wavelengths.  There is also a significant wavelength variation in where the retardance spectrum crosses the theoretical {\it net-zero induced retardance} value of 180$^\circ$.  The right hand graphic of Figure \ref{fig:group_model_nlsp_retardance} shows the resulting thicknesses from our simple two-layer retardance fitting technique. There is significant scatter with a strong anti-correlation between thinner top layer and thicker bottom layer. The bottom layer varies in thickness by $\pm$10.2 nm representing a roughly 10\% thickness variation.  The top layer fit varies by $\pm$3.5 nm but this represents a 40\% variation.  Of course, our simple two-layer model does not necessarily reflect the complexity in a coating that could contain many more layers than we are modeling.  But this variation is representative of the variation in retardance measured for many samples.  We can take this measured range of parameters and show the uncertainties in the DKIST system model given this possible range of coating behaviors.  We also can provide a more accurate system model given that we can use these measured values to put a more representative coating specific to each mirror into the system model.

\begin{wraptable}{r}{0.52\textwidth}
\vspace{-3mm}
\caption{DKIST Telescope Mirror Model}
\label{table:mirror_coating_formula_for_groups}
\centering
\begin{tabular}{l r r r r l}
\hline
\hline
Optic		& AOI			& Ax		& ZnS		& Al$_2$O$_3$	 	& Note		\\
\hline
\hline
{\it M1}	& 14.0$^\circ$		& X		& --			& 4.0				& Al$_2$O$_3$	 over Al				\\
M2 		& 11.7$^\circ$		& X		& 7.700 		& 106.6			& posA, 18th \\
\hline
M3 		& 45.0$^\circ$		& X		& 8.800		& 102.8			& 19th 	\\
M4		& 1.8	$^\circ$		& Y		& 6.075		& 113.2			& pos3, 18th	\\
\hline
M5		& 15.0$^\circ$		& Y		& 8.250		& 99.4			& 20th \\
M6		& 30.0$^\circ$		& Y		& 11.000		& 99.1			& 17th \\
\hline
M7  		& 45.0$^\circ$  		& Y		& 5.025   		& 118.0	    		& 17th 	\\
M8  		& 5.3$^\circ$  		& X		& 8.850  		& 109.5	  		& Use CN16BB21	\\
M9  		& 10.0$^\circ$  		& X		& 11.975 		& 103.1      		& Use CN PW1	\\
M10 		& 15.0$^\circ$  		& X		& 9.550  		& 103.2	   		& DM, posU 24th	\\
\hline
\end{tabular}
\end{wraptable}
\vspace{-2mm}

In Table \ref{table:mirror_coating_formula_for_groups} we show the two-layer model fits and the incidence angle assumed for a simple mirror group model. We use the brute-force method of fitting for minimum retardance difference between measurements and model using a model grid with a step size of 0.025 nm in ZnS and 0.1 nm in Al$_2$O$_3$ running from 0 nm to 250 nm thickness. The first column shows the chief ray incidence angle in degrees for each mirror. The second column shows the tilt axis in the local coordinates of the beam propagating to the mirror in the Zemax optical model. This column is meant to show that mirror groups do not necessarily share a plane of incidence and thus the Mueller matrix for each mirror group may not follow the simple equation for a single mirror. 

\begin{figure}[htbp]
\begin{center}
\vspace{-1mm}
\hbox{
\hspace{-0.6em}
\includegraphics[height=6.3cm, angle=0]{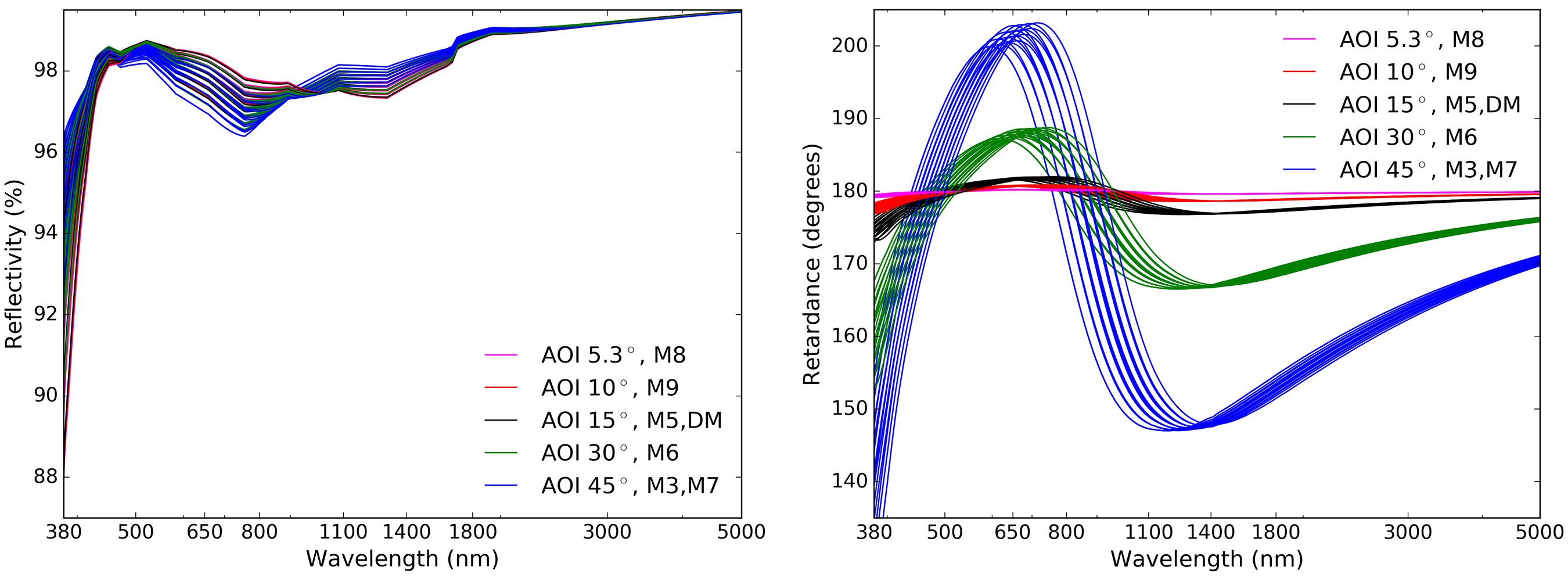}
}
\caption[] 
{\label{fig:mirror_model_reflectivity_retardance} Reflectivity and retardance tolerance analysis on the individual DKIST mirrors M1 through M10 at the appropriate incidence angles.  The left graphic shows reflectivity ranges at incidence angles noted in the legend. As we do not fit reflectivity, this range is representative but can be less accurate. The right hand graphic shows the retardance model and these curves follow the measured retardance variation shown above. }
\vspace{-5mm}
\end{center}
\end{figure}

In the DKIST system, the mirrors are crossed at many telescope pointings when the tilt axes are 90$^\circ$ away from each other. In this case, the diattenuation and retardance subtracts, leading to a lower net magnitude in particular for the case of M7 and M8 through M10. We replace M8 and M9 values with other DKIST silver coated mirrors as these optics (along with M7 and M10) are part of the modulation matrix combined with FIDO optics and instrument relay optics. The third column of Table \ref{table:mirror_coating_formula_for_groups} shows the brute-force fit thickness of the top ZnS layer using the RefractiveIndex.info refractive index.  The fourth column shows the brute-force fit thickness of the Al$_2$O$_3$ layer when using the Boidin reference from RefractiveIndex.info for the refractive index. The last column shows notes on the selection of the fit parameters. We include the DKIST primary mirror (M1) in this Table as the first row as the bare aluminum metal coating does form an oxide layer (Al$_2$O$_3$). We show an example fit to M1 coating samples in Appendix \ref{sec:sub_DKIST_M1_alum_mirror}. We also have several samples from different spatial positions in the chamber as well as repeat measurements on multiple days.  We show the data set used for the fit as well as spatial position of the sample when available.

These models are representative of the measured range of variation as expected with normal coating manufacturing tolerances. As an example of the expected range of system performance, we can take our DKIST silver two-layer model and vary the thickness of each layer independently for each coating shot following our procurement schedule.  We can make a grid of coating layer thickness models where the top ZnS layer is varied from 6 nm to 12 nm in steps of 2 nm thickness while the bottom Al$_2$O$_3$ layer varies from 95 nm to 110 nm in steps of 5 nm thickness. These thickness choices follow our layer thickness fits to retardance measurements of the many witness samples in the DKIST silver coating shots as well as the DL-NIRSP and Cryo-NIRSP shots from the same vendor with the same coating formula.

The mirrors that were coated in the same coating chamber shot will be assumed to have the same coating formula. However, each coating shot will vary independently from every other shot. Spatial variation is certainly present and measured in our samples through various vendors, but outside the scope of this paper. For the sake of simplicity with reasonable model accuracy, we will use the silver refractive indices that represents reflectivity at visible wavelengths in the reflectivity calculations. We use the silver refractive indices that better match retardance for calculating retardance and diattenuation. We note that the diattenaution in this model does not match the measurements to better than a fraction of a percent.  Diattenuation was not included in the fit error metric, nor were the refractive indices of the silver fit. This gives model simplicity and the range of polarization variation is similar for a variety of refractive index choices.

In Figure \ref{fig:mirror_model_reflectivity_retardance} we show reflectivity ranges for the various DKIST mirrors as functions of incidence angle in the left hand graphic computed with our Berreman package. The right hand graphic shows the retardance ranges predicted for the individual DKIST mirrors down stream of the calibration unit. The primary and secondary mirrors form their own group which is static in time and fit separately as modeled in HS17 \cite{2017JATIS...3a8002H}. The different colors represent different incidence angles for the various mirrors denoted in the legend.

\begin{figure}[htbp]
\begin{center}
\vspace{-1mm}
\hbox{
\hspace{-0.6em}
\includegraphics[height=6.3cm, angle=0]{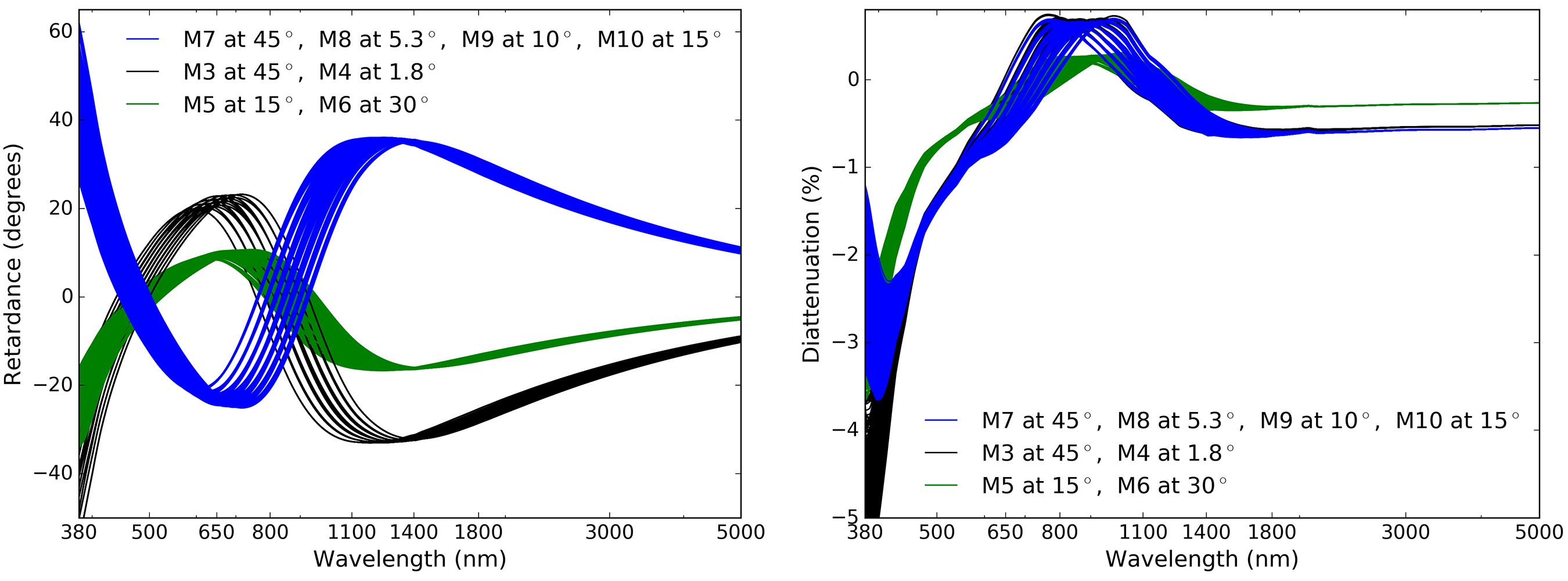}
}
\caption[] 
{ \label{fig:group_model_retardance_diattenuation} Group model retardance (left) and diattenuation (right) tolerance analysis. The mirrors are grouped appropriately and model Mueller matrices are multiplied. Black shows the M3:M4 group.  Green shows the M5:M6 group.  Blue shows the M7:M9:M9:M10 group where M10 is the AO system deformable mirror.  }
\vspace{-5mm}
\end{center}
\end{figure}

Continuing the tolerance analysis, we combine mirrors into groups and compute polarization performance by multiplying the appropriate mirror Mueller matrices at the appropriate incidence angles. The simulated layer thickness variations on any one mirror are compounded with variations on every other mirror to evaluate the worst-case possibilities. Figure \ref{fig:group_model_retardance_diattenuation} shows this group model and the tolerance analysis.  The left hand graphic shows the retardance for the three mirror groups. The M5:M6 group is shown in green with relatively low retardance but significant variation in the tolerances because both mirrors are at significant incidence angles (15$^\circ$, 30$^\circ$). The M3:M4 group is shown in black with moderate retardance dominated by the 45$^\circ$ incidence of M3.  

The M7 through M10 group is shown in blue. This group of mirrors is fixed on the coud\'{e} table and will become part of the instrument modulation matrix. We group the mirrors independently for now here as they impact the expected variation of the instrument modulation matrix. The mirrors also impart some variation with instrument field of view on the polarization calibrations as the pupil is demagnified by a factor of 20x giving rise to incidence angle variation of roughly 1$^\circ$ when observing over the 5 arc minute field of view. Figure \ref{fig:group_model_retardance_diattenuation} shows how typical manufacturing tolerances can create variability in the expected polarization properties of the telescope optical path.

This M7:M10 group will impart significant complexity to the azimuth-elevation dependent behavior of the telescope diattenuation and polarizance. M7 has significant retardance at a 45$^\circ$ incidence angle and there will be differences imparted to the first row and columns of the Mueller matrix. Equation \ref{eqn:two_mirrors_combined} shows the theoretical form of the Mueller matrix for two mirrors that do not share a plane of incidence. The first mirror is tilted along the vertical plane becoming a Q diattenuator with retardance between U and V components.  The second mirror is tilted along the 45$^\circ$ plane which creates U diattenuation and retardance between Q and V components.  The resulting Mueller matrix has polarizance in the first column that is not equal to the diattenuation in the first row.  With the elevation and azimuth axes providing such rotations to each group of mirrors, we expect significant differences for DKIST in the first row and column of the Mueller matrix.

\vspace{-4mm}
\begin{equation}
{\bf M}_{toFIDO} = \  {\bf M}_{10}\  {\bf M}_9\  {\bf M}_8\  {\bf M}_7\ \ \ \ \     {\bf \mathbb{R}}(-Az-TA)\ \ \    {\bf M}_6\  {\bf M}_5\ \  {\bf \mathbb{R}}(-El)\ \  {\bf M}_4\  {\bf M}_3\  {\bf M}_2\  {\bf M}_1\ \  {\bf \mathbb{R}}(El)  \ \ \  {\bf \mathbb{R}}(Az+TA)
\label{eqn:MM_group_model_MMcoude}
\end{equation}
\vspace{-4mm}
\begin{equation}
{\bf \mathbb{M}}_{ij} =
 \left ( \begin{array}{rrrr}
T_1			& D_1		& 0				& 0			\\
D_1 			& T_1		& 0				& 0			\\
0			& 0			& c_\gamma	 	& s_\gamma	\\
0			& 0			& -s_\gamma		& c_\gamma	\\ 
 \end{array} \right ) 
 \left ( \begin{array}{rrrr}
T_2			& 0			& D_2			& 0			\\
0 			& c_\beta		& 0				& s_\beta		\\
D_2			& 0			& T_2	 		& 0			\\
0			& -s_\beta		& 0				& c_\beta		\\ 
\end{array} \right )  = 
\left ( \begin{array}{rrrr}
T_1 T_2			& D_1 c_\beta			& T_1 D_2		&  D_1 s_\beta			\\
T_2 D_1 			& T_1 c_\beta			& D_1 D_2		&  T_1 s_\beta			\\
 D_2 c_\gamma	& -s_\gamma s_\beta	& T_2 c_\gamma	&  s_\gamma c_\beta	\\
-D_2 s_\gamma	& -c_\gamma s_\beta	&-T_2 s_\gamma	& c_\gamma c_\beta		\\ 
\end{array} \right ) 
\label{eqn:two_mirrors_combined}
\end{equation}

\begin{wrapfigure}{l}{0.65\textwidth}
\centering
\vspace{-3mm}
\begin{tabular}{c} 
\hbox{
\hspace{-1.4em}
\includegraphics[height=8.2cm, angle=0]{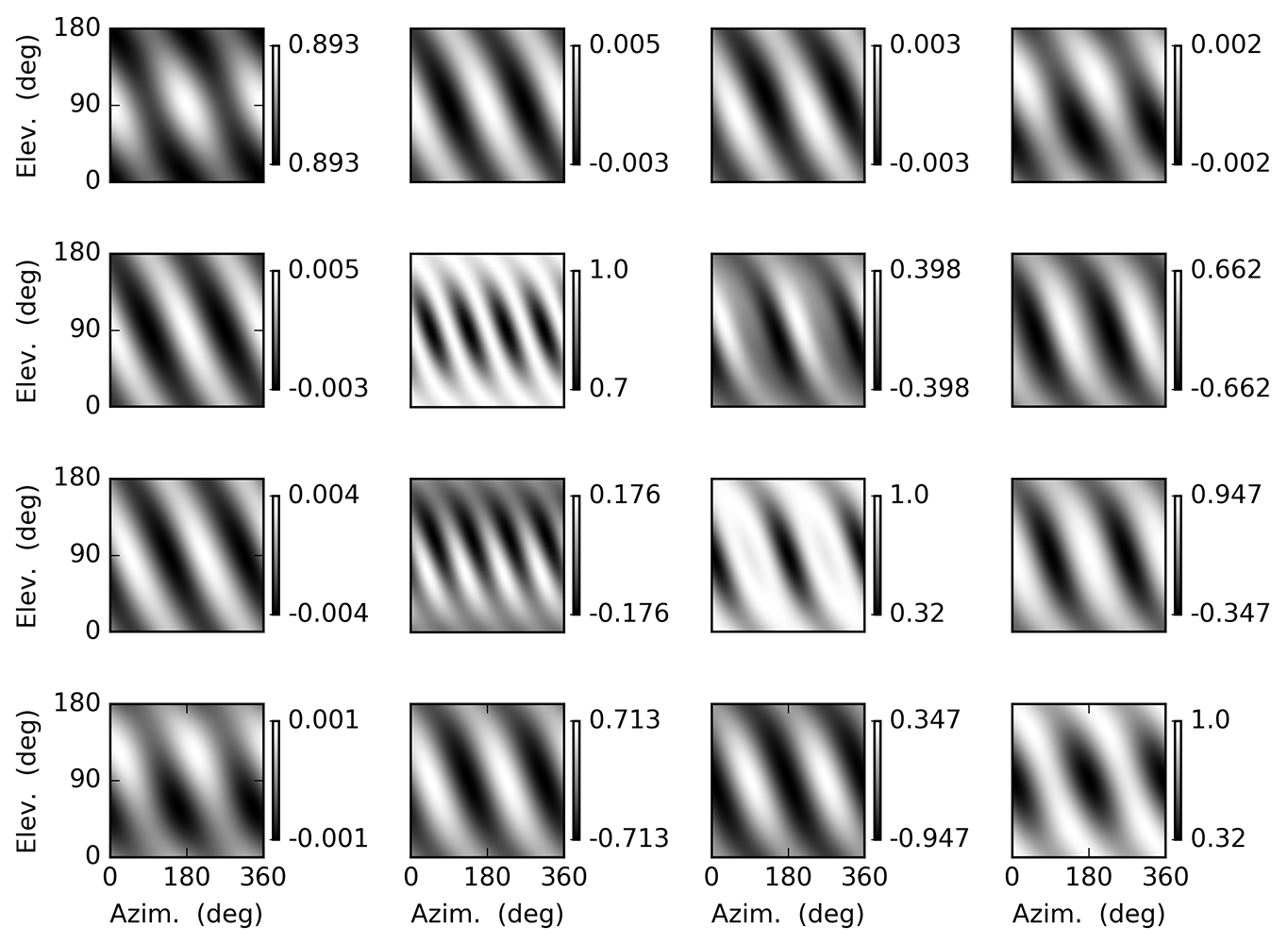}
}
\end{tabular}
\caption[] 
{\label{fig:group_model_MM_1565}  The Mueller matrix for the combined optics of M1 through M10 as combined in the system group model at 1565 nm wavelength. Each element shows azimuth from 0$^\circ$ to 360$^\circ$ and elevation angle 0$^\circ$ to 180$^\circ$.}
\vspace{-3mm}
\end{wrapfigure}

We can implement the group model by combining the Mueller matrices derived for each mirror. We get the diattenuation and retardance using the Berreman calculus with the coating layer thicknesses in Table \ref{table:mirror_coating_formula_for_groups} and the theoretical equation for a flat fold mirror at a single incidence angle from Equation \ref{eqn:flat_mueller_matrix}. We then apply the geometrical rotations between the groups of mirrors as the telescope rotates in azimuth, elevation and coud\'{e} laboratory table angle. In Equation \ref{eqn:MM_group_model_MMcoude}, we explicitly show the positive and negative rotation matrices as required to rotate the groups of mirrors into the appropriate frame. We combine the azimuth and coud\'{e} laboratory table angle as the laboratory freely rotates to arbitrary orientation.

Figure \ref{fig:group_model_MM_1565} shows an example of such a group model calculation. We have normalized all elements by the [0,0] element except the [0,0] element itself as in Equation \ref{eqn:MM_IntensNorm}. As we showed in Equation \ref{eqn:two_mirrors_combined} for two mirrors at 45$^\circ$ relative rotation between incidence planes, the dominant impact of the geometry is to create oscillations in azimuth and elevation as the mirror properties add when sharing planes of incidence and subtract when having orthogonal planes of incidence. We run this calculation over a full 360$^\circ$ of azimuth + table angle to show the symmetry of the combined optical system. We also run the calculation from 0$^\circ$ to 180$^\circ$ in elevation axis to show a complete cycle of parallel and perpendicular mirror pair incidence planes. The telescope mount structure is not capable of this articulation, but this simulation shows the symmetry of the underlying coating model behavior as well as the geometric QU rotation with azimuth and elevation.

We note that there are very slight changes in system transmission seen in the [0,0] matrix element. The color scale labels in Figure \ref{fig:group_model_MM_1565} round to three decimal places for clarity.  The transmission ranges from 89.26684\% to 89.26595\% or a range of only about 9 parts per million. This tiny effect is easily ignored.  The $IV$ and $VI$ elements are roughly a factor of two different. We now have the telescope mirrors M1 though M10 modeled with as-built coating formulas and the associated azimuth-elevation-table-angle behavior modeled. Next we consider the dichroics and feed optics of individual instruments to show the polarization behavior of the rest of the optical train to instrument modulators.

\subsection{Instrument Feed Optics: ViSP Reflectivity \& Polarization With Many Layers}
\label{sec:FeedOptics_visp}

The Visible Spectropolarimeter (ViSP) has three enhanced protected silver mirrors between the last FIDO optic reflection and the spectrograph entrance slit. The original specification had very high reflectivity mirrors using 29 dielectric layers and a physical thicknesses of 3 microns. Thick, many-layer designs can have very narrow spectral features that impact polarization calibration and heightened sensitivity to various manufacturing and environmental conditions.  Though the ViSP team has stripped and re-coated their optics, we consider what would have been the polarization properties of these feed optics.

\begin{figure}[htbp]
\begin{center}
\vspace{-1mm}
\hbox{
\hspace{-0.8em}
\includegraphics[height=6.3cm, angle=0]{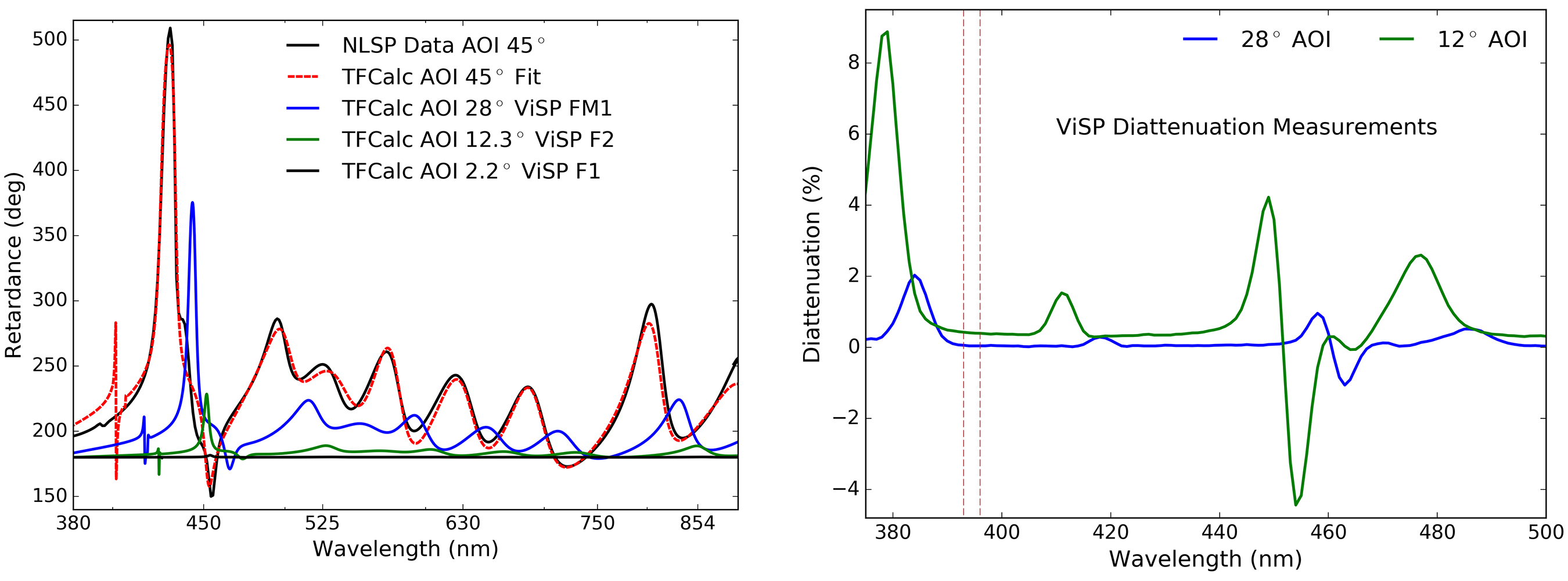}
}
\caption[] 
{ \label{fig:nutek_mirrors_nlsp_tfcalc_ret_ioi_diat} The left hand graphic shows retardance measured with NLSP for the ViSP feed mirror coating witness sample at 45$^\circ$ incidence from 380 nm to 900 nm as the oscillating black curve. The TFCalc model fit to that NLSP retardance is shown as the dashed red line.  With this successful TFCalc fit, we then predict the retardance for the ViSP mirrors FM1, F1 and F2 at the appropriate incidence angles. The right hand graphic shows vendor measured reflected diattenuation at the appropriate ViSP incidence angles in the 380 nm to 520 nm bandpass where the mirrors show very strong changes. }
\vspace{-7mm}
\end{center}
\end{figure}

\begin{wraptable}{l}{0.25\textwidth}
\vspace{-3mm}
\caption{ViSP Feed}
\label{table:visp_opt}
\centering
\begin{tabular}{l l l}
\hline
\hline
Optic		& AOI 		& Run	\\
\hline
FM1		& 28$^\circ$	& ViSP1	\\
F1		& 2.2$^\circ$	& ViSP1	\\
F2		& 12.3$^\circ$	& ViSP1	\\
Slit$_f$	& 5.4$^\circ$	& ECI	\\
Slit$_b$	& 5.4$^\circ$	& ECI	\\
\hline
\hline
\end{tabular}
\vspace{-4mm}
\end{wraptable}

For the baseline ViSP feed mirrors coatings, we were given a coating design file for the coating with roughly 29 layers of dielectric material over silver. Figure \ref{fig:nutek_mirrors_nlsp_tfcalc_ret_ioi_diat} shows the NLSP measurements of a witness sample retardance measurement at 45$^\circ$ incidence along with the TFCalc model fit to that data set in the left hand graphic. With this TFCalc model fit to retardance at 45$^\circ$ over wavelengths 380 nm to 800 nm, we can then predict performance for the baseline ViSP mirrors before stripping at appropriate incidence angles outlined in Table \ref{table:visp_opt}. ViSP FM1 at 28$^\circ$ incidence is shown in blue. The ViSP powered optic F1 is shown in black at 2.2$^\circ$ incidence. The ViSP powered optic F2 is shown in green at 12.3$^\circ$ incidence.

The polarization introduced by these three mirrors combine with the diattenuation and retardance introduced by the tilted broad-band anti-reflection coated glass substrate of the slit mask to create static elements of the ViSP system modulation matrix. The modulator is mounted behind the slit mask. Table \ref{table:visp_opt} shows the three ViSP feed optics between FIDO and the ViSP modulator. The ViSP slit masks are deposited on a glass substrate mounted at 5.4$^\circ$ incidence angle to the beam.  Both sides of this substrate are coated with a broad-band anti-reflection coating with reflectivity averaging 1.0\% over the 380 nm to 900 nm bandpass.  Each side has slightly different coating behavior with both coatings having spectral oscillations ranging from 0.2\% to 1.5\% across the bandpass from shot-to-shot variation. Given the low incidence angle and high transmission, the retardance and diattenuation contributions are very small from this optic.

\begin{figure}[htbp]
\begin{center}
\vspace{-0mm}
\hbox{
\hspace{-0.6em}
\includegraphics[height=6.3cm, angle=0]{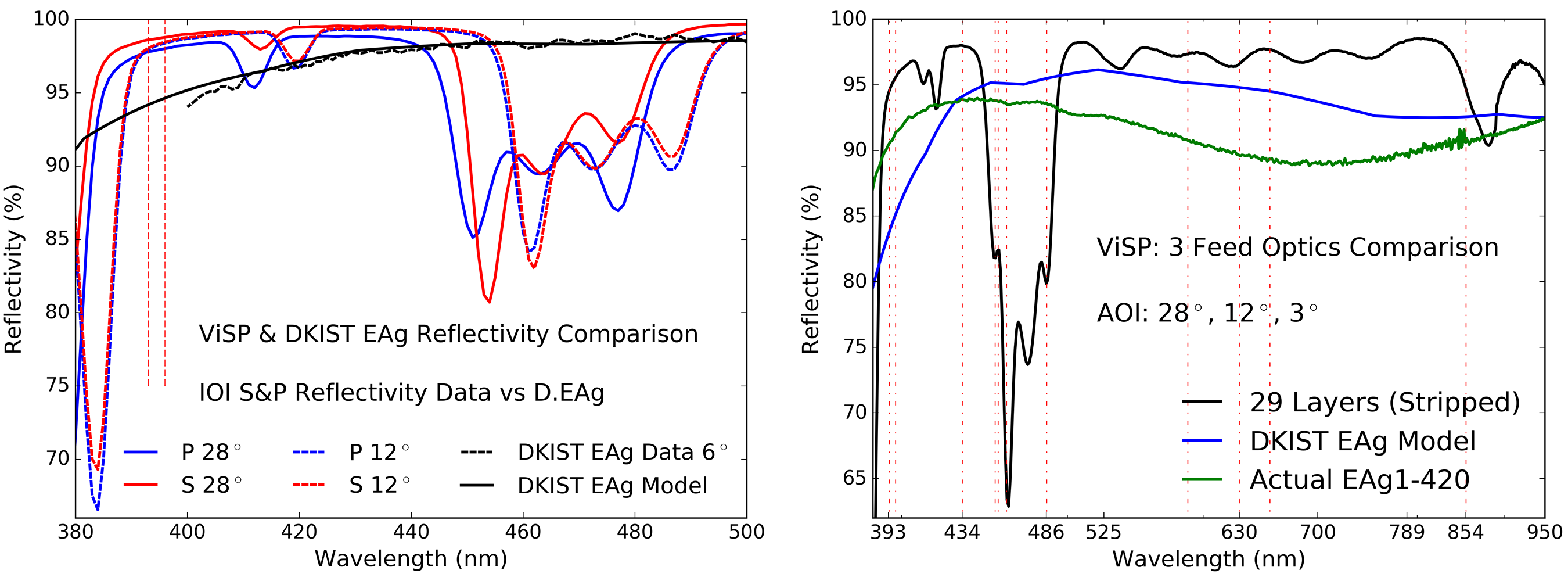}
}
\caption[] 
{ \label{fig:visp_reflectivity_3fold_mirrors} Left shows the S and P reflectivity measured by the vendor for a witness sample from the single coating shot covering all three ViSP fold mirrors. The solid lines show incidence angle 28$^\circ$ for ViSP FM1 and dashed lines show 12$^\circ$ for ViSP F2. The black lines show both data (dashed) and models (solid) of the DKIST silver for comparison.  Right shows the reflectivity of the combined ViSP mirrors with each mirror at the appropriate incidence angle.  Black shows the highly enhanced coating with the measured reflectivity issue for the three combined optics.  Blue shows the nominal DKIST silver formula (D.EAg) modeled for each optic at the appropriate incidence angle.  Green shows the stripped and re-coated mirrors as measured using the formula EAg1-420, the same as for the FIDO mirrors. }
\vspace{-6mm}
\end{center}
\end{figure}

\begin{wraptable}{l}{0.17\textwidth}
\vspace{-3mm}
\caption{Ex. Lines}
\label{table:visp_lines}
\centering
\begin{tabular}{l l}
\hline
\hline
$\lambda$	& Name 	\\
(nm)		&		\\
\hline
393.4	& Ca	II K	\\
396.9	& Ca	II H	\\
434.1	& H$_\gamma$	\\
453.6	& Ti I		\\
455.4	& Ba II	\\
460.7	& Sr I	\\
486.1	& H$_\beta$	\\
587.6	& He I		\\
630.2	& Fe I		\\
656.3	& H$_\alpha$	\\
854.2	& Ca II		\\
\hline
\hline
\end{tabular}
\vspace{-4mm}
\end{wraptable}

Figure \ref{fig:visp_reflectivity_3fold_mirrors} shows the parallel (P) and perpendicular (S) polarization state reflectivity of the ViSP fold mirrors measured at the appropriate incidence angles for the ViSP fold mirrors in the left hand graphic, measured several months after applying the original (now stripped) coating. The ViSP FM1 is at 28$^\circ$ incidence plotted as the solid lines. The ViSP F2 mirror is at 12.3$^\circ$ incidence and is the dashed lines. As expected, when the incidence angle increases, the complex spectral pattern generally shifts to the blue and changes morphology. The DKIST enhanced silver mirror coating is shown as a comparison using black lines. Solid shows a model while a dashed line shows reflectivity data for one of the DKIST mirrors (M10). Two dashed vertical lines show the Ca II spectral lines of interest for ViSP at 393.4 nm and 396.9 nm which drove the coating design. 

We note that the band of reflectivity at 80\% to 90\% was not present in the original test coating metrology. A manufacturing error gave rise to the actual mirror coatings behaving differently than the preliminary test coatings. The TFCalc models fit to vendor data on the ViSP enhanced silver coating failed to reproduce the relatively low reflectivity curves measured by the vendor in the 450 nm to 490 nm bandpass. However, the model does show that the wavelength shift from incidence angles of 12.3$^\circ$ to 2.2$^\circ$ are very small.  We do not have vendor reflectivity data at 2.2$^\circ$ incidence, but we can approximate the F1 mirror as the average of S and P reflectivity curves at 12.3$^\circ$ with zero diattenuation and zero retardance. Using this assumption, we can make a reflectivity model of the three combined ViSP feed mirrors as the black curve in the right hand graphic of Figure \ref{fig:visp_reflectivity_3fold_mirrors}. The system throughput would have generally been around 96\%. We show common atomic spectral lines in Table \ref{table:visp_lines} intended for ViSP observations that are noted as vertical dashed lines in some Figures of this section. However, there is a drop to roughly 70\% to 75\% throughput that will impact observations of spectral lines such as Ba II at 455.4 nm, Sr I at 460.7 nm and H$_\beta$ at 486.1 nm shown as vertical dashed lines. Additionally, there is a drop to 90\% throughput for the Ca II near infrared line at 854.2 nm. Given this metrology result, these mirrors were stripped and re-coated in October 2018. 

The new system throughput after stripping and recoating the three ViSP feed mirrors are shown as the green curves in the right hand graphic of Figure \ref{fig:visp_reflectivity_3fold_mirrors}. These curves are generated using spectrophotometry on witness samples for the two appropriate coating shots.  The blue curve shows the model prediction for the nominally specified DKIST enhanced protected silver. As shown above, the as-coated performance for this coating can vary substantially so we only show the nominal modeled performance. Each coating shows a few percent throughput loss at various wavelengths, but the polarization curves are very smooth, following the two-layer coating models shown in previous sections.

\begin{figure}[htbp]
\begin{center}
\vspace{-1mm}
\hbox{
\hspace{-0.8em}
\includegraphics[height=6.3cm, angle=0]{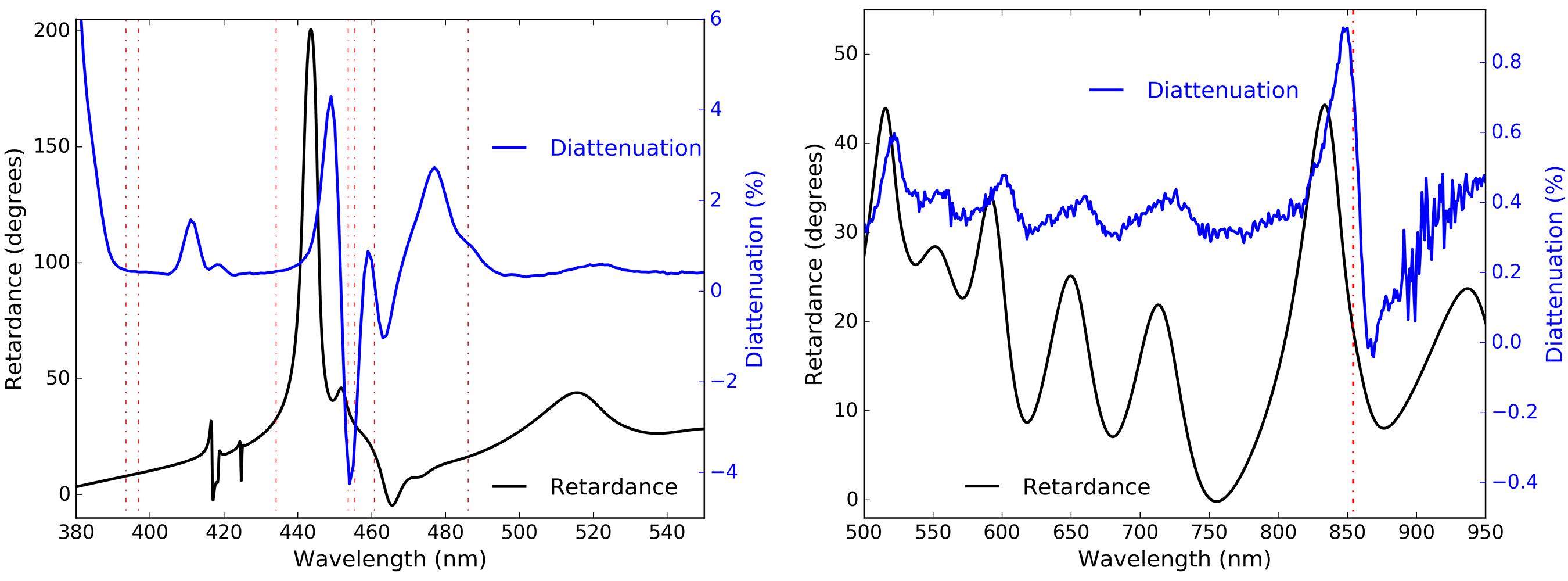}
}
\caption[] 
{ \label{fig:visp_3fold_pol_sys} Retardance and diattenuation from the combination of ViSP feed optics FM1, F1 and F2 before stripping of this 29 layer coating.  Left shows the 380 nm to 550 nm wavelength region. Retardance varies by almost 200$^\circ$ around 445 nm wavelength as seen by the black curve on the left hand Y axis. The diattenuation in blue using the right hand blue Y axis, varying by $\pm$4\% across adjacent wavelengths.  Right shows longer wavelengths on reduced Y scales.}
\vspace{-7mm}
\end{center}
\end{figure}

Figure \ref{fig:visp_3fold_pol_sys} shows the polarization model for the three ViSP feed optics combined at the appropriate incidence angles for the older (baselined) 29 layer coatings. The Mueller matrices were created for each optic and multiplied in sequence to create the model. The reflectivity and diattenuation from the vendor data were used while the retardance was predicted from the TFCalc model fit to our NLSP retardance measurements. The left hand graphic of Figure \ref{fig:visp_3fold_pol_sys} shows wavelengths between 380 nm and 550 nm a different vertical scale than the right graphic covering 500 nm to 950 nm. Around 450 nm there is a change of retardance of about one full wave over roughly 15 nm bandpass as the retardance goes from zero to over half wave and back. The spectral change in retardance in this bandpass goes over 60$^\circ$ retardance per nm of bandpass. This instrument would have to be calibrated at high spectral resolving power to account for the rapid spectral changes in this bandpass. Figure \ref{fig:visp_3fold_pol_sys} also includes vertical lines that denote ViSP spectral lines of interest. Fortunately, the 434.1 nm H$_\gamma$ line and 453.1 nm Ti I line avoid this coating spectral retardance feature. There is a band of relatively high diattenuation at slightly longer wavelengths of 445 nm to 490 nm with magnitudes of over $\pm$4\%.

The right hand graphic of Figure \ref{fig:visp_3fold_pol_sys} shows the three mirrors at longer wavelengths. The retardance oscillates between 0$^\circ$ and 45$^\circ$ with an oscillation period much larger than the nm scale ViSP bandpass. There is a small diattenuation feature around the 854.2 nm Ca II spectral line. Thus we could have expected some mild continuum polarization gradients in calibrations. We show in Appendix \ref{sec:sub_DKIST_visp_thor_Kmirrors} further confirmation of spectrally narrow reflectivity and polarization artifacts using our own NLSP metrology in an image-rotator (K-cell) type configuration even at incidence angles of 11$^\circ$.  Though these coatings may have offered high reflectivity, when high spectral resolving power and polarization are considered, these coatings could have caused substantial calibration complexity.

\clearpage

\subsection{DKIST \& FIDO Configured for DL-NIRSP On-Disk Observations}
\label{sec:fido1}

\begin{wrapfigure}{r}{0.44\textwidth}
\centering
\vspace{-5mm}
\begin{tabular}{c} 
\hbox{
\hspace{-1.0em}
\includegraphics[height=3.1cm, angle=0]{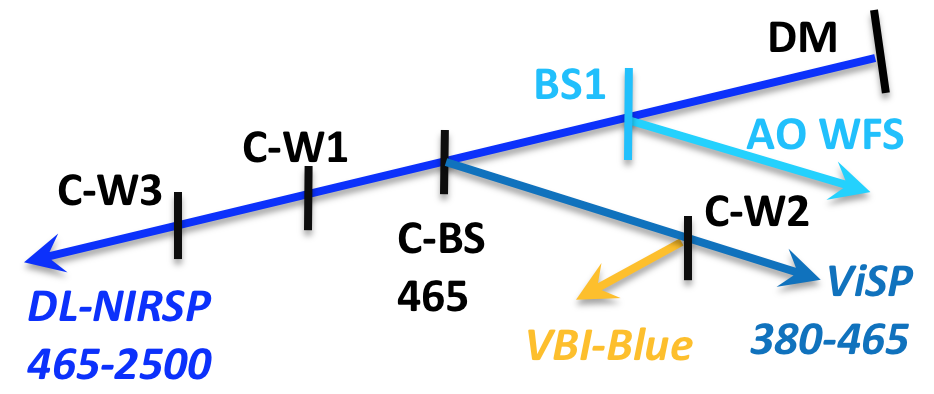}
}
\end{tabular}
\caption[] 
{\label{fig:fido_config1_cartoon}  The FIDO layout for Configuration 1. }
\vspace{-3mm}
\end{wrapfigure}

In this section we show example calculations of the DL-NIRSP system modulation matrix and group model terms given the uncertainties in coatings. We use the F/ 24 configuration and describe the impact of incomplete knowledge of coating properties  in computing expected Mueller matrices for the optics. In Table \ref{table:dl_fido_config1} we show which beam splitters and mirrors would be installed at which stations. We choose a setup with maximum transmission in the wavelength range 465 nm to 1800 nm allowing DL-NIRSP to cover the full wavelength range within its capabilities. The DL-NIRSP sees high transmission through the windows and the dichroic beam splitter C-BS-465. In the last two stations, CL3 and CL4, we put in a window with anti-reflection coatings on both sides. 

\begin{wraptable}{l}{0.22\textwidth}
\vspace{-3mm}
\caption{FIDO Config1}
\label{table:dl_fido_config1}
\centering
\begin{tabular}{l l}
\hline
\hline
Station		& Optic		\\
\hline
CL2			& C-BS-465	\\
CL3			& C-W1		\\
CL4			& C-W3		\\
\hline
CL2a		& C-W2		\\
\hline
\hline
Camera		& $\lambda$ (nm)		\\
\hline
DL-VIS		& 530		\\
DL-NIR1		& 1083		\\
DL-NIR2		& 1565		\\
\hline
ViSP1		& 393		\\
ViSP2		& 455		\\
\hline		
VBI-Blue		& 393		\\
\hline
\hline
\end{tabular}
\vspace{-4mm}
\end{wraptable}

As a demonstration, we chose to install the FIDO window with an uncoated front surface C-W2 in the optical station CL2a.  This window would send the $\sim$4\% Fresnel surface reflection to VBI-Blue so that instrument could perform its function as a context imager at wavelengths of 380 nm to 465 nm for the ViSP spectrograph slit. The beam transmitted through C-W2 would contain significant flux for wavelengths shorter than 465 nm due to the dichroic reflection off C-BS-465. ViSP can be configured to observe wavelengths shorter than this, subject to the constraint on physical spacing between the camera arms. With the anti-reflection coated window C-W1 in station CL3, the beam going to VTF and the VBI-red systems would have very little flux rendering these systems undesirable for use in this configuration. In Figure \ref{fig:fido_config1_cartoon}, we show a simplified cartoon of the FIDO setup and the wavelength ranges for the various instruments.

In the lower half of Table \ref{table:dl_fido_config1} we also show an example of some camera configurations possible with this setup. For this particular observing setup, the ViSP and the blue arm of VBI receiving 4\% of the flux would both be able to simultaneously observe in a few relevant channels.

\begin{wrapfigure}{r}{0.52\textwidth}
\centering
\vspace{-4mm}
\begin{tabular}{c} 
\hbox{
\hspace{-0.8em}
\includegraphics[height=6.5cm, angle=0]{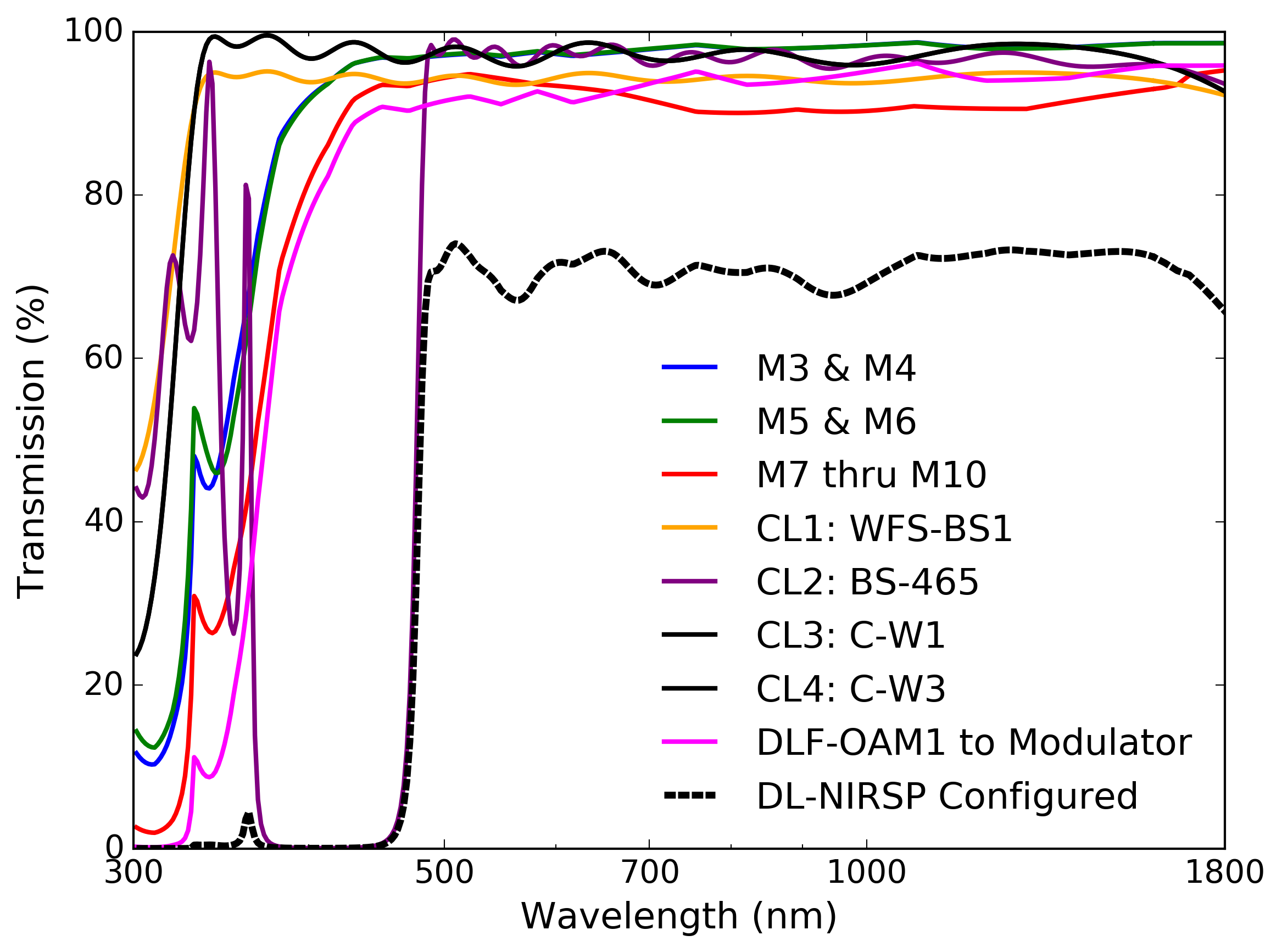}
}
\end{tabular}
\caption[] 
{\label{fig:dl_transmission_FIDO465_windows}  The DL-NIRSP system transmission in this configuration for all mirror groups beyond M1 \& M2: DKIST M3 to M10, BS1, FIDO optics and instrument feed mirrors. }
\vspace{-3mm}
\end{wrapfigure}

We choose two example wavelengths for the Ca II UV line at 393 nm and the Ba II line at 455 nm as a convenient second wavelength.  The ViSP is flexibly configured so several alternate choices are possible.  We list the three cameras in DL-NIRSP with three example wavelengths. We chose the shortest wavelength presently planned for the visible channel at 530 nm.  We note that alternates include 587 nm, 630 nm, 789 nm, 854 nm which can be easily observed in this FIDO configuration.  Similarly the two near infrared cameras could also observe 1075 nm, 1080 nm or 1430 nm. 

In Figure \ref{fig:dl_transmission_FIDO465_windows}, we show the transmission functions of all mirror groups and also all coud\'{e} optics. We use the silver refractive indices that are closer to the measured mirror reflection values for this calculation.  The WFS-BS1 optic is assumed to have a bare fused silica front surface reflection computed directly from the Fresnel equations.  The back surface is coated with WBBAR1 as described in Section \ref{sec:wbbar1}.  We perform the same calculation for the FIDO windows C-W1 and C-W3 where both surfaces are coated with WBBAR1.  The FIDO dichroic C-BS-465 utilizes the transmission from our best fit TFCalc model on the front surface and the WBBAR1 coating on the back surface. 

\begin{wrapfigure}{r}{0.56\textwidth}
\centering
\vspace{-5mm}
\begin{tabular}{c} 
\hbox{
\hspace{-0.6em}
\includegraphics[height=6.99cm, angle=0]{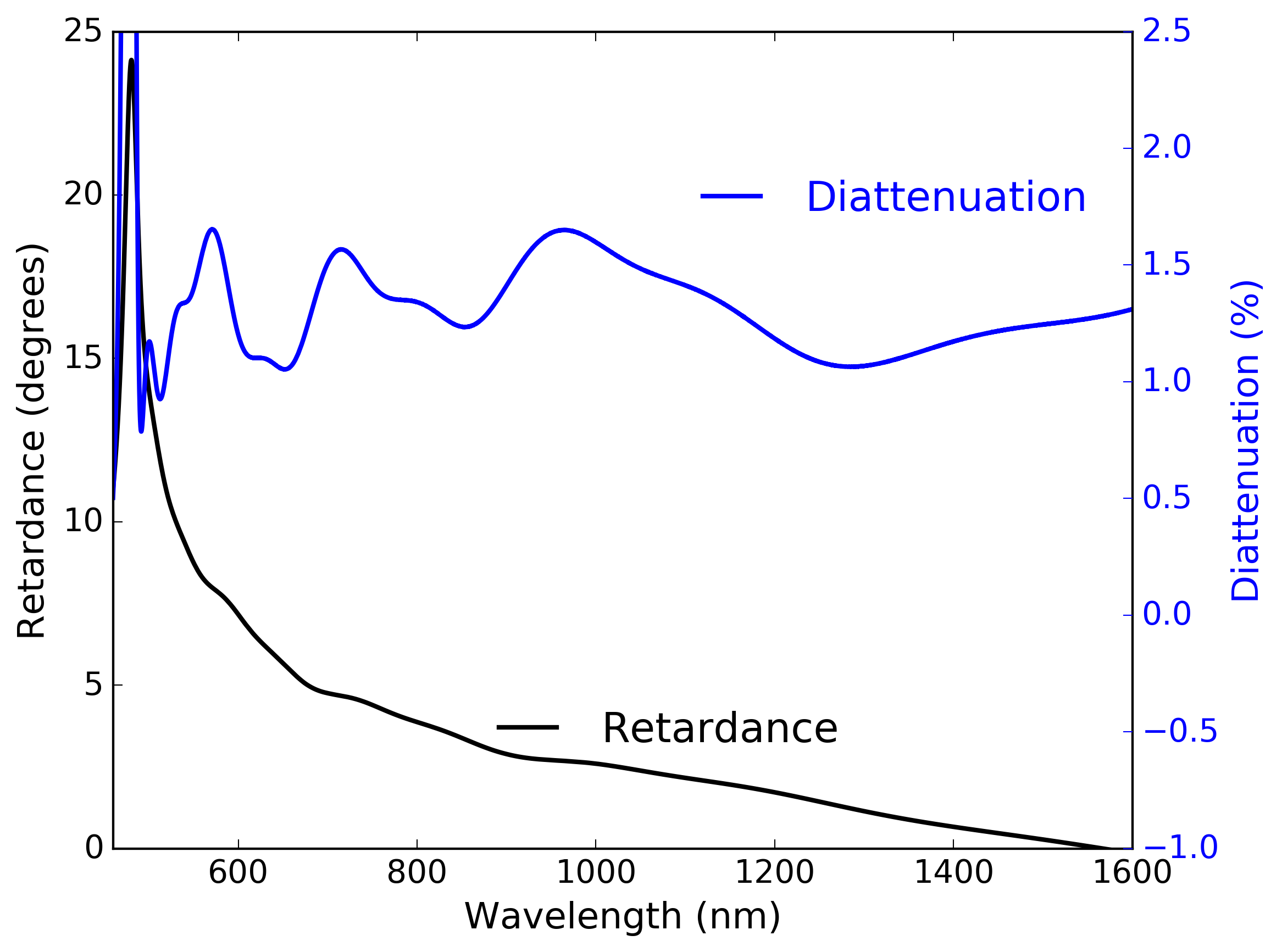}
}
\end{tabular}
\caption[] 
{\label{fig:fido_config1_ret_dia}  The retardance in black and diattenuation in blue caused by the transmissive optics WFS-BS1, C-BS-465, C-W1 and C-W3 combined as installed for FIDO configuration 1. }
\vspace{-3mm}
\end{wrapfigure}

We show in Figure \ref{fig:fido_config1_ret_dia} the retardance and diattenuation for propagation through the FIDO optics in configuration 1 as well as the WFS-BS1. The diattenuation is a relatively small 1\% to 2\% across the wavelength range transmitted to DL-NIRSP.  The retardance is 10$^\circ$ at 530 nm wavelength caused almost entirely by the front surface dichroic coating of C-BS-465.

The diattenuation in the DL-NIRSP modulation matrix is a roughly equal combination of the transmissive optics in FIDO and the WFS from Figure \ref{fig:fido_config1_ret_dia} along with the three high incidence angle fold mirrors within the instrument relay optics.  For the retardance, the three high incidence mirrors in the DL-NIRSP optics dominate with magnitudes up to 100$^\circ$ varying strongly across the visible and near infrared wavelength region.  The FIDO optics are less than 10$^\circ$ at most wavelengths, and are a very small portion of the system retardance. The DKIST telescope mirror retardance from Figure \ref{fig:group_model_retardance_diattenuation} is also significant at roughly one third this magnitude, and are also time dependent.  Diattenuation of the DKIST mirror groups has a magnitude similar to the combined FIDO optics and DL-NIRSP feed optics before the modulator. 

In summary, the telescope, FIDO optics and DL-NIRSP feed mirrors give roughly 70\% throughput from 525 nm to 1600 nm wavelength.  The DL-NIRSP feed optics are the dominant source of retardance at magnitudes up to 100$^\circ$. The diattenuation is roughly equal combinations of feed mirrors and FIDO optics in this configuration but with combined magnitudes less than several percent.


\clearpage
\subsection{DKIST \& FIDO Configured for Multi-Instrument Spectropolarimetry}
\label{sec:fido_many}

We consider in this section a configuration for the FIDO optics that operates all cameras in all three AO-assisted DKIST polarimetric instruments. 

\begin{wraptable}{l}{0.22\textwidth}
\vspace{-3mm}
\caption{FIDO Config 2}
\label{table:dl_fido_config3maxpol}
\centering
\begin{tabular}{l l}
\hline
\hline
Station		& Optic		\\
\hline
CL2			& C-BS-555	\\
CL3			& C-BS-680	\\
CL4			& C-W3		\\
\hline
CL2a		& C-W2		\\
\hline
CL3a		& C-BS-643	\\
\hline
\hline
\end{tabular}
\vspace{-4mm}
\end{wraptable}

\begin{wrapfigure}{r}{0.52\textwidth}
\centering
\vspace{-4mm}
\begin{tabular}{c} 
\hbox{
\hspace{-0.9em}
\includegraphics[height=3.8cm, angle=0]{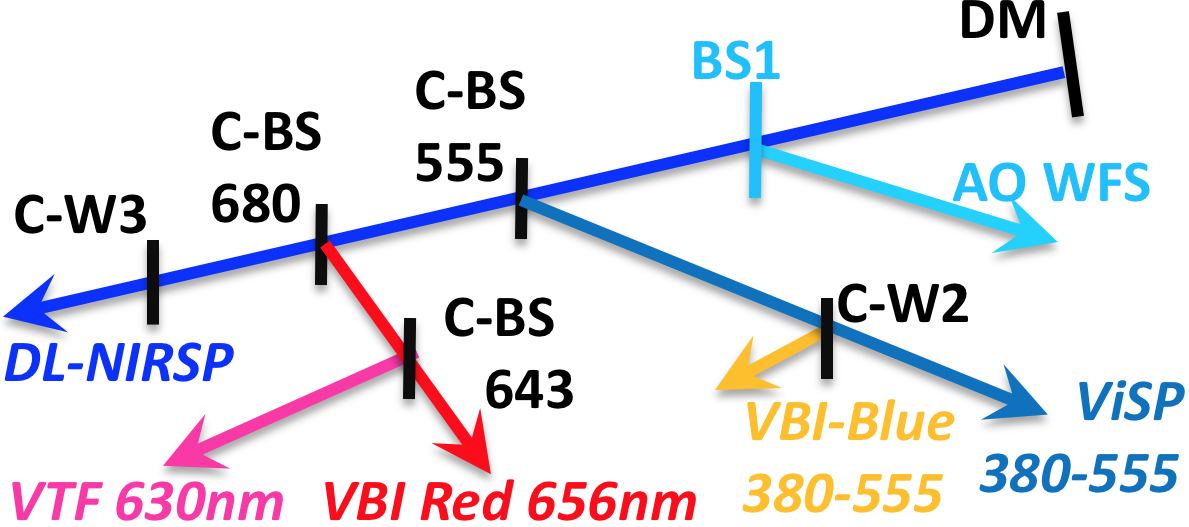}
}
\end{tabular}
\caption[] 
{\label{fig:fido_config3maxpol_cartoon}  The FIDO layout for a configuration where all polarimetric cameras are operated simultaneously. }
\vspace{-2mm}
\end{wrapfigure}

In Figure \ref{fig:fido_config3maxpol_cartoon}, we show a schematic for which FIDO dichroic beam splitters and windows will be installed in which stations. The WFS-BS1 sends the 4\% Fresnel reflection from the uncoated fused silica window to the AO system. In station CL2 we mount dichroic C-BS-555 where we reflect wavelengths shorter than 555 nm toward ViSP and VBI-blue.  In station CL3 we mount dichroic C-BS-680 to reflect wavelengths between 555 nm and 680 nm towards VTF and VBI-red.  The final station CL4 has the window C-W3 with broad band anti-reflection coefficients on all surfaces for high transmission. We show the beam splitters and stations in Table \ref{table:dl_fido_config3maxpol}.

In the station CL2a, we install the window C-W2 which sends the 4\% Fresnel reflection from the uncoated front surface towards VBI-blue. This camera operates as a context imager for ViSP with limited flux. ViSP can be configured to operate three cameras over a diverse range of spectral lines.  In Table \ref{table:dl_fido_config3maxpol_cameras}, we arbitrarily choose the three ViSP camera wavelengths to cover the Ca II H and K lines at 393 nm, the H$_\beta$ line at 486 nm and the photospheric Fe I line at 525nm. In station CL3a, we install the dichroic C-BS-643 which would transmit 643 nm to 680 nm towards the VBI-red camera. The reflected beam would allow VTF to observe in the 555 nm to 643 nm wavelength range.  It has one filter for the H$_\alpha$ line at 656 nm in this wavelength range.

\begin{wraptable}{l}{0.22\textwidth}
\vspace{-3mm}
\caption{Cameras: 2}
\label{table:dl_fido_config3maxpol_cameras}
\centering
\begin{tabular}{l l}
\hline
\hline
Camera		& $\lambda$ (nm)		\\
\hline
ViSP1		& 393		\\
ViSP2		& 486		\\
ViSP3		& 525		\\
\hline
VTF			& 630		\\
\hline
DL-VIS		& 854		\\
DL-NIR1		& 1083		\\
DL-NIR2		& 1565		\\
\hline		
VBI-Blue		& 393		\\
VBI-Red		& 656		\\
\hline
\hline
\end{tabular}
\vspace{-4mm}
\end{wraptable}

In Table \ref{table:dl_fido_config3maxpol_cameras}, we list the VTF as observing this 630 nm line. We note that VTF technically operates 3 separate cameras. Two cameras are synchronized and operated as a dual-beam spectropolarimeter as the etalons scan the spectral line in steps of a few picometers. These two cameras can be compared to the spectrograph instrument calibrations as these cameras scan in wavelength to sample the spectral line.  The third camera is also synchronized with the two polarimetric cameras. This third camera is set to observe continuum wavelengths adjacent to the spectral line. These three cameras all operate at wavelengths very close to 630 nm and are treated as one wavelength range, similar to spectrograph calibrations for our purposes here. 

A highly transparent FIDO window C-W3 is installed in station CL4. The DL-NIRSP would thus receive all wavelengths from 680 nm to the cutoff of the Infrasil transmission near 3000 nm wavelength.  The anti-reflection coatings and dichroics all become fairly reflective for wavelengths longer than 1800 nm. We configure DL-NIRSP in Table \ref{table:dl_fido_config3maxpol_cameras} for observation at 854 nm using the visible wavelengths camera.  We set the two near infrared cameras to 1083 nm and 1565 nm.

\clearpage

\begin{wrapfigure}{r}{0.56\textwidth}
\centering
\vspace{-1mm}
\begin{tabular}{c} 
\hbox{
\hspace{-0.8em}
\includegraphics[height=6.99cm, angle=0]{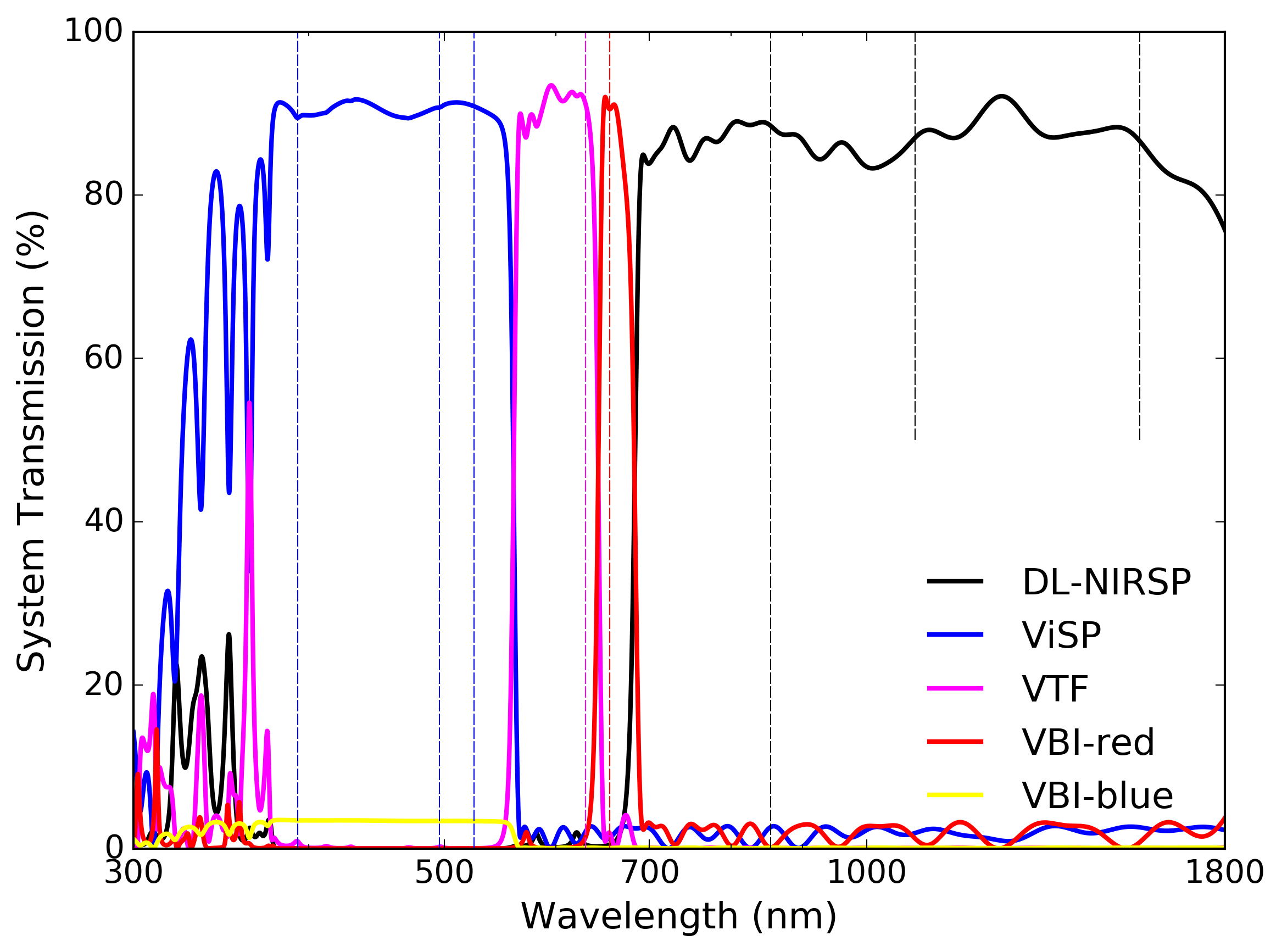}
}
\end{tabular}
\caption[] 
{\label{fig:fido_config3_transmission_coude_optics}  The flux through combined optics from BS1 the last FIDO element as combined and installed for FIDO configuration 3. The optical throughput delivered to the first feed optic of each sub-system is shown in different colors.}
\vspace{-3mm}
\end{wrapfigure}

Figure \ref{fig:fido_config3_transmission_coude_optics} shows the BS1 and FIDO optics throughput to the first instrument optic. We show throughput in the UV bandpass as this is important for calculations of optical degradation in polymer optics and the oils used in the crystal modulators. The ViSP feed is shown in blue receiving roughly 90\% of the flux from 390 nm to 540 nm wavelength. The VBI blue channel is shown in yellow receiving only a few percent of the flux over a similar bandpass. The VTF feed is shown in magenta with roughly 90\% of the flux in the 565 nm to 630 nm wavelength. The VBI red system is shown in red and receives 90\% of the flux at 656 nm wavelength to observe the H$_\alpha$ line.  The DL-NIRSP is shown in black receiving light at wavelengths longer than 690 nm. From Figure \ref{fig:fido_config3_transmission_coude_optics}, we can also see that no instrument sees more than a few percent of the flux at longer wavelengths outside the intended bandpass.  The VTF, DL-NIRSP and VBI-red channels do receive some UV flux at varying wavelengths depending on the imperfect UV reflectivity of the combined optics.

\begin{figure}[htbp]
\begin{center}
\vspace{-0mm}
\hbox{
\hspace{-0.6em}
\includegraphics[height=6.3cm, angle=0]{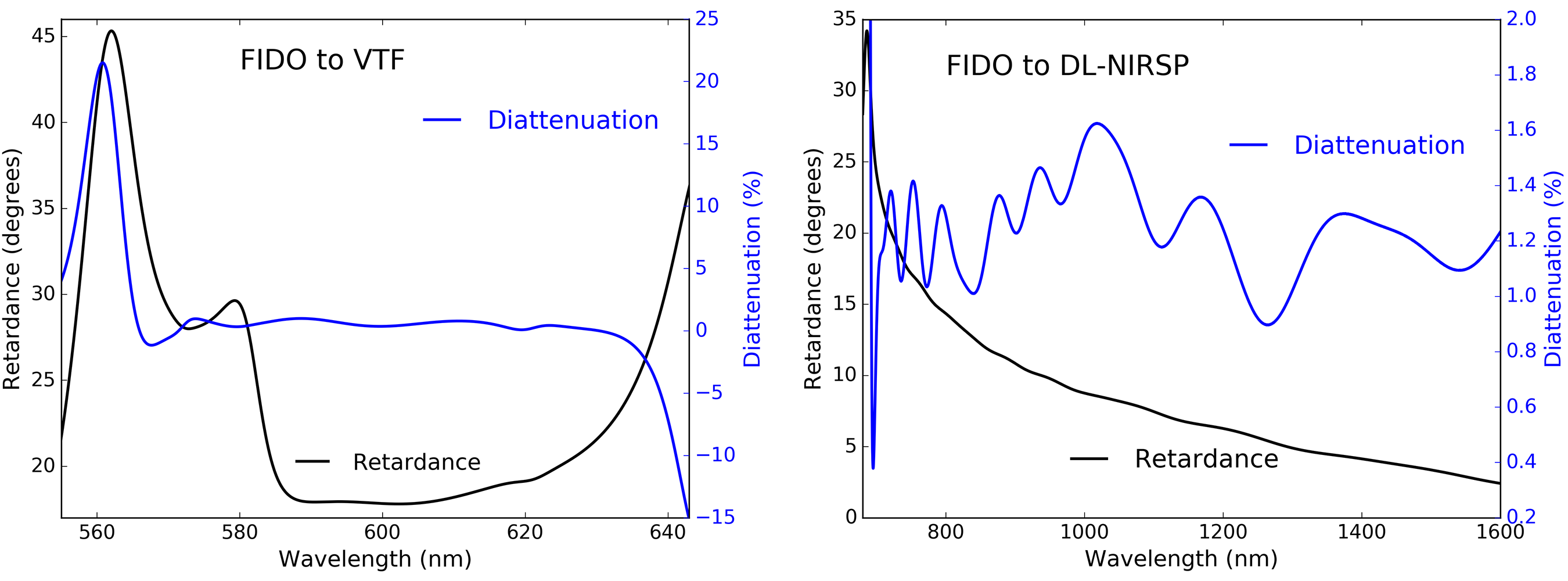}
}
\caption[] 
{ \label{fig:fido_config3_ret_dia} The retardance in black and diattenuation in blue for the beam through WFS-BS1 and FIDO optics C-BS-555, C-BS-643, C-BS-680 and windows C-W2 and C-W3 as combined for FIDO configuration 3. The left graphic shows the combined polarization response of the beam to VTF in the band for observation at 630 nm.  The right graphic shows the full  wavelength range available to DL-NIRSP in this configuration from 700 nm to 1600 nm. }
\vspace{-5mm}
\end{center}
\end{figure}

Figure \ref{fig:fido_config3_ret_dia} shows the retardance and diattenuation in the beam propagated through BS1 and the FIDO optics as configured for this use case. The left graphic shows the appropriate wavelength range for the beam to the first lens of the VTF.  In this configuration, the VTF would only be observing the 630 nm spectral line. The right hand graphic shows the beam sent to the DL-NIRSP first fold mirror DL-FM1.  We listed three wavelengths of 854 nm, 1083 nm and 1565 nm.  However, this configuration could easily support other DL-NIRSP configurations for any wavelength longer than 690 nm.

The VTF mirrors have not yet been coated but the nominal specified coating is the same as the GREGOR telescope mirrors.  We show our NLSP measurements and two-layer model fits to this GREGOR coating in Section \ref{sec:sub_VTF_GREGOR_mirrors}. The VTF feed mirror polarization response will essentially come from the two combined fold mirrors at 45$^\circ$ incidence.  The $\sim$20$^\circ$ retardance and zero diattenuation from FIDO in Figure \ref{fig:fido_config3_ret_dia} will likely be only a small contribtion to the modulation matrix properties.  For DL-NIRSP, we draw the same conclusions as in the prior Section \ref{sec:fido1}.

\subsection{Summary: Mueller matrices of DKIST Instruments: FIDO \& Coud\'{e} Optics}

In Section \ref{sec:system_group_model}, we have presented models of the DKIST mirrors grouped together between rotation axes for the on-axis beam. We showed how typical manufacturing tolerance errors can impact reflectivity, retardance and diattenuation. Variation directly follows the layer thickness variation derived from two layer model fits to our NLSP measurements. We then showed examples of the additional polarization effects from mirror combinations in two DKIST instruments: ViSP and DL-NIRSP.  In the case of DL-NIRSP, the nominal DKIST silver formula coated on several high incidence angle mirrors creates significant variability in the expected polarization properties in response to manufacturing variability even though all mirrors were coated in the same run. Additionally, this instrument did not choose the nominal DKIST silver coating for some of their mirrors, complicating the modeling of the entire system. However, the DL-NIRSP mirrors measured in Appendix \ref{sec:sub_DKIST_DL_EMF_mirrors} showed their alternate coating is broadly similar to the DKIST silver in magnitude and spectral smoothness of polarization properties. For the ViSP instrument optics of Section \ref{sec:FeedOptics_visp}, the original specification of a highly enhanced 29-layer coating on the feed mirrors would have significantly complicated the spectral behavior. We estimated reflectivity, diattenuation and retardance using a combination of vendor reflectivity and diattenuation, vendor modeling and NLSP retardance measurements. We were able to predict the Mueller matrix of these combined optics had they not been stripped and re-coated with one of our nominal FIDO enhanced protected silver coatings. We note that the Cryo-NIRSP uses an all reflective feed by inserting a mirror (M9a) in front of the AO deformable mirror (DM) at a low incidence angle of 9$^\circ$ and with this same FIDO EAg coating. The two other Cryo-NIRSP feed mirrors are at very low incidence angles of 4$^\circ$ and 1$^\circ$. Thus, the polarization properties of the Cryo-NIRSP feed optics are substantially simplified compared to DL-NIRSP.  In addition, the VTF feed mirrors are essentially two fold mirrors around 45$^\circ$. They are clocked with respect to the coud\'{e} floor, introducing both $QV$ and $UV$ retardance in the frame of their analyzer.  But this is similar to DL-NIRSP, predictable with witness sample data, and easily calibrated. With this information, and a few more coating samples delivered in the near future, we have the Mueller matrix of every metal-coated optic in the telescope and on the coud\'{e} floor.

We then showed examples of how the dichroics of the FIDO could be configured for two use cases. The first configuration of Section \ref{sec:fido1} shows maximum transmission to DL-NIRSP with relatively simple and spectrally smooth Mueller matrices.  In Section \ref{sec:fido_many}, we add the complexity of operating polarimeters in all AO-assisted instruments simultaneously.  FIDO is configured for ViSP from 380 nm to 555 nm, the VTF at 630 nm and DL-NIRSP 680 nm to 1800 nm. The predicted throughput from the combined reflections and transmissions are shown along with the retardance and diattenuation expected for each system.  This process can be used for all DKIST instruments and updated in the future to reflect the as-built coating performance once the FIDO dichroics are procured and we have measured those as-coated witness samples in the near future.

\clearpage

\section{Summary}
\label{sec:summary}

The DKIST project must deliver a stable, polarization-calibrated beam to instruments in the coud\'{e} laboratory. The optical properties of coatings on each optic must be well known in order to predict performance as a function of field angle, telescope configuration and wavelength at DKIST spectral resolving powers of roughly 100,000.  The altitude and azimuth articulation of the telecope, coud\'{e} laboratory table rotation, adaptive optics beam splitter along with the windows, mirrors and dichroics of the Facility Instrument Distribution Optics (FIDO) all require measurement and modeling at spectral resolving powers above 10,000 to detect narrow spectral features in many-layer coatings. Dichroics and highly enhanced mirrors can have complex behavior over the $<$1 nm bandpasses of DKIST instrument suite as we showed above. The internal optics to each DKIST polarimeter also contribute significantly to the expected polarization performance of the system. We explored here the metrology tools and system models for DKIST and several instruments using coating witness samples, vendor models, and vendor data. We have measured at least one sample for every mirror within the telescope and coud\'{e} relay optics and presented several examples of instrument partner mirrors, beam splitter dichroics and window anti-reflection coatings. We extensively used our National Solar Observatory laboratory spectropolarimeter (NLSP) that simultaneously covers visible and near infra-red wavelengths outlined in Section \ref{sec:nlsp_data}. For this paper, we consider only the on-axis beam properties for DKIST noting that the field-dependence for the telescope is as we assessed previously \cite{2017JATIS...3a8002H}. In this prior work, we only used a single nominal silver and aluminum metal coating prescription.  With this new work, we can now identify realistic coatings for all mirrors, anti-reflection coatings, and dichroics that can easily be exported to optical modeling tools for use in predicting performance on powered optics and in articulated systems, as we showed in 2017 \cite{2017JATIS...3a8002H}. We presented an assessment of mirror polarization data from many samples measured with NLSP in Section \ref{sec:mirror_coating_fits}. We achieve statistical signal-to-noise ratios over 10,000 at spectral resolving power of a few hundred along with very high temporal stability over months. The reflective configuration is highly stable. Retardance measurement stability is better than 1$^\circ$ retardance after remounting of reflective samples over a year timespan with a complete optical realignment. 

We adapted the Berreman calculus\cite{1972JOSA...62..502B, 2014btfp.book.....M} to use in predicting and fitting a wide variety of polarization behavior of coatings.  Here we presented adaptation of the scripts to fit NLSP Mueller matrix elements for diattenuation and retardance for enhanced protected metallic mirror coatings. We show refractive index curves and associated materials thicknesses used to model diattenuation and retardance with high quality fits across visible and near-infrared wavelengths. This lab spectropolarimeter will also be critical for the DKIST project when assessing our upcoming FIDO dichroic and anti-reflection coatings. We assessed polarization performance on several DKIST anti-reflection coating witness samples in Section \ref{sec:wbbar1}. We also compared our results with vendor-provided models and reflectivity data from Infinite Optics. The data and model matches were quite good when vendor-provided materials and refractive index data were used in the models. 

We then assessed anti-reflection coatings and dichroic coatings that use many dielectric layers to create wavelength ranges of high reflection and high transmission in Section \ref{sec:dichroics}. We presented here example dichroic formulas and witness-sample measurements from Infinite Optics. We measured several witness samples with NLSP and compared the results to designs in the industry standard TFCalc software package as well as verified the output in our Berreman calculus scripts. The measurements of transmission, retardance and diattenuation agree quite well with the measurements. Fitting for the as-built coating layer thicknesses using transmission data was demonstrated with TFCalc. Polarization predictions using these revised coating designs gives good matches to our NLSP measurements. We showed examples of polarization properties and reflectivity changing over wavelength ranges smaller than a nanometer. These narrow features are a property of many-layer coatings and were highlighted in Section \ref{sec:sub_dichroic_spikes} and measured with spectral resolving power greater than 20,000. These narrow features can be significantly mitigated through design. We presented FIDO coating designs that follow this strategy for DKIST.

System level models for the DKIST telescope group model and some FIDO configurations were shown in Section \ref{sec:system_group_model}. The impact of manufacturing tolerances was assessed to show that metrology of witness samples is critical for accurate prediction of system performance. The polarization properties of several DKIST instruments were also assessed to create predictions of the instrument Mueller matrices. An example of a 29-layer protected silver mirror was shown for ViSP along with the spectral consequences for using such a coating in the instrument feed optics in Section \ref{sec:FeedOptics_visp}. Spectral features as narrow as the nanometer scale instrument bandpasses are present in this coating. These metrology efforts resulted in stripping and re-coating of the ViSP feed mirrors with a new, slightly less reflective but polarimetricaly smooth coating presented in Appendix \ref{sec:sub_DKIST_M9_ViSP_FiDO}. 

In Appendix  \ref{sec:DKIST_mirror_path}, we show the current coating status of every surface between the primary mirror and the instrument modulators. Many dichroics are not yet coated, but essentially all mirrors except the instrument VTF are now complete. In Appendix \ref{sec:refractive_index_and_fitting}, we show data on several commercial mirrors and witness samples used in DKIST along with fits to a range of two-layer coating material models. These largely fell into two groups of high-index over low-index and low-index over high-index, as is in standard coating recipe textbooks. Our models ignore several aspects of real coatings including use of strippable layers, fabrication processes, textured growth, refractive index variation with depth and layer, etc. However, the simple polarization models presented here for coating performance with wavelength and incidence angle can be quite useful for simulating system polarization performance. Using thin film modeling software such as TFCalc or optical modeling in Zemax or other ray-tracing programs can give predictions and assess manufacturing tolerances, even if the materials modeled do not correspond to those of the actual coating. 

There are several opportunities for extension of this work. We only present NLSP data at 45$^\circ$ incidence angles but are not limited to this single configuration. Adding variable incidence angles as well as common image rotator (K-cell, K-mirror) type reflectivity configurations shown in Appendix \ref{sec:refractive_index_and_fitting} provides a straightforward path for future work on fitting additional coating properties. We identified several variables with large impact such as metal layer complex refractive index and modified refractive index spectral behavior as well as assessing coating polarization model fidelity over a wide range of incidence angles (Section \ref{sec:mirror_fit_limitations} and Appendix \ref{sec:refractive_index_and_fitting}). Future work includes expanding the spatial variation of coatings described in Section \ref{sec:FIDO_spatial_variation} and the field-of-view dependence in our prior publication HS17\cite{Harrington:2018bt} to predict the polarization calibration limits of the system across the diverse wide-field scanning techniques intended for use at DKIST.

\section{Acknowledgements}

This work was supported by the DKIST project. The DKIST is managed by the National Solar Observatory (NSO), which is operated by the Association of Universities for Research in Astronomy, Inc. (AURA) under a cooperative agreement with the National Science Foundation (NSF). This research made use of Astropy, a community-developed core Python package for Astronomy (Astropy Collaboration, 2013). We acknowledge the community effort devoted to the development of the following open-source packages used in this work: numpy(numpy.org), matplotlib(matplotlib.org). We thank Thomas Kentischer, Lucia Kleint, Manolo Collados and the KIS staff for mirror samples and discussions about GREGOR mirror polarization. We thank Kwangsu Ahn at Big Bear Solar Observatory for assistance understanding their polarimeter optical performance and for the BBSO mirror samples.  We thank David Elmore for his assistance, guidance and insight into the long history of work on the DKIST project.  We thank Tom Schad, Sarah Jaeggli and Christian Beck for their ongoing discussions of polarization performance at various telescopes. This research has made use of NASA’s Astro-physics Data System Bibliographic Services.


\appendix

\section{DKIST Coating Status: Mirrors, Dichroics \& Windows}
\label{sec:DKIST_mirror_path}

Tables \ref{table:mirror_coating_formula_list} and \ref{table:mirror_coating_formula_list_instr} show the current status of DKIST mirrors and coatings in the DKIST optics and within the initial polarimetric instruments. We show the name of the various optics in column one. Column two shows their incidence angle, or a range of angles for powered optics. The third column shows the focal ratio (F/) for the beam at that optic. Optics denoted (OA) are off axis mirrors. The fourth column shows the coating name.  Column five names the coating vendor where available.  Column six shows the coating run number as many optics were combined into single coating shots.  The seventh column shows current status of witness samples and the quantity we have from each run. D.EAG denotes the specified enhanced protected silver within the DKIST project. Otherwise, EAg is short-hand for Enhanced protected silver (Ag).  Vendors include Infinite Optics, Inc. (IOI), Zygo Corporation and Dynasil's Evaporated Metal Films (EMF). The Air Force Research Labs (AFRL) has a coating chamber adjacent to DKIST on the summit of Haleakala used for the DKIST primary mirror. We also have samples from Tafelmaier used by the GREGOR solar telescope with the intended formula to be applied to optics not yet coated within the DKIST VTF instrument. In some cases, we did not receive sufficient documentation to specifically attribute a coating run to a specific optic.  For DL-NIRSP, we were able to test some of the small flat mirrors directly in NLSP. For the beam splitters in FIDO and the AO system, we show front side and back side coatings as -f and -b respectively.  We do not list the three FIDO windows described in Section \ref{sec:wbbar1} that will be coated with WBBAR1 or left uncoated.  We also do not list the five dichroic coatings of FIDO described in Section \ref{sec:dichroics} as they also have not yet been coated. We have currently completed several coating stress, uniformity and repeatability tests for the dichroic coating formulas and anticipate a more thorough study in the coming year. Bold font highlights coatings not yet performed, changes of vendors or special cases in which we do not have clearly identified samples. In some cases, we test the mirror itself.

\begin{table}[htbp]
\vspace{-0mm}
\caption{DKIST Mirrors, Incidence angles \& Coating Status}
\label{table:mirror_coating_formula_list}
\centering
\begin{tabular}{l l l l l l l}
\hline
\hline
Optic		& AOI		& Power		& Coating	 	& Provider		& Run		& Sample?		\\
\hline
\hline
M1		& 7.13 - 20.56	& F/2 OA		& Al			& AFRL		& 			& Yes (2)			\\
M1 spare	& 7.13 - 20.56	& F/2 OA		& Al			& AFRL		& 			& Yes (2)			\\
M2		& 6.03 - 17.27	& F/13 OA		& D.EAG		& --			& 13BE18		& Yes (2)			\\
M3		& 42.81 - 47.19	& Flat		& D.EAG		& --			& 14BE04 	& Yes 			\\
M4		& 0.93 - 2.57	& F/53 OA		& D.EAG		& --			& 15BA35		& Yes (2+1)		\\
M5		& 14.47 - 15.53	& Flat		& D.EAG		& --			& 12BD18		& Yes		\\
M5 spare	& 14.47 - 15.53	& Flat		& D.EAG		& --			& 12BD19		& Yes	 	\\
M6		& 29.47 - 30.53 & Flat		& D.EAG		& --			& 14BE05		& Yes		\\
M6 spare	& 29.47 - 30.53 & Flat		& D.EAG		& --			& 14BE04		& Yes	 	\\
M7 		& 44.47 - 45.53 & Flat		& D.EAG		& --			& 16BD16		& Yes		\\
M8		& 5.06 - 5.60	& F/53 OA		& bAG99		& EMF		& Unlabeled	& Yes (3)		\\
M9		& 10			& Flat		& EAg1-450 	& IOI			& 9-3095		& Yes (3)		\\
\hline
\hline
M10	DM	& 15			& Flat		& D.EAG		& --			& 15BA23		& Yes (4)	\\
WFS-BS1-f& 15		& Flat		& none		& --			& 			& Uncoated		\\
WFS-BS1-b& 14.5		& Flat		& WBBAR1	& IOI			& 10-0233		& {\bf Coming}  		\\
\hline
\hline
FIDO 	& Mirror		&			&			&			&			&		\\
\hline
C-M1	& 15			& Flat		& EAg1-420 	& IOI			& 6-7766		& Yes, 3		\\
C-M2	& 15			& Flat		& EAg1-450 	& IOI			&			& {\bf Not Yet Coated}		\\
\hline \hline 
FIDO 	& Dichroic		&			&			&			&			&		\\
\hline
C-BS-465-f& 15		& Flat		& Dich465 	& IOI			&			& {\bf Not Yet Coated}		\\
C-BS-465-b& 14.5		& Flat		& WBBAR1 	& IOI			&			& {\bf Not Yet Coated}		\\
C-BS-555-f& 15		& Flat		& Dich555 	& IOI			&			& {\bf Not Yet Coated}		\\
C-BS-555-b& 14.5		& Flat		& WBBAR1 	& IOI			&			& {\bf Not Yet Coated}		\\
C-BS-643-f& 15		& Flat		& Dich643 	& IOI			&			& {\bf Not Yet Coated}		\\
C-BS-643-b& 14.5		& Flat		& WBBAR2 	& IOI			&			& {\bf Not Yet Coated}		\\
C-BS-680-f& 15		& Flat		& Dich680 	& IOI			&			& {\bf Not Yet Coated}		\\
C-BS-680-b& 14.5		& Flat		& WBBAR2 	& IOI			&			& {\bf Not Yet Coated}		\\
C-BS-950-f& 15		& Flat		& Dich950 	& IOI			&			& {\bf Not Yet Coated}		\\
C-BS-950-b& 14.5		& Flat		& WBBAR2 	& IOI			&			& {\bf Not Yet Coated}		\\
\hline
FIDO 	& Window		&			&			&			&			&		\\
\hline
C-BS-W1-f& 15			& Flat		& WBBAR1 	& IOI			&			& {\bf Not Yet Coated}		\\
C-BS-W1-b& 14.5		& Flat		& WBBAR1 	& IOI			&			& {\bf Not Yet Coated}		\\
C-BS-W2-f& 15			& Flat		& none	 	& 			&			& Uncoated				\\
C-BS-W2-b& 14.5		& Flat		& WBBAR1 	& IOI			&			& {\bf Not Yet Coated}		\\
C-BS-W3-f& 15			& Flat		& WBBAR1 	& IOI			&			& {\bf Not Yet Coated}		\\
C-BS-W3-b& 14.5		& Flat		& WBBAR1 	& IOI			&			& {\bf Not Yet Coated}		\\
\hline
\hline
\end{tabular}
\vspace{-4mm}
\end{table}

\clearpage

\begin{table}
\vspace{-3mm}
\caption{DKIST Instrument Mirrors, Incidence angles \& Coatings}
\label{table:mirror_coating_formula_list_instr}
\centering
\begin{tabular}{l l l l l l l}
\hline
\hline
Optic		& AOI		& Power		& Coating	 	& Provider		& Run		& Sample?		\\
\hline
\hline
		&			&			&			&			&			&		\\
\hline \hline
{\bf ViSP}		& 		 	& 			& 			&			&			&		\\
\hline \hline
V-FoldM1	& 28			& Sph		& EAg1-420	& IOI			& 6-7767		& Yes, 3	 \\
V-Feed1	& 2.2			& Sph		& EAg1-420	& IOI			& 6-7766		& Yes, 3	 \\
V-Feed2	& 12.3		& Sph		& EAg1-420	& IOI			& 6-7767		& Yes, 3	 \\
Slit glass-f	& 5.4			& Flat		& BBAR		& ECI		& Unknwn		& No, graphed	\\
Slit glass-b& 5.4		& Flat		& BBAR		& ECI		& Unknwn		& No, graphed	\\
V-FoldM2	& 47.7		& Flat		& EAg		& RMI		& Y31071216	& No		\\ 
\hline
Modulator	& 0			& Flat		& MgF$_2$	& MLO		& 			& Yes	 \\	
\hline 
V-FoldM3	& 45			& Flat		& EAg		& RMI		& Y31121216	& No		\\
V-FoldM4	& 45			& Flat		& EAg		& RMI		& Y31121216	& No		\\
\hline
{\it ViSP}	& {\it Old}		& {\it Stripped}	& 29 Layer	& 			& 			& Yes, 2	\\
\hline \hline
		&			&			&			&			&			&		\\
		&			&			&			&			&			&		\\
\hline \hline
{\bf Cryo-NIRSP}&			&			&			&			&			&		\\
\hline \hline
M9a		& 9			& Flat		& EAg	 	& {\bf Zygo}	& G4194725	& Yes, 5 (Unif)		\\
Scan		& 4			& Flat		& D.EAG		& --			& 16BB07		& Yes, FM1, CN-OPT-0001	\\
Foc		& 1.13		& OAP		& D.EAG		& --			& 16BB21		& Yes, FM2, CN-OPT-0002	\\
\hline
Modulator	& 0			& Flat		& None		& MLO		& --			& --			\\
\hline
Slit 		&			&			&			&			&			&	\\
Filter		&			&			&			&			&			&	\\
SM2 Fold1& 5.5		& F/18 Flat	& D.EAG		& --			& 16BB07		& Yes, CN-OPT-0003		\\
SM3 Col	& 5.5			& F/18 OAH	& D.EAG		& --			& 16BB21		& Yes, CN-OPT-0004		\\
SM4	Fold2& 7			& Flat		& D.EAG		& --			& 16BB07		& Yes, CN-OPT-0005		\\
Grating	& Var		& Flat		& Al			& Newport		& --			& 					\\
SM5 Cam	& 7.7			& F/8 OAE	& D.EAG		& --			&{\it 16BD15?}	& Yes, CN-OPT-0006	\\
\hline \hline
		&			&			&			&			&			&		\\
		&			&			&			&			&			&		\\
\hline \hline
{\bf DL-NIRSP}& 		 	& 			& 			&			&			&		\\
\hline \hline
DL-FM1	& 45			& Flat		& EAg		& {\bf Zygo}	& G4194726	& Yes, 1 (\& 5 test -28)		\\
DLF-OAM1& 7			& OAh		& D.EAG		& --			& 16BE17		& Yes, F00-102, 400mm	\\
DLF-FM2	& 45			& Flat		& D.EAG		& --			& 16BB22		& {\bf No}	But Verify DL-207	 \\	
DLF-FM3	& 45			& Flat		& D.EAG		& --			& 16BB22		& {\bf No} But Verify DL-207	 \\	
DLF-FSM	& 3.8			& Sph		& D.EAG		& --			& {\bf Unknown}& {\bf No} F00-105, 220mm	\\
\hline
{\bf F/ 24 \& F/ 8} &			&			& 			&			&			&	 	\\
\hline
DLF-MF24& 4.4		& OAe		& D.EAG		& --			& 16BE16		& Yes, F00-106, 250mm \\	
DLF-M4	& 49.7		& Flat		& D.EAG		& --			& 16BB22		& {\bf No} But Verify DL-207	 \\	
\hline
Modulator	& 0			& Flat		& BBAR		& IOI			& 12-6523		& Yes  \\	
\hline
{\bf F/62}		&			& 			&			&			&			& 	\\
\hline
DLF-MF62-1& 3.3		& OAe		& EAg		& {\bf Unkn}	& {\bf Unknown}&  {\bf No}, F00-108, 230mm \\	
DLF-MF62-2& 4.9		& OAe		& D.EAG		& --			& 16BE17		& Yes, F00-109 \\	
DLF-M4	& 45.9		& Flat		& D.EAG		& --			& 16BB22		& {\bf No}	But Verify DL-207	\\	
\hline
Modulator	& 0			& Flat		& BBAR		& IOI			& 12-6523		& Yes \\	
\hline
\hline
		&			&			&			&			&			&		\\
		&			&			&			&			&			&		\\
\hline \hline
VTF		& SAMP		&			& EAG		&Tafelmaier	&			& Yes	\\
\hline \hline
\end{tabular}
\vspace{-4mm}
\end{table}

\clearpage

\section{NLSP Calibration \& Optical Stability}
\label{appendix:nlsp_cal}

\begin{figure}[htbp]
\begin{center}
\vspace{-1mm}
\hbox{
\hspace{-0.8em}
\includegraphics[height=10.9cm, angle=0]{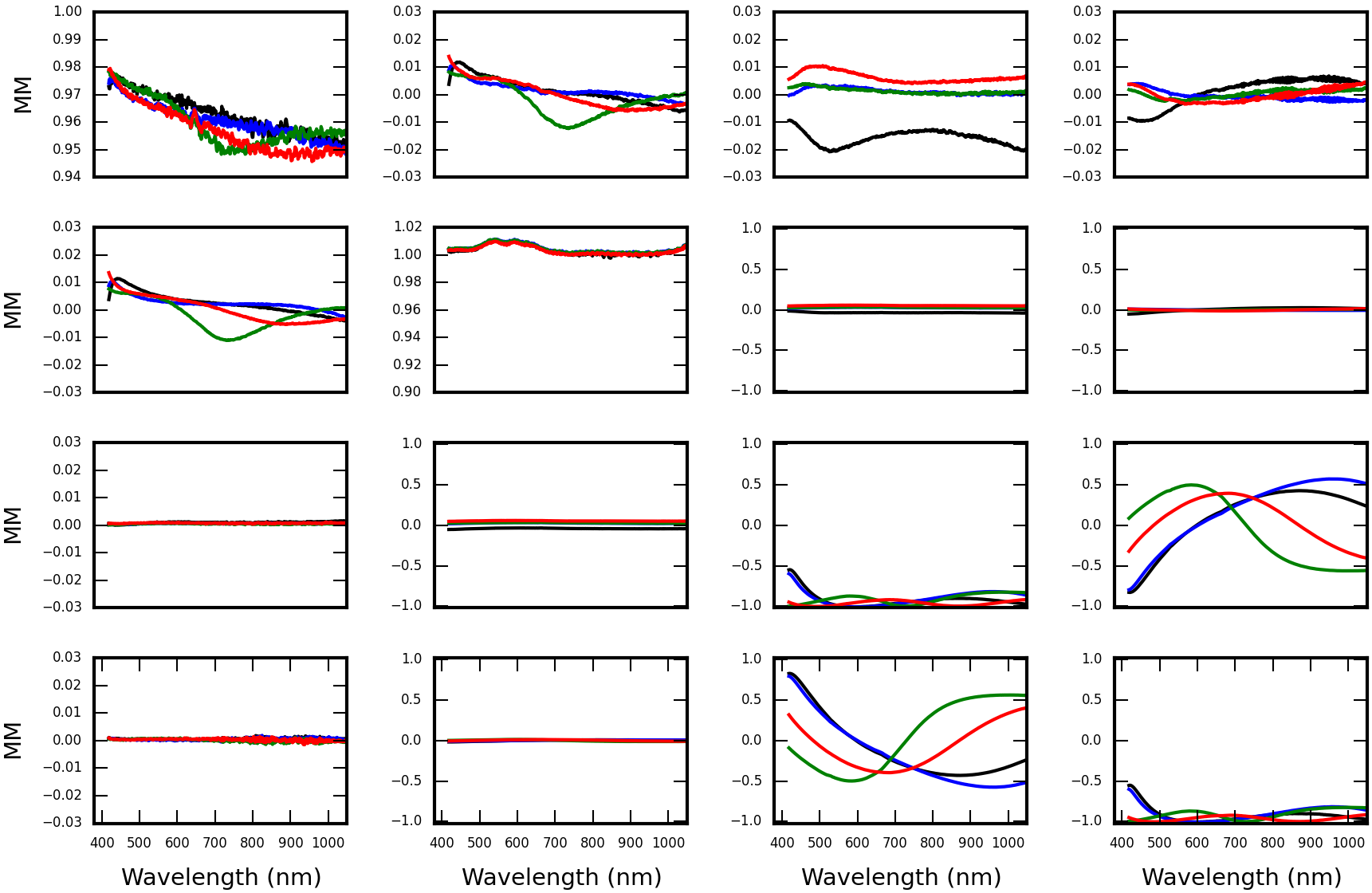}
}
\caption[] 
{ \label{fig:nlsp_measured_mueller_matrix}  The NLSP measured Mueller matrix for a DKIST enhanced protected silver mirror sample as well as the three Infinite Optics mirror samples tested for the DKIST FIDO optics. Reflection was at 45$^\circ$ incidence angle and we show only visible spectrograph data here for clarity. Red shows the DKIST enhanced protected silver mirror. Black shows the IO EAG300 5-5033 sample. Green shows the IO EAG700 8-6282 sample. Blue shows the IO EAG450 8-6898 sample.  The data has been normalized as in Equation \ref{eqn:MM_IntensNorm}. }
\vspace{-5mm}
\end{center}\end{figure}

The NLSP measured Mueller matrix using the visible spectrograph for four samples are shown in Figure \ref{fig:nlsp_measured_mueller_matrix}. A DKIST silver witness sample from one coating of the telescope feed optics is shown in red. Witness samples from Infinite Optics tested as part of the FIDO mirror process are also shown in black, green and blue. These witness samples used a few different materials and were provided for polarization comparison between NLSP, the Infinite Optics metrology equipment and theoretical calculations. We note that we use the industry standard Thin Film Calculator (TFCalc), the Zemax optical design software as well as our Python-based Berreman calculus scripts, which all agree to numerical precision when we have checked against known coatings and crystals\cite{Harrington:2018bt,Harrington:2017jh,2017JATIS...3a8002H}.

\begin{figure}[htbp]
\begin{center}
\vspace{-3mm}
\hspace{-0.8em}
\hbox{
\includegraphics[height=6.3cm, angle=0]{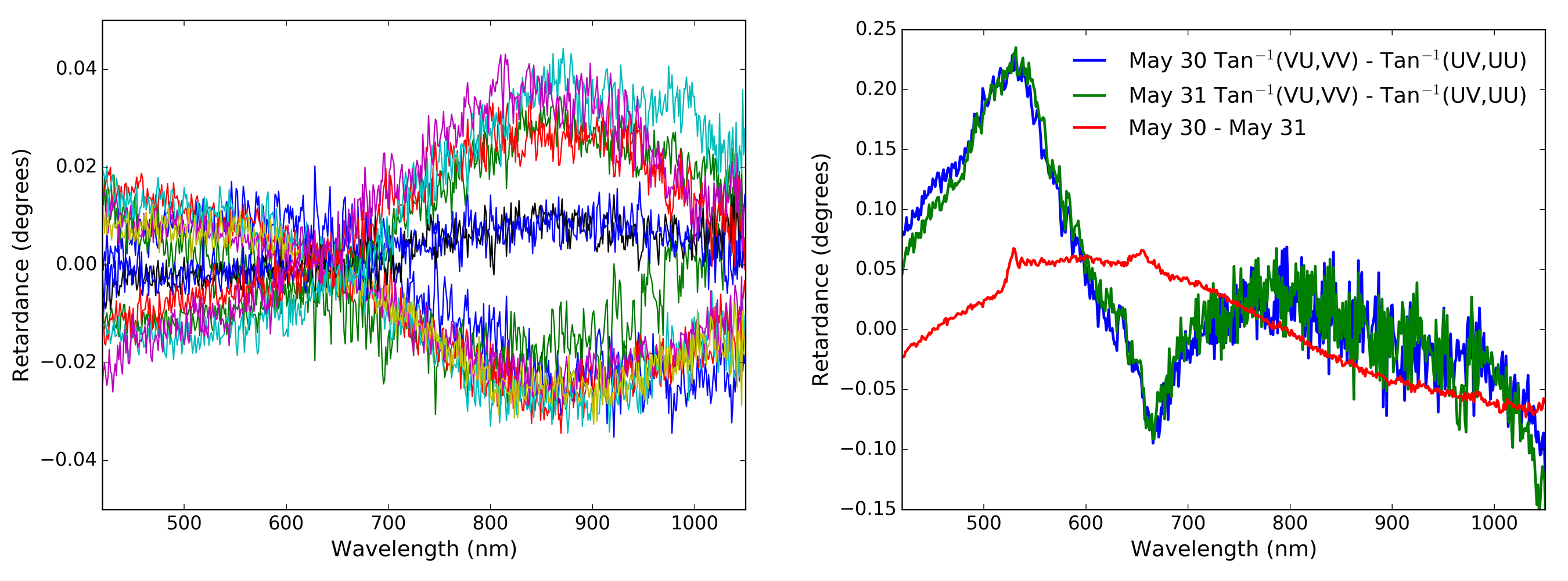}
}
\caption[] 
{ \label{fig:quantum_retardance_change_with_time} The left panel shows retardance changes between data sets taken on the same day. Variation is less than 0.05$^\circ$ retardance.  The right panel shows a comparison of retardance measurements taken on May 30th and May 31st.  The change between days is shown as the red curve and is less than 0.06$^\circ$ peak to peak. However, the biggest systematic difference is in the Mueller matrix being not entirely physical.  For retardance, the $UU, UV, VU, VV$ elements should all be $\sin$ or $\cos$ of the retardance magnitude. As such, any estimate computed as an inverse tangent of the appropriate Mueller matrix elements should give an identical retardance value. The retardance values estimated using $\arctan(VU,VV)$ does not agree with $\arctan(UV,UU)$ at amplitudes of up to 0.2$^\circ$ retardance. This is a small disagreement but it does show that systematic errors are a few times larger than repeatability errors. }
\vspace{-4mm}
\end{center}
\end{figure}

The NLSP reflective setup gives retardance results consistent within roughly one degree when perturbed by remounting samples over days to a year. We frequently re-mount samples to repeat measurements using various calibrations as a test of our systematic error levels. As an example, Figure \ref{fig:quantum_retardance_change_with_time} shows retardance measurements made after unmounting and remounting the sample 24 hours later. The left hand graphic shows the difference between the fit retardance values. Different curves show the changes between various inverse tangent methods of computing retardance from the Mueller matrix. The beam footprints are not identical on the optic, nor is the optical alignment guaranteed to be exact. The individual estimates of retardance agree between subsequent remounting at levels of 0.05$^\circ$ or below. In the right hand plot, we see the difference between the retardance estimates using different Mueller matrix elements in an inverse tangent computation. The theoretical Equation \ref{eqn:flat_mueller_matrix} says the various inverse tangents of the various $UV$ elements should produce identical retardance values. However, we see variation at amplitudes of $\pm$0.15$^\circ$ retardance with an offset of roughly 0.05$^\circ$ on average as the green and blue curves in the right hand graphic of Figure \ref{fig:quantum_retardance_change_with_time}. We also include in that graphic the difference between May 30 and May 31 measurements as the red curve, which is a factor of a few less than the error between the different inverse tangent estimates. The alignment procedures produce highly repeatable measurements at levels less than one tenth of a degree.

Strong arguments can be made about requiring any data-derived Mueller matrix  to be physical by imposing a process  to relate the measured matrix to be the nearest physical Mueller matrix using an appropriate distance metric \cite{Givens:1993cl, Kostinski:1993uh, Chipman:2006iu,Twietmeyer:2005bz,2003SPIE.5158..184C,1990JOSAA...7..693G,Chipman:2010tn,1996SPIE.2873....5C,Noble:2011wx,2012OExpr..20...17N,2012ApOpt..51..735N,2000SPIE.4133....1C}. Our matrix measurements are stable and produce retardance measurements that vary by less than 0.2$^\circ$ when comparing different inverse tangent estimates of the UV, VU, VV and UU elements. For the sake of simplicity, we choose the average of the inverse tangent estimates of retardance and proceed with the analysis. As seen in the right graphic of Figure \ref{fig:quantum_retardance_change_with_time}, the two estimates we chose sample all four $U$ and $V$ Mueller matrix cross talk elements. We average the two retardance estimates as a very straightforward, simple procedure.

\begin{figure}[htbp]
\begin{center}
\vspace{-0mm}
\hbox{
\hspace{-0.7em}
\includegraphics[height=6.3cm, angle=0]{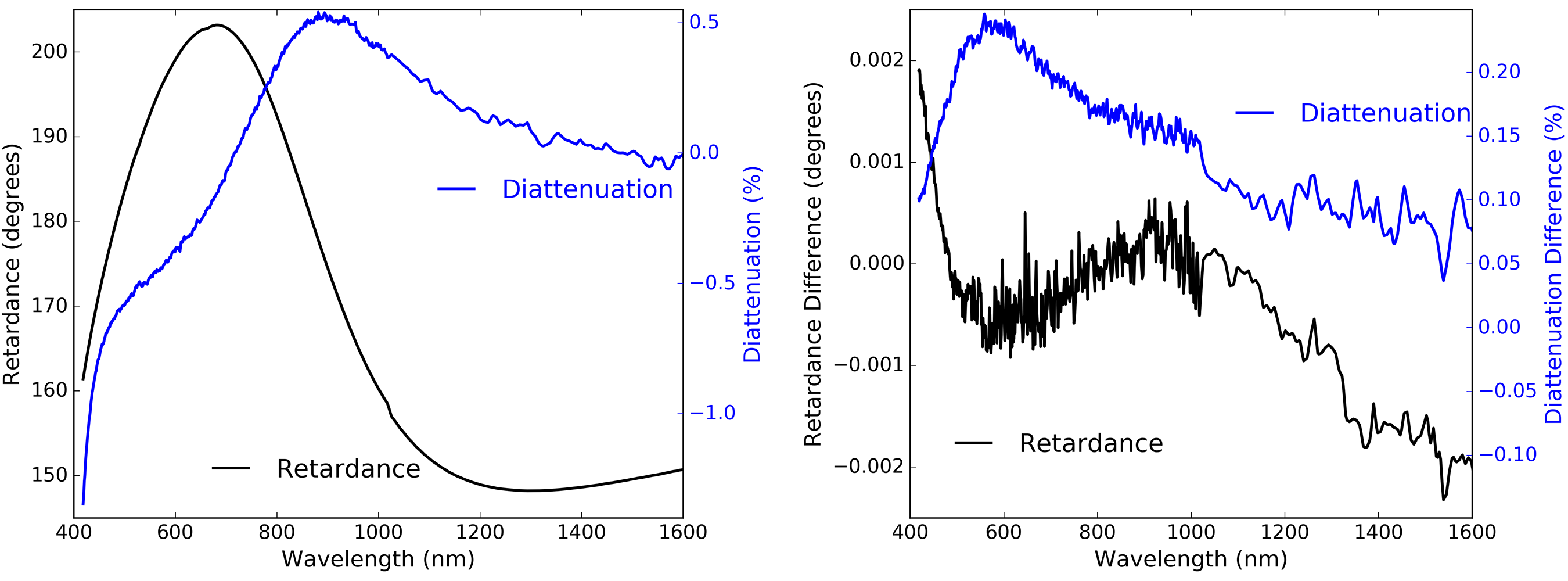}
}
\caption[] 
{\label{fig:quantum_may31_nlsp_data}   The left graphic shows NLSP processed retardance and diattenuation measured at 45$^\circ$ incidence angle with both VIS and NIR spectrographs for a DKIST enhanced protected silver witness sample. NLSP data was recorded from 380 nm to 1650 nm. The VIS and NIR spectrograph data sets showed good overlap and were subsequently stitched together at a cross-over wavelength of 1020 nm. The right hand graphic shows the difference between retardance and diattenuation when internal calibrations from three months prior were used instead of same-day calibrations. Retardance is in black on the left hand y axis. Diattenuation is in blue on the right hand y axis.}
\vspace{-5mm}
\end{center}
\end{figure}

Retardance and diattenuation measurements are very stable over months with statistical signal to noise ratios over 10,000. Figure \ref{fig:quantum_may31_nlsp_data} shows an example measurement of retardance and diattenuation for an enhanced protected silver mirror coating witness sample in May of 2017 in the left graphic. The retardance and diattenuation are derived from the NLSP measured Mueller matrix of a DKIST enhanced protected silver witness sample reflecting at 45$^\circ$ incidence angle. A slight discontinuity can be seen at 1020 nm wavelength as well as a change in the statistical noise properties. This is the wavelength we have chosen to switch from visible to near infrared data sets. The black curve shows the retardance in the range of 150$^\circ$ to over 200$^\circ$ where a retardance free mirror would give a phase change of 180$^\circ$. The blue curve shows the diattenuation in the range -1.4\% to +0.5\%.  

The right hand graph of Figure \ref{fig:quantum_may31_nlsp_data} shows the difference between retardance and diattenuation when internal calibrations are used from May to calibrate a July data set, compared with calibrations taken within 24 hours of measurement in July. The input retardance ranges over 50$^\circ$ over this bandpass but the results are repeatable to better than 0.002$^\circ$ peak to peak. Diattenuation results vary more but are still stable. We see -1\% to +0.5\% diattenuation in this sample but using different calibrations the results are offset by roughly 0.15\% and range spectrally from this offset by 0.2\%.

\begin{figure}[htbp]
\begin{center}
\vspace{-0mm}
\hspace{-0.8em}
\hbox{
\includegraphics[height=6.3cm, angle=0]{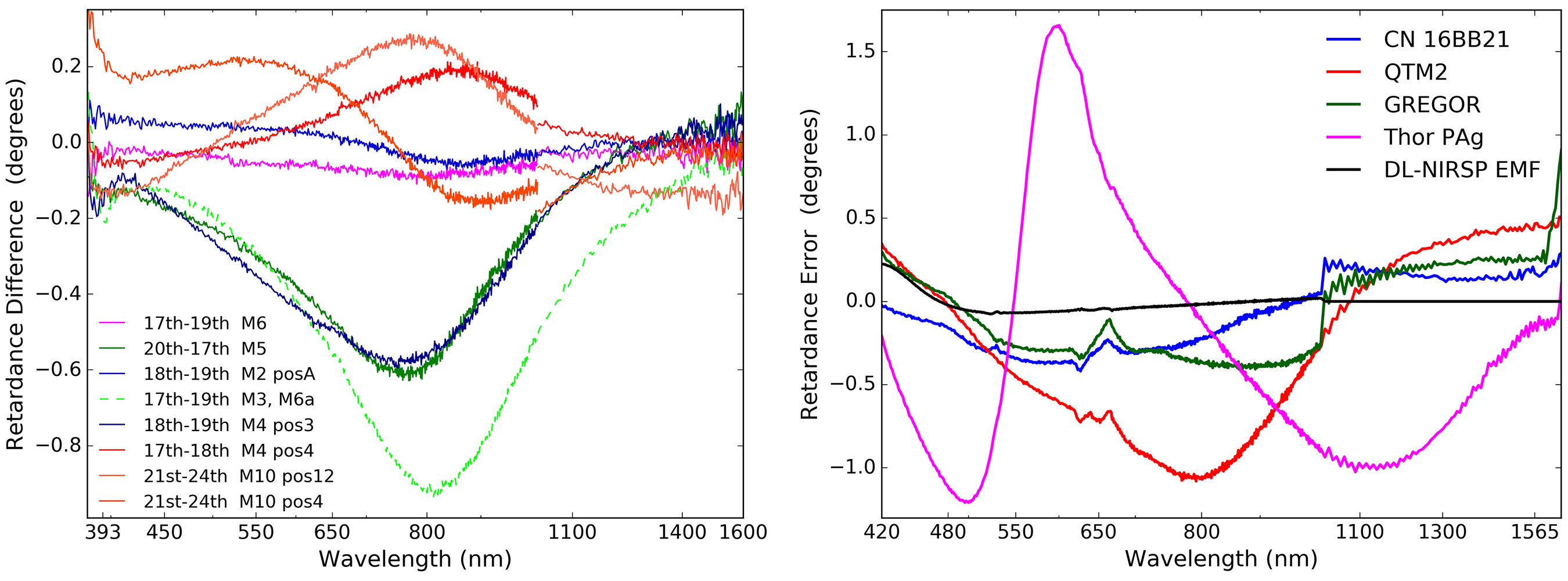}
}
\caption[] 
{ \label{fig:new_retardance_repeatability_with_time} The difference between retardance measurements over time. The left panel shows retardance changes between several DKIST EAg data sets taken in the September 2018 run. Variation is less than 1$^\circ$ retardance. The right panel shows a similar repeatability test but comparing retardance variation over a year between samples of many different silver coating formulas. }
\vspace{-2mm}
\end{center}
\end{figure}

Though the NLSP reflective system calibrations are stable, a larger effect is seen when mounting the same sample in a different mount after an optical re-alignment of the system. The left hand graphic of Figure \ref{fig:new_retardance_repeatability_with_time} shows the difference between retardance measurements when samples were remounted with a day or few in September 2018. The optical re-alignment procedure was performed and data analysis done with identical calibration files. The difference in retardance is always below 1$^\circ$ with most values within $\pm$0.5$^\circ$. The right hand graphic of Figure \ref{fig:new_retardance_repeatability_with_time} shows how the retardance varies between summer 2017 and September 2018 for samples from various telescopes and vendors.  In this case, we observe variations up to 1.5$^\circ$ magnitude.  The optical alignment and derived retardance results are stable to degree magnitude over timescales of days to years.

\begin{figure}[htbp]
\begin{center}
\vspace{-1mm}
\hspace{-0.7em}
\hbox{
\includegraphics[height=6.3cm, angle=0]{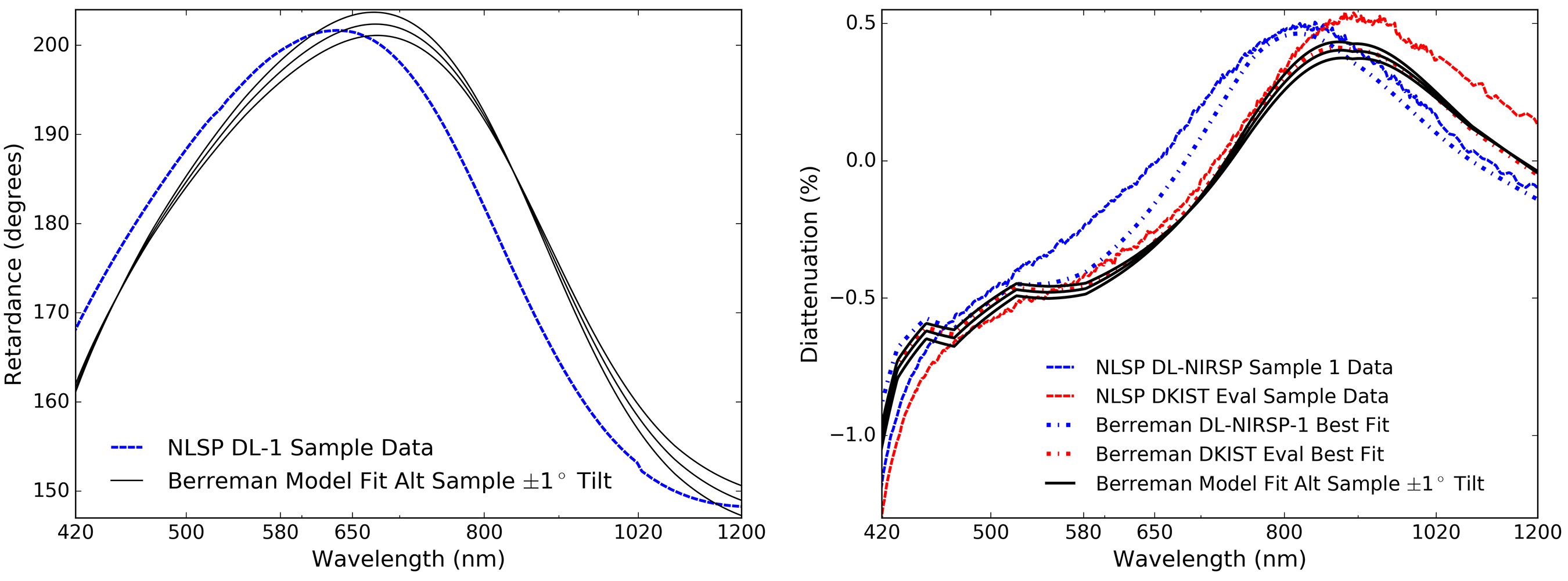}
}
\caption[] 
{ \label{fig:nlsp_tilt_sensitivity} The left hand graphic shows a comparison of our retardance measurement sensitivity to tilt. Two significantly different retardance curves were selected:  A DKIST evaluation sample and the DL-NIRSP 1 sample. The NLSP data for the DL-1 sample is shown in dashed blue. The Berreman models shown in black represent the best fit coating formula to retardance of the DKIST evaluation sample with change in predicted retardance from a $\pm$1$^\circ$ mounting tilt error.  The black curves are significantly displaced in wavelength showing tilt does not explain the shot-to-shot variations in our measurements. The right hand graphic shows a similar tilt sensitivity analysis for diattenuation. }
\vspace{-4mm}
\end{center}
\end{figure}

We show an example of the robustness of the NLSP retardance measurements to sample tilt and mounting errors in the left panel of Figure \ref{fig:nlsp_tilt_sensitivity}. We compare predictions of the best fit two-layer coating Berreman models to measurements. The blue curve shows the DL-NIRSP-1 sample retardance measurement that is significantly different from the DKIST sample.  The black curves show how the best fit Berreman model predictions change for tilts $\pm$1$^\circ$ about the nominal 45$^\circ$ incidence angle. When new samples are mounted in NLSP, there is some mechanical error in the manual positioning of the sample depending on what kinematic mounts are used.  The Berreman models in Figure \ref{fig:nlsp_tilt_sensitivity} change magnitude at certain wavelengths, but the black curves generally do not shift wavelength.  In particular, the wavelength of the theoretical 180$^\circ$ retardance is almost completely insensitive to incidence angle errors. We also show example diattenuation differences as well as tilt sensitivity in the right hand graphic of Figure \ref{fig:nlsp_tilt_sensitivity}.  We show the same NLSP measurements for the DL-NIRSP-1 sample in blue along with the tilt sensitivity of the best-fit two layer Berreman model in black.  We additionally show the NLSP measurements for the DKIST evaluation sample in red. Diattenuation is not presently fit in our current coating model so we do not anticipate a high quality match.

\begin{wrapfigure}{r}{0.55\textwidth}
\centering
\vspace{-3mm}
\begin{tabular}{c} 
\hbox{
\hspace{-0.95em}
\includegraphics[height=6.95cm, angle=0]{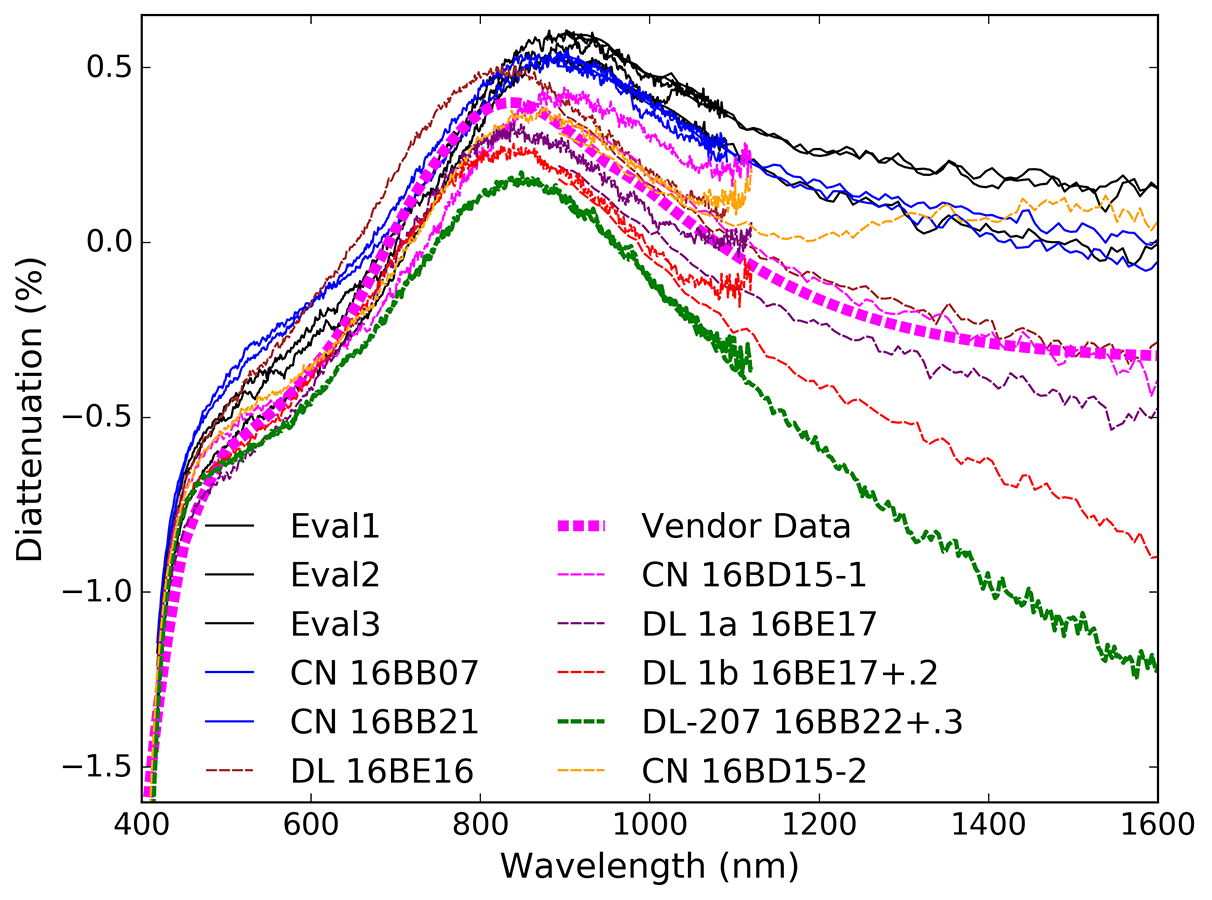}
}
\end{tabular}
\caption[] 
{\label{fig:quantum_compiled_nlsp_data_diatt}  Diattenuation measured for NLSP samples  is strongly rising from -3\% at 393 nm wavelengths to +0.5\% around 850 nm, then decreasing towards values of -0.5\% to +0.2\% at 1650nm. Vendor data is shown as thick dashed magenta.}
\vspace{-3mm}
\end{wrapfigure}

The diattenuation measurements were not repeatable at the level of roughly 0.2\%. In addition to a constant offset of 0.15\% there were spectral differences of $\pm$0.1\%. Thus we consider the systematic errors on diattenuation to be at magnitudes of at least this level.  However, the diattenuation tilt sensitivity of Figure \ref{fig:nlsp_tilt_sensitivity} is much smaller than these magnitudes as seen by the black curves. In addition, the two separate two-layer coating models show that both spectral shifts and magnitude changes are expected. The right hand plot of Figure \ref{fig:nlsp_tilt_sensitivity} shows the diattenuation models over-plotted as the dot-dashed lines. The spectral shifts between blue and red curves is largely explained by thickness variations of the two dielectric layers in our coating model. The diattenuation measurements seems to be more sensitive to the optical alignment of the samples and is the main limiting error at this time.  In Figure \ref{fig:quantum_compiled_nlsp_data_diatt} we show the diattenuation for several DKIST silver samples recorded in the same campaign shown above in Figure \ref{fig:quantum_compiled_nlsp_data}. We also show the nominal diattenuation prediction derived from vendor S and P reflectivity as the dot-dash magenta line.  We do expect coatings to vary significantly run to run as shown in our tolerance analysis but the offsets of some fraction of a percent in Figures \ref{fig:quantum_compiled_nlsp_data_diatt} and \ref{fig:quantum_may31_nlsp_data} are significant and so far unpredictable.

\begin{figure}[htbp]
\begin{center}
\vspace{-1mm}
\hspace{-1.0em}
\hbox{
\includegraphics[height=6.25cm, angle=0]{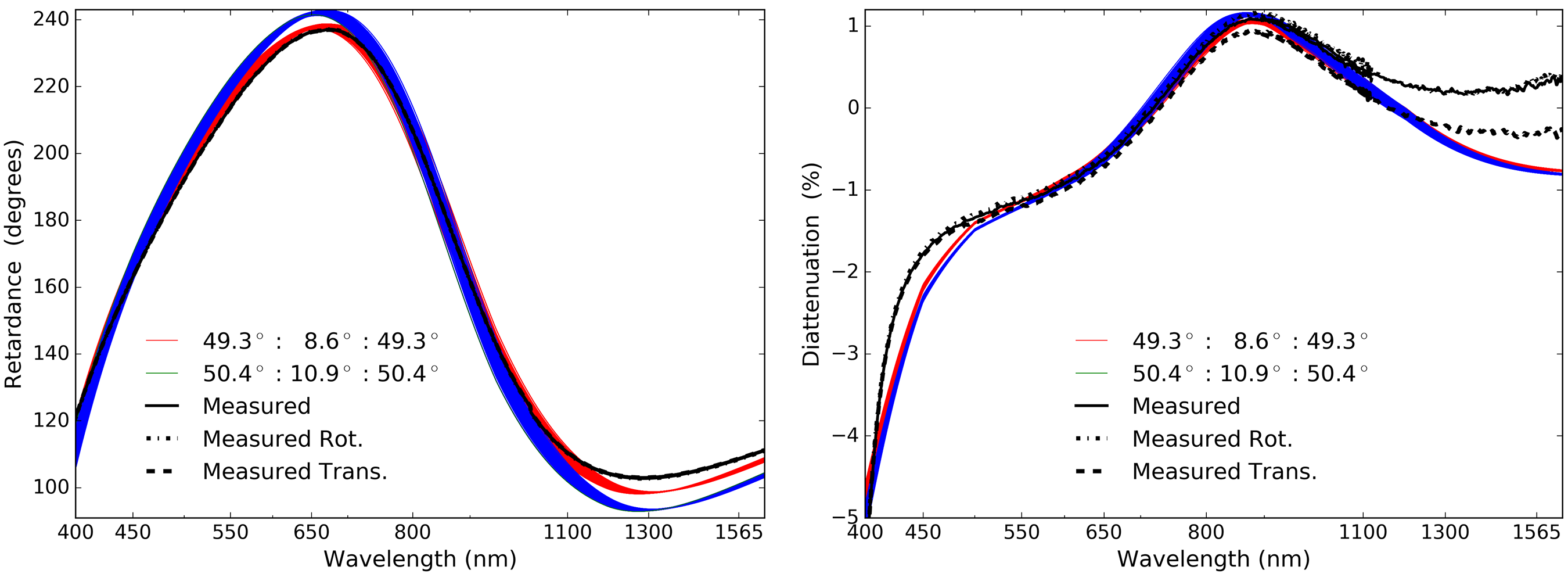}
}
\caption[] 
{ \label{fig:k_cell_quantum} The K-cell retardance measurements and model for the DKIST enhanced silver mirrors. Individual colors show families of coating model predictions following the perturbation of $\pm$0.5 nm in the top layer and $\pm$1.5 nm in the bottom layer (3$^2$ models for 3 mirrors). Different colors represent different incidence angles. The black lines show NLSP measurements with some perturbations in K-cell orientation and spatial position. }
\vspace{-4mm}
\end{center}
\end{figure}

To verify that we can predict system-level polarization for DKIST with combinations of multiple mirrors, we introduced a three-mirror K-cell (image rotator) type setup into our lab spectropolarimeter. We combine three fold mirrors which nominally preserve the beam translation and tilt to verify the model predictions using the better-calibrated transmission arm of NLSP for testing a combination of DKIST enhanced silver mirrors. In Figure \ref{fig:k_cell_quantum} we show the K-cell data for three mirrors at incidence angles of roughly 50$^\circ$, 10$^\circ$ and 50$^\circ$.  Each mirror has different properties that are included in the model. We take the nominal retardance-only two layer coating model fits as the starting point.  The first mirror is 11.975 nm ZnS over 103.1 nm Al$_2$O$_3$. The second mirror is 8.925 nm ZnS over 109.4 Al$_2$O$_3$.  The third mirror is 9.975 nm ZnS over 101.4 nm Al$_2$O$_3$. We show the impact of fitting errors by allowing the top layer to vary by $\pm$0.5 nm and the bottom layer to vary by $\pm$1.5 nm independently for each mirror in the K-cell. This gives rise to 3 thicknesses for 2 layers in 3 mirrors giving 18 total models per K-cell configuration.

We also show the impact of incidence angle by allowing a 5 mm error in measuring the spatial positioning of the beam along with the requirement that the K-cell be aligned geometrically.  We measured the long axis of the K-cell to be 83 mm while the short axis was 26 mm. The equations for a perfectly aligned K-cell give the small interior angle as the arctangent of the two distances.  The triangles must all sum to 180$^\circ$ interior angles and thus the exterior fold angle is (90$^\circ$ + interior angle) / 2. We estimate that the smallest fold angles would be 49.3$^\circ$:8.6$^\circ$:49.3$^\circ$. The predicted 3-mirror retardance is shown by the red curves on the left in Figure \ref{fig:k_cell_quantum}. Diattenuation is shown on the right.  If we add the 5 mm perturbation to increase the angles, we would get 50.4$^\circ$:10.9$^\circ$:50.4$^\circ$ as shown by the blue curves in Figure \ref{fig:k_cell_quantum}.  

Each family of colored curves in Figure \ref{fig:k_cell_quantum} shows the range of coating formula layer thicknesses. Clearly, the impact of incidence angle is stronger than the coating layer thickness for this geometric measurement uncertainty.  The black curves show the retardance derived from the NLSP Mueller matrix measurements. We tried a few different optical alignment perturbations by rotating and translating the sample. The data matches the smaller incidence angle predictions best while slightly under-predicting retardance around 1400 nm wavelength. Given the close match, we have reasonable confidence that we can predict the system model retardance to similar tolerances.

We have shown that NLSP can repeatably measure retardance to values less than 0.05$^\circ$ when remounting identical samples without a perturbation in the optical alignment.  The retardance calculations from the $UU, UV, VU$ and $VV$ Mueller matrix elements are self-consistent to better than 0.25$^\circ$ magnitudes within a single data set. When remounting and re-calibrating the same sample on different days with a possibly perturbed optical alignment, we compute the same retardance within roughly 0.5$^\circ$ with spectral dependence well above statistical noise limits. The same samples repeatedly measured over a year show magnitude variation of up to $\pm$1.5$^\circ$ depending on the quality of the optical alignment.  When using system calibrations taken months apart, the diattenuation changes by 0.15\% in absolute offset with some spectral dependence at 0.1\% magnitudes.

We also showed examples of reflective polarization measurements for several DKIST enhanced protected silver samples and how the measurement setup is insensitive to tilt errors in the sample mounting. Measured variation within witness samples is very significant compared to NLSP systematic error limits.  The reflective arm of NLSP is very capable of measuring sample retardance and diattenuation.

\section{Mirror Examples \& Coating Model Fitting}
\label{sec:refractive_index_and_fitting}  

\begin{figure}[htbp]
\begin{center}
\vspace{-1mm}
\hbox{
\hspace{-0.2em}
\includegraphics[height=10.8cm, angle=0]{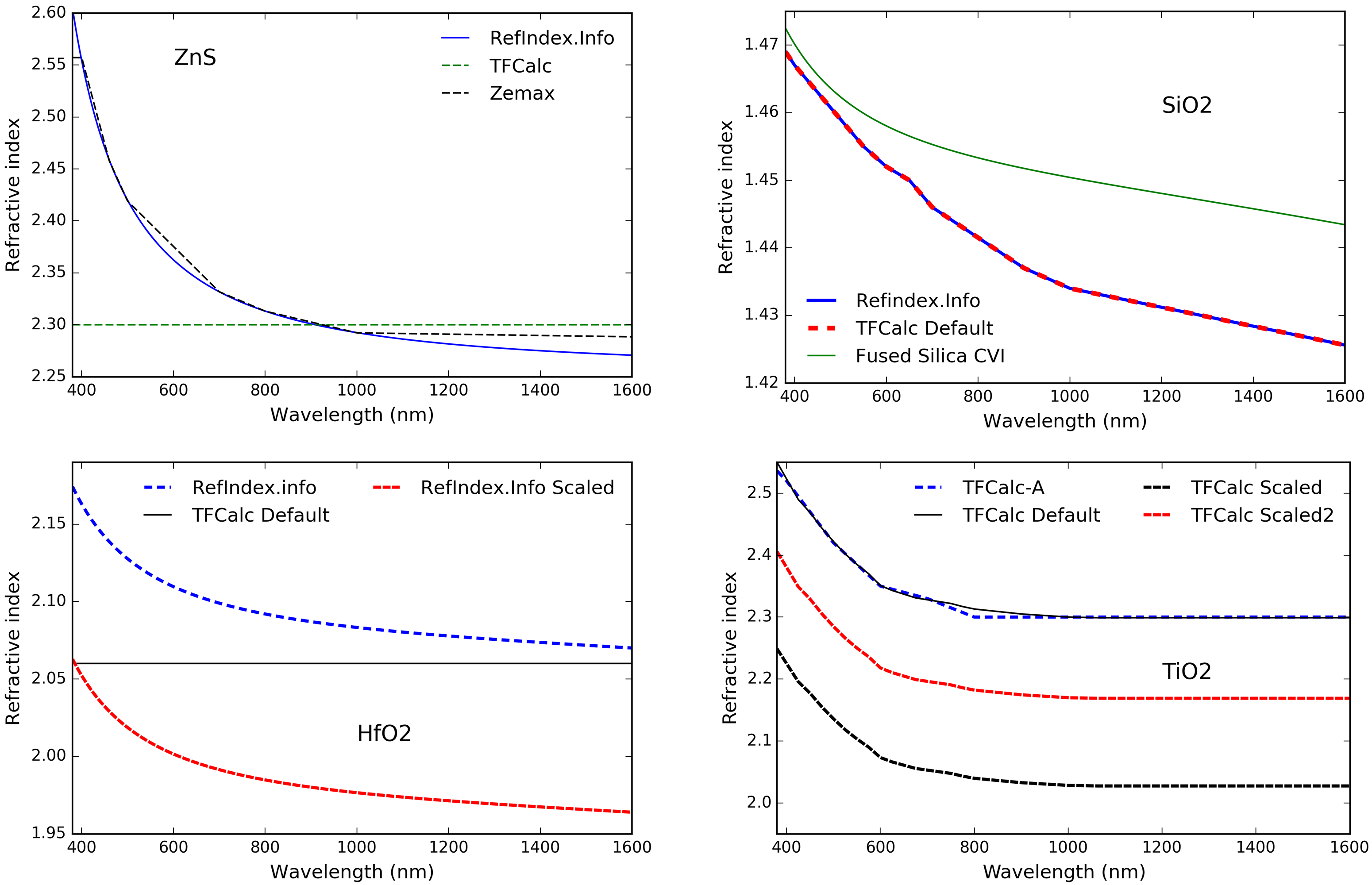}
}
\caption[Refractive Index] 
{  The refractive indices for common coating materials.  Common RefractiveIndex.info website data is shown along with data from the Zemax coating file provided with version 16.5, 2016 and the TFCalc default material files where applicable. 
\label{fig:refractive_index_many}  }
\vspace{-8mm}
\end{center}
\end{figure}

Though common software programs provide some nominal refractive indices for common materials, the data often poorly captures wavelength dependence. The uncertainty in wavelength dependence of materials properties is one of the major limitations in predicting polarization performance significantly better than 1$^\circ$ retardance. Figure \ref{fig:refractive_index_many} shows example refractive index curves gathered from the common coating modeling tools in TFCalc and Zemax, as well as online references in the RefractiveIndex.info database.

The upper left plot shows Zinc Sulfide (ZnS) where the RefractiveIndex.info database gives a dispersion formula of Equation \ref{eqn:SellmeierZnS} based on two references. The black curve shows the Zemax coating file likely performs a linear interpolation of this curve with wavelengths of 400, 460, 500, 700, 800, 1000 and 2000 nm. The TFCalc software package gives only a single refractive index of 2.3 at a wavelength of 550 nm.

\begin{wrapfigure}{r}{0.59\textwidth}
\vspace{-0mm}
\begin{equation}
n^2 = 8.393 + \frac{0.14383}{\lambda^2-0.2421^2} + \frac{4430.99}{\lambda^2-36.71^2}
\label{eqn:SellmeierZnS}
\end{equation}
\begin{equation}
n^2 - 1 = \frac{0.6961663\lambda^2}{\lambda^2-0.0684043^2} + \frac{0.4079426\lambda^2}{\lambda^2-0.1162414^2} + \frac{0.8974794\lambda^2}{\lambda^2?9.896161^2}
\label{eqn:SellmeierCVIsio2}
\end{equation}
\vspace{-4mm}
\end{wrapfigure}

The upper right graphic shows SiO$_2$ for the RefractiveIndex.info database in solid blue as well as the TFCalc coating properties in dashed red with good agreement. Interpolation wavelengths for both curves are 300, 350, 400, 450, 500, 550, 600, 650, 700, 900, 1000, 2000nm. Equation \ref{eqn:SellmeierCVIsio2} is also shown in green as published in the RefractiveIndex.info database and is in the CVI Melles Griot catalog. In the lower two graphics, we show common coating materials of HfO$_2$ and TiO$_2$. A technique used in fitting refractive indices is to scale or offset the refractive index equation by some constant.  For HfO$_2$ and TiO$_2$ we show curves that are scaled by 5\% to 20\%.



Often, vendors for lower cost off-the-shelf parts will only provided limited information about the polarization performance of the mirrors, if any. A common scenario is to be given a theoretical model for reflectivity and possibly diattenuation with coarsely sampled, roughly-interpolated spectral data. When attempting to fit coating performance models to these data sets, we've often found that the models are easily reproduced with publicly available refractive index information with additional adjustments or interpolation. As an example, we show here models for a coating which was described as a single dielectric layer protecting aluminum metal. The coating was described by the vendor as a {\it quarter wave of silicon monoxide protecting aluminum}. We also had explicit follow up communications that silicon dioxide was not correct, and that silicon monoxide was the correct model. With nominal refractive indices for SiO and Al, we completely failed to reproduce the mirror performance, both modeled and measured. However, with SiO$_2$ as the dielectric, and using TFCalc default values we get an exact match to the model provided by the vendor at visible wavelengths. At infrared wavelengths, we model significantly different reflectivity.  As is quite common, the measured performance was significantly different from the model, especially in polarization properties.

\begin{figure}[htbp]
\begin{center}
\vspace{-2mm}
\hbox{
\hspace{-0.8em}
\includegraphics[height=6.3cm, angle=0]{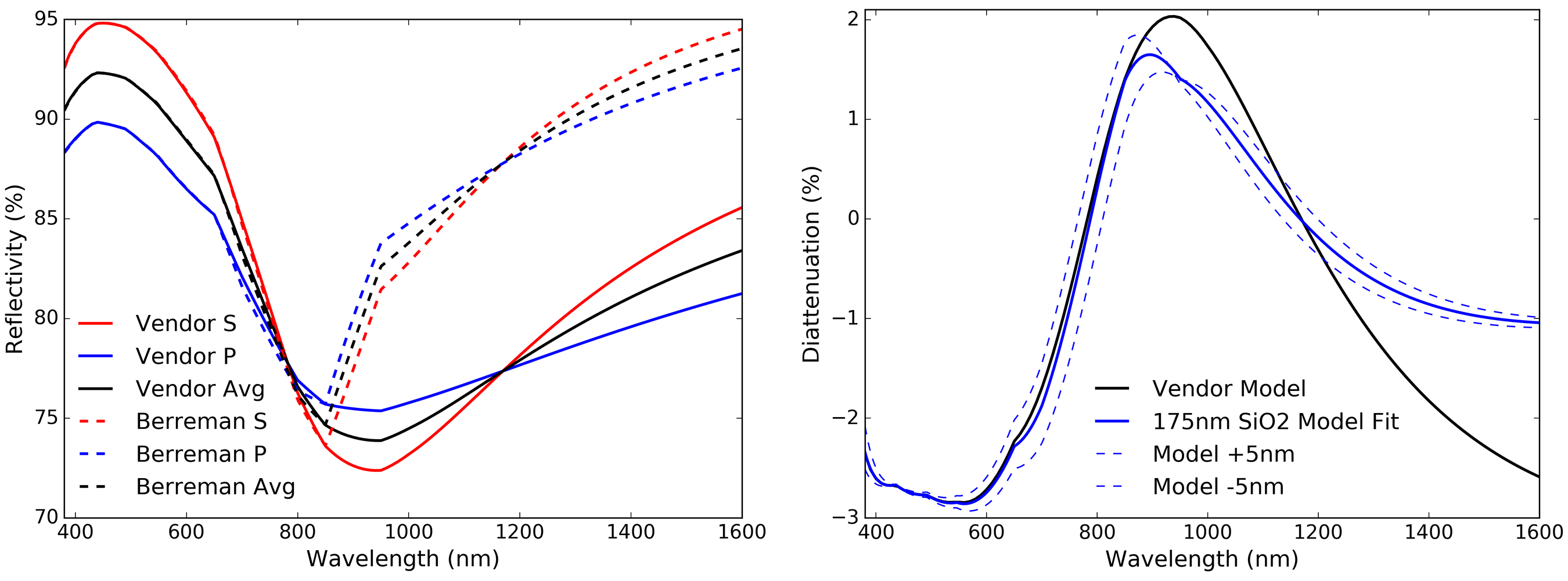}
}
\caption[] 
{ \label{fig:edmun_model_ref_dia} Commercial mirror models and corresponding Berreman fits. Left shows reflectivity fits to protected aluminum mirror and our Berreman model fits using common material formulas. Right shows diattenuation. For this mirror, the coating description was a quarter wave of silicon monoxide protecting aluminum. We find 170 nm of SiO$_2$ with the CVI formula protecting aluminum using the default TFCalc refractive index values provides an exact numerical match at short wavelengths, including interpolation points.}
\vspace{-6mm}
\end{center}
\end{figure}

Figure \ref{fig:edmun_model_ref_dia} shows our Berreman calculations along side the vendor provided data at a 45$^\circ$ incidence angle.  In the left hand graphic, we see the reflectivity for S as red, P as blue and the average polarization as black. Solid lines represent the vendor provided model and dashed lines show our Berreman model with 175 nm of SiO$_2$ protecting Al. In the right graphic of Figure \ref{fig:edmun_model_ref_dia}, we see the vendor provided diattenuation in black and our Berreman models in blue. For wavelengths from 380 nm to 850 nm the models match almost exactly, including the linear interpolation between coarsely sampled points. However, for wavelengths longer than 850 nm, the model predictions diverge from our Berreman computations. In the diattenuation plot in the right graphic of Figure \ref{fig:edmun_model_ref_dia} the dashed blue lines show a change in coating thickness of 5 nm. We see that the fit is quite accurate at short wavelengths and that these changes in coating thickness do not significantly improve the long wavelength fit while obviously degrading the short wavelength match between models. 

Astronomical systems need accurate knowledge of coating performance at all wavelengths of interest for a range of incidence angles.  We show examples below of a variety of coatings used in DKIST and other solar telescopes.

\subsection{DKIST VTF \& GREGOR Enhanced Protected Silver Commercial Mirrors}
\label{sec:sub_VTF_GREGOR_mirrors}

The Kiepenheuer Institute for Solar Physics (KIS) provided us GREGOR telescope witness samples for test in summer 2017. These witness samples are also the same enhanced protected silver coating formula nominally proposed for the Visible Tunable Filter (VTF) instrument KIS is constructing for installation on DKIST.  The KIS staff were informed that the coating could be modeled as a high index dielectric layer over a low index dielectric layer on top of a silver metal layer similar to the models presented here and in our prior work \cite{Harrington:2017ejb, 2017JATIS...3a8002H}.

\begin{figure}[htbp]
\begin{center}
\vspace{-2mm}
\hbox{
\hspace{-0.8em}
\includegraphics[height=6.3cm, angle=0]{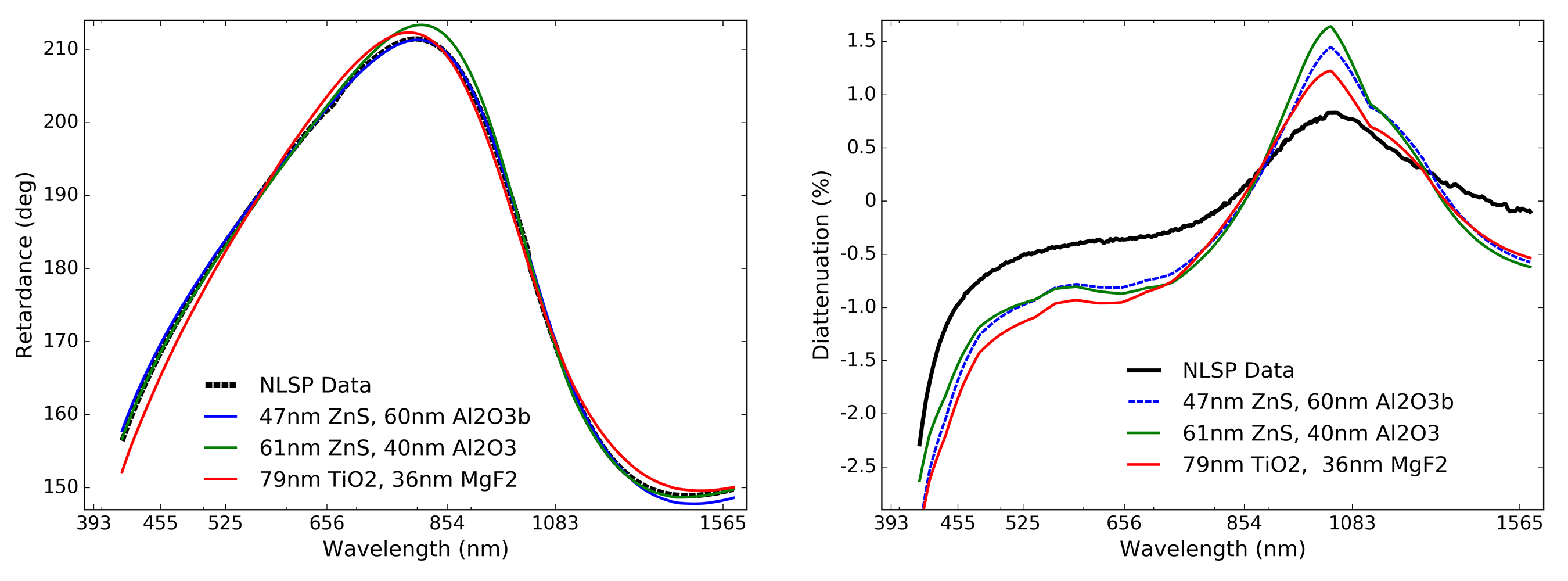}
}
\caption[] 
{\label{fig:gregor_nlsp_data} NLSP measurements and corresponding Berreman model fits. Left shows retardance fits to an enhanced protected silver mirror used at the GREGOR solar telescope along with some model fits using common materials. Right shows diattenuation along with predictions for the retardance-fit Berreman models. Three different materials combinations are shown that were more successful at fitting retardance than several other material combinations.}
\vspace{-6mm}
\end{center}
\end{figure}

Four repeated data sets were collected at 45$^\circ$ incidence angle. Figure \ref{fig:gregor_nlsp_data} shows the retardance and diattenuation properties of the Mueller matrix derived from measurements. The $IQ$ / $II$ diattenuation Mueller matrix element is shown in the right hand graphic. The inverse tangent of the $VV$ and $UV$ elements is used as a proxy for the linear retardance component. Agreement between visible and near-infrared spectrographs is quite good as seen by the overlap in the 950 nm to 1050 nm wavelength range. The extreme edges of each data set are removed for this analysis by stitching together the data sets at 1020 nm wavelength. We ignore longer wavelengths measured by the visible spectrograph and we ignore shorter wavelength data measured by the near-infrared spectrograph. 

We attempted to fit Berreman models of many material choices in two-layer configurations to the retardance measurements. The three best fits were selected for Figure \ref{fig:gregor_nlsp_data}.  All three models represent relatively high refractive index materials on top of a relatively moderate refractive index material.  The retardance behavior is similar to the DKIST enhanced protected silver in that there are two wavelengths around blue and near infrared where the retardance is zero.  If only the visible-wavelength data set is used, the best fit model layer thickness changes slightly. As an example, the best-fit layer thicknesses are (62nm, 60nm) when using only VIS spectrograph data as opposed to (65nm, 56nm) when using the full VIS+NIR range for a HfO$_2$ over Al$_2$O$_3$ formula model.   




%

\subsection{Commercial Mirror Samples: Big Bear Solar Observatory \& Newport PAg}
\label{sec:sub_BBSO_mirrors}

We also aim to help create a telescope model comparison for the Big Bear Solar Observatory (BBSO). Many observatories use commercial off-the-shelf (COTS) optics. It is straightforward to create polarization performance models. We were given witness samples from the BBSO aluminum coated mirrors as well as a commercial off-the-shelf protected metal coated mirror from Newport used at BBSO. This Newport coating was used on several mirrors as part of the BBSO optical train. The BBSO uses several other commercial mirrors in their optical train and knowledge of each coating is required to create a detailed polarization model. 

\begin{figure}[htbp]
\begin{center}
\vspace{-0mm}
\hbox{
\hspace{-0.4em}
\includegraphics[height=5.5cm, angle=0]{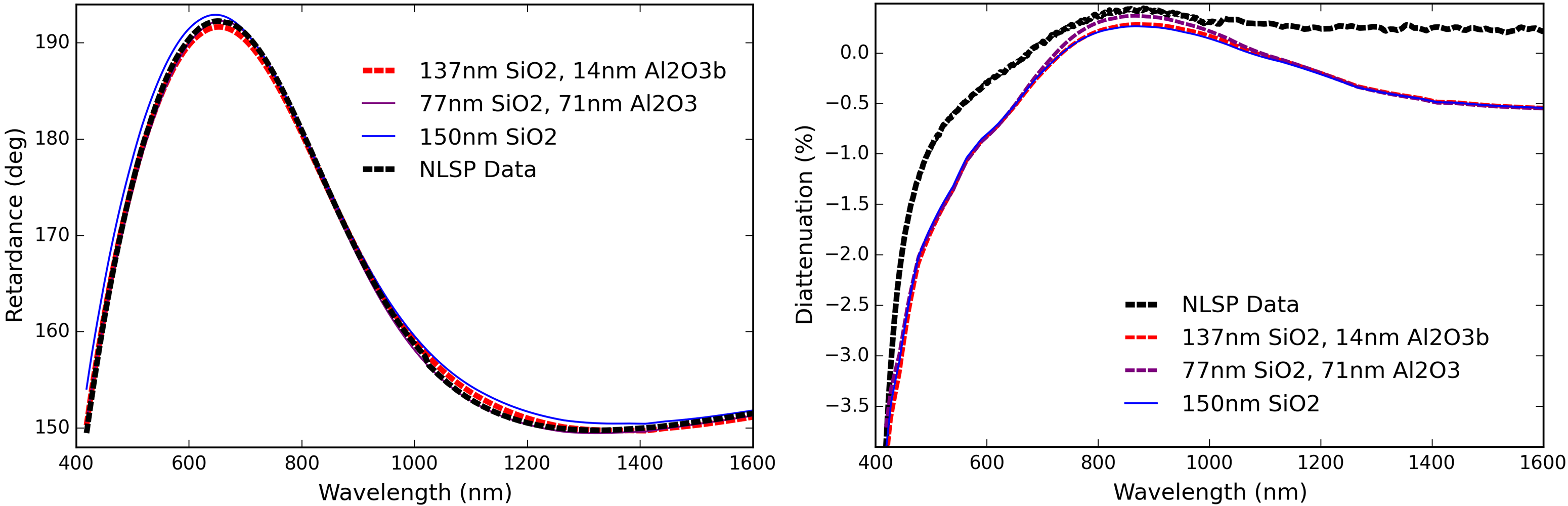}
}
\caption[] 
{\label{fig:newport_nlsp_data_ret_dia}  Commercial Newport enhanced protected silver. Left shows retardance fits and some model fits using common materials. Right shows measured diattenuation along with the Berreman models for diattenuation (not fit).}
\vspace{-4mm}
\end{center}
\end{figure}

We measured the Mueller matrix of the Newport mirror at 45$^\circ$ incidence angle with NLSP. The linear retardance in the $UV$ plane as well as the $IQ$ diattenuation was derived using the Mueller matrix. The resulting stitched data sets and Berreman models are shown in Figure \ref{fig:newport_nlsp_data_ret_dia}. There are two models that fit the Newport coating equally well. The first fit is 137 nm thickness of the SiO$_2$ model coated over 14 nm thickness of the Al$_2$O$_3$ model using the Boidin et al. \cite{Boidin:2016kx} refractive index formula using index 1.67 at 850 nm wavelength.  A similarly good model fit is seen for 77 nm thickness of the SiO$_2$ model coated over 71 nm thickness of the Al$_2$O$_3$ model with refractive index of 1.55 at 850 nm wavelength.  Both SiO$_2$ models have refractive index 1.55 at 850 nm wavelength. The 0.5$^\circ$ retardance step barely visible at 1020 nm is the stitching wavelength where VIS spectrograph data is concatenated with NIR spectrograph data.

For each vendor the BBSO used, we requested information on the coating prescription. As expected, only limited and incomplete information could be obtained, if any was even available. This is clearly insufficient for creating polarization models of reasonable fidelity at observatories with powered optics at varying incidence angles.

\subsection{Commercial Mirror Samples: DKIST \& A Thor Labs Protected Silver K-Cell}
\label{sec:sub_DKIST_thor_mirrors}

\begin{wrapfigure}{l}{0.55\textwidth}
\centering
\vspace{-2mm}
\begin{tabular}{c} 
\hbox{
\hspace{-1.0em}
\includegraphics[height=6.99cm, angle=0]{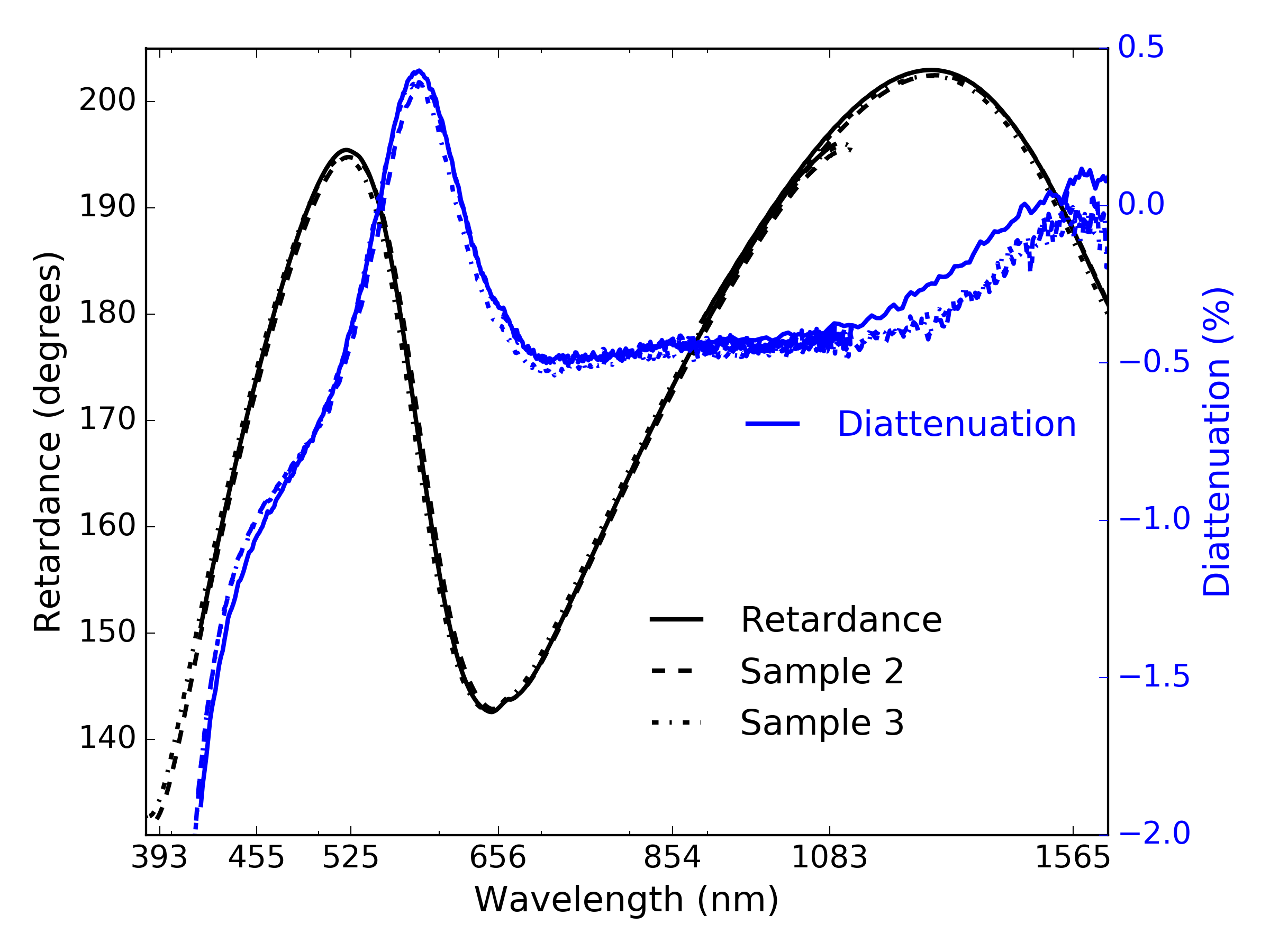}
}
\end{tabular}
\caption[] {\label{fig:thorAG_nlsp_3data_sets_data_ret_dia}  NLSP measurements of three Thor Labs protected silver mirrors. Black shows retardance using the left hand Y axis.  Blue shows diattenuation on the right hand Y axis. The first sample was measured 7 months before the second two, and the diattenuation was shifted by 0.25\% for the curves to match.}
\vspace{-3mm}
 \end{wrapfigure}

We tested three Thor Labs protected silver mirrors we use in our DKIST laboratory. These Thor Labs mirrors have a different wavelength dependence of retardance and diattenuation than several of the previous samples reported in the main paper. This difference is likely caused by substantially thicker coating layers as suggested below by our fitting process outlined above. In Figure \ref{fig:thorAG_nlsp_3data_sets_data_ret_dia} we show the retardance in black and diattenuation in blue using the right hand Y axis measured with NLSP in the reflective configuration for each of the three samples. The retardance curves cross 180$^\circ$ at four wavelengths over the DKIST instrument wavelength range. The diattenuation for this sample has a somewhat strong peak near 600 nm. One mirror was measured in July of 2017 while the other two mirrors were measured in February of 2018. 

The three mirror samples were all procured at the same time. While it is possible that they were all coated in the same shot, there is no guarantee of their coating heritage. We measured significant differences in both the diattenuation and the retardance. The diattenuation measured in the first mirror was spatially offset by 0.25\% in Figure \ref{fig:thorAG_nlsp_3data_sets_data_ret_dia} to match values at near infrared wavelengths. As we showed above for the DKSIT silver mirrors in Figure \ref{fig:nlsp_tilt_sensitivity}, there are significant impacts to diattenuation for even small variations in the layer thicknesses of the dielectrics in the model. Our 0.25\% offset is a likely a combination of real systematic difference between measurements as outlined for NLSP in prior sections and also real variations between these commercial mirrors.

\begin{wrapfigure}{r}{0.55\textwidth}
\centering
\vspace{-4mm}
\begin{tabular}{c} 
\hbox{
\hspace{-1.0em}
\includegraphics[height=6.99cm, angle=0]{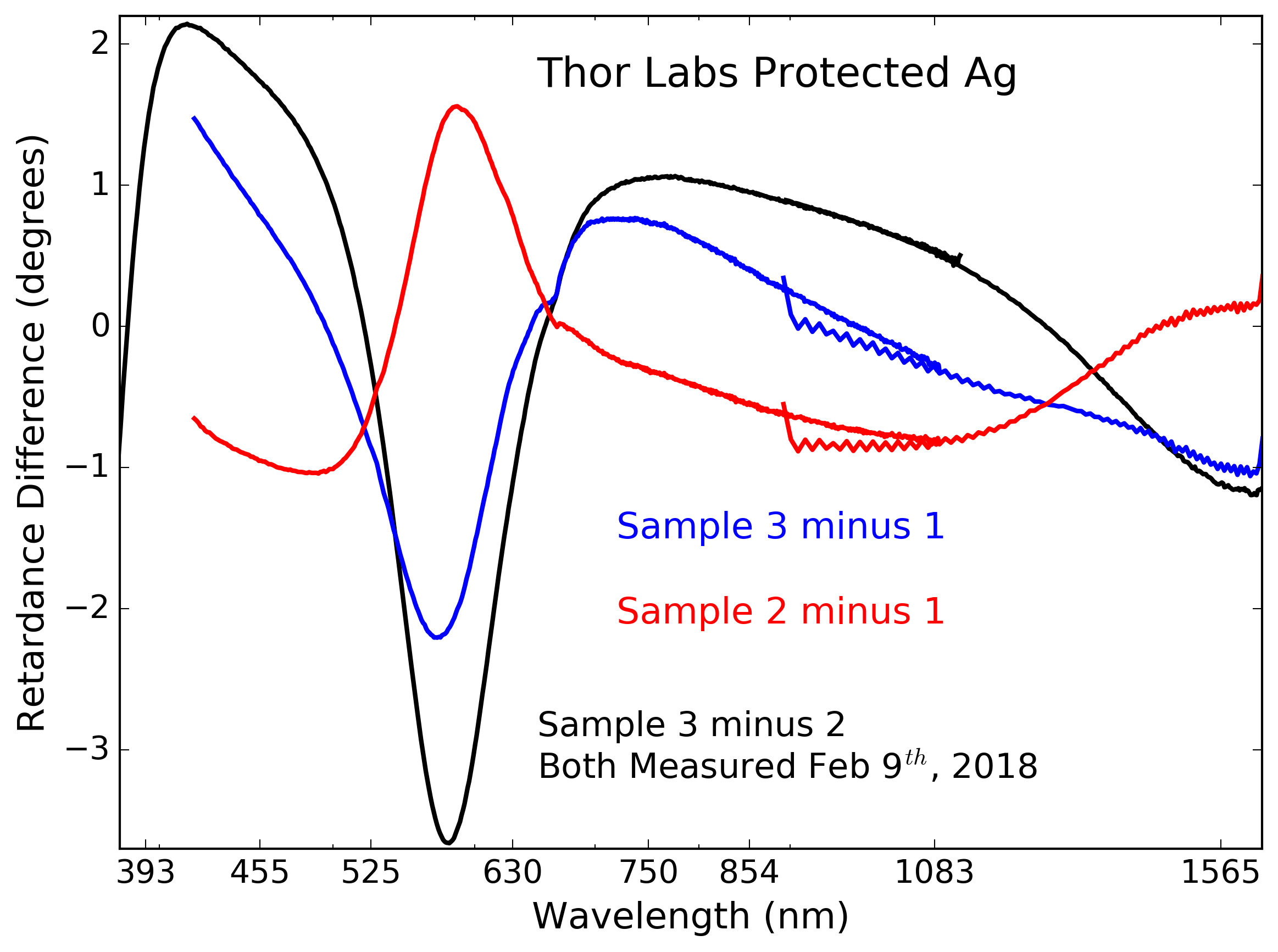}
}
\end{tabular}
\caption[] {\label{fig:thorAG_nlsp_ret_dif}  The difference in retardance between Thor Labs protected silver mirrors. Sample number 2 and 3 measured with NLSP on the same day. Sample 1 was measured in July 2017. Differences are more than an order of magnitude larger than systematic errors.}
\vspace{-2mm}
 \end{wrapfigure}

The retardance curves for all three samples match to within 3$^\circ$ but there are easily detectable difference between nominally identical samples. We show in Figure \ref{fig:thorAG_nlsp_ret_dif} the difference between mirror retardance for Thor Labs mirrors number 3 and 2 in black. Both mirrors were measured on the same day with NLSP at 45$^\circ$ incidence angle.  We also show the difference between sample 3 and sample 1 in blue as well as sample 2 and sample 1 in red.

We note that repeatability tests shown above in Figures \ref{fig:quantum_retardance_change_with_time} and \ref{fig:quantum_may31_nlsp_data} showed that after remounting and aligning a mirror, we can reproduce a measurement to better than 0.05$^\circ$.  Estimates of retardance computed by using different combinations of the $UV$ Mueller matrix elements agree to better than 0.3$^\circ$. Both of these systematic errors are an order of magnitude smaller than the difference measured here between Thor Labs mirrors. 

\begin{figure}[htbp]
\begin{center}
\vspace{-0mm}
\hbox{
\hspace{-0.95em}
\includegraphics[height=6.3cm, angle=0]{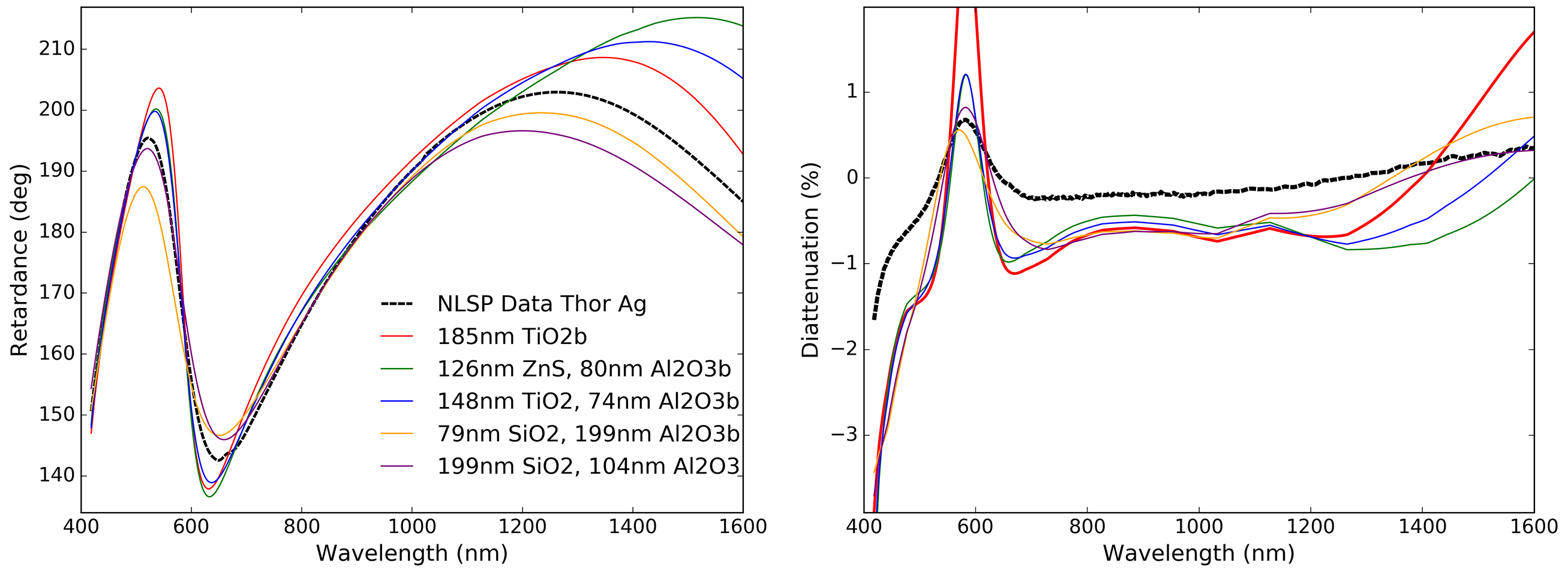}
}
\caption[]  { \label{fig:thorAG_nlsp_data_ret_dia} Commercial Thor Labs protected silver mirror and corresponding fits. Left shows retardance fits and some model fits using common materials. Right shows measured diattenuation along with the Berreman models for diattenuation (not fit). For this mirror the dielectric layers are much thicker and the fits to retardance data are significantly worse. }
\vspace{-6mm}
\end{center}
\end{figure}

With these data sets, we can attempt to fit various coating models to the retardance measurements. Figure \ref{fig:thorAG_nlsp_data_ret_dia} shows the retardance and diattenuation measurements as the dashed black lines. The prediction from the best fit to retardance from our two layer Berreman coating models are shown as solid lines of varying color.  Retardance is shown on the left with the diattenuation on the right. The dielectric layer thicknesses found in the fit are shown in the legend of the retardance graph. 

We attempted to fit a large range of two-layer coating models but found only a few reasonably reproduced the retardance curves.  These better fitting models generally used relatively high refractive index material curves such as ZnS, TiO$_2$ from TFCalc or vendor references and Al$_2$O$_3$ from either Zemax or Boidin et al \cite{Boidin:2016kx} references. The red curve shows a coating model is a single layer of TiO$_2$ at 185 nm physical thickness using the Boidin et al \cite{Boidin:2016kx} refractive index curves from refractiveindex.info.  The blue curve shows another model as layer of amorphous SiO$_2$ at 199 nm physical thickness coated on top of 104 nm of Al$_2$O$_3$. In all models fit, the layers ended up with thicknesses significantly larger than 150 nm. The two layer coating models described in the main text all had total thickness around or less than 100 nm. All models diverge at longer wavelength where it is possible that both the metal and the refractive index curves for the dielectrics have higher error.

The right hand graphic of Figure \ref{fig:thorAG_nlsp_data_ret_dia} shows the diattenuation.  All models consistently underestimate the diattenuation for blue and near infrared wavelengths.  The spectral feature around 600 nm shows a diversity of results between the models.  We do not expect diattenuation to be modeled well as this parameter is not included in the fit. We showed above in Section \ref{sec:mirror_fit_limitations} that the metal complex refractive index can strongly influence the fit and can be highly variable between coating process and vendors.

\begin{wraptable}{l}{0.23\textwidth}
\vspace{-3mm}
\caption{K-Cells}
\label{table:thor_Kcell_aoi}
\centering
\begin{tabular}{l l l}
\hline
\hline
Name		& $\theta_{1,3}$	& $\theta_2$	\\
\hline
Narrow		& 50.4			& 10.8		\\
Middle		& 53.4			& 17.4		\\
Wide			& 57.6			& 25.3		\\
\hline
\hline
\end{tabular}
\vspace{-4mm}
\end{wraptable}

We combined the three Thor Labs mirrors to create a sample where the beam enters and exits in translation and tilt exactly following the unperturbed incoming beam. This setup is often called an image rotator, derotator, K-mirror or K-cell. Common designs use relatively high incidence angles near 60$^\circ$ for the outer two mirrors and a smaller angle for the inner mirror.  In our setup, we were able to place these round one-inch diameter mirrors into kinematic mounts with incidence angles near 50$^\circ$ to 60$^\circ$ for the outer mirrors and 10$^\circ$ to 17$^\circ$ on the inner mirror.  Table \ref{table:thor_Kcell_aoi} shows the incidence angles estimated for the two outer mirrors ($\theta_{1}$ \& $\theta_{3}$) as well as the lower incidence angle inner mirror.

\begin{figure}[htbp]
\begin{center}
\vspace{-2mm}
\hbox{
\hspace{-0.8em}
\includegraphics[height=6.3cm, angle=0]{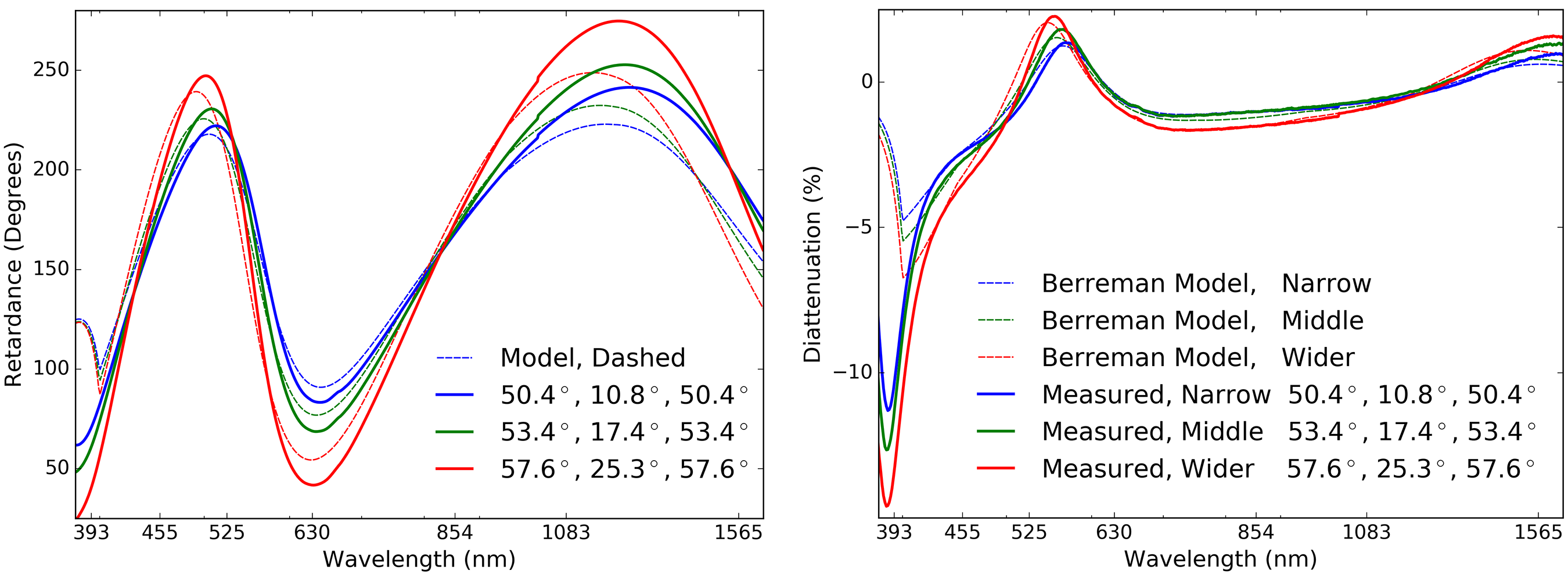}
}
\caption[] { \label{fig:thorAG_nlsp_KCell_data_ret_dia} Transmission measurements from a 3-mirror K-cell using our three commercial Thor Labs protected silver mirrors. Left shows retardance measurements and a Berreman prediction using our best fit two layer coating model. Right shows measured diattenuation along with the Berreman models for diattenuation (not fit). Note the sharp change in behavior for wavelengths short of 400 nm is caused by a strong change in the silver refractive index complex component in the default TFCalc values used here.}
\vspace{-6mm}
\end{center}
\end{figure}

Using a HeNe laser with a $\sim$3 mm beam, we aligned the optical footprint location on each mirror as well as the individual mirror tilts. We ensured that the exit beam roughly matched the unperturbed beam in height and spot location over a $\sim$1 meter path with and without the K-cell translated into the beam. As a further alignment step, our NLSP software provides tools to show the symmetry of detected flux upon 180$^\circ$ rotation of the NLSP retarders as the various optics are rotated in 10$^\circ$ steps through 360$^\circ$. This diagnostic shows photometric errors over 0.5\% if the beam is not well centered on the spectrograph fibers. We performed additional tilt adjustment of each mirror after aligning for translation and height in the NLSP beam to maximize flux through the system (minimize vignetting on the fiber) as well as to ensure the $\sim$4 mm NLSP beam was centered on the optics. 

In Figure \ref{fig:thorAG_nlsp_KCell_data_ret_dia}, we show the NLSP measured retardance and diattenuation as solid colored lines. The narrower K-cell is shown as blue. The moderate K-cell configuration is shown in green. A wider K-cell is shown in red. Dashed lines show a two-layer Berreman coating model for the three mirrors operating in series at the appropriate incidence angles. We only show one of our better fit two-layer models as 200 nm of SiO$_2$ over 100 nm of Al$_2$O$_3$ coated over silver modeled using the default TFCalc values. There is a sharp change in the predictions for wavelengths short of 400 nm. This is caused by a strong change in the silver refractive index complex component in the default TFCalc values interpolated here.  The value of k changes from 1.93 at 400 nm to 5.85 at 375 nm wavelength. We use linear interpolation, which accounts for the sharp change in behavior at 400 nm.

\begin{wrapfigure}{r}{0.55\textwidth}
\centering
\vspace{-3mm}
\begin{tabular}{c} 
\hbox{
\hspace{-1.05em}
\includegraphics[height=6.99cm, angle=0]{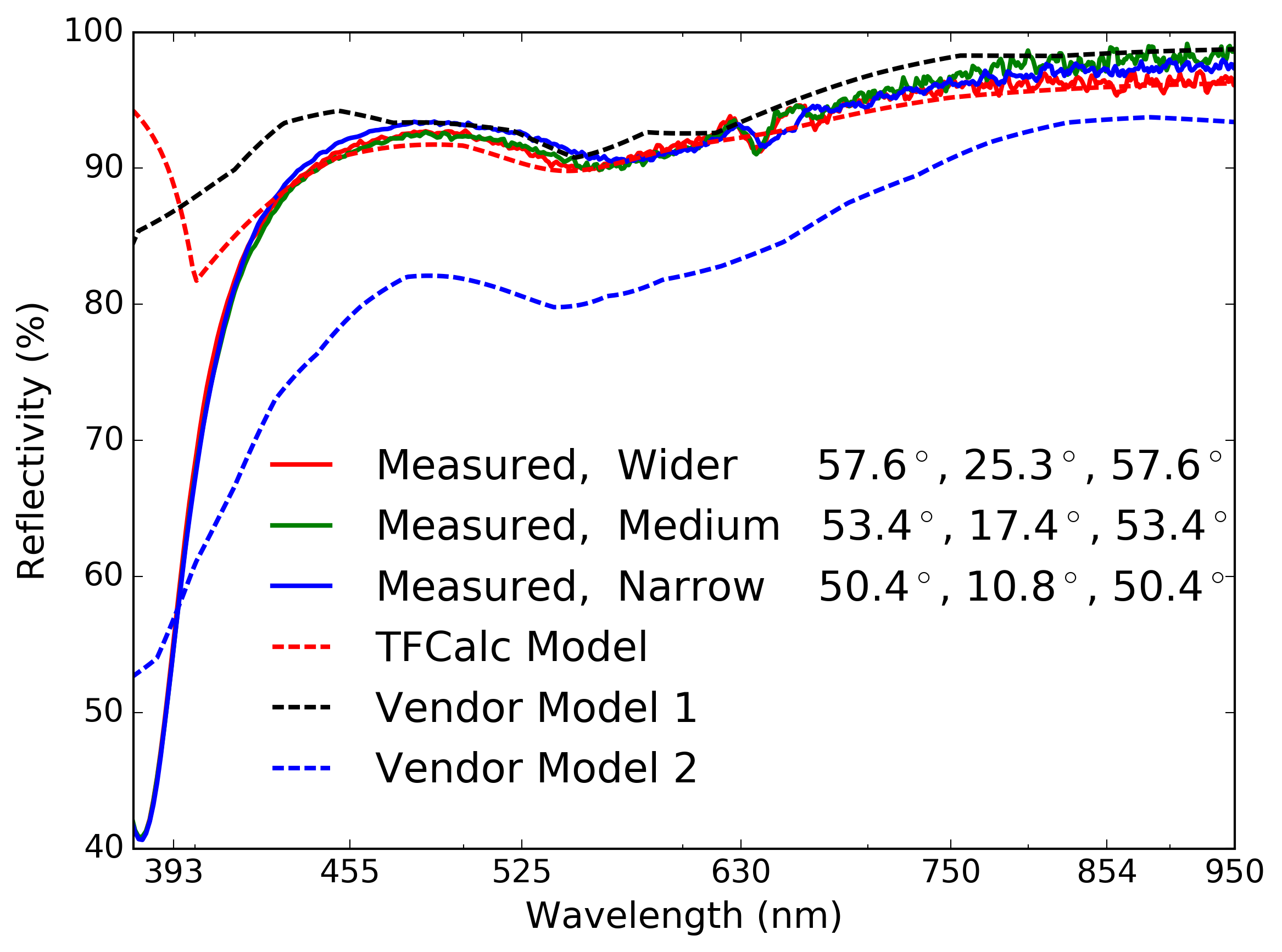}
}
\end{tabular}
\caption[] {\label{fig:thorAG_kcell_refl}  The combined three-mirror reflectivity of the wider K-cell with Thor Labs silver mirrors.  Dashed lines show two layer coating models with 200 nm of SiO$_2$ over 100 nm of Al$_2$O$_3$ protecting silver.  The complex refractive index of the silver varies between models, giving rise to strong changes in predicted reflectivity.}
\vspace{-2mm}
 \end{wrapfigure}

This issue with TFCalc default metal refractive index and all other metal refractive index values, is drastically seen by the failure of the Berreman models to match the measured reflectivity through the K-cell.  We show the three K-cell reflectivity measurement of the wider configuration as solid red in Figure \ref{fig:thorAG_kcell_refl}. The Berreman model using the default TFCalc values is shown in Figure \ref{fig:thorAG_kcell_refl} as a red dashed line. There is reasonably good agreement in spectral shape and throughput for wavelengths longer than 420 nm.  At short wavelengths however, the high complex index of k=5.85 in the TFCalc silver creates reflectivity values over 50\% higher than observed.  We additionally show two different sets of vendor values as black and blue.  These values respectively over and under estimate the throughput of this 3 mirror system at longer wavelengths. Both silver refractive index values over-estimate reflectivity at the shortest wavelengths.

These models are not expected to match in detail as this COTS mirror has an unknown coating in an unknown process. Additional work is required to adjust the refractive index of the dielectric and the metal to match this particular optic. The two outer mirrors are at relatively high incidence with the inner mirror at small incidence angles. With all folds are in roughly the same plane, we expect the retardance to be somewhat larger than double the magnitude about 180$^\circ$ than that of a single mirror at 45$^\circ$. This is easily seen by a comparison between the single mirror of Figure \ref{fig:thorAG_nlsp_data_ret_dia} and the three mirrors in Figure \ref{fig:thorAG_nlsp_KCell_data_ret_dia}.   

One major benefit of testing K-mirror type setups is that a much larger range of incidence angles can be tested. The polarization predictions need to be valid from near normal incidence to 45$^\circ$ to account for the full range of DKIST reflections.  By building this type of setup, we can further constrain our simple coating models to ensure accurate predictions are derived from these coatings as we have applied to telescopes such as DKIST and AEOS \cite{Harrington:2017ejb,2017JATIS...3a8002H}.

\subsection{Image Rotator K-Cell: A NLSP Sample With Both ViSP \& Thor Silver Mirrors}
\label{sec:sub_DKIST_visp_thor_Kmirrors}

The K-cell type sample can be useful for deriving reflectivity of individual mirrors using the well-calibrated transmission arm of NLSP.  Given the various measurement techniques applied to the ViSP many-layer silver mirrors, we confirm the reduced transmission bands as well as the non-negligible polarization impacts at low incidence angles. We took the Thor Labs K-cell in the narrow configuration and replaced the low-incidence middle mirror with the ViSP witness samples.  This allows extraction of the polarization and reflectance properties at an incidence angle around 11$^\circ$, similar to the ViSP F2 mirror at 12.3$^\circ$ incidence. 

\begin{figure}[htbp]
\begin{center}
\vspace{-2mm}
\hbox{
\hspace{-0.7em}
\includegraphics[height=6.35cm, angle=0]{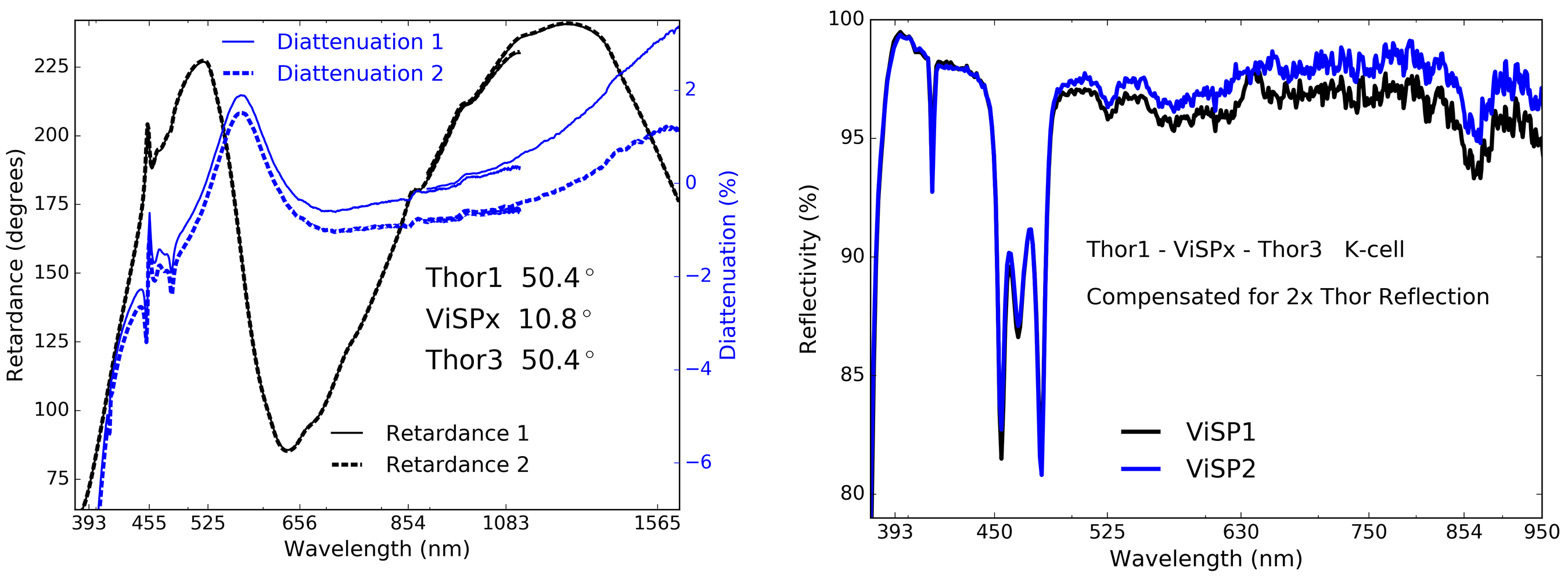}
}
\caption[] { \label{fig:thor_and_visp_KCell_refl_ret_dia} The left graphic shows retardance and diattenuation for a K-cell made of two Thor Labs protected silver mirrors at high incidence angles and a ViSP witness sample at low incidence angles.  Right shows the reflectivity of the ViSP sample after compensating for the two Thor Labs mirror reflections. The ViSP sample adds narrow spectral features seen around 450 nm wavelength and substantial drops in reflectivity in two separate bandpasses.}
\vspace{-6mm}
\end{center}
\end{figure}

We show in the left hand graphic of Figure \ref{fig:thor_and_visp_KCell_refl_ret_dia} the retardance and diattenuation of the combined 3-mirrors in the narrow configuration.  The smooth, large magnitude structures of Figure \ref{fig:thorAG_nlsp_KCell_data_ret_dia} are reproduced by the two high-incidence Thor Labs silver mirrors at 50.4$^\circ$ indicence. However, we now see significant narrow spectral features in both retardance and diattenuation around 450 nm to 500 nm wavelength.  Additional small amplitude ripples are also seen in the retardance curve caused by the many-layer enhanced silver coating design. We attribute these narrow {\it spikes} and broader {\it ripples} to the ViSP many-layer mirrors. 
 
In the right hand graphic of Figure \ref{fig:thor_and_visp_KCell_refl_ret_dia}, we extract the reflectivity of the ViSP samples by dividing out the reflectivity of the two Thor Labs mirrors derived in the last section. We use the cube-root of the Thor Labs K-cell of Figure \ref{fig:thorAG_kcell_refl} as an approximation of an individual Thor Labs mirror. The various incidence angles tested did not show an appreciable difference in reflectivity.  Dividing the measured K-cell reflectivity by this Thor Labs reflectivity curve twice results in a reflectivity curve for the ViSP witness sample that very well correlates with the vendor data presented in the main paper. The black and blue reflectivity curves show two separate witness samples that each had different storage histories over roughly 9 months between coating and measurement. Though their storage conditions varied, the reflectivity and polarization properties between the two samples are remarkably consistent.  In testing at the vendor site immediately after coating, the low reflectivity band between 450 nm and 500 nm was not present.

\subsection{Protected Silver Mirrors: DKIST Feed Optics With Zygo Silver}
\label{sec:sub_DKIST_Zygo_mirrors}

\begin{wrapfigure}{r}{0.61\textwidth}
\centering
\vspace{-6mm}
\begin{tabular}{c} 
\hbox{
\hspace{-0.9em}
\includegraphics[height=7.7cm, angle=0]{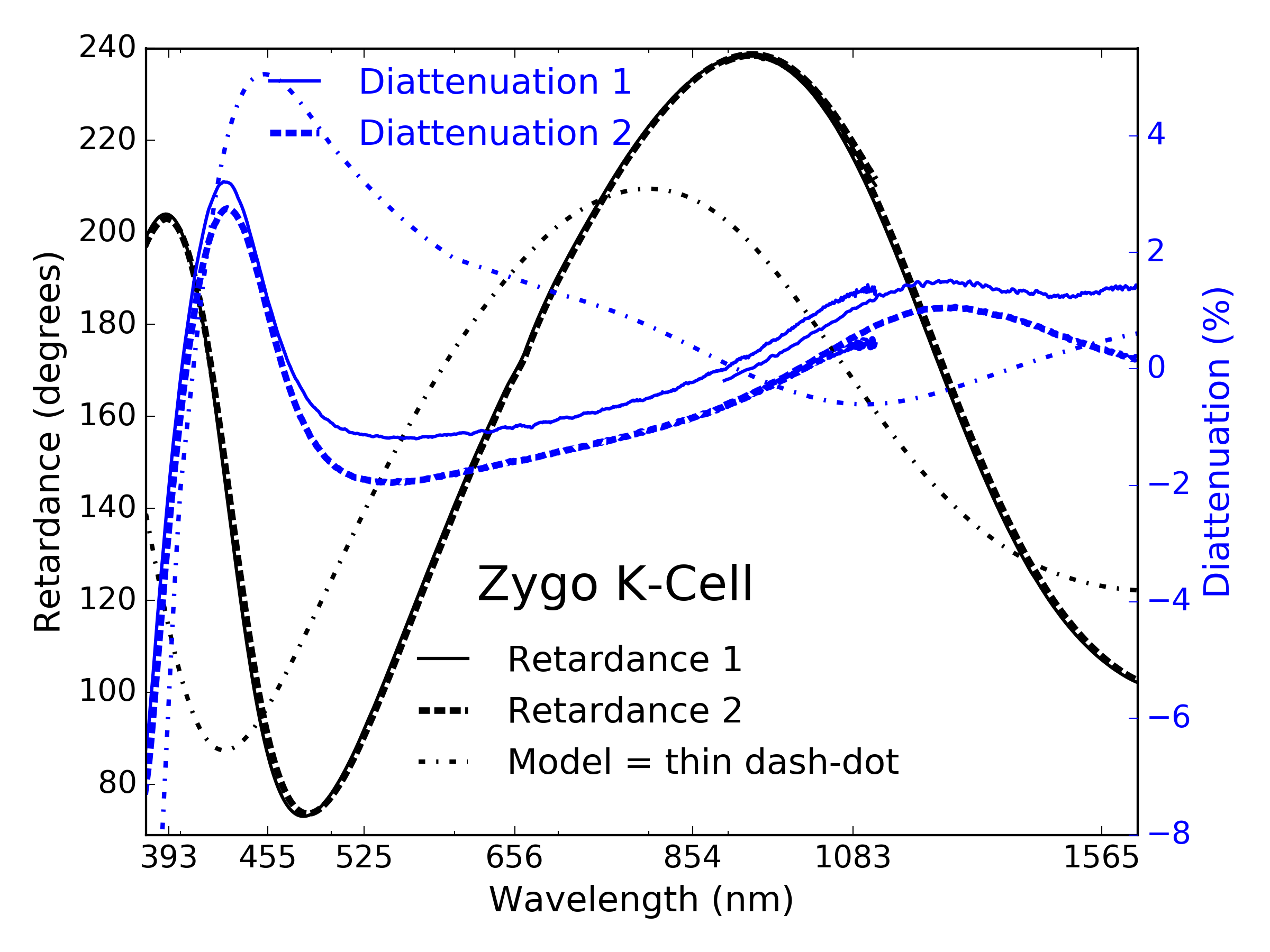}
}
\end{tabular}
\caption[] {\label{fig:zygo_silver}  An enhanced protected silver from Zygo in a K-cell configuration.  Black shows retardance while blue shows diattenuation using the right hand Y axis. A model is a thin dash-dotted line.}
\vspace{-0mm}
 \end{wrapfigure}

DKIST has used an enhanced protected silver from Zygo for the feed mirror M9a as well as one instrument feed mirror (DL-FM1) inside the DL-NIRSP instrument relay optics. Figure \ref{fig:zygo_silver} shows the retardance and diattenuation measured in a K-cell with witness samples from Zygo. Samples were roughly 8 years old and stored in laboratory conditions. These coatings use an ion-assisted deposition process and are described as having a {\it few layers} over silver.  The samples were stored in a laboratory environment that is roughly 50\% humidity in the summer and a bit drier in the winter.   The K-cell sample was measured at the same 50$^\circ$ - 11$^\circ$ - 50$^\circ$ configuration.  The two different curves of Figure \ref{fig:zygo_silver} show the difference when we replaced the first fold at 50$^\circ$ incidence with a separate sample from a separate coating shot. There is only a small difference in retardance with a fraction of a percent change in diattenuation.

\begin{figure}[htbp]
\begin{center}
\vspace{-1mm}
\hbox{
\hspace{-0.9em}
\includegraphics[height=6.3cm, angle=0]{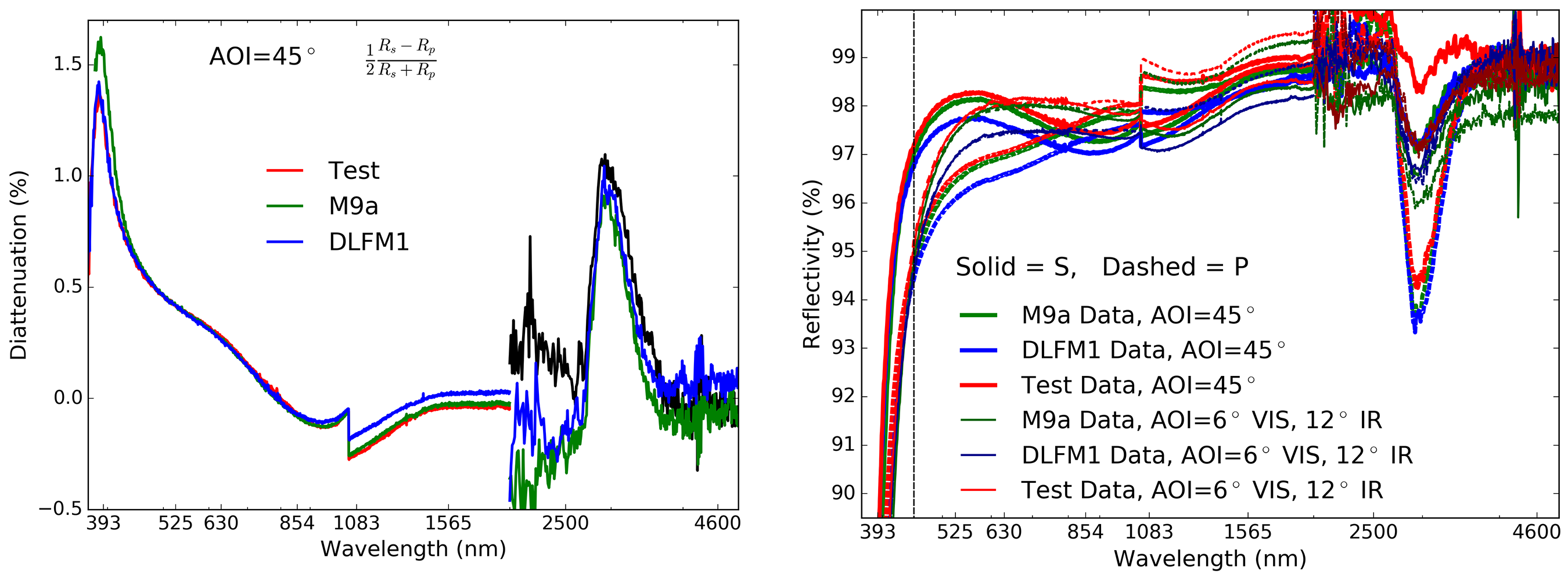}
}
\caption[] {\label{fig:zygo_eag_m9a_dlfm1_refl} Diattenuation of S- and P- polarization states measured by Zygo at 8$^\circ$ and 45$^\circ$ incidence at left. Reflectivity of S- and P- polarization states measured by Zygo at 8$^\circ$ and 45$^\circ$ incidence at right. Colors show different coating runs.}
\vspace{-6mm}
\end{center}
\end{figure}

In fall of 2018 the M9a and DL-FM1 optics were coated. Figure \ref{fig:zygo_eag_m9a_dlfm1_refl} shows the reflectivity and diattenuation. The test run is shown in red. The M9a coating run is shown in green.  The DL-FM1 coating run is shown in blue. These coatings were not blue-enhanced as they feed infrared optimized instruments working at wavelengths longer than 450 nm. The diattenuation is less than half percent for wavelengths longer than 500 nm.  There is a fairly prominent absorption feature around 3$\mu$m wavelength and a corresponding change in diattenuation. Both curves show discontinuities around 1000 nm wavelength from the change in spectrophotometric equipment, also illustrating the level of systematic error present in the data. The reflectivity curves show there are differences of roughly 1\% between coating runs.  \\

\clearpage

\subsection{Protected Silver Mirrors: DKIST VBI \& BBSO With Edmund Optics}
\label{sec:sub_DKIST_VBI_mirrors}

\begin{wrapfigure}{r}{0.53\textwidth}
\centering
\vspace{-2mm}
\begin{tabular}{c} 
\hbox{
\hspace{-1.1em}
\includegraphics[height=6.99cm, angle=0]{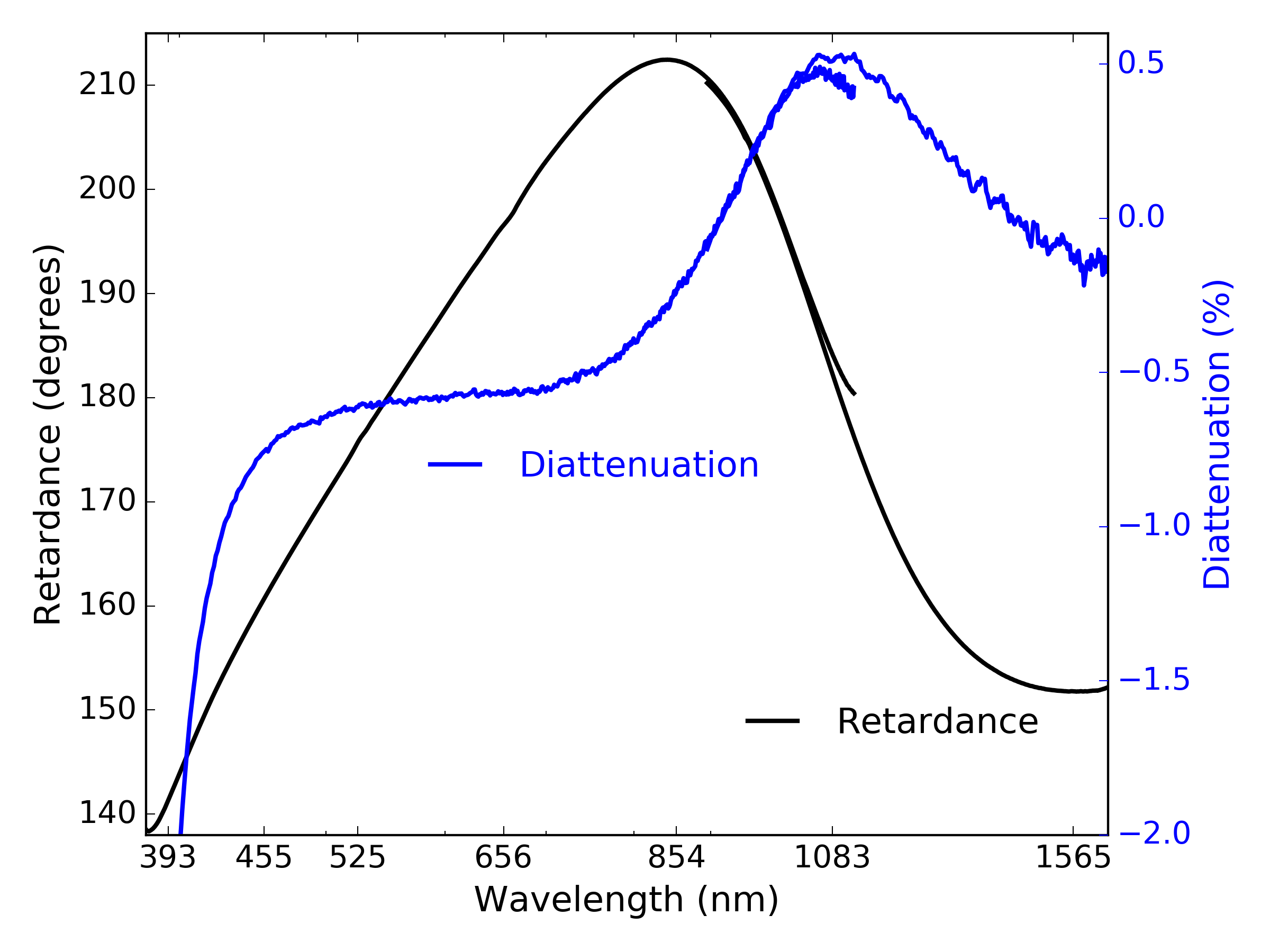}
}
\end{tabular}
\caption[] {\label{fig:edmund_vbi_silver}  An Edmund Optics protected silver used in VBI-Blue (and possibly similar to BBSO mirrors).  Black shows retardance and blue shows diattenuation on the right hand Y axis.}
\vspace{-4mm}
 \end{wrapfigure}

We show additional examples of polarization properties derived from NLSP Mueller matrix measurements for commercial protected silver mirrors relevant to DKIST and BBSO. The Visible Broadband Imager (VBI) instrument in DKIST uses an Edmund Optics off-the-shelf protected silver mirror. We took the actual six inch diameter mirror from the VBI optical path and measured the Mueller matrix at 45$^\circ$ incidence in the center of the optic. The BBSO optical path also contains a few mirrors that are a commercial Edmund Optics protected silver coating with polarization properties possibly similar to this mirror. Figure \ref{fig:edmund_vbi_silver} shows the retardance and diattenuation derived from the measurements.  Diattenuation is rapidly increasing in magnitude for wavelengths shorter than 420 nm.  For visible and near infrared wavelengths, the diattenuation stays below 1\% magnitude with a few sign changes. The retardance is broadly similar to the two-layer simple models we've presented in this paper with retardance crossing the theoretical 180$^\circ$ magnitude at wavelengths around 525 nm and 1100 nm.

\subsection{Protected Silver Mirrors: DKIST DL-NIRSP \& EMF}
\label{sec:sub_DKIST_DL_EMF_mirrors}

\begin{wrapfigure}{l}{0.55\textwidth}
\centering
\vspace{-2mm}
\begin{tabular}{c} 
\hbox{
\hspace{-1.2em}
\includegraphics[height=6.99cm, angle=0]{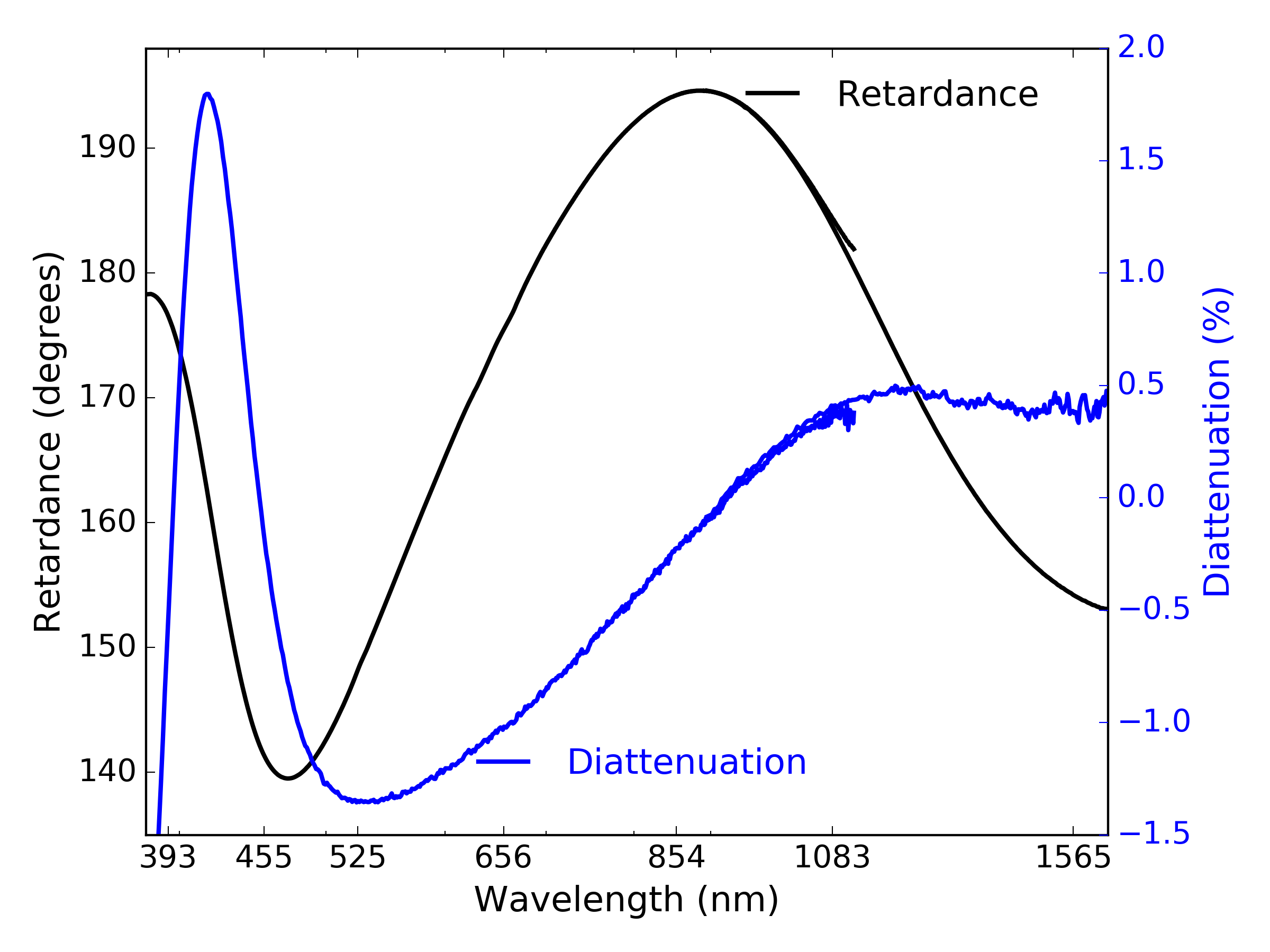}
}
\end{tabular}
\caption[] {\label{fig:emf_dl_silver}  A coating by EMF used on two of the DL-NIRSP spectrograph mirrors as well as the spherical steering mirror feed optic. Black shows retardance and blue shows diattenuation on the right hand Y axis.}
\vspace{-2mm}
 \end{wrapfigure}

The DL-NIRSP instrument contracted a vendor who ultimately procured some coatings that were not the specified DKIST silver. Once we discovered this alternate coating was present on several mirrors in the DL-NIRSP optics, we directly measured one of the flat DL-NIRSP mirrors that had used this alternate coating. Figure \ref{fig:emf_dl_silver} shows measurement of a coating on a DL-NIRSP spectrograph mirror. Diattenuation values for the DL-NIRSP mirror are slightly higher, but still below 2\% magnitude. This coating shows much stronger oscillations of diattenuation at short wavelengths.  However, the DL-NIRSP only observes wavelengths longer than 525 nm and this coating has less than 1\% diattenuation for wavelengths longer than 700 nm.  The retardance also crosses the theoretical 180$^\circ$ magnitude at effectively three wavelengths:  380 nm, 700 nm and 1080 nm. The reflectivity, retardance and diattenuation is acceptable for DKIST purposes, especially since only one DL-NIRSP feed mirror uses this coating and only at 3.8$^\circ$ incidence angle.

\subsection{Protected Silver Mirrors: DKIST M8 \& EMF blue-optimized AG99}
\label{sec:sub_DKIST_M8_EMF_mirrors}

DKIST also utilized EMF for one of the coud\'{e} feed mirrors: M8.  We used a blue-modified flavor of the AG99 formula to ensure we did not compromise the 393 nm wavelength throughput.  Metrology was recorded at 8$^\circ$ incidence while the M8 off axis paraboloa is mounted at roughly 5$^\circ$ incidence for the chief ray. 

\begin{wrapfigure}{l}{0.55\textwidth}
\centering
\vspace{-3mm}
\begin{tabular}{c} 
\hbox{
\hspace{-1.0em}
\includegraphics[height=6.99cm, angle=0]{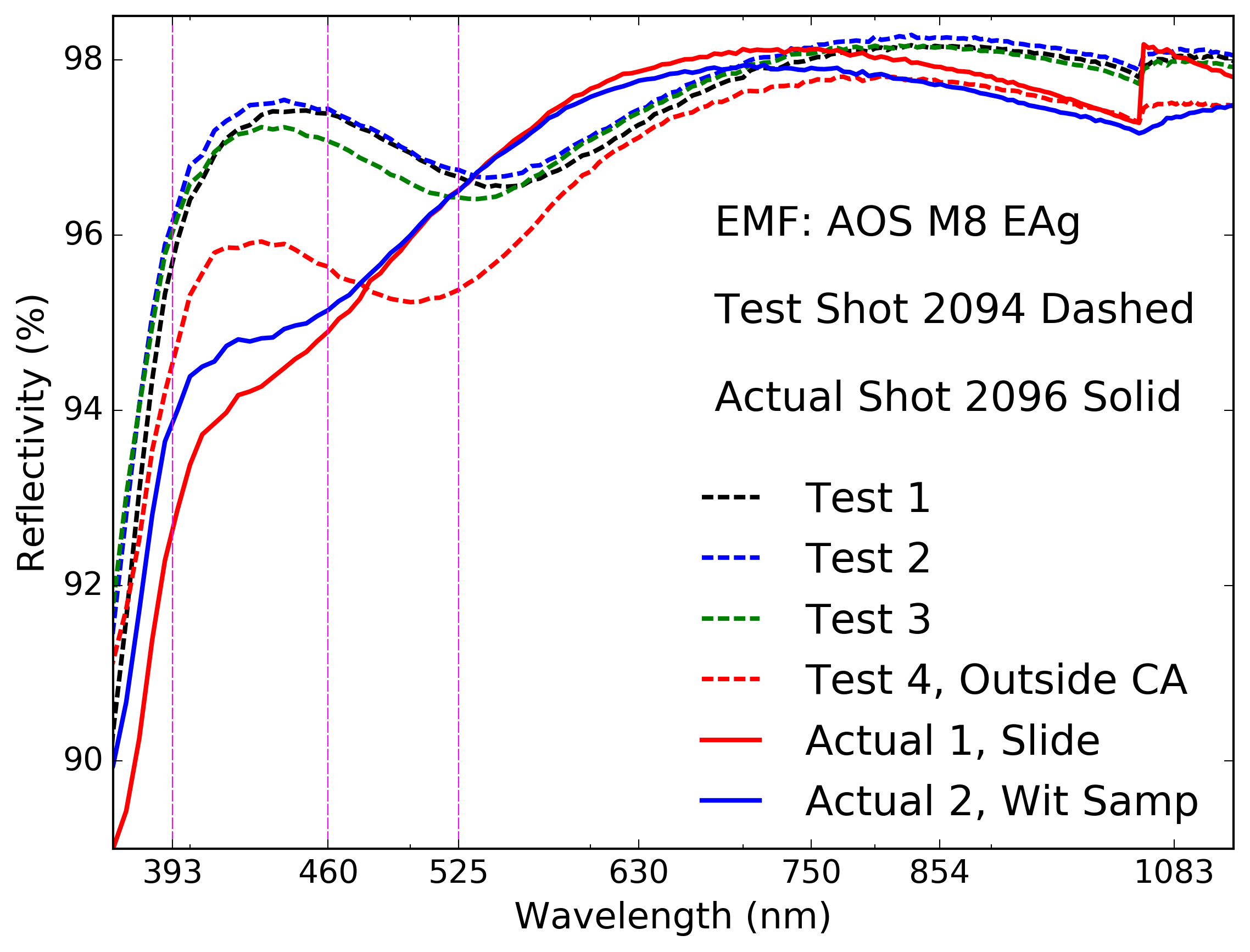}
}
\end{tabular}
\caption[] {\label{fig:emf_M8_silver_meas}  The blue enhanced AG99 coating by EMF used on DKIST M8. Test run data shown as dashed lines. Solid lines show witness samples adjacent to the M8 optic during the coating run.}
\vspace{-3mm}
 \end{wrapfigure}

Figure \ref{fig:emf_M8_silver_meas} shows the spectrophotometry from EMF at 8$^\circ$ incidence for two coating shots. The test coating shot number 2094 is shown as dashed lines while the coating shot deposited on the actual DKIST M8 is shown in solid lines.  As is typical, the test data shows spectral reflectivity variation of up to 3\% at some wavelengths. The test shot run 2094 had three samples distributed within the clear aperture shown as black, blue and green dashed lines.  There was a fourth sample located outside the clear aperture of the M8 optic that shows somewhat similar spectral behavior but with significant deviation at short wavelengths. The two solid lines of the actual coating shot show tests on a glass slide and also a standard witness sample coupon.  As the M8 optic was coated, these samples are by definition outside the clear aperture of the optic and are only representative of the actual M8 coating within the limits of the coating spatial uniformity as deposited in this chamber. The test run data shows a similar drop in blue reflectivity though with some differences in the spectral oscillations.

\begin{figure}[htbp]
\begin{center}
\vspace{-0mm}
\hbox{
\hspace{-0.8em}
\includegraphics[height=6.3cm, angle=0]{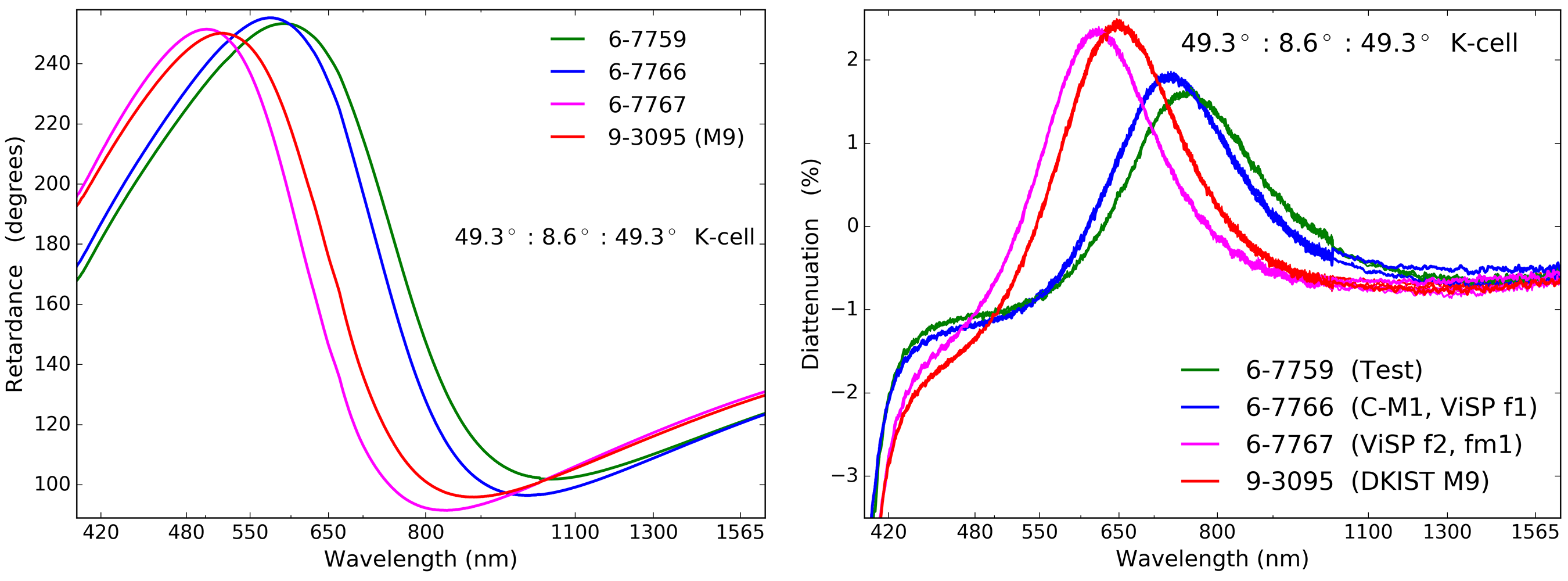}
}
\caption[] { \label{fig:M9_CM1_ViSP_Feed_IOI} The left graphic shows retardance, right shows diattenuation. The chamber 6 test shot 7759 is shown in green, shot 7766 in blue, shot 7767 in magenta and the chamber 9 shot for DKIST M9 in red. There are actually 12 red curves and 9 curves of all other colors as we repeated each optical alignment 3 or 4 times, and repeated each test at each alignment 3 times.  Some enhanced statistical noise is seen in the diattenuation graph at right.}
\vspace{-6mm}
\end{center}
\end{figure}


\clearpage 

\subsection{Protected Silver Mirrors: DKIST C-M1 \& M9,  ViSP Mirrors F1, F2, FM1}
\label{sec:sub_DKIST_M9_ViSP_FiDO}

\begin{wraptable}{l}{0.28\textwidth}
\vspace{-3mm}
\caption{IOI EAg Runs}
\label{table:ioi_eag_opt_coating}
\centering
\begin{tabular}{l l l}
\hline
\hline
Run		& Name 			& AOI			\\
\hline
6-7759	& Samples		& --				\\
6-7766	& FIDO C-M1		& 15$^\circ$		\\
6-7766	& ViSP F1			& 2.2$^\circ$		\\
6-7767	& ViSP FM1		& 28$^\circ$		\\
6-7767	& ViSP F2			& 12.3$^\circ$		\\
9-3095	& DKIST M9		& 10$^\circ$		\\
\hline
\hline
\end{tabular}
\vspace{-4mm}
\end{wraptable}

We had several mirrors coated with an Infinite Optics enhanced protected silver with formulas EAg1- 420 and -450.  We list in Table \ref{table:ioi_eag_opt_coating} the coating run number associated with each of the optics.  The FIDO mirror C-M1 was in the same shot as the first ViSP feed mirror (FM1).  The second two ViSP mirrors were in a second shot. A separate coating shot was done for DKIST M9.  

We obtained 3 witness samples from each coating shot containing DKIST optics. With these, we were able to create a K-cell image rotator type setup to use as a sample in the transmissive arm of NLSP.   Figure \ref{fig:M9_CM1_ViSP_Feed_IOI} shows the retardance on the left and diattenuation on the right for this sample.

\begin{wrapfigure}{l}{0.55\textwidth}
\centering
\vspace{-3mm}
\begin{tabular}{c} 
\hbox{
\hspace{-1.0em}
\includegraphics[height=6.99cm, angle=0]{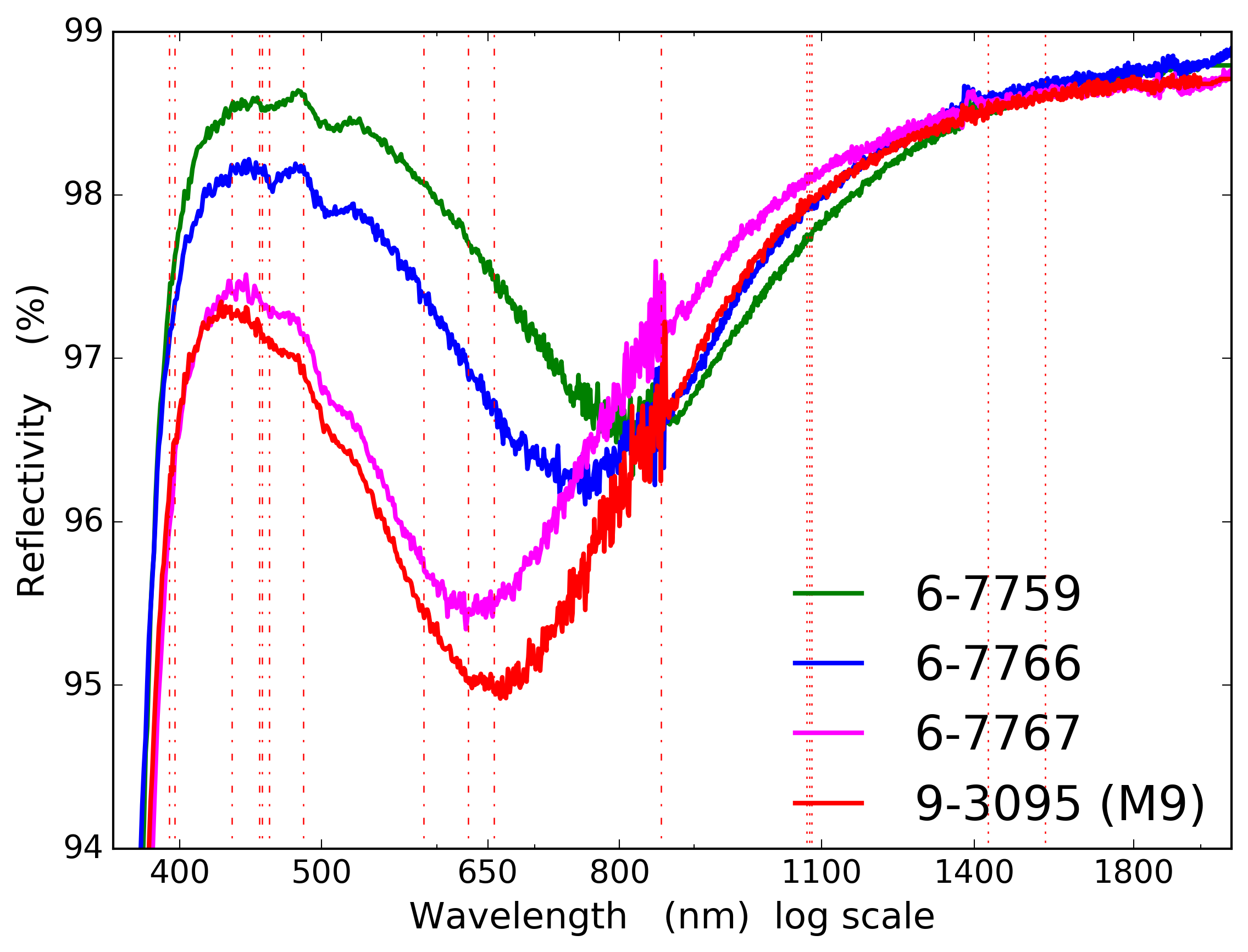}
}
\end{tabular}
\caption[] {\label{fig:ioi_eag420_M9_CM1_ViSPfeed}  The reflectivity of each coating by Infinite Optics measured at 8$^\circ$ incidence. Vertical dashed lines show spectral lines of interest.}
\vspace{-3mm}
 \end{wrapfigure}

We used the narrow K-cell setup with approximate incidence angles of 49$^\circ$ on the outer two samples and 8.6$^\circ$ on the inner sample.  We only had two samples for the chamber 6 test shot 7759 so we substituted a DKIST sample from M10 spatial position U.  At the low incidence angle of 8.6$^\circ$, we expect retardance impact of less than a few degrees when modeling this system.  The four curves represent the same materials but with different thicknesses of each material.  All of these coatings are slightly thinner than the DKIST specified coating with zero retardance values at correspondingly shorter wavelengths. 

Figure \ref{fig:ioi_eag420_M9_CM1_ViSPfeed} shows the reflectivity measured at IOI from 300 nm to 2200 nm. We follow the same color scheme as Figure \ref{fig:M9_CM1_ViSP_Feed_IOI}. One sample was measured to 5000 nm wavelength with reflectivity slowly increasing above 99\% as predicted in the coating model.  All coatings have $>$96\% reflectivity at the 393 nm Ca K line.  \\

\begin{figure}[htbp]
\begin{center}
\vspace{-0mm}
\hbox{
\hspace{-0.8em}
\includegraphics[height=6.3cm, angle=0]{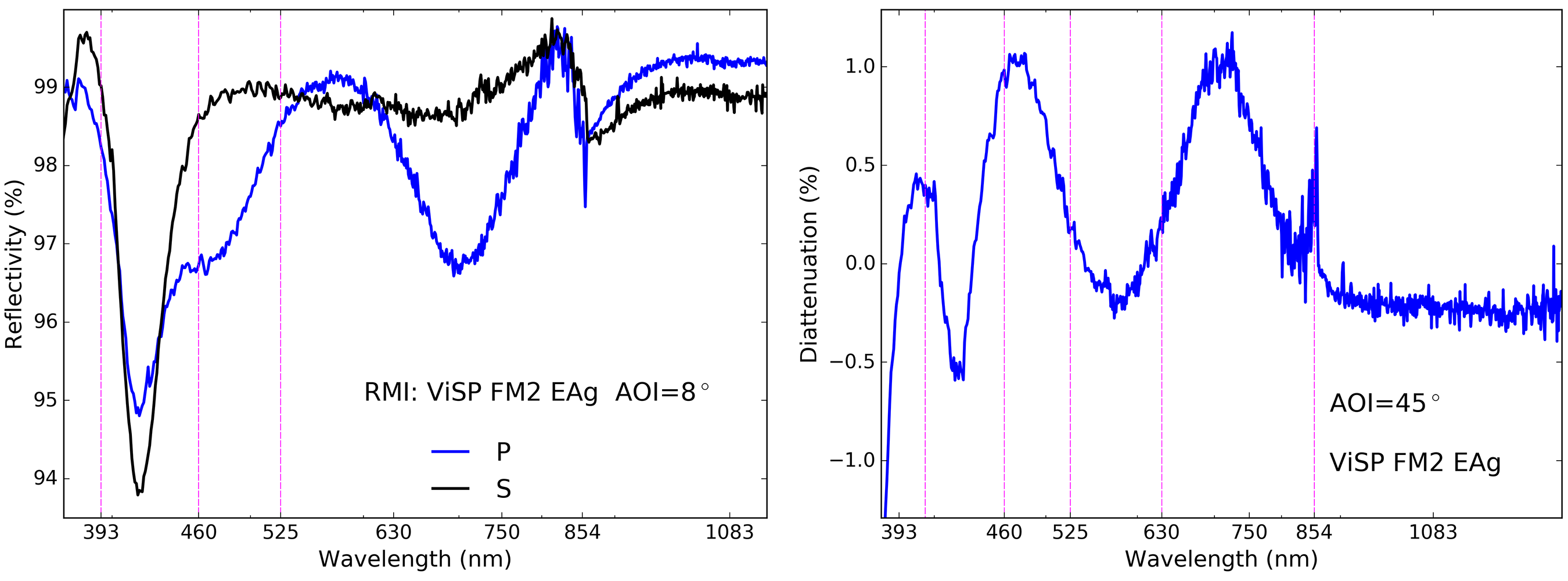}
}
\caption[] {\label{fig:ViSP_RMI_specphot}  Spectrophotometry for S (black) and P (blue) polarization states at 45$^\circ$ incidence shown at left. Diattenuation derived from S\&P reflectivity at right. Vertical dashed lines show typical solar observing wavelengths.}
\vspace{-6mm}
\end{center}
\end{figure}

\subsection{Protected Silver Mirrors: ViSP Fold Mirrors FM2, FM3, FM4 \& RMI EAg}
\label{sec:sub_visp_rmi_eag}

The ViSP feed optics include a fold mirror (FM2) coated with a Rocky Mountain Instruments enhanced protected silver. The optic is mounted in the F/ 32 diverging beam after the slit reflecting at an incidence angle of 47.7$^\circ$ towards the modulator. This optic sees footprints of only a few millimeters and is included in the system modulation matrix as it is ahead of the modulator. In Figure \ref{fig:ViSP_RMI_specphot} we show the reflectivity at left and diattenuation at right for 45$^\circ$ incidence. The reflectivity is mostly over 96\% with an exception around 420 nm wavelength.  The vertical dashed lines show spectral channels of interest. The diattenuation oscillates spectrally with six zero crossings in the nominal ViSP bandpass. The diattenuation is never more than 1\% though the sign and magnitude changes along with the spectral oscillations.

\subsection{Bare Aluminum Mirrors: DKIST Primary \& Test Mirror coated at AFRL}
\label{sec:sub_DKIST_M1_alum_mirror}

\begin{wrapfigure}{l}{0.56\textwidth}
\centering
\vspace{-3mm}
\begin{tabular}{c} 
\hbox{
\hspace{-0.8em}
\includegraphics[height=6.95cm, angle=0]{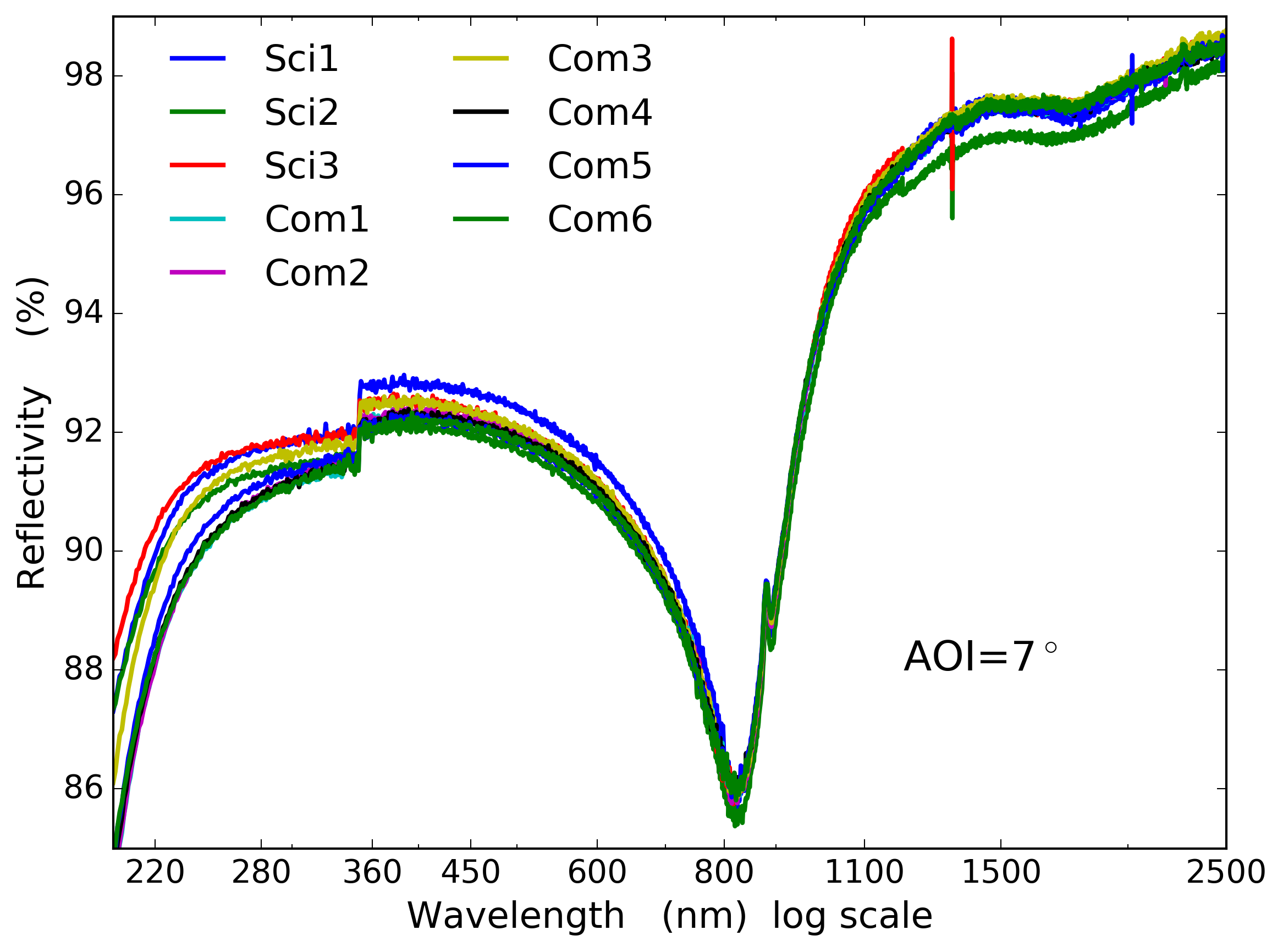}
}
\end{tabular}
\caption[] {\label{fig:M1_alum_coating_data_thick}  The reflectivity measured at 7$^\circ$ incidence angle for three samples coated with the actual M1 science grade mirror and another  6  samples measured during coating of the full sized M1 commissioning mirror.}
\vspace{-3mm}
 \end{wrapfigure}

The DKIST primary mirror (M1) is coated at the Air Force Research Labs (AFRL) coating chamber located adjacent to DKIST on the summit of Haleakala, Maui. We also have a full sized commissioning mirror as a replica of M1. Both were coated by AFRL with bare aluminum. Reflectivity of multiple samples in each coating run  measured by a Varian Cary 5000 at Gemini in Hilo are shown in Figure \ref{fig:M1_alum_coating_data_thick}.

\begin{figure}[htbp]
\begin{center}
\vspace{-0mm}
\hbox{
\hspace{-0.7em}
\includegraphics[height=6.3cm, angle=0]{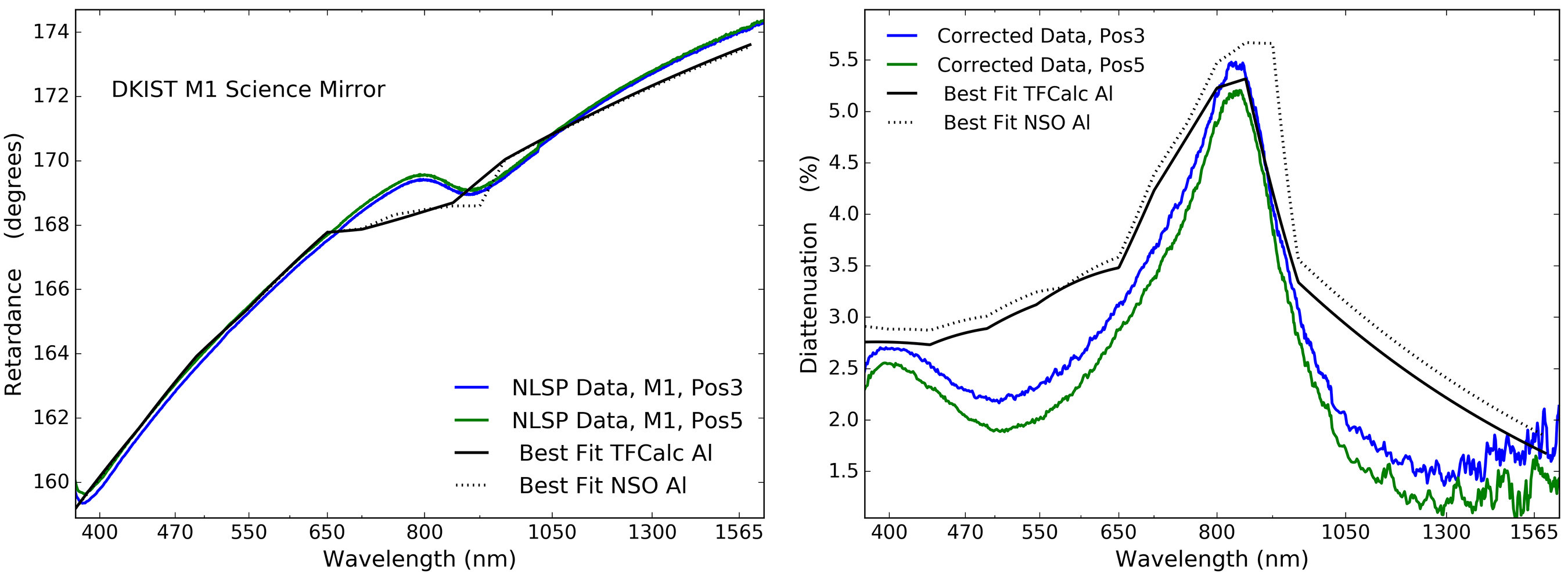}
}
\caption[] { \label{fig:M1_alum_coating_model_fit} The left graphic shows retardance, right shows diattenuation of the DKIST M1 science mirror witness sample from spatial position 3 measured in reflection at 45$^\circ$ incidence angle. Blue shows the NLSP data.  Green shows spatial position 5.  The solid black line shows the Berreman code fit when using the TFCalc refractive indices for aluminum. Dashed black shows the fit with an internal NSO aluminum refractive index interpolation.}
\vspace{-6mm}
\end{center}
\end{figure}

We tested two witness samples from the first coating on the science mirror. Figure \ref{fig:M1_alum_coating_model_fit} shows the retardance on the left and the diattenuation on the right. We ran our two-layer fitting routines using the Berreman calculus in our Python scripts. However, for this run we let the thickness of the aluminum metal layer be the second thickness variable on a grid from 40 nm to 200 nm thickness in steps of 10 nm.  The top layer was the aluminum oxide using the Boidin refractive index curves running from 0 to 10 nm in steps of 0.5 nm.  We find a coating model of 2.5 nm Al$_2$O$_3$ over 50 nm aluminum as the best fit when using the TFCalc aluminum metal refractive indices. We also note that we have corrected the diattenuation in Figure \ref{fig:M1_alum_coating_model_fit} for a linear offset to highlight the often crude spectral sampling of typical coating models. We find the same 2.5 nm of oxide but a thinner 40 nm metal layer when using an internal NSO aluminum metal interpolation.

Coatings on large mirrors are expected to be spatially variable. We note that we have interferrometric testing of larger samples distributed throughout the chamber during testing showing that the coating was physically between 90 nm and 150 nm thick across the 4 meter aperture so neither of these metal thickness fits are close to a direct thickness measurement. A recent study on polarization aberrations and the impact on the Habex system by Breckenridge et al.\cite{Breckinridge:2018hv} shows spatial variation in the retardance in reflection for a 3.75m diameter mirror coated at the University of Arizona in Figure 19 \cite{Breckinridge:2018hv}. This form birefringence measurement at magnitudes of 0.002 radians retardance required a special setup {\it developed, built and measured by. B Daugherty\cite{Breckinridge:2018hv}}. As we have detailed above, getting correct values for the optical constants of the coated aluminum is critical for matching the data with high spectral accuracy. The Breckenridge et al.\cite{Breckinridge:2018hv} work shows spatial variation is present and measurable across large area mirrors. Given the factor of $\sim$2 thickness variation in the DKIST aluminum coating, we certainly anticipate spatial variations in the mirror at some undetermined magnitude due to the varying properties of the aluminum across the mirror.


\section{Antireflection Coatings: Spectral Oscillations }
\label{sec:BBAR_Coatings}

We provide more details here on the anti-reflection coatings WBBAR1 and WBBAR2 described above in Section \ref{sec:wbbar1}. We have multiple coating runs over more than a year in several coating chambers at IOI. We show how the general magnitude and incidence angle behavior is very repeatable, but that spectral oscillations are always present impacting estimates of coating behavior at individual spectral channels typical of solar spectropolarimeters.

\begin{figure}[htbp]
\begin{center}
\vspace{-1mm}
\hbox{
\hspace{-0.7em}
\includegraphics[height=6.35cm, angle=0]{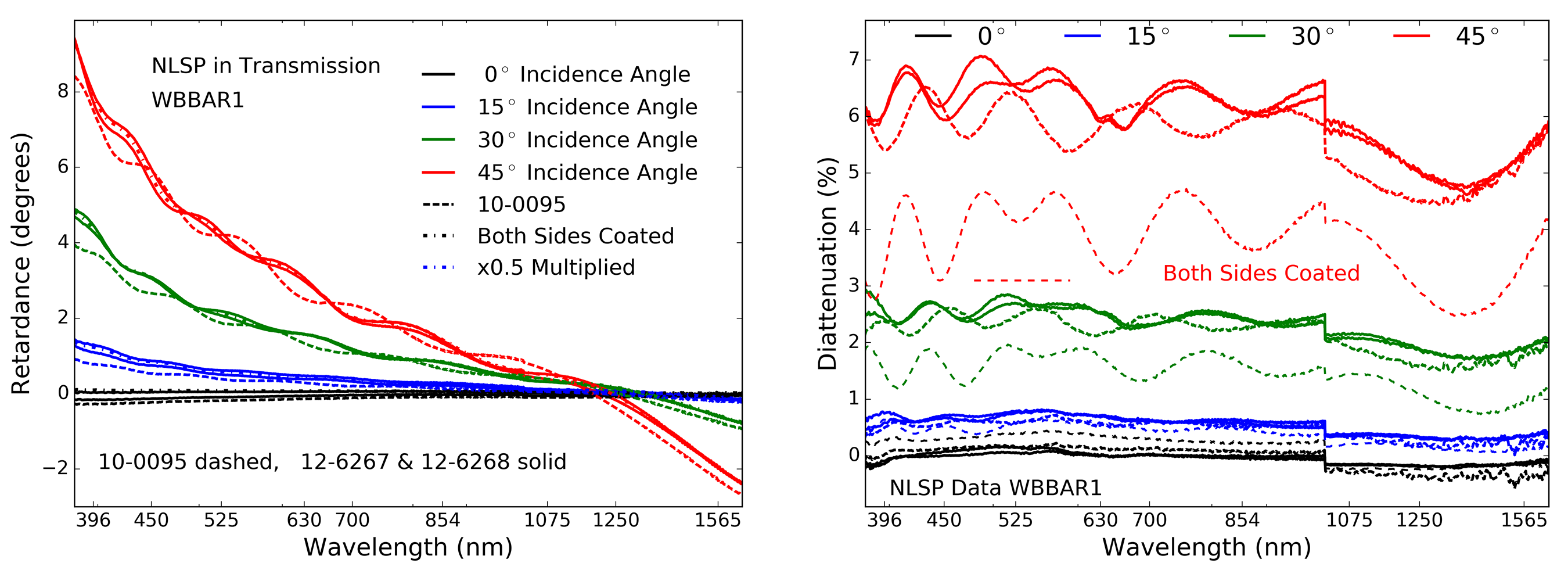}
}
\caption[] 
{ \label{fig:IO_BBAR_Transmission_Retardance_Diattenuation_CompareMany} Transmission retardance (left) and diattenuation (right) from NLSP measurements of Infinite Optics WBBAR1 on all samples with a common design: 10-0095, 12-6267, 12-6268.  The solid lines show measurements of samples 12-6267 and 12-6268. Dashed lines show the test shot 10-0095. The double-side coated sample with both 12-6267 and 12-6268 is shown with long dashes and has been divided by 2. The the net retardance scales appropriately as the uncoated sample back surface reflection does not introduce retardance while the coated back surface reflection roughly doubles the sample net retardance. The diattenuation is significantly impacted as the back surface reflection is now also coated and significantly less diattenuating than uncoated surfaces. }
\vspace{-6mm}
\end{center}
\end{figure}

We did two test coating shots of the WBBAR1 formula in chambers 7 and 10 and then coated both sides of an Infrasil window intended for use in one of the DKIST calibration polarizer assemblies in chamber 12. Both Infrasil window runs were coated sequentially in chamber 12 with side one coated in shot 6267 and the second side in shot 6268.  We had a 1.1 mm thick Infrasil 301 sample coated on both sides for photo-thermal and spectropolarimetric assessment in NLSP.

\begin{wrapfigure}{r}{0.57\textwidth}
\centering
\vspace{-4mm}
\begin{tabular}{c} 
\hbox{
\hspace{-1.2em}
\includegraphics[height=7.4cm, angle=0]{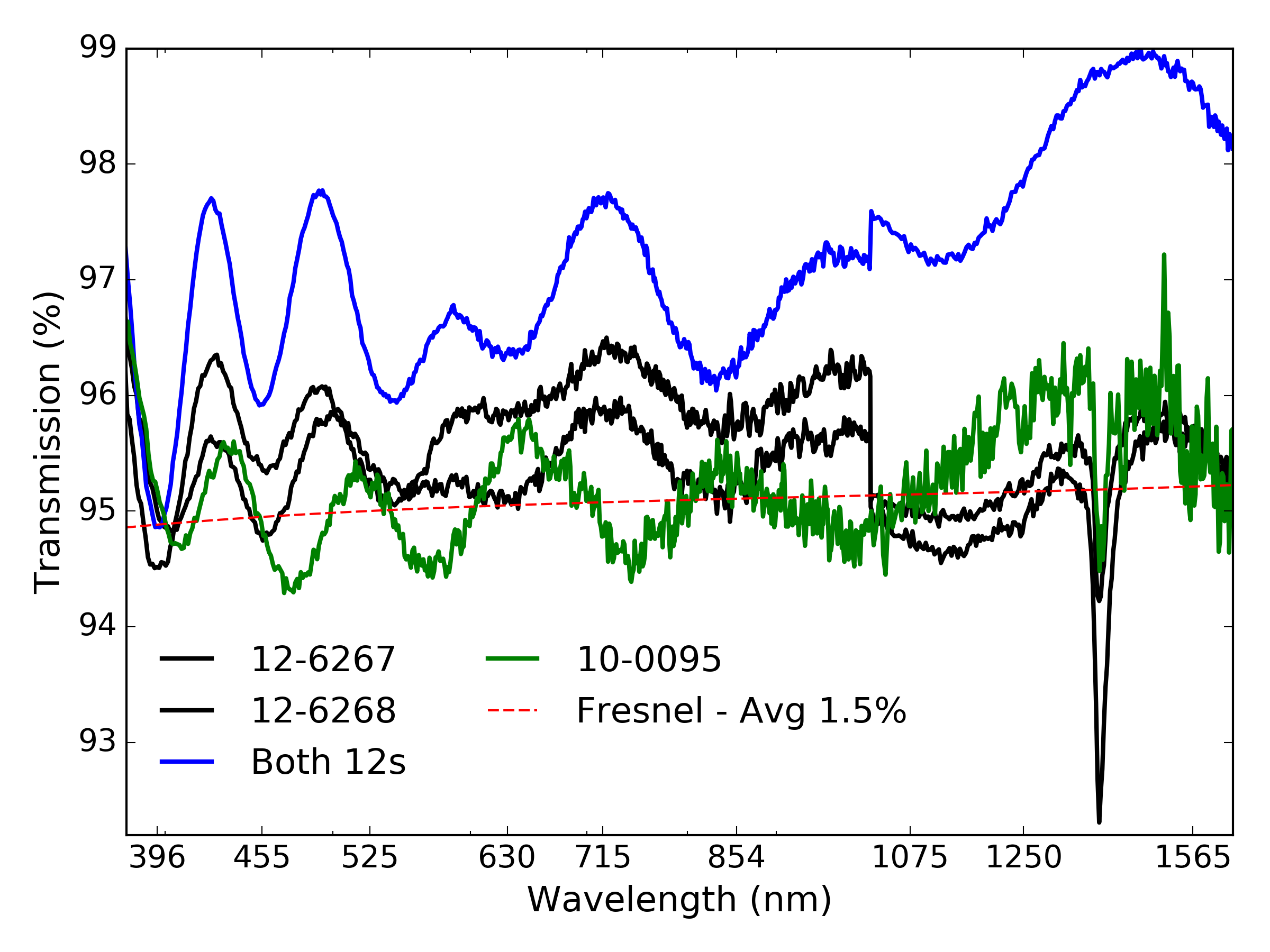}
}
\end{tabular}
\caption[] 
{\label{fig:IO_BBAR_Transmission_CompareMany}  The transmission in  NLSP at normal incidence of the WBBAR1 coating on a Heraeus Infrasil 301 fused silica substrate. The test shot 10-0095 is shown in green. The two sides of a DKIST window were coated sequentially and are shown in black for runs 12-6267 and 12-6268.  The combined two-surface coated sample transmission is shown in blue with larger oscillations. }
\vspace{-4mm}
 \end{wrapfigure}

Figure \ref{fig:IO_BBAR_Transmission_Retardance_Diattenuation_CompareMany} shows the retardance and diattenuation versus incidence angle for all WBBAR1 samples tested in NLSP. The two solid lines show excellent repeatability for the sequential shots 12-0267 and 12-0268.  The dashed line shows the preliminary shot 10-0095 has very similar performance but with slightly different spectral oscillations. This is expected for a different shot in a different chamber.  Additional comparison with the Infrasil sample that is coated on both sides is shown with the dot-dash lines in both graphics where results have been divided by two.

The legend shows a few colors to note that the double-side coated sample had the results divided by 2 for each measurement at each incidence angle. The retardance in transmission should scale as 2x but the transmission diattenuation will not.  The diattenuation of the uncoated back surfaces adds significantly more polarization than a 2-side coated surface.  We do not include the preliminary sample 7-4246 as the design changed slightly in response to a sensitivity analysis. We also note the offset in diattenuation between the visible and infrared spectrographs spliced at 1020nm is due to the optical misalignment caused by the translation of the beam through a tilted sample.

\begin{wraptable}{r}{0.23\textwidth}
\vspace{-3mm}
\caption{Photothermal}
\label{table:wbbar1_photothermal}
\centering
\begin{tabular}{c c c}
\hline
\hline
$\lambda$	& Side1	& Side2	\\
nm		& ppm	& ppm	\\
\hline
355		& 650	& 650 	\\
532		& 115	& 65 		\\
690		& 42		& 32		\\
785		& 28		& 21		\\
830		& 27		& 15		\\
1064		& 10		& 5		\\
\hline
\hline
\end{tabular}
\vspace{-3mm}
\end{wraptable}

Figure \ref{fig:IO_BBAR_Transmission_CompareMany} shows the NLSP-measured transmission functions.  At normal incidence (0$^\circ$) we see reasonable agreement between NLSP metrology showing $\sim$1\% surface reflectivity hence 99\% transmission on a single surface combined with the $\sim$3.8\% Fresnel reflection loss from the uncoated sample back surface. With a $\sim$1.5\% WBBAR1 average reflection loss, the transmission of a one-side coated sample should oscillate about an average of 94.7\% transmission with peaks between 94.2\% and 96.2\% (max transmission, WBBAR1 reflectivity at zero). This is shown in Figure \ref{fig:IO_BBAR_Transmission_CompareMany} as the thin dashed red line.  The two-side coated Infrasil sample is shown as the blue line with much higher transmission expectations, oscillating about 97.0\%.  The narrow spectral absorption spike around 1300 nm is caused by the fused silica used by IOI in their standard test samples.

In Table \ref{table:wbbar1_photothermal} we show the absorption in parts per million at wavelengths from the ultraviolet to near infrared measured at Stanford Photothermal Solutions. We obtained low values, similar to our low-absorption isotropic MgF$_2$ coatings assessed as part of our prior calibration retarder thermal modeling. \cite{Harrington:2018cx}  Absorption at 355 nm wavelength was roughly 0.65\% or 650 ppm in both coatings.  The side two coating then shows absorption rapidly dropping to less than 100 ppm at visible wavelengths and less than 30 ppm at near infrared wavelengths. This coating will not significantly contribute to the heating of most DKIST optics, which see roughly 80 Watts of optical power after the 2.8 arc minute field stop at Gregorian focus.

\begin{figure}[htbp]
\begin{center}
\vspace{-0mm}
\hbox{
\hspace{-0.7em}
\includegraphics[height=6.3cm, angle=0]{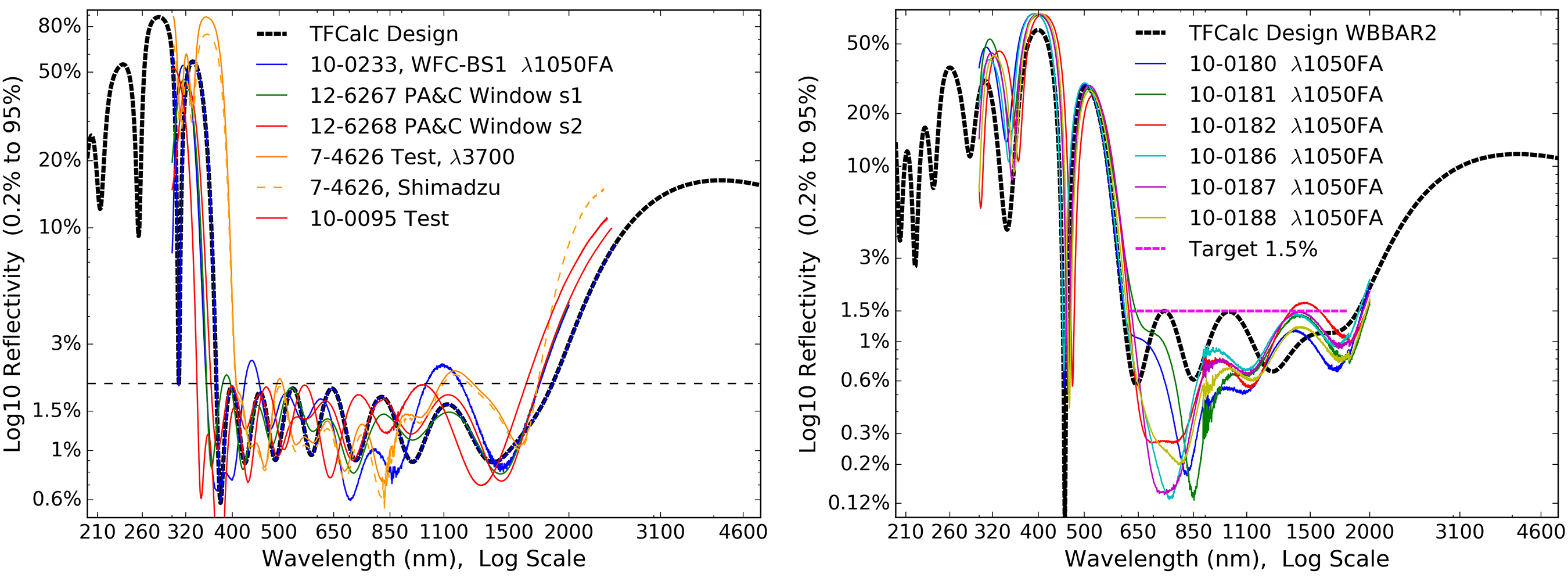}
}
\caption[] {\label{fig:wbbar_compare_1n2} Reflectivity measured for a single-surface reflection in multiple test coatings for the DKIST wide wavelength range anti reflection coatings: WBBAR1 at left and WBBAR2 at right. IOI measured the reflectivity of samples at their facility using either Shimadzu or Lambda ($\lambda$) spectrophotometers with various model numbers.   }
\vspace{-6mm}
\end{center}
\end{figure}

The spectral oscillations are not repeatable run to run as a small fraction of a nanometer thickness layer variation can shift the oscillations while still preserving overall coating performance.  We show examples of multiple repeated WBBAR coating runs in Figure \ref{fig:wbbar_compare_1n2}. The left graphic shows the wider  WBBAR1 formula and the $<$2.0\% absolute reflectivity spec with an average less than 1.5\%. The right hand graphic shows six sequential coating runs of the WBBAR2 formula done as part of our coating stress testing. The absolute reflectivity value spec was $<$1.5\% in the 630 nm to 1800 nm wavelength range with an average less than 1.1\%. Spectrophotometry was  performed at IOI with their Lambda1050 FA system.

The DKIST throughput estimates also crucially depend on knowing the anti-reflection coatings on the many lenses and windows internal to the instruments. Coatings are often optimized for the specific bandpasses in each specific camera. The three ViSP camera arms can be configured to cover any ViSP wavelength and coatings are optimized for 380 nm to 950 nm with a goal to not significantly degrade performance at 1083 nm. In DL-NIRSP, a series of dichroic beam splitters limit camera 1 to  wavelengths shorter than 900nm, the second camera to 950 nm to 1300 nm range and the third camera to the 1400 nm to 1800 nm range.

\begin{wrapfigure}{r}{0.56\textwidth}
\centering
\vspace{-3mm}
\begin{tabular}{c} 
\hbox{
\hspace{-1.0em}
\includegraphics[height=7.1cm, angle=0]{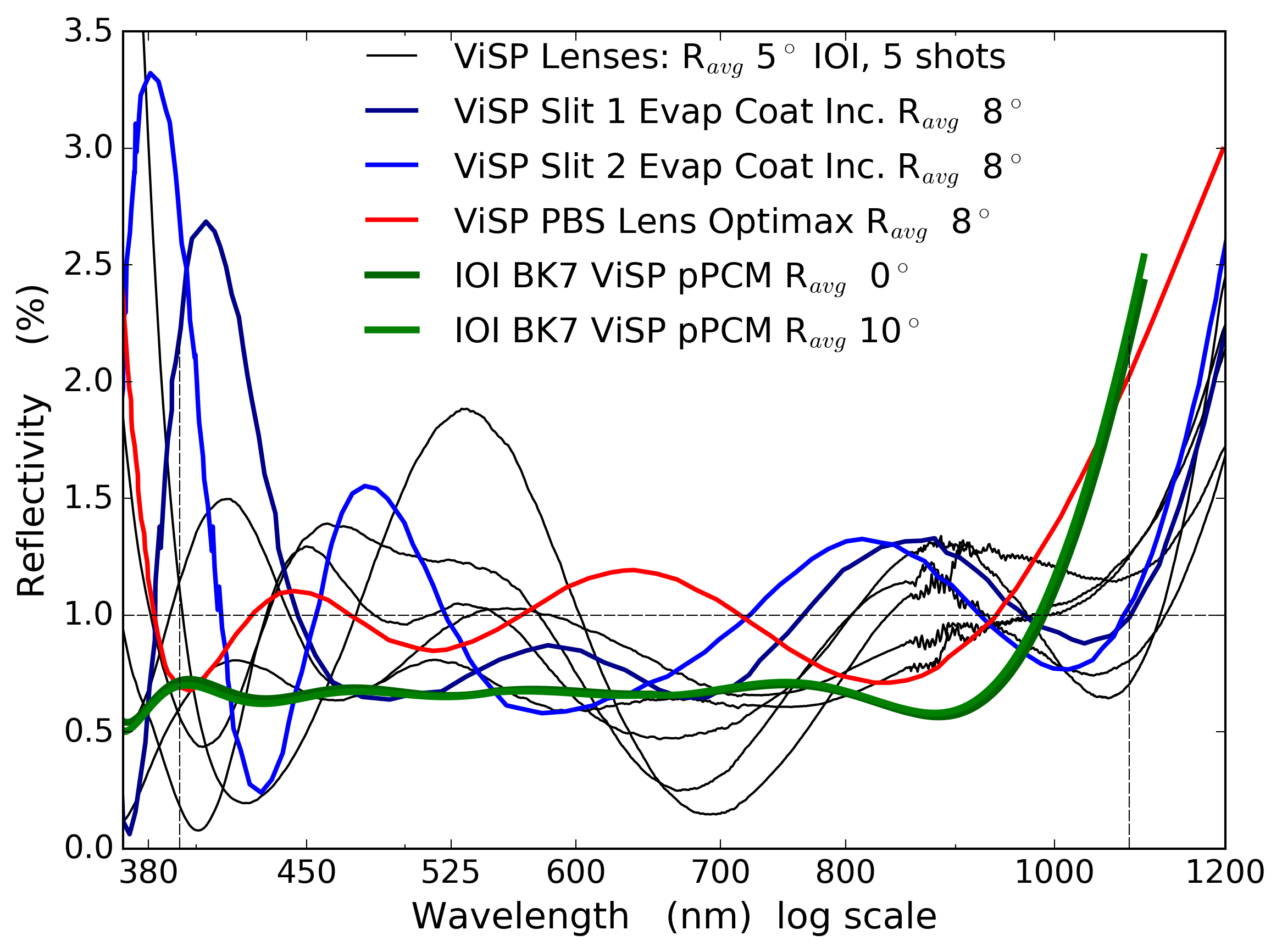}
}
\end{tabular}
\caption[] 
{\label{fig:ViSP_bbars_measured} Spectrophotometry measured by Evaporated Coatings Inc (ECI) for the ViSP slit entrance and exit interfaces at 8$^\circ$ incidence in blue. Optimax spectrophotometry at incidence angle 8$^\circ$ for the telecentric lens air-glass interfaces before bonding to the ViSP polarizing beam splitter (PBS).  The upcoming polycarbonate modulator for ViSP from Meadowlark Optics will have BBAR coated BK7 windows with the design from IOI shown in green. }
\vspace{-4mm}
 \end{wrapfigure}

Figure \ref{fig:ViSP_bbars_measured} shows the transmission measurements of the various coatings on the ViSP lenses. There were five coating shots at IOI to cover the lenses in the three ViSP cameras shown in black. The ViSP slit substrate has antireflection coatings on both sides from Evaporated Coatings Inc. (ECI) shown in light and dark blue. At shorter wavelengths these coatings combine to reflect 5\% of the light whereas longer wavelengths can have losses less than 1.5\%.  The spectral oscillations in the coatings can represent a significant uncertainty in the system flux budget. The polarizing beam splitter prisms were also coated by Optimax as shown in red. We also overlay a coating design from IOI we intend to use in an upcoming upgrade of the ViSP modulator. The specification is shown as the dashed black lines with an absolute value less than 1.0\% at any wavelength between 390 nm and 950 nm. The performance at the 1083 nm spectral line is still significantly below an uncoated Fresnel reflection loss. 

We have compiled here examples of various antireflection coatings and their properties in transmission. Significant retardance of a few degrees can be observed depending on the complexity of the coating but the diattenuation of coated tilted surfaces is significantly reduced. The spectral oscillations are common and should be directly measured to ensure an accurate throughput estimate. For DKIST, we also have oils and polycarbonate layers in various optics. The UV and IR bandpasses can see significant throughput changes depending on the oscillations of the coatings. Knowing the coating behavior into the UV impacts lifetime and damage estimates for these kinds of polarimetric optics common in solar telescopes.

\section{Mueller matrix formalisms: (R$_s$, R$_p$, $\delta$)  or (X, $\tau$)}
\label{sec:app_Mueller}

We summarize here the Mueller matrix terms and conventions for relating reflectivity and retardance to Mueller matrix elements \cite{Chipman:2014ta,Chipman:2018ug,Chipman:2010tn}. Many solar telescopes perform calibrations following an (X,$\tau$) style Mueller matrix where X relates to diattenuation and $\tau$ is retardance\cite{Skumanich:1997eh, Makita:1991ur, Kiyohara:2004eva, Hanaoka:2009dd, 2005A&A...443.1047B, 2005A&A...437.1159B, Capitani:1989kx}.  We show how to relate this solar formalism to a reflectivity and phase formalism common in optics where strict equality between conventions allows us to compare reflectivity, diattenuation and retardance \cite{Chipman:2014ta,Chipman:2018ug,Chipman:2010tn}.

We adopt a standard notation where the S and P polarization states represent incoming linear polarization states parallel and perpendicular to the plane of incidence. Their reflectivity is denoted as R$_s$ and R$_p$ respectively, and their average is denoted as R$_{avg}$. Retardance is denoted as $\delta$ which is the same as $\tau$ in the solar convention. In Equation \ref{eqn:flat_mueller_matrix} we show a common definition of the Mueller matrix for a mirror folded along the +Q plane. The $II$ element is the average reflectivity. The retardance ($\delta$) is a term in the UV rotation matrix in the lower right quadrant.

\vspace{-6mm}
\begin{normalsize}
\begin{equation}
\label{eqn:flat_mueller_matrix}
{\bf M}_{ij} =
 \left ( \begin{array}{rrrr}
 R_{avg}	 				& \frac{R_s - R_p}{2}  		& 0		& 0		\\
 \frac{R_s - R_p}{2}   	& R_{avg}						& 0		& 0		\\
 0 						& 0							& \sqrt{R_p R_s} C_\delta		& \sqrt{R_p R_s} S_\delta	\\
 0					 	& 0							& -\sqrt{R_p R_s} S_\delta		& \sqrt{R_p R_s} C_\delta	\\ 
 \end{array} \right )  = R_{avg}
 \left ( \begin{array}{rrrr}
 1	 	& \Delta  	& 0		& 0		\\
 \Delta   	& 1		& 0		& 0		\\
 0 		& 0		&  \frac{\sqrt{R_p R_s}}{R_{avg}}C_\delta	& \frac{\sqrt{R_p R_s}}{R_{avg}}S_\delta	\\
 0		& 0		& -\frac{\sqrt{R_p R_s}}{R_{avg}}S_\delta	& \frac{\sqrt{R_p R_s}}{R_{avg}}C_\delta	\\ 
\end{array} \right ) 
\vspace{-2mm}
\end{equation}
\end{normalsize}

In the normalized Mueller matrix the $\frac{IQ}{II}$ and $\frac{QI}{II}$ terms are a normalized reflectivity difference ratio (R$_s$-R$_p$)/(R$_s$+R$_p$) often denoted with a capital delta ($\Delta$). The lower right UV rotation matrix terms are modified by the scale factor $\frac{\sqrt{R_p R_s}}{R_{avg}}$. This term is above 0.999 for mirrors with diattenuation less than 10\% as are all mirrors considered for DKIST.

\vspace{-4mm}
\begin{normalsize}
\begin{equation}
\label{eqn:flat_mueller_matrix_X}
{\bf M}_{ij} =
\frac{R_p}{2} \left ( \begin{array}{rrrr}
 1 + \frac{R_s }{R_p} 	 	& 1 - \frac{R_s}{R_p}  		& 0		& 0		\\
1- \frac{R_s }{R_p}   			& 1 +  \frac{R_s }{R_p} 		& 0		& 0		\\
 0 						& 0							&  2\sqrt{ \frac{R_s}{R_p}} C_\delta		& 2\sqrt{ \frac{R_s}{R_p}} S_\delta	\\
 0					 	& 0							& -2\sqrt{ \frac{R_s}{R_p}} S_\delta		& 2\sqrt{ \frac{R_s}{R_p}} C_\delta	\\ 
\end{array} \right )   =
\frac{R_p}{2}\left ( \begin{array}{rrrr}
 1 + X^2	 & 1-X^2  	& 0		& 0		\\
 1 - X^2   	& 1+X^2	& 0		& 0		\\
 0 		& 0		&  2XC_\delta	& 2XS_\delta	\\
 0		& 0		& -2XS_\delta	& 2XC_\delta	\\ 
\end{array} \right ) 
\vspace{-4mm}
\end{equation}
\end{normalsize}

\begin{wrapfigure}{r}{0.35\textwidth}
\centering
\vspace{-6mm}
\begin{equation}
R_{avg}\Delta = \frac{R_p}{2} (1-X^2)
\end{equation}
\begin{equation}
\label{eqn:x_to_IQ}
X^2 = 1 - \frac{R_s + R_p}{R_p} \Delta
\end{equation}
\begin{equation}
X^2 = 1 - \frac{2 IQ}{(II + IQ)}
\end{equation}
\begin{equation}
\frac{IQ}{II} = \Delta = \frac{1-X^2}{1+X^2} = \frac{R_p - R_s}{R_p + R_s}
\end{equation}
\begin{equation}
X^2 = \frac{ 1 - \Delta } {1 + \Delta}
\end{equation}
\vspace{-7mm}
\end{wrapfigure}

A reflectivity ratio denoted X can be computed from the IQ or QI elements of the intensity-normalized Mueller matrix (IQ/II or QI/II) and is typically near 1.  Divide out one of the polarized reflectivities and denote the upper 2x2 submatrix in terms of intensity reflection coefficients X = $\sqrt{ \frac{R_s}{R_p}}$ as in Equation \ref{eqn:flat_mueller_matrix_X}. The retardance is denoted as tau ($\tau$).  Other systems using this formalism include The Advanced Stokes Polarimeter (ASP) at the Dunn Solar Telescope (DST) \cite{Skumanich:1997eh}, The Hida Domeless Solar Telescope  \cite{Makita:1991ur, Kiyohara:2004eva, Hanaoka:2009dd}, the German Vacuum Tower Telescope \cite{2005A&A...443.1047B} and the Polarimetric Littrow Spectrograph \cite{2005A&A...437.1159B} and the solar tower in Arcetri \cite{Capitani:1989kx}. We equate the two Mueller matrices and use IQ element to solve for X.  Equation \ref{eqn:flat_mueller_matrix} gives the average mirror reflectivity ($R_{avg}$) times the normalized IQ element ($\Delta$) giving the product: $R_{avg}\Delta$. This is equated to the Mueller matrix in Equation \ref{eqn:flat_mueller_matrix_X} with intensity coefficient ($R_p$/2) times the normalized IQ element (1-X$^2$).  Equation \ref{eqn:x_to_IQ} gives the relationship for X in terms of measured Mueller matrices and the relevant intensity scaling coefficients. We substituted 2$R_{avg}$ = $R_s + R_p$.  Note that in the limit of small diattenuation ($R_s + R_p$)/$R_p$ is approximately 2 but the scale factor between matrix conventions is $R_p$/2.  If we compute the normalized Mueller matrix elements, we can equate the X values to the reflectivity ratios. A diagonalized convention for the Mueller matrix with a total transmission term outside the matrix and II=1 is shown in Equation \ref{eqn:flat_mueller_matrix_cb_norm}. 

The overall throughput term II is scaled by $\frac{R_p (1+X^2)}{2}$ for every matrix in our group model. As diattenuation is low, X$\sim$1 and thus the overall throughput is close to 1. We also can solve for X using the terms in in Equation \ref{eqn:flat_mueller_matrix_cb_norm}. The relations between (X,$\tau$) and optical properties such as throughput and polarized reflectivities R$_s$ and R$_p$ are useful when comparing polarization calibrations and throughput estimates from these varying conventions.

\vspace{-4mm}
\begin{normalsize}
\begin{equation}
\label{eqn:flat_mueller_matrix_cb_norm}
{\bf M}_{ij} =
 \frac{R_p (1+X^2)}{2}\left ( \begin{array}{rrrr}
 1	 	& \frac{1-X^2}{1 + X^2}  	& 0		& 0		\\
 \frac{1 - X^2}{1 + X^2}   	& 1	& 0		& 0		\\
 0 		& 0		&  \frac{2X}{1 + X^2}C_\tau	& \frac{2X}{1 + X^2} S_\tau	\\
 0		& 0		& \frac{-2X}{1 + X^2} S_\tau	& \frac{2X}{1 + X^2} C_\tau	\\ 
\end{array} \right )  = R_{avg}
 \left ( \begin{array}{rrrr}
 1	 	& \Delta  	& 0		& 0		\\
 \Delta   	& 1		& 0		& 0		\\
 0 		& 0		&  \frac{\sqrt{R_p R_s}}{R_{avg}}C_\delta	& \frac{\sqrt{R_p R_s}}{R_{avg}}S_\delta	\\
 0		& 0		& -\frac{\sqrt{R_p R_s}}{R_{avg}}S_\delta	& \frac{\sqrt{R_p R_s}}{R_{avg}}C_\delta	\\ 
\end{array} \right ) 
\vspace{-2mm}
\end{equation}
\end{normalsize}


\bibliography{ms_ver03} 			
\bibliographystyle{spiebib}		

\end{document}